\documentclass[a4paper,12 pt,twoside]{book}
\usepackage[T1]{fontenc}
\usepackage[english]{babel}
\usepackage{cite}
\usepackage{amssymb}
\usepackage{amsmath}
\usepackage{textcomp}
\usepackage{txfonts}
\usepackage{mathrsfs}
\usepackage{braket}
\usepackage{indentfirst}
\usepackage{appendix}
\usepackage{graphicx}
\usepackage{bbold}
\usepackage[hidelinks]{hyperref}
\usepackage{xcolor}
\usepackage[centertableaux]{ytableau}

\usepackage{adjustbox}

\usepackage{version}
\usepackage{fancyhdr}
\usepackage{appendix}
\usepackage{multirow}

\usepackage{soul} %scrittura barrata
\usepackage{float}

\newcommand{\nordg}[1]{:\! #1 \!:_g} 
\newcommand{\nord}[1]{:\! #1 \!:}

\newcommand{\comm}[2]{\left[#1,#2\right]}
\newcommand{\acomm}[2]{\left\{#1,#2\right\}}
\DeclareMathOperator{\Tr}{Tr}
\DeclareMathOperator{\tr}{tr}

\newcommand{\trp}{\Tr_{\cR}^\prime}

\newcommand{\adj}{\mathrm{adj}}
\newcommand{\bone}{\mathbf{1}}
\newcommand{\ii}{\mathrm{i}}
\makeatletter
\newcommand*{\letterdef@}{}
\newcommand*{\letterdef}[3]{%
	\def\letterdef@##1{\expandafter\newcommand\csname #1\endcsname{#2{##1}}}%
	\@tfor\@tempa :=#3\do{\expandafter\letterdef@\expandafter{\@tempa}}}
\makeatother
\letterdef{c#1} {\mathcal}{ABCDEFGHIJKLMNOPQRSTUVWXYZ} % \cX = \mathcal{X}
\letterdef{rm#1}{\mathrm} {dDeimM} % \rmX = \mathrm{X} for X in {dDmM}

\numberwithin{equation}{section}
\newcommand{\vev}[1]{{\left\langle #1 \right\rangle}}
\newcommand{\norm}[1]{{\left| #1 \right|}}

\newcommand{\beq}{\begin{equation}}
\newcommand{\eeq}{\end{equation}}

\newcommand{\bsT}{\bar{\mathsf{T}}}
\newcommand{\sT}{\mathsf{T}}

\newcommand{\bs}{\bar{\sigma}}

\renewcommand{\a}{\alpha}
\renewcommand{\b}{\beta}
\renewcommand{\c}{\gamma}
\renewcommand{\d}{\delta}
\newcommand{\pa}{\partial}
\newcommand{\g}{\gamma}

\newcommand{\D}{\Delta}
\newcommand{\e}{\epsilon}
\newcommand{\z}{\zeta}

\renewcommand{\l}{\lambda}

\newcommand{\m}{\mu}

\newcommand{\n}{\nu}

\newcommand{\s}{\sigma}

\renewcommand{\t}{\tau}

\newcommand{\tmb}[1]{{\mbox{\tiny{#1}}}}

\newcommand{\cusp}{\text{cusp}}

\newdimen\tableauside\tableauside=1.0ex
\newdimen\tableaurule\tableaurule=0.4pt
\newdimen\tableaustep
\def\phantomhrule#1{\hbox{\vbox to0pt{\hrule height\tableaurule
			width#1\vss}}}
\def\phantomvrule#1{\vbox{\hbox to0pt{\vrule width\tableaurule
			height#1\hss}}}
\def\sqr{\vbox{%
		\phantomhrule\tableaustep
		\hbox{\phantomvrule\tableaustep\kern\tableaustep\phantomvrule\tableaustep}%
		\hbox{\vbox{\phantomhrule\tableauside}\kern-\tableaurule}}}
\def\squares#1{\hbox{\count0=#1\noindent\loop\sqr
		\advance\count0 by-1 \ifnum\count0>0\repeat}}
\def\tableau#1{\vcenter{\offinterlineskip
		\tableaustep=\tableauside\advance\tableaustep by-\tableaurule
		\kern\normallineskip\hbox
		{\kern\normallineskip\vbox
			{\gettableau#1 0 }%
			\kern\normallineskip\kern\tableaurule}%
		\kern\normallineskip\kern\tableaurule}}
\def\gettableau#1 {\ifnum#1=0\let\next=\null\else
	\squares{#1}\let\next=\gettableau\fi\next}
\newcommand{\Yfund}{\tableau{1}}
\newcommand{\Ysymm}{\tableau{2}}
\newcommand{\Yasymm}{\tableau{1 1}}

{\left\lbrace\begin{array}{@{}l@{}}}%
{\end{array}\right.}

\makeatletter
\def\cleardoublepage{\clearpage\if@twoside \ifodd\c@page\else
\hbox{}
\thispagestyle{empty}
\newpage
\if@twocolumn\hbox{}\newpage\fi\fi\fi}
\makeatother

\begin{document}

%\begin{titlepage}
\iffalse
%%%%%%%
\thispagestyle{empty}
\begin{center}
\begin{large}
Universit\`a degli Studi di Torino \\
{\bf Scuola di Dottorato} \\
\end{large}
\end{center}
\hrulefill

\vspace{2cm}
\centerline {\includegraphics[width=4cm]{logo.jpg}}
\vspace{5cm}

\large{\bf Wilson loops as defects in $\mathcal{N}=2$ conformal field theories}

\vspace{2cm}

\large{\bf Francesco Galvagno}
%%%%%%%%%%%%%%%%%%%%%%%%%%%%%%%%%%%%%%%%%%%%%%%%%%%%%%%%%%
\newpage
%\pagenumbering{arabic}
\begin{center}
\begin{large}
Universit\`a degli Studi di Torino \\
{\bf Scuola~di~Dottorato}
\end{large}
\end{center}
\hrulefill
\begin{center}
\begin{large}
%{\bf Tesi di Dottorato di Ricerca in Scienza ed Alta Tecnologia} \\
{\bf Dottorato in Fisica ed Astrofisica}
\end{large}
\end{center}

\vspace{2cm}
\centerline{\Large{\bf Wilson loops as defects in $\mathcal{N}=2$ conformal field theories}}

\vspace{8cm}
\large{\bf Francesco Galvagno}

\vspace{1cm}
\large{\bf Tutor: Marco Bill\`o}
%%%%%%
\fi

%\iffalse
\begin{titlepage}
\thispagestyle{empty}
\centerline {\Large{\textsc{ UNIVERSIT\`A DEGLI STUDI DI TORINO}}}
\vskip 20 pt
\centerline {\Large{\textsc Scuola di Dottorato}}
%\vskip 20 pt
%\centerline {{\textsc SCUOLA DI SCIENZE DELLA NATURA}}
\vskip 20 pt
\centerline {\Large{\textsc Dottorato in Fisica e Astrofisica}}
\vskip 60 pt
%\begin{tabular}{ccc}
\centerline {\includegraphics[width=4cm]{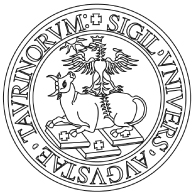}}
%\end{tabular}
\vskip 2.7cm
\centerline {\Large {\bf Wilson loops as defects in $\mathcal{N}=2$ conformal field theories}}
\vskip 1.7cm
\noindent Candidate: Francesco Galvagno 
\hfill  {Advisor: Prof. Marco Bill\`o}
%\vskip 2.7cm
%\centerline{ANNO ACCADEMICO 2018-2019}
\end{titlepage}
%\fi

\thispagestyle{empty}
\cleardoublepage
\pagestyle{headings}

\frontmatter

\cleardoublepage
\frontmatter
\cleardoublepage
\pagestyle{fancy}

\fancyhf{}
\fancyhead[LE,RO]{\thepage}
\fancyhead[LO]{\nouppercase{\rightmark}}
\fancyhead[RE]{\nouppercase{\leftmark}}
\renewcommand{\headrulewidth}{0.5pt}
\renewcommand{\footrulewidth}{0pt}
\addtolength{\headheight}{2.5pt}

\tableofcontents
\pagestyle{fancy}
\newpage
\thispagestyle{empty}
\phantom{.}

\chapter*{Acknowledgements}
\addcontentsline{toc}{chapter}{Acknowledgements}
\chaptermark{Acknowledgements}
First of all, I am extremely grateful to my supervisor, Marco Bill\`o, and to Alberto Lerda, for their constant support in each step of my PhD, for having introduced me to these fascinating topics and taught me their special approach to scientific research. 

Special thanks are due to my friends and collaborators, Lorenzo Bianchi and Paolo Gregori, for the fruitful collaborations and especially for their precious advices and exchanges about the future. I also thank my more recent collaborators, Matteo Beccaria and Azeem Hasan, for sharing their knowledge and skills.

I am grateful to Elli Pomoni and Jorge G. Russo for having accepted to read this thesis, and for the enlightening discussions that opened new directions in my research.

Afterwards, I have to thank all the Turin Openspace's people, for creating such a community spirit inside a working ambient, and the many PhD students and postdocs I met during these years: a particular thought to Luca Ciambelli, Edoardo Lauria, Lorenzo Menculini, Pierluigi Niro, Michelangelo Preti, Paolo Soresina, Stefano Speziali and C\'{e}line Zwikel, for the moments of real friendship we enjoyed.

Finally, I am immensely grateful to my family and all my friends for their love and for being so close during the difficult moments. This thesis is entirely dedicated to all of you.

\renewcommand{\chaptermark}[1]{\markboth{#1}{Overview}}
\renewcommand{\sectionmark}[1]{\markright{\thesection\ #1}} 
\chapter*{Overview and main results}
\addcontentsline{toc}{chapter}{Overview and main results}
\chaptermark{Overview and main results}

In theoretical particle physics and in the context of fundamental interactions the leading role is played by Quantum Field Theories (QFTs), which have been consistently harmonized in a gauge theory dubbed Standard Model. The mathematical building blocks, \textit{i.e.} the fields, are organized in a Lagrangian formalism and associated to fundamental particles, providing an excellent agreement with the experimental results involving electromagnetic, weak and strong nuclear interactions. In this framework non-abelian gauge theories still represent one of the most challenging puzzle. Indeed their high energy behavior is under control thanks to the mechanism of \textit{asymptotic freedom}, which means that the strength of the interaction goes to zero for large energies. By contrast, their behavior at low energies is far more difficult to treat: experimental evidences show that interactions become strong enough to prevent complete separation of the quarks bound in a hadron (mechanism of \textit{confinement}), but the full theoretical understanding of this phenomenon has remained elusive.

When interactions are weak, their effects can be approximated through the perturbation theory, systematically developed using the Lagrangian formulation. However, the calculation of higher order effects in the perturbation theory is rather cumbersome and a number of additional effects cannot be seen within perturbation theory. Understanding the strong coupling behavior of gauge theories appears a hopeless task from this point of view, since it would require a complete resummation of both perturbative and non-perturbative contributions. Therefore, new techniques and different theories need to be introduced.\\
Some progress has recently been made on this issue: a way to deepen our knowledge about non-perturbative methods in QFTs is by applying new constraints coming from additional symmetries.\\

The main example of this procedure is the introduction of supersymmetry (SUSY), which organizes bosons and fermions in supermultiplets, associating a "superpartner" to each particle. Supersymmetry is not realized in nature at accessible energies but it represents a powerful tool to drastically simplify computation in field theories: it implies that quantum corrections from bosonic degrees of freedom partially cancel against similar contributions from fermionic ones, leading to systems sufficiently constrained to give some exact results. A powerful example of this mechanism is supersymmetric localization, which is able to reduce the path integral to a finite dimensional integral. 

Another effective method to obtain highly constrained results in QFTs by means of the implementation of further symmetries is conformal symmetry. Its physical relevance derives from enabling QFTs to enjoy scale invariance promoted also at the quantum level. Conformal invariance has emerged in many contexts of high energy physics, as a signal of asymptotic freedom of QCD, or in its extremely powerful two-dimensional version, where it enjoys an infinite number of symmetry generators. Recently Conformal Field Theories (CFTs) have led to remarkable achievements also in case of the insertion of conformal defects, which are intrinsically interesting for phenomenological interests, but also can act as probes to explore topological sectors of special theories.

The natural playground where to explore non-perturbative effects and where to apply the just outlined symmetries is represented by String Theory, where it is possible to realize gauge theories through the insertion of D-branes, as well as to build consistent quantum field theories including the gravitational interaction, realizing a proper quantum gravity theory. Within this framework, one of the most important results in recent theoretical physics was born, namely the AdS/CFT correspondence \cite{Maldacena:1997re,Gubser:1998bc,Witten:1998qj}. It relates  a ten-dimensional supergravity (SUGRA) theory on a AdS$_5$ $\times$ $S^5$ space and a four-dimensional gauge theory endowed with both conformal symmetry an extended supersymmetry ($\cN=4$) degree, in the limit of large value of the rank of the gauge group.\\

The correspondence has passed many convincing tests, and one of the most important observables in this context is represented by the Wilson loop \cite{Maldacena:1998im}. The Wilson loop operator was introduced in gauge theories in order to study the confinement mechanism of quarks. Indeed it represents a gauge invariant object, whose vacuum expectation value (vev) measures the interaction potential between a $q\bar q$ pair. As a physical operator it represents the phase factor picked up by the quarks moving along a closed loop $C$, so it is an example of non-local operator. In the AdS/CFT context the Wilson loop vev has a well established dual object: the vev $\vev{W(C)}$ corresponds to the partition function for a string moving in AdS$_5$ $\times$ $S^5$ with boundary on $C$ \cite{Berenstein:1998ij}. The precise match between the Wilson loop vev computed using field theory techniques in a $\cN=4$ SYM theory at the strong coupling limit and the string partition function in the classical supergravity approximation represents one of the best examples of the validity of AdS/CFT duality.

Precisely the Wilson loop operator plays a major role also in the present thesis, thanks to its capacity to relate many aspects of theoretical physics, even beyond the AdS/CFT correspondence. In supersymmetric theories with extended SUSY, the Wilson loop vacuum expectation value has been computed exactly using supersymmetric localization, reducing the partition function to a finite dimensional matrix model on a four sphere \cite{Pestun:2007rz}. Such achievement has opened the doors to several examples of exact results in supersymmetric theories, especially those with a Lagrangian formulation, where the power of localization enables to obtain information about field theoretical observables. On the other hand, in the context of CFTs, the Wilson loop represents the simplest example of non-local operator, which behaves as a conformal defect. Therefore it represents the reference object also for many applications of defect conformal symmetry to generic (also non-Lagrangian) theories, for example inside the conformal bootstrap program.\\

From these observations the main goal of the present work follows: considering a $\cN=2$ Lagrangian theory which is conformal at the quantum level (namely the beta function coefficient vanishes for all the perturbative orders), we want to exploit the interplay between the extended supersymmetry achievements, especially regarding exact results thanks to localization techniques, and the defect conformal theory results. In particular, after the insertion of an extended probe (the Wilson loop) which explicitly affects the vacuum of the theory, we study the residual symmetry pattern and we constrain special classes of observables using DCFT techniques. Then we exploit the localization approach and derive matrix model techniques to perform explicit computations on the field theory side.

The fruitful combination of the these two general techniques will be clearer throughout this dissertation.

\section*{Outline of the thesis}
\noindent This thesis is based on the original work presented in the following papers \cite{Billo:2018oog, Billo:2019fbi, Bianchi:2019dlw}:
\begin{itemize}
\item
M.~Bill\`o, F.~G., P.~Gregori and A.~Lerda, \\
  ``Correlators between Wilson loop and chiral operators in $ \mathcal{N}=2 $ conformal gauge theories,'' JHEP {\bf 1803} (2018) 193,
[arXiv:1802.09813 [hep-th]].
\item
M.~Bill\`o, F.~G., and A.~Lerda,\\
 ``BPS Wilson loops in generic conformal N = 2 SU(N)
SYM theories,'' 
JHEP {\bf 1908} (2019) 108, 
[arXiv:1906.07085 [hep-th]].

\item
L.~Bianchi, M.~Bill\`o, F.~G., and A.~Lerda,\\
 ``Emitted radiation and geometry,''
 JHEP {\bf 01} (2020) 075, 
[arXiv:1910.06332 [hep-th]].
\end{itemize}
\noindent The thesis introduces and describes the content of these papers without a pure chronological order, but rather trying to follow the conceptual path across the publications. The main original contributions can be found in Chapters \ref{chap:3}, \ref{chap:5} and \ref{chap:6}.\\

Part I is devoted to a collection of introductory material and to the computation of the vacuum expectation value of a 1/2 BPS Wilson loop in a generic $\cN=2$ superconformal theory with gauge group SU$(N)$.\\
In particular, Chapter \ref{chap:1} contains a fast introduction to the physical set up. The additional symmetries with respect to a generic QFT, supersymmetry and conformal invariance, are described with the goal of introducing some concepts which are needed in the following. Next, we give the notion of Wilson loop operator in a gauge theory and its supersymmetric extension.\\
In Chapter \ref{chap:2} we provide additional (and more specific) tools. We describe all the tools for perturbative computations, namely the Lagrangians and the Feynman rules for extended SUSY theories, using the $\cN=1$ superspace formalism. We reserve a specific attention to the description of the ``difference theory'' between $\cN=2$ and $\cN=4$. Afterwards, we review the methods to place a $\cN=2$ theory with rigid supersymmetry on a curved space: this is useful both to place $\cN=2$ Lagrangian theories on four-dimensional spheres and ellipsoids and to derive the matrix model which comes out the localization computation. All these ingredients are the main building blocks for the computations of the following Chapters.\\
Chapter \ref{chap:3} contains the explicit computation of the Wilson loop vev in a $\cN=2$ theory with SU$(N)
$ gauge group, both at finite N and in the large-N limit, using the interacting matrix model provided
by localization results. We single out some families of theories for which the Wilson loop
vev approaches the $\cN=4$ result in the large-N limit, in agreement
with the fact that they possess a simple holographic dual. At finite N and in the generic
case, we explicitly compare the matrix model result with the field-theory perturbative
expansion up to a four-loop order, finding
perfect agreement. Organizing the Feynman diagrams as suggested by the structure of the
matrix model turns out to be very convenient for this computation. \\[1mm]

Part II is devoted to the computation of specific classes of observables inside the vacuum affected by the presence of this extended probe, following the approach of Defect Conformal Field Theories, combined with the localization achievements.\\
Chapter \ref{chap:4} is a collection of background material for study correlation functions in presence of conformal defect. After reviewing the explicit symmetry breaking pattern and the restricted conformal group, we construct and list the correlation functions of bulk and defect operators, concentrating on classes of correlators whose kinematics is completely fixed by the residual conformal invariance. We study energy-momentum conservation in presence of the external probe, in particular for a one-dimensional defect, which corresponds to the case in analysis in this thesis.\\
In Chapter \ref{chap:5} we study the first example of correlation function in presence of a Wilson loop, namely correlators of chiral primary operators. The residual conformal symmetry fixes the kinematic factor of the one-point functions, while the presence of supersymmetry still allows to exploit the localized matrix model to compute the one-point coefficient in terms of the coupling dependence. The comparison between the $S^4$ matrix model and the flat space perturbative computations is performed, up to a disentangling procedure from the sphere to the plane. \\
Chapter \ref{chap:6} contains a roadmap of the results in four dimensional superconformal theories for the emitted energy from a charged particle in an accelerated motion, represented by the Wilson loop. The Bremsstrahlung radiation can be related to the one-point function of the stress-energy tensor in presence of the Wilson loop. In $\cN=4$ this observable is related to the chiral one-point function, so perfectly captured by supersymmetric localization on the four sphere, producing an exact result which links many observables of the theory. We analyze the same computation in the less symmetric $\cN=2$ SCFT case, and we prove that a stress tensor supermultiplet insertion is related to a small variation of the sphere geometry, which becomes an ellipsoid. Therefore, using general properties of the geometric background and of residual conformal symmetry, we provide an exact formula for the Bremsstrahlung radiation in terms of a matrix model on a squashed sphere, and we derive a perturbative structure of the result, which can be organized in a transcendentality expansion.

Finally Part III contains all the technical material collected in three Appendices.

\mainmatter
\pagestyle{fancy}\addtolength{\headwidth}{0pt}
\renewcommand{\chaptermark}[1]{\markboth{\thechapter.\ #1}{}}
\renewcommand{\sectionmark}[1]{\markright{\thesection \ #1}{}}
\cfoot{}
\rhead[\fancyplain{}{\leftmark}]{\fancyplain{}{\thepage}}
\lhead[\fancyplain{}{\thepage}]{\fancyplain{}{\rightmark}}
\part{Wilson loops in supersymmetric gauge theories}

\chapter{Introduction}\label{chap:1}
In this Chapter we provide an introduction to the real core of the thesis, with the special goal of outlining the general philosophy, namely the study of Quantum Field Theories endowed with additional symmetries: this top-down approach is an ideal way to compute observables of general interest for theoretical physics, and in general it represents a functional playground to understand the behavior of more realistic theories.

The plan of the Chapter is the following. In Section \ref{sec1:susy} we give a brief introduction to supersymmetric theories, and in particular we concentrate on the extended SUSY case, whose field content and some specific features are outlined.
Section \ref{X3} is an introduction on conformal symmetry. In this case also we simply point out the main concepts we will need in the following; a special Subsection is devoted to the embedding formalism, which we will strongly use to constrain DCFT correlators. In Section \ref{sec1:SCFT} we describe superconformal algebra and a special class of operators belonging to superconformal representations, namely the chiral primaries. Finally in Section \ref{sec1:WL} we describe the role of Wilson loops in gauge theories, also in their supersymmetric extension.

\section{Supersymmetry}\label{sec1:susy}
Supersymmetry relates the two classes of elementary particles, namely of integer spin (bosons) with particles of half-integer spin (fermions). For a detailed introduction to this broad topic see for example \cite{Sohnius:1985qm,Wess:1992cp,West:1990tg}. Historically the introduction of SUSY invariance has represented  a way out of the Coleman-Mandula no-go theorem \cite{Coleman:1967ad}, which states that the only possible continuous symmetries of a consistent QFT are Poincar\`e $\times$ Internal Symmetries $G$. The Poincar\`e algebra reads:
\begin{subequations}\label{Poinctot}
\begin{align}
[P_{\mu},P_{\nu}]& =  0~, \label{Poinca}\\
[M_{\mu \nu},M_{\rho \sigma}]& = i(\eta_{\mu \sigma} M_{\nu \rho} + \eta_{\nu \rho} M_{\mu \sigma} - \eta_{\mu \rho} M_{\nu \sigma} - \eta_{\nu \sigma} M_{\mu \rho})~,\label{Poincb}\\
[M_{\mu \nu}, P_{\rho}]& = i(\eta_{\rho \nu} P_{\mu} - \eta_{\rho \mu} P_{\nu})~,\label{Poincc}
\end{align}
\end{subequations}
while internal global symmetry algebra generated by $T_a$ and with structure constants $f_{abc}$ is given by:
\begin{equation}\label{internalsym}
\comm{T_a}{T_b}=if_{abc}T_c~.
\end{equation}

The Coleman-Mandula theorem has been evaded in 1975 by Haag, Lopuszanski and Sohnius \cite{Haag:1974qh} by weakening one of its assumptions: the only possible consistent extension is to include a set of $ \mathcal{N} + \mathcal{N} $ anticommuting $fermionic$ generators, $ ( Q_{\alpha}^{\cI}, \bar{Q}_{\dot{\alpha}}^{\cI} ) $, where $ \cI = 1, \dots, \mathcal{N}$ and $ \alpha, \dot{\alpha} = 1,2 $ are spinorial indices.  This implies that supersymmetry is not an internal symmetry, but an extension of Poincar\`e space-time symmetries. Since the SUSY generators transform bosons into fermions (and viceversa); thus this symmetry naturally mixes radiation with matter. \\
The Super-Poincar\`e algebra adds the following commutation rules to the previous \eqref{Poinctot} and \eqref{internalsym}:
\begin{subequations}
\label{SUSYalgebra}
\begin{align}
\left[P_{\mu},Q_{\alpha}^{\cI}\right] &=  0~, \label{1.1.1a}\\
\left[M_{\mu \nu}, Q_{\alpha}^{\cI}\right] &= i\left(\sigma_{\mu \nu}\right)_{\alpha}^{\beta}Q_{\beta}^{\cI}~, \label{1.1.1d} \\
\left\{Q_{\alpha}^{\cI}, \bar{Q}_{\dot{\beta}}^{\cJ} \right\} &= 2 \sigma^{\mu}_{\alpha \dot{\beta}} P_{\mu} \delta^{\cI \cJ}~, \label{1.1.1e}\\
\left\{Q_{\alpha}^{\cI}, Q_{\beta}^{\cJ}\right\} &= \epsilon_{\alpha \beta} Z^{\cI \cJ}\label{1.1.1f}~,\\
\comm{Q_{\alpha}^{\cI}}{T_c}&=(B_c)^\cI_\cJ Q_{\alpha}^{\cJ}\label{1.1.1g}~.
\end{align}
\end{subequations}
%where $ Z^{I J} = - Z^{J I} $ are bosonic generators commuting with all the other generators, called central charges. 
We refer to Appendix \ref{app:Notations} for all our notations and conventions. Some important remarks about \eqref{SUSYalgebra}:
\begin{itemize}
\item
Equation (\ref{1.1.1d}) follows from the fact that $ Q_{\alpha}^{I} $ and $ \bar{Q}_{\dot{\alpha}}^{I} $ are spinors of the Lorentz group, and thus they can rise/lower the spin by half unit.
\item
Equation (\ref{1.1.1e}) implies that the square of two SUSY transformations is a translation. In theories with local supersymmetry this relation expresses the invariance under a general coordinate transformation, namely a supergravity theory.
\item
Equation (\ref{1.1.1f}) introduces new objects, $ Z^{I J} = - Z^{J I} $ that are bosonic generators commuting with all the other generators, therefore they are central charges. They span an invariant subalgebra of the internal symmetry group $G$.
\item
From (\ref{1.1.1g}) we see that $ Q $'s carry a representation of $G$. Since they are spinors, it follows that the largest possible internal symmetry group which can act on the $ Q $'s is $ U(\mathcal{N})_R $ and this is called the \textit{R-symmetry group}\footnote{For $\cN=4$ the R-symmetry group is SU$(4)_R$ only, we will motivate this in Section \ref{sec1:SCFT}}.
\end{itemize}

Then the prescription for building a supersymmetric theory follows from the standard field theoretical approach: we study the irreducible representations of the super-Poincar\`e algebra. We get a set of supermultiplets, whose components correspond to the common fields. The number $ \mathcal{N} $ of SUSY generator cannot be arbitrarily large: indeed, any supermultiplet contains particles with spin at least large as $ \cN/4 $. Therefore $ \mathcal{N} $ can be at most as large as 4 for gauge theories (maximal spin 1).

\subsection{Theories with extended supersymmetry}\label{subsec1:extendedSUSY}
Throughout the present work we will deal with theories with extended supersymmetry, \textit{i.e.} $\cN=2,4$, especially due to their special behavior at the quantum level. In this Subsection we start by introducing the supermultiplets content in terms of fundamental fields and their Lagrangian formulation. 
\subsubsection*{Field contents}
\begin{itemize}
\item\textbf{$\cN=4$}\\
This is tha maximally supersymmetric gauge theory. There is a unique multiplet, the vector superfield:
\begin{itemize}
\item
Vector: $V=\left(A_\m,\lambda^\cI_\a,\phi^u\right)$,\\
made of a gauge fields, four Weyl spinors ($\cI=1,\dots 4$) and six real scalars ($u=1,\dots 6$), all in the adjoint representation of the gauge group.
\end{itemize}  
 R-symmetry allows the four spinors to transform in the fundamental of SU$(4)_{R} $, the six real scalars in the rank 2 antisymmetric representation, which is the fundamental of SO$(6)$.
\item \textbf{$\cN=2$}\\
We have two superfields with the following (on-shell) degrees of freedom:
\begin{itemize}
\item Vector: $V=\left(A_\m,\lambda^\cI_\a,\phi^u\right)$, \\
with a gauge fields, two Weyl spinors (here the R-symmetry index runs over $\cI=1,2$) and two real scalars ($u=1,2$), all in the adjoint representation of the gauge group.
\item Hypermultiplet: $H=\left(q,(\psi_{q})_\a,\tilde{q},(\psi_{\tilde q})_\a\right)$,\\
with two complex scalars and two Weyl fermions, in a generic representation $\cR$ of the gauge group.
\end{itemize}
\end{itemize}

\subsubsection*{Lagrangians}
We write the Lagrangian for the pure Yang-Mills part (involving the vector multiplet only) for both the theories:
\begin{equation} \label{a16} \begin{split}
L^{\mathcal{N}=2,4}_{SYM} = \hspace{0.2 cm} \Tr~ &\biggl\{-\frac{1}{4}F_{\mu \nu}F^{\mu \nu} + \frac{\theta}{32\pi^{2}}g^{2}F_{\mu \nu}\tilde{F}^{\mu \nu} - i \bar{\lambda}^{\cI}\bar{\sigma}^{\mu}D_{\mu}\lambda_{\cI} + D_{\mu}\phi^{u}D^{\mu}\phi_{u} + \\
+& g C^{\cI\cJ}_{u}\lambda_{\cI}[\phi^{u},\lambda
_{\cJ}] + g\bar{C}_{u\cI\cJ}\bar{\lambda}^{\cI}[\phi^{u},\bar{\lambda}^{\cJ}] + \frac{g^{2}}{2}[\phi^{u},\phi^{v}]^{2} \biggr\}
\end{split}
\end{equation}
where $ C^{\cI\cJ}_{u} $ are the structure constants of the R-symmetry groups U$(2)_{R} $ and SU$(4)_{R} $ which rotate the supersymmetry generators.\\
No matter can be inserted in a $\cN=4$ theory in the usual
sense, due to the strong constraints of the larger amount of supersymmetries. In the $\cN=2$ case, instead, the SU$(2)_R$ symmetry action prevents the hypermultiplet part to have self-interactions, so all the interactions with the matter part turn out to be gauge interactions \footnote{An example of an interactive $\cN=2$ Lagrangian can be found in Subsection \ref{subsecn:susy}, where we will describe $\cN=2$ theories on ellipsoids. To get the flat space version it is sufficient to switch off all the background fields listed in \eqref{offsm}}.\\
We will deepen our analysis on the extended supersymmetric Lagrangians throughout Chapter \ref{chap:2}, where we will define the set up to perform perturbative calculations.

\subsubsection*{Quantum behavior}
We mention the most important advantage of a supersymmetric theory: it makes quantum corrections much better behaved with respect to ordinary field theories. These results about UV properties of supersymmetric theories can be summarized in terms of non-renormalization theorems, in practice the constraints imposed by the supersymmetry charges strongly fix the renormalization factors:
\begin{itemize}
\item
The $ \mathcal{N}=4 $ theory enjoys the maximal constrains from supersymmetry: the renormalization factors are trivial, and thus $ \mathcal{N}=4 $ SYM is \textit{perturbatively finite}; in other words it does not exhibit ultraviolet divergences \cite{Grisaru:1980nk}. That's a remarkable result, since having a vanishing gauge coupling $ \beta $-function at the full quantum level implies $ \mathcal{N}=4 $ SYM is also \textit{superconformal invariant} \cite{Sohnius:1981sn,Mandelstam:1982cb,Brink:1982wv}. We will exploit this feature in the following. 
\item
In $ \mathcal{N}=2 $ the fact that the supermultiplets are made of $ \mathcal{N}=1 $ vector and $ \mathcal{N}=1 $ matter multiplet, the presence of a non-abelian R-symmetry SU$(2)_{R} $ and the non-renormalization rules for the interaction terms make a $ \mathcal{N}=2 $ theory \textit{one-loop exact} in perturbation theory, \textit{i.e.} the gauge coupling $ \beta $-function gets only one loop contributions \cite{Howe:1983wj}.    
\end{itemize}

\section{Conformal symmetry} \label{X3}
Conformal invariance arises as a bosonic extension of Poincar\'e group, introducing the addition of scale invariance. Such enhanced spacetime symmetry constrains many observables of the theories, such as correlation functions, in their spacetime dependence. In the following of this Section we will keep the dimension $d$ of the spacetime as generic, but our goal is to introduce conformal symmetry only for $d>2$. Again for a more detailed analysis see for example \cite{DiFrancesco:1997nk,Ginsparg:1988ui, Rychkov:2016iqz}.\\

Given a d-dimensional space with a metric $ g_{\mu \nu}(x) $, in general it transforms under diffeomorphisms $ x^{\mu} \mapsto x'^{\mu} $ as a rank-two tensor:  
\begin{equation}\label{c1}
g'_{\mu \nu} (x')= \frac{\partial x^{\alpha}}{\partial x'^{\mu}}\frac{\partial x^{\beta}}{\partial x'^{\nu}}g_{\alpha \beta}(x).
\end{equation}

The conformal group is the subgroup of coordinate transformations that leaves the metric tensor invariant up to a local scale change ($ x \rightarrow x' = \lambda x $):
\begin{equation}\label{c2}
g'_{\mu \nu} (x')= \Omega (x) g_{\mu \nu}(x),
\end{equation}
where $\Omega (x) \equiv \lambda^{2}(x) > 0  $. Taking an infinitesimal coordinate transformation ($ \lambda =1- \epsilon$)
\begin{equation} \label{c14}
   x^{\mu}= x'^{\mu} + \epsilon^{\mu}(x) + \mathcal{O}(\epsilon^{2}) \hspace{0.2cm}, \hspace{0.2 cm} \epsilon \rightarrow 0
\end{equation} 
and asking a general coordinate transformation to satisfy (\ref{c2}), we obtain the differential equation satisfied by $ \epsilon^{\mu} $ \footnote{Notice how this equation simplifies for $d=2$, see \cite{Belavin:1984vu} for further details}:
\begin{equation} \label{c3}
 \partial_{\mu} \epsilon_{\nu} + \partial_{\nu} \epsilon_{\mu} = \frac{2}{d} (\partial_{\alpha} \epsilon^{\alpha})\eta_{\mu \nu}. 
\end{equation}

The solutions to this equation are the parameters associated to the generators which describe such transformations:
\begin{equation} \label{c5}
\begin{split}
\textrm{Translations} \hspace{0.5 cm} &P_{\mu} = -i\partial_{\mu} \\
\textrm{Rotations} \hspace{0.5 cm} &M_{\mu \nu} = i(x_{\mu}\partial_{\nu}-x_{\nu}\partial_{\mu}) \\
\textrm{Dilatations} \hspace{0.5 cm} &D = -ix^{\mu}\partial_{\mu} \\
\textrm{Special~Conformal~Transformations} \hspace{0.5 cm} &K_{\mu} = -i(2x_{\mu}x^{\nu}\partial_{\nu}-x^{2}\partial_{\mu}) \\
\end{split}
\end{equation}
which satisfy the following algebra, which extends \eqref{Poinctot}:
\begin{equation} \label{c23}
\begin{split}
&[K_{\mu}, P_{\nu}] = 2i(\eta_{\mu \nu}D-L_{\mu \nu})\, ,\\
&[P_{\rho},M_{\mu \nu}]=i(\eta_{\rho \mu}P_{\nu}- \eta_{\rho \nu}P_{\mu})\, ,\\
&[K_{\rho},M_{\mu \nu}]= i(\eta_{\rho \mu}K_{\nu}- \eta_{\rho \nu}K_{\mu})\, ,\\
&[M_{\mu \nu},M_{\rho \sigma}] = i(\eta_{\mu \sigma} M_{\nu \rho} + \eta_{\nu \rho} M_{\mu \sigma} - \eta_{\mu \rho} M_{\nu \sigma} - \eta_{\nu \sigma} M_{\mu \rho})\, , \\
&[D,P_{\mu}]=iP_{\mu}\, , \\
&[D,K_{\mu}]=-iK_{\mu}\, , \\
&[P_{\mu}, P_{\nu}]=[K_{\mu}, K_{\nu}]=[D,M_{\mu \nu}]=0. 
\end{split}
\end{equation}
We outline some of the main features of conformal invariance:
\begin{itemize}
\item
The total number of generators, respectively
\begin{equation}
d + \frac{1}{2}d(d-1)+ 1+ d = \frac{1}{2}(d+1)(d+2),
\end{equation}
we can see that there exists an isomorphism between the conformal group and SO$(1,d+1) $, namely the Lorentz group on $ \mathbb{R}^{1,d+1} $.
In the next Subsection we will review and exploit the isomorphism with SO$(1,d+1) $, since it admits a realization acting on $ \mathbb{R}^{d} $ stereographically projected to the sphere $ S^{d} $ and embedded in the light cone of $ \mathbb{R}^{1,d+1} $.
\item
For a generic metric $ g_{\mu \nu} $, satisfying (\ref{c2}), the infinitesimal transformation (\ref{c14}) generates a small variation $ \delta g_{\mu \nu} = 2\epsilon \eta_{\mu \nu} $. Using a definition of the stress tensor $ T_{\mu \nu} $ as an operator measuring response to changing the metric, the Hamiltonian of the theory changes by
\begin{equation}
   \Delta H = \int d^{d}x T_{\mu \nu} \delta g^{\mu \nu} \propto \int d^{d}x T_{\mu}^{\mu}.
   \end{equation}
Hence a fundamental condition for a conformal invariant theory is the \textit{tracelessness} of the stress tensor.
\item
Since $ [D,P_{\mu}] \neq 0 $, dilatation symmetry implies that the mass spectrum is either continuous or all masses are zero. Thus conformal theory cannot be interpreted in terms of particles: no $S$-matrix exists and the only observables are the correlation functions.
\item
We note that $ \{ K_{\mu}, M_{\mu \nu}, D \} $ generate a subalgebra which, when exponentiated, corresponds to the stability group of the transformations that leave the point $ x=0 $ invariant. Indeed, choosing a classical multicomponent field $ \Phi(x) $ belonging to an irreducible representation of the Lorentz group, we have:
\begin{equation} \label{c9}
\begin{split}
&M_{\mu \nu} \Phi (0) = S_{\mu \nu} \Phi (0), \\
&D \Phi (0) = -i \Delta \Phi (0), \\
&K_{\mu} \Phi (0) = 0,
\end{split}
\end{equation}
The first relation defines the spin of the field $\Phi$, where $ S_{\mu \nu}$ forms a finite dimensional representation of the Lorentz group.\\
The second relation defines the \textit{conformal dimension} of the field $ \Phi $, which specifies its behavior under a dilatation $x\rightarrow \lambda x$:
\begin{equation}
\Phi(x)\rightarrow\lambda^\Delta\Phi(\lambda x)
\end{equation} 
To understand the last relation of (\ref{c9}), we translate it outside the origin using $ P_{\mu} \Phi (x) = -i \partial_{\mu} \Phi (x) $. From the conformal algebra \eqref{c23}, we read that $P_\m$ raises the scaling dimension of the field, whereas $K_\m$ lowers it. Since in unitary CFT there is a lower bound on the dimensions of the field, we can define all the fields annihilated by $K_\m$ as \emph{conformal primary} operators, or simply primaries. By acting with $P_\m$ we can construct the whole tower of operators with dimension $\geq \Delta$, defined \emph{descendants} of $\Phi$. 
\item
A final remark which is important to to realize is that conformal invariance at the quantum level does not follow from classical conformal invariance. Indeed, a quantum field theory does not make sense without a regularization prescription that introduce a scale in the theory; this scale breaks the conformal symmetry, except at the renormalization-group fixed point. Hence conformal symmetry in a QFT, namely the tracelessness of the stress tensor also at the quantum level, is associated to the vanishing of the Callan-Symanzik $ \beta $ function. We will find the concept of conformal invariance at the quantum level throughout all the the corp of this thesis.
\end{itemize} 

\subsection{Correlation functions and OPE} 
We briefly outline how conformal invariance constrains the observables of a theory, namely correlation functions.  We consider correlators between primary fields only, since any descendant can be simply written as a convergent series of differential operators acting on primaries.
  
The one-point function of a generic operator $ \phi_\Delta(x) $ is constrained by covariance under translations, which implies that it must be a constant, scale invariance fixes it to be zero. Then in a CFT vacuum expectation values (v.e.v) must vanish: 
\begin{equation}
\langle \phi_\Delta(x) \rangle =0~.
\end{equation} 
The two point function  is constrained by translational and rotational invariance to depend on $ r_{12} \equiv \norm{x_{1}-x_{2}} $, while scale invariance and special conformal transformations impose that correlators with different conformal dimension must vanish. Apart from a field normalization constant $C$, which can depend on the couplings of the theory, a two point function is fixed to:
\begin{equation} \label{c20}
 \langle \phi_{1}(x_{1})  \phi_{2}(x_{2}) \rangle = \frac{C\,\delta_{\Delta_{1},\Delta_{2}}}{r_{12}^{\Delta_{1}+\Delta_{2}}}. 
 \end{equation} 
From similar constraining arguments, we write the result for the three point function:
\begin{equation} \label{c21}
  \langle \phi_{1}(x_{1})  \phi_{2}(x_{2})  \phi_{3}(x_{3}) \rangle = \frac{C_{123}}{r_{12}^{\Delta_{1}+\Delta_{2}-\Delta_{3}}r_{23}^{\Delta_{2}+\Delta_{3}-\Delta_{1}}r_{13}^{\Delta_{1}+\Delta_{3}-\Delta_{2}}}.
  \end{equation}
And here conformal invariance leaves the structure constants $ C_{ijk} $ undetermined.

Wilson's idea \cite{Wilson:1969zs} of Operator Product Expansion (OPE) produces further constraints to higher order correlators of the theory.
The idea states that we should be able to replace a product of two local quantum operators, in the limit where they are very close to each other, by an asymptotic series of operators.
In a CFT we have no dimensional constants and we are allowed to classify all local conformal operators into primaries and descendants, so we can write the product as a sum just over primaries $ \phi_\Delta $:
\begin{equation}\label{c22}
 \phi_{i}(x_{i})\phi_{j}(x_{j}) \stackrel{x_{i} \rightarrow x_{j}}{=} \sum_{\phi} f_{ij\phi}(x_{ij}^{2}) C_{\phi} (x_{i}-x_{j}, \partial_{j}) \phi(x_{j}). 
 \end{equation}

This formula is the starting point for many CFT techniques, such as the conformal bootstrap, that we will not use in the present thesis.
    
\subsection{Embedding formalism} \label{X4}
As already remarked, the conformal group on $ \mathbb{R}^{d} $ is isomorphic to the Lorentz group SO$(1,d+1) $. We can prove it, following the embedding formalism,  mainly due to P. Dirac \cite{Dirac:1936fq}, which provides a realization of conformal transformations as linear coordinate transformations on the light-cone of $ \mathbb{R}^{1,d+1} $. We also review this method since it represents the best way to realize the constraints on correlation functions coming from conformal symmetry.
 
We define embedding coordinates in $ \mathbb{R}^{1,d+1} $
\begin{equation}
P^{\cM} = (P^{0},x^{\mu},P^{d+1}),
\end{equation}
where $ P^{0} $ is the time-like direction. Using now light-cone coordinates $ P^{\pm}=P^{0}\pm P^{d+1} $ we identify the generators (\ref{c5}) with the generators of the Lorentz group $ J_{\cM \cN} = - J_{\cN \cM} $ as follows:
\begin{equation}
J_{\mu \nu}=M_{\mu \nu}~, \hspace{0.8 cm} J_{\mu +}=K_{\mu}~, \hspace{0.8 cm}
J_{+ -}=D~, \hspace{0.8 cm} J_{\mu -}=P_{\mu}~.
\end{equation}
And now, comparing with (\ref{c23}), they correctly satisfy:
\begin{equation}
[J_{\cM \cN},J_{\cR \cS}]=\ii(\eta_{\cM \cS}J_{\cN \cR}+\eta_{\cN \cR}J_{\cM \cS}-\eta_{\cM \cR}J_{\cN \cS}-\eta_{\cN \cS}J_{\cM \cR})~.
\end{equation}
While $ d $-dimensional representations of the conformal group are non-trivial, we find a natural action on the $ \mathbb{R}^{1,d+1} $ space:
\begin{equation}
X^{\cM} \rightarrow X'^{\cM} = \Lambda^{\cM}_{\cN} X^{\cN}, \hspace{0.7 cm} \Lambda^{\cM}_{\cN} \in \mathrm{SO}(1,d+1)~.
\end{equation}
To get an action on $ \mathbb{R}^{d} $, one has to get rid of two extra coordinates:
\begin{itemize}
\item
One dimension is eliminated by working on the $ \mathbb{R}^{1,d+1} $ null cone $ P^{2} = 0 $, which is invariant under the action of the Lorentz group.
\item
We get down to $ d $ dimension by declaring the light-cone to be projective: $P\sim \lambda P$, $\lambda\in \mathbb{R}^+$. In general we have the freedom to take a section such that the induced metric is Euclidean.
The most common choice is to map $x\in \mathbb{R}^d$ to a null point $P_x\in\mathbb{R}^{1,d+1}$ in the so called Poincar\`e section
\begin{equation}\label{Poincarè}
x^\m \rightarrow P^\cM =\left(P^{+}, P^{-}, P^{\mu} \right) = (1, x^{2}, x^{\mu})~.
\end{equation}
This section is not preserved by the action of a generic element $g\in$ SO$(1,d+1)$, but it is possible to define an action $\tilde{g}$ on the section by rescaling back the point: writing
\begin{equation}
g\,P^\cM = g(x)(1, x^{\prime 2}, x^{\prime \mu})~,
\end{equation}
we can define $\tilde{g}$ such that $\tilde{g}x=x^\prime$, and $\tilde{g}$ is precisely a conformal transformation.
\end{itemize}

Fields on the light cone must coincide with the $d$-dimensional field on the euclidean section \eqref{Poincarè}. Therefore, we define a SO$(1,d+1)$ tensor field $F_{\cM_1,\dots\cM_\ell}(P)$ with the following properties:
\begin{itemize}
\item
Defined on the cone $P^2=0$;
\item
Homogeneous of degree $-\Delta$: $F_{\cM_1,\dots\cM_\ell}(\lambda P) = \lambda^{-\Delta}F_{\cM_1,\dots\cM_\ell}(P), ~~\lambda>0$;
\item
Symmetric, traceless and transverse: $(P\cdot F)_{\cM_2,\dots\cM_\ell}=0$.
\end{itemize}
Then projecting $F$ to the Poincar\`e section \eqref{Poincarè} automatically defines a symmetric tensor field on $\mathbb{R}^d$:
\begin{equation}
f_{\m_1,\dots,\m_\ell} = \frac{\partial P^{\cM_1}}{\partial x^{\m_1}}\dots \frac{\partial P^{\cM_\ell}}{\partial x^{\m_\ell}}F_{\cM_1,\dots\cM_\ell}(P)~.
\end{equation}
Using this prescription we have a precise $f\leftrightarrow F$ correspondence (see \cite{Costa:2011mg} for further details) and conformal invariance of the result is guaranteed.\\
This formalism helps the computation of correlation functions, especially for spin operators and in presence of a conformal defects, now we simply give a hint of how it works.

\subsubsection*{Examples of correlation functions}

The two-point function of scalar primaries with scaling dimension $ \Delta $ on the light cone is directly constrained by the SO$(1,d+1)$ Lorentz group:
\begin{equation} \label{c13}
\langle \phi(X) \phi(Y) \rangle = \frac{C}{(X \cdot Y)^{\Delta}},
\end{equation}
where $ C $ is the same constant as \eqref{c20}. The above is the most general $\mathrm{SO}(1,d+1)$ invariant expression consistent with scaling, since $ X^{2} = Y^{2} = 0 $ cannot appear. To obtain the two-point function in the physical space, we project $ X $ and $ Y $ on the section \eqref{Poincarè}, \textit{i.e.}
\begin{equation}\label{1.2.24}
X=\left(X^{+}, X^{-}, X^{\mu} \right) = (1, x^{2}, x^{\mu}), \hspace{0.7 cm} Y=\left(Y^{+}, Y^{-}, Y^{\mu} \right) = (1, y^{2}, y^{\mu})~,
\end{equation}
and we get the same expression as \eqref{c20}.\\
Similarly we provide an example for vector fields. In this case the most general expression invariant under Lorentz symmetry is:
\begin{equation}
     \langle \phi_{\cM_{1}}(X) \phi_{\cM_{2}}(Y) \rangle = \frac{\eta_{\cM_{1} \cM_{2}} - \frac{X_{\cM_{1}}Y_{\cM_{2}}}{X \cdot Y}}{(X \cdot Y)^{\Delta}}~,
     \end{equation}
where at the numerator we recognize the usual transverse tensorial structure.
The projection to the section \eqref{1.2.24} returns:
\begin{equation}
\langle \phi_{\m}(x) \phi_{\n}(y) \rangle= \frac{\delta_{\mu \nu}-\frac{2(x-y)_{\mu}(x-y)_{\nu}}{(x-y)^{2}}}{(x-y)^{2 \Delta}}~.     
     \end{equation}
In Chapter \ref{chap:4} we will see how this formalism can be sharpened for higher order tensor fields and in presence of a conformal defect.

\section{Superconformal symmetry}\label{sec1:SCFT}
We introduced the two possible extensions of the Poincar\`e group, \textit{i.e.} supersymmetry and conformal invariance. It is possible to combine these two symmetry groups generating an enhanced superconformal invariance. 

We unify the super-Poincar\`e \eqref{SUSYalgebra} and the conformal \eqref{c23} algebras, but in order to close it another spinorial generator $S^\cI_\a$ is needed. This additional class of generators commute with special conformal transformations $K_\m$, so that it plays the same role of $Q^\cI_\a$ for the translation operator $P_\m$. The full classification of superconformal algebras was given by Nahm \cite{Nahm:1977tg}, in this case we simply concentrate on the four dimensional case.

The $\cN$-extended superconformal algebra is dubbed SU$(2,2|\cN)$ and includes \eqref{SUSYalgebra}, \eqref{c23} and the following relations:
\begingroup
\allowdisplaybreaks
\begin{align}\label{Sconfalgebra}
\acomm{Q^\cI_\a}{S^\cJ_\b} &=2 \epsilon_{\a\b}\delta^{\cI\cJ} D-\ii (\s^{\m\n})_\a^\c \epsilon_{\c\b}\delta^{\cI\cJ}M_{\m\n}-4\ii \epsilon_{\a\b}\delta^{\cI\cJ}R~, \notag \\
\acomm{S^\cI_\a}{S^\cJ_\b}&= 2\delta^{\cI\cJ}\s^\m_{\a\dot \a}K_\m~,\hspace{0.7cm}\comm{Q^\cI_\a}{P_{\m}} =\comm{S^\cI_\a}{K_{\m}} =0~, \notag \\
\comm{S^\cI_\a}{M_{\m\n}} &= (\s^{\m\n})_\a^\b S^\cI_\b ~,\hspace{0.7cm}\comm{S^\cI_\a}{D}=-\frac{1}{2}S^\cI_\a~, \hspace{0.7cm}
\comm{S^\cI_\a}{P_{\m}} =-\ii \s^\m_{\a\dot \a}\bar Q^{\cI\dot\a}~,\notag \\
\comm{Q^\cI_\a}{M_{\m\n}} &= (\s^{\m\n})_\a^\b Q^\cI_\b ~,\hspace{0.7cm}\comm{Q^\cI_\a}{D}=\frac{1}{2}Q^\cI_\a~, \hspace{0.7cm}
\comm{Q^\cI_\a}{K_{\m}} =\ii \s^\m_{\a\dot \a}\bar S^{\cI\dot\a}~,\notag \\
\comm{Q^\cI_\a}{R}&= -\ii \left(\frac{4-\cN}{4\cN}\right)Q^\cI_\a~,\hspace{1cm}\comm{S^\cI_\a}{R}=\ii \left(\frac{4-\cN}{4\cN}\right)S^\cI_\a~.
\end{align}
\endgroup
The operator $R$ is the generator of the U$(1)$ factor of the U$(\cN)_R$ R-symmetry group. We see that the $\cN=4$ case is special because $R$ commutes with $Q^\cI_\a$ and $\bar Q^{\cI\dot\a}$; that's why the R-symmetry group in this case is SU$(4)$ and not the full U$(4)$ as one would expect.

We act with this enhanced algebra in a similar way we did in the conformal case. There primaries were the operators annihilated by $K_\m$, while $P_\m$ generated the full tower of descendants. In a superconformal case the corresponding roles are played by $S^\cI_\a$ and $Q^\cI_\a$ respectively. Therefore we define a \emph{superconformal primary} as the operator satisfying:
\begin{equation}
\comm{S^\cI_\a}{\Phi} = \comm{\bar S^{\cI\dot a}}{\Phi}=0~,
\end{equation}
then the action of $Q^\cI_\a$ and $\bar Q^{\cI\dot\a}$ generates the tower of superconformal descendants. A general superconformal primary is labeled by the quantum numbers $(\Delta,j_l,j_r,s,r)$ associated to dilatations, Lorentz and the R-symmetry $su(2)_R\times u(1)_R$.

Among the superconformal primaries we select a special class of operators that will play a central role in the following, namely the Chiral Primary Operators (CPOs), denoted as $\cO$. These operators are superconformal primaries which are also annihilated by all the Poincar\`e charges with a certain chirality
\begin{equation}
\comm{\bar Q^{\cI\dot\a}}{\cO} = 0~,
\end{equation}
together with the anti-chiral primaries $\bar\cO$ annihilated by $Q^\cI_\a$.
This structure provides a lot of constraints, so that CPOs enjoy many interesting properties:
\begin{itemize}
\item
Unitarity of the CFT and the anticommutator $\acomm{Q^\cI_\a}{S^\cJ_\b}$ defined in \eqref{Sconfalgebra} imply that:
\begin{align}\label{CPOcond}
\cO:~\Delta&=\frac{R}{2}~, ~~~~~~~j_r=s=0~,\notag\\
\bar\cO:~\Delta&=-\frac{R}{2}~, ~~~~~~~j_l=s=0~.
\end{align}
Moreover for SCFTs with a Lagrangian description all chiral primaries must be Lorentz scalar, so with also $j_r=0$. The important point is that the conformal dimension of CPOs is completely determined by their R-charge $R$.
\item
In $\cN=2$ conformal theories, CPOs parametrize the Coulomb branch of vacua of the SCFT, where $su(2)_R$ is preserved and $u(1)_R$ is broken.
\item
The OPE of chiral primaries is non-singular due to the unitarity bound $\Delta\geq R/2$. Therefore CPO generate the so-called chiral ring:
\begin{equation}\label{chiralring}
\cO_I(x)\cO_J(0)= \sum_K C_{IJ}^K \cO_K(x)~,
\end{equation}
where $C_{IJ}^K$ are the CFT structure constants. For Lagrangian $\cN=2$ theories the chiral ring is freely generated, namely there exists a finite-dimensional basis such that any element of the ring can be written as a linear combination of the basis elements. The number of generators of the chiral ring is the dimension of the Coulomb branch of the SCFT.
\end{itemize}
We will deal with chiral primaries in Chapter \ref{chap:5}, where we will compute their correlation functions for $\cN=2$ theories.

\section{Wilson loop in gauge theories}\label{sec1:WL}
We conclude this introductory Chapter by introducing the main ingredient, namely the Wilson loop operator.

The high energy behavior of non abelian gauge theories like QCD is under control thanks to the mechanism of asymptotic freedom: it is possible to treat the theory using the standard perturbation theory approach, inserting dynamical quarks as light degrees of freedom in the action. However, at low energies QCD is confining, namely light quarks are not free in the vacuum, but they appear as quark-antiquark pairs $q\bar{q}$. In this case we need to introduce a different tool to face this regime. The idea is to insert external heavy quarks that are no more dynamical degrees of freedom, but in such a way that the distance between a $q\bar{q}$ pair is fixed in time. The physical object that defines and measures the interaction potential $V_{q\bar{q}}$ is dubbed \emph{Wilson loop}, introduced by Wilson \cite{Wilson:1974sk} in 1974.

The Wilson loop represents the phase factor picked up by an external quark moving along a closed path and is the most general gauge invariant observable. It is defined as the  traced holonomy of the gauge connection $A_{\mu}(x)$ which take values in the gauge algebra $A_{\mu}(x)= A_{\mu}^a(x)T^a$:
\begin{equation}
\label{WLQCD}
		W(C)=\frac{1}{\mathrm{dim}_\cR}\Tr_\cR \: \mathcal{P}
		\exp \left\{\ii g \oint_C dx^{\mu}  \,A_{\mu}(x)\right\}~,
\end{equation}
where dim$_\cR$ is the dimension of the representation $\cR$, $g$ is the Yang-Mills coupling and $C$ is the path, parametrized by the vector $x^\m$. The symbol $\cP$ denotes the path-ordering exponential. For example if we parametrize the curve as $x^\m =x^\m(\tau)$ where $\tau\in[0,\ell]$, and we choose $\tau_1>\tau_2$:
\begin{align}\label{pathordering}
\mathcal{P}\exp \left\{\ii g \oint_C dx^{\mu}  \,A_{\mu}(x)\right\}=\mathbb{1} &+\ii g \int_0^\ell d\tau \Big[\dot x^{\mu}  \,A_{\mu}(x)\Big]\notag \\
&-g^2\int_0^\ell d\tau_1 \int_0^{\tau_1} d\tau_2 \Big[\dot x_1^{\mu}\dot x_2^{\mu}~A_{\mu}(x_1)A_{\mu}(x_2)\Big]+\dots
\end{align}
In practice, the path-ordering keeps track of the non-commutativity of the gauge algebra matrices, by imposing to preserve the order in performing the multiple integrals in the expansion \eqref{pathordering}. In this way it ensures gauge invariance.

The Wilson loop is the most general gauge invariant operator and is related to many important observables \cite{Giles:1981ej,Migdal:1984gj}. We briefly mention here its relationship with the quark-antiquark potential and its role in AdS/CFT, while throughout the text we will analyze its connection with many other physical quantities.

\subsection{Interquark potential from Wilson loops}
We briefly describe the role of the Wilson loop for describing confinement mechanism, since it helps to understand the real physical meaning of this operator. We consider a rectangular loop, which describes a $q\bar q$ pair created at $t=0$ and stretched at a certain distance $R$ in the spatial direction and for a time length equal to $T$.
\begin{figure}[htb]
	\begin{center}
		\includegraphics[scale=0.8]{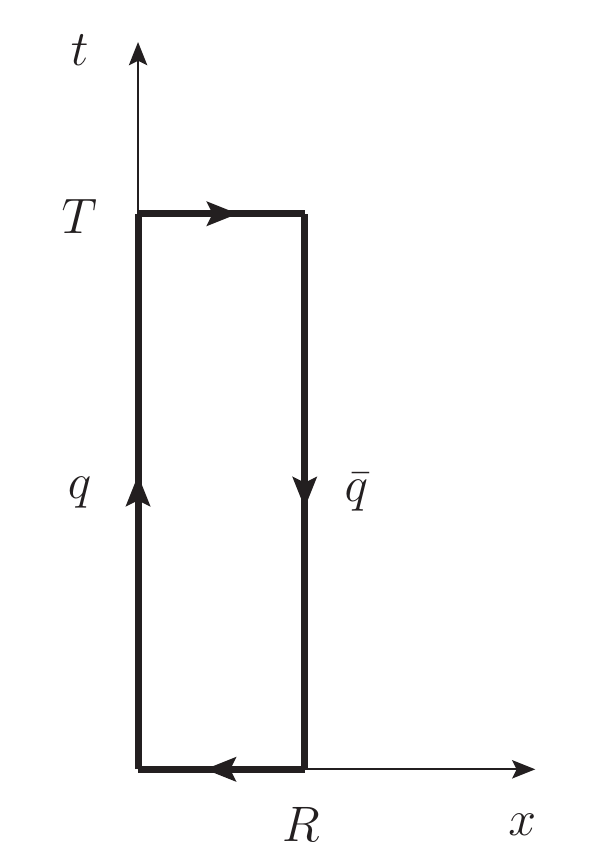}
	\end{center}
	\caption{Rectangular Wilson loop}
	\label{fig:WLrectangular}
\end{figure}
It is possible to prove that the vacuum expectation value of the Wilson loop in Figure \ref{fig:WLrectangular} is related to the quark-antiquark potential as:
\begin{equation}\label{Wqbarq}
\vev{W(C)} \propto e^{-V_{q\bar q}(R)T}~, ~~~T\to\infty~.
\end{equation}
We can understand this as follows.\\
Adding static quarks to a gauge theory corresponds to insert an additional source term to the action coupled to the gauge field. Schematically:
\begin{equation}
\int dt j^\m A_\m = \int dt~ g\Big[ A_0(x_q)-A_0(x_{\bar q})  \Big]
\end{equation}
where $g A_0(x_q)$ is the potential at the position of $q$. Therefore for large values of $t$, $\vev{W(C)}$ parametrizes the $q\bar q$ potential. Now we see how equation \eqref{Wqbarq} is a good criterion for confinement.\\
The statement of confinement is equivalent to the presence of a constant force that resists when one tries to stretch the $q\bar q$ away: the two quarks cannot be separated to an infinite distance with finite energy. So we have a linear potential
\begin{equation}\label{confphase}
V_{q\bar q}(R) \sim \sigma R~,
\end{equation}
where $\s$ is the ``string tension'', which represents a confined flux tube. This string is not a fundamental object, but rather an effective description of confinement.
The unconfined phase, instead, corresponds to the Coulomb static potential
\begin{equation}\label{deconfphase}
V_{q\bar q}(R) \sim \frac{\a}{R}~,
\end{equation}
which is indeed the typical behavior of the electrodynamic case.\\
We can observe the how these two regimes affect the Wilson loop vev. In a confined theory like QCD we get:
\begin{equation}
\vev{W(C)} \propto e^{\sigma R T} = e^{-\sigma A}~,
\end{equation}
this behavior is precisely known as area law, since the vev scales with the area of the loop. In a deconfined case like QED with external quarks we find:
\begin{equation}
\vev{W(C)} \propto e^{-\a\,\frac{T}{R}}~.
\end{equation}
The Wilson loop vev here depends on the scale invariant quantity $T/R$. So we distinguish two regimes, which clarify how the Wilson loop vev is interpreted as the order parameter of confinement.
\begin{align}
&\lim_{R\to\infty} \vev{W(C)}\sim e^{-\s A}\rightarrow 0 \hspace{1.5cm} \mathrm{confined~phase} \notag \\
&\lim_{R\to\infty} \vev{W(C)}\sim e^{-\mathrm{const}}\rightarrow \neq 0 \hspace{1cm} \mathrm{deconfined~phase}
\end{align}

\subsection{Supersymmetric Wilson loops}
The Wilson loop has played a crucial role in AdS/CFT context \cite{Maldacena:1998im,Berenstein:1998ij}. Here we make a brief recap of the Maldacena construction and derive the expression of supersymmetric invariant Wilson loop (often dubbed Maldacena-Wilson loops), since we will strongly need its properties in the following.

Following the AdS/CFT conjecture, there exists a correspondence between type IIB string theory compactified on AdS$_5\times S^5$ and a four-dimensional $\cN=4$ theory. The explicit realization involves the insertion of D3-branes, whose open string sector realizes a $\cN=4$ theory on the  D3-brane worldvolume. The closed string description reduces to a string theory on AdS$_5\times S^5$. Within this set up, we can introduce the concept of supersymmetric Wilson loop.

\begin{figure}[htb]
	\begin{center}
		\includegraphics[scale=0.8]{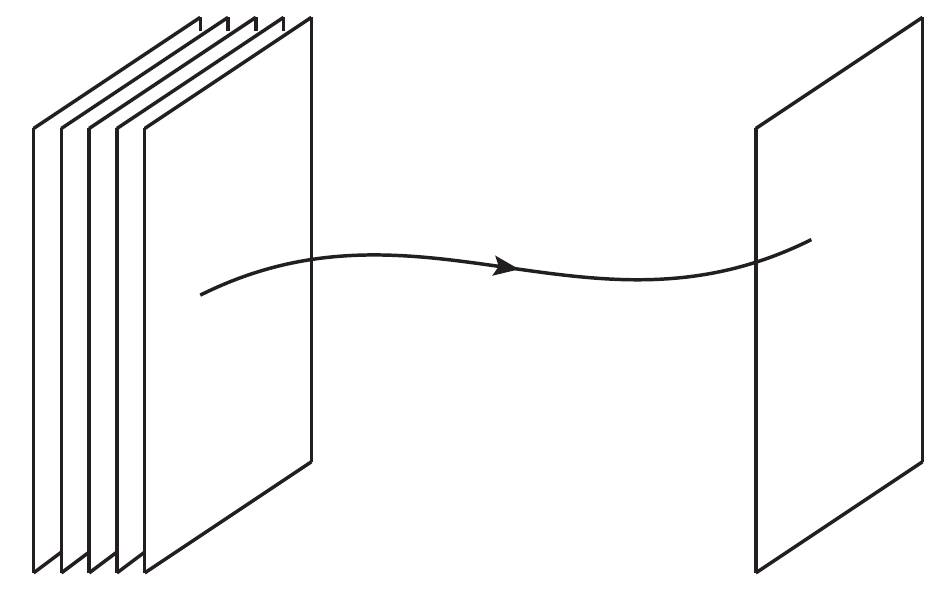}
	\end{center}
	\caption{The single brane separated from the stack of $N$ branes acts as a probe for the theory living on the stack worldvolume.}
	\label{fig:WLbranes}
\end{figure}

Following Polchinski's idea \cite{Polchinski:1995mt}, U$(N)$ gauge theories are realized on a set of $N$ D-branes situated at the same point. Strings with endpoints on these branes are massless and give rise to gauge fields $A_\m^a$. According to the Maldacena construction the insertion of D$3$ branes generates a $\cN=4$ SYM theory on their world volume. The way to insert a Wilson loop inside a $\cN=4$ theory is to start from a stack of $N+1$ D$3$-branes, which realize a U$(N+1)$ gauge group, and to separate one of them from the rest, see Figure \ref{fig:WLbranes}.
This procedure generates a Higgs mechanism that breaks the gauge group to U$(N)\times$U$(1)$. The strings stretched along the $N$ branes and the single brane have a state in the fundamental representation of the unbroken U$(N)$ gauge group. This state has a mass which is proportional to the distance between the probe and the stack:
\begin{equation}
M=\frac{1}{2\pi\a'}r~.
\end{equation}
Taking $r\to\infty$ we get an infinite mass state.
Therefore, from the U$(N)$ theory point of view, the massive string state acts as a source term for the gauge fields.

Note that from the 10-d perspective this realization consists in placing the $\cN=4$ gauge theory at infinity in the transverse direction. The infinitely heavy quark is obtained having a string stretched in this 5-dimensional space, which in the AdS$_5$/CFT$_4$ framework corresponds to a 5-dimensional AdS space. The Wilson loop is a boundary condition for the string, and the string worldsheet stretches between the contour $C$ at infinity down to a finite point in AdS, forming a surface. Therefore from the string perspective $\vev{W(C)}$ can be computed as the partition function for a string with boundary on $C$
\begin{equation}
\vev{W(C)}= Z_{\mathrm{string}}[C]\sim e^{-\frac{1}{2\pi\a'}A}~,
\end{equation}
where $A$ is the area of the worldsheet. 

In this string perspective we need to keep in mind that the massive string which acts as a probe for the 4-dimensional gauge theory is also situated on the five-sphere S$_5$, parametrized by coordinates $n^u$, $u=1,\dots,6$. The corresponding fields are the 6 scalars $\phi^u$ of $\cN=4$ SYM\footnote{This is also a key point of the correspondence: scalars of $\cN=4$ transform under the SO$(6)$ group, which is the R-symmetry group from the SUSY point of view, but also the isometry group of the five sphere}. Therefore we expect the massive state to be a source for a generalized \emph{supersymmetric Wilson loop} in the fundamental representation:
\begin{equation}
	\label{WLsusy}
		W(C)=\frac{1}{N} \tr \: \mathcal{P}
		\exp \left\{g \oint_C d\tau \Big[\ii \,A_{\mu}(x)\,\dot{x}^{\mu}(\tau)
		+n^u(\tau)\phi_u(x)|\dot{x}|\Big]\right\}
\end{equation}
where $x^\m(\tau)$ is a parametrization of the loop and $n^u$ is a 6-dimensional unit vector. \\
The object \eqref{WLsusy} preserves some of the supersymmetry charges. We act with a supersymmetry variation starting from the SUSY variations of the bosonic fields:
\begin{align}
\delta_\epsilon A_\m &= \bar \lambda\, \Gamma_\m \epsilon~, \notag \\
\delta_\epsilon \phi_u &= \bar \lambda \,\Gamma_u \epsilon~,
\end{align}
where $(\Gamma_\m,\Gamma_u)$ are ten-dimensional matrices and the SUSY transformation parameter is a 10-dimensional Majorana-Weyl spinor. Then we have:
\begin{equation}
\delta_\epsilon W[C]= \frac{1}{N} \tr \: \mathcal{P}
g \oint_C d\tau \bar{\lambda} \Big[\ii \,\Gamma_{\mu}\,\dot{x}^{\mu}
		+n^u\Gamma_u|\dot{x}|\Big]\epsilon\,
		\exp \left\{g \oint_C d\tau \Big[\ii \,A_{\mu}\,\dot{x}^{\mu}
		+n^u\phi_u|\dot{x}|\Big]\right\}~.
\end{equation}
The supersymmetry preserving condition is then:
\begin{equation}\label{WLsusycondition}
\Big[\ii \,\Gamma_{\mu}\,\dot{x}^{\mu}
		+n^u\Gamma_u\,|\dot{x}|\,\Big]\,\epsilon=0
\end{equation}
The request to have global supersymmetry, \textit{i.e.} a constant $\epsilon$, induces a constraint on $x_\m(\tau)$ and $n^u(\tau)$. The number of linearly independent $\epsilon$'s satisfying \eqref{WLsusycondition} determines the number of conserved supercharges. This analysis has been pursued in a series of papers \cite{Zarembo:2002an, Drukker:2006zk,Drukker:2007dw,Drukker:2007qr,Drukker:2007yx}. Throughout the full thesis we will concentrate on the highest supersymmetric case, namely $1/2$ BPS loops, which can be realized in two possible ways.
\begin{itemize}
\item \textbf{Straight line}:
We choose $n^u$ not depending on $\tau$. The only solution in this case is $C$ equal to a straight line. With a parametrization such that $|\dot{x}|=1$, we differentiate \eqref{WLsusycondition} with respect to $\tau$ and we get:
\begin{equation}\label{susyconditionline}
\ii \Gamma^\m \ddot x_\m \epsilon=0~.
\end{equation}
This implies that $\ddot x=0$ is the only solution. This means that in this case $W[C]$ commutes with all the Poincar\`e supercharges $Q_\cI$.
\item \textbf{Circular loop}: it is possible to move from the straight line to a circular loop using a conformal transformation (specifically an inversion). We will see how this operation affects the computation of the vacuum expectation value. In this case the Wilson loop is still $1/2$ BPS but it preserves a combination of Poincar\`e and conformal supercharges. 
\end{itemize}
We also mention another class, which is not maximally supersymmetric, the \textbf{latitude loop}. We can take the loop to be a non maximal circle, \textit{i.e.} a latitude of $S_2$ embedded in $\mathbb{R}^4$ and parametrized by the latitude angle $\theta_0$. This specific class of operators, for $\cos\theta_0\neq 0$ are 1/4 BPS. We will meet a Wilson loop operator belonging to this class in Section \ref{sec:6N4Brem}.

A similar classification can be performed in the less supersymmetric $\cN=2$ case, the main difference lies in the number of scalars (in that case $u=1,2$, as seen before) belonging to the vector multiplet. We will treat the $1/2$ BPS Wilson loop vev in $\cN=2,4$ theories in Chapter \ref{chap:3}.

\chapter{$\mathcal{N}=2$ Superconformal field theories}\label{chap:2}
This Chapter is focused on $\cN=2$ superconformal field theories in four dimensions, which represent the main framework of the present thesis. We will not review all the several features that $\cN=2$ SCFTs enjoy (see the reviews\cite{Bilal:1995hc, DiVecchia:1998ky} and the collection \cite{Teschner:2014oja} for a detailed analysis of the recent results in this context). Our goal is rather to collect a series of ingredients that we will use throughout the thesis. 

Section \ref{sec:FTactions} contains all the formalism we will use to perform perturbative computations on flat space $\mathbb{R}^4$. In particular, to fully exploit the presence of extended supersymmetry, we write the Lagrangians and Feynman rules following the $\cN=1$ superspace formalism \cite{Gates:1983nr}. \\
From Section \ref{sec2:susycurved} we move to the description of $\cN=2$ theories on curved space. After an introduction on the general formalism, in Section \ref{sec2:ellips} we discuss the construction of $\cN=2$ theories on four dimensional ellipsoids. This enables us to introduce the idea of supersymmetric localization on compact spaces (in Section \ref{sec2:localization}) and to review the shape of the $\cN=2$ matrix models that will guide all the perturbative computations in the following.
\section{$\cN=2$ Lagrangian on $\mathbb{R}^4$ in $\cN=1$ superspace formalism}\label{sec:FTactions}
We write the Lagrangian for a generic conformal $\cN=2$ theory in a four-dimensional Euclidean flat space with SU($N$) gauge group. We follow the $\cN=1$ superspace formalism, which is an important and well established way to construct interacting Lagrangians, and is one of the main tools to be used in this thesis. We will mainly follow the conventions and notations from Wess and Bagger \cite{Wess:1992cp}, but we will also present a new way to compute superFeynman diagrams.

\begin{figure}[H]
\begin{center}
    \includegraphics[scale=0.7]{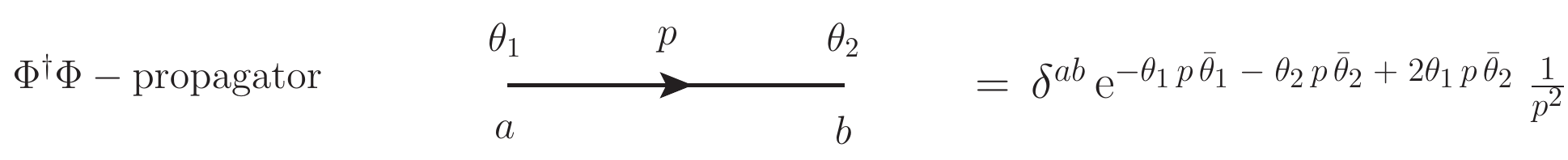}\\ \vspace{0.2cm}
	\hspace{-2.6cm}\includegraphics[scale=0.7]{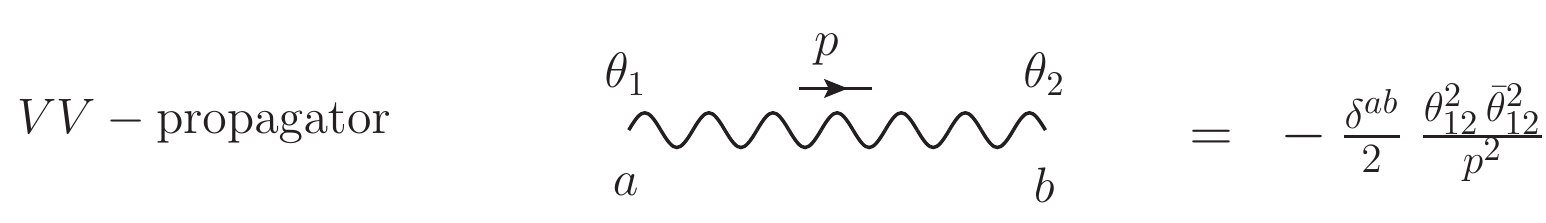} \\[0.1cm]
	\includegraphics[scale=0.6]{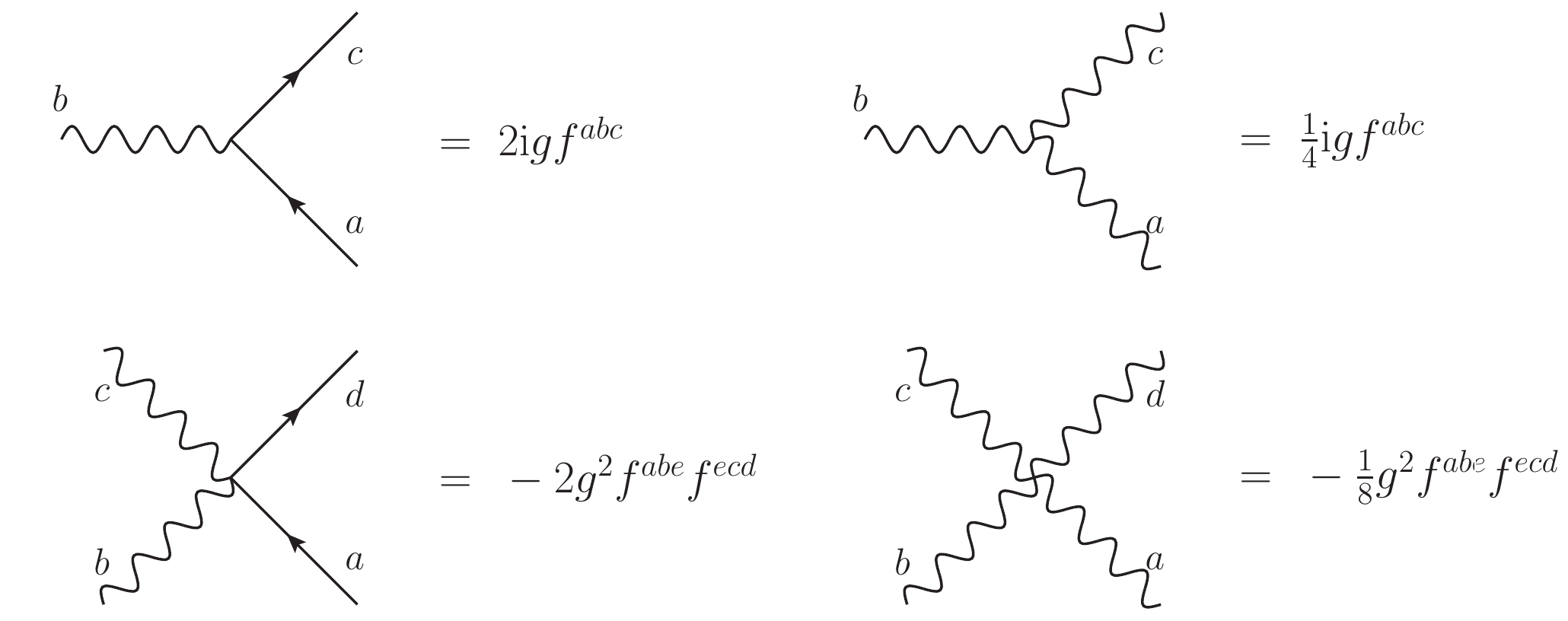}
	\caption{Feynman rules for the gauge part  of the $\cN=2$ theory that are relevant
	for our calculations.}
	\label{fig:Feyngauge}
	\end{center}
\end{figure}

$\cN=2 $ theories contain both gauge fields, organized in a $\cN=2$ vector multiplet which determines the pure Yang-Mills (YM) part, and matter fields, organized in hypermultiplets. Therefore the generic action will be given by the sum
\begin{equation}\label{Stot}
S_{\cN=2}=S_{\mathrm{YM}}+S_{\mathrm{matter}}
\end{equation}
In terms of $\cN=1$ fields, the $\cN=2$ vector multiplet is a combination of a vector $V$ with a chiral $\Phi$ multiplet, transforming in the adjoint representation of the gauge group. Hypermultiplets in a generic representation $\cR$ of the gauge group are made of two chiral multiplets, $Q$ in the representation $\cR$, $\widetilde{Q}$ in the conjugate representation $\bar{\cR}$. Schematically:
\begin{align}\label{N2fields}
\mathrm{Vector}_{(\cN=2)} &= \big(V, \Phi \big) ~~\mathrm{adj~of~SU}(N) \notag \\
\mathrm{Hyper}_{(\cN=2)} &= \big(Q, \widetilde{Q} \big)~~\mathrm{representations~} \cR, \bar{\cR}  ~\mathrm{of~SU}(N)~.
\end{align}
Comparing \eqref{N2fields} with the field content described in Subsection \ref{subsec1:extendedSUSY}, we can recover the fundamental field content of $\cN=2$ theories \footnote{Notice that the scalar fields $\phi^{1,2}$ of the vector multiplet here appear as $\varphi,~ \bar\varphi$, scalar components of the chiral/antichiral multiplets $\Phi,~\Phi^\dagger$. See also \eqref{fibarfi} for our normalization}.\\
We build separately the Lagrangians for the gauge and the matter part. The gauge theory action is:
\begin{align}\label{Sgauge1}
S_{\mathrm{YM}} =& \frac{1}{8g^2} \left(\!\int\!d^4x\,d^2\theta\, \tr(W^\a W_\a)+\mathrm{h.c.}\right) +2\!\int\!d^4x\,d^2\theta\,d^2\bar{\theta}\, \tr\left( e^{-2gV}\Phi^\dagger e^{2gV}\Phi\right) \notag \\
&-\frac{\xi}{4}\!\int\!d^4x\,d^2\theta\,d^2\bar{\theta}\, \tr\left( \bar{D}^2 V D^2 V\right)~,
\end{align}
where $g$ is the gauge coupling and $W_\a$ is a chiral field corresponding to the super field strength of V:
\begin{align}
W_\a=-\frac{1}{4}\bar{D}^2\left(e^{-2gV}D_\a e^{2gV}\right)~.
\end{align}
See Appendix \ref{app:Notations} for our notation for the covariant derivatives and spinor indices. Some important remarks:
\begin{itemize}
\item
From \eqref{Sgauge1} after the $\theta$ integration, it is possible to obtain the expression for the Yang-Mills action in terms of fundamental fields, namely \eqref{a16}.
\item
The last term in \eqref{Sgauge1} specifies the gauge fixing, and the Fermi-Feynman gauge ($\xi=1$) will be our preferred choice.
\item
$S_{\mathrm{YM}}$ should also contain a ghost contribution, which we omit since it is not relevant in any of our calculations \footnote{The ghosts contributions cancel in the first perturbative orders of the difference theory, which we introduce in Subsection \ref{subsec::difference}}.
\item
With our conventions, the total action $S$ is negative defined, so it appears in the path integral as $\exp[S]$.
\end{itemize} 
We expand the superfields $V=V^aT^a$, $\Phi= \Phi^aT^a$ in terms of the generators $T^a$ in the representation $\cR$ of the gauge group, and then the action \eqref{Sgauge1}, expanded up to second order, becomes:
\begin{align}
	\label{Sgauge}
		S_{\mathrm{YM}}&=\!\int\!d^4x\,d^2\theta\,d^2\bar{\theta}\,\Big(
		- V^a\square V^a+\Phi^{\dagger a}\Phi^a
		+\frac{\ii}{4}g f^{abc}\,\big[\bar{D}^2(D^{\alpha} V^a)\big]\, V^b\, (D_{\alpha} V^c) 
		\nonumber\\
		& \qquad\qquad 
		-\frac{1}{8}\,g^2 f^{abe}f^{ecd}\, V^a (D^{\alpha} V^b) (\bar{D}^2 V^c)(D_{\alpha} V^d) \nonumber\\
		& \qquad\qquad+ 2\,\ii g f^{abc}\,\Phi^{\dagger a}V^b\Phi^c
		- 2g^2 f^{abe}f^{ecd}\,\Phi^{\dagger a}V^b V^c \Phi^d
		+\cdots\Big)~,
\end{align} 
where the dots stand for higher order vertices of the schematic form $g^k\: \Phi^{\dagger}V^k \Phi$ with $k\geq 3$. 
Here $f^{abc}$ are the structure constants of SU$(N)$. 
The Feynman rules following from this action are displayed in Figure \ref{fig:Feyngauge}.

\begin{figure}[H]
\begin{center}
	\includegraphics[scale=0.7]{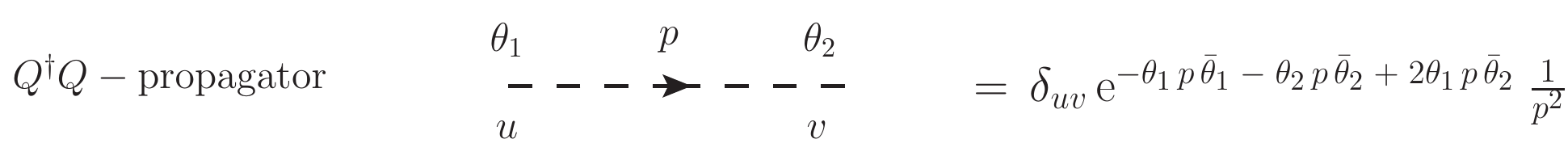}\\ \vspace{0.2cm}
	\includegraphics[scale=0.7]{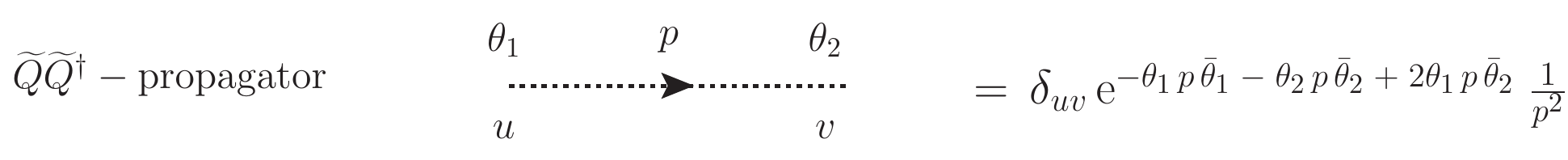} \\
	\includegraphics[scale=0.6]{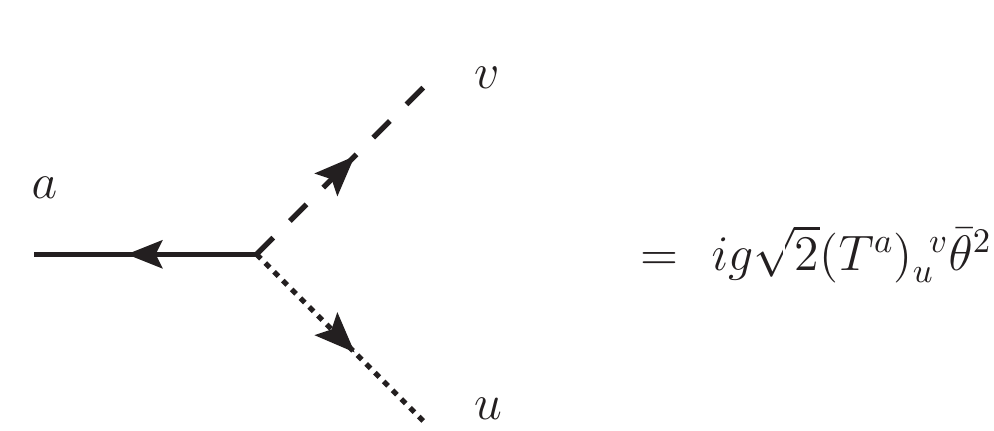}\hspace{1.2cm}
	\includegraphics[scale=0.6]{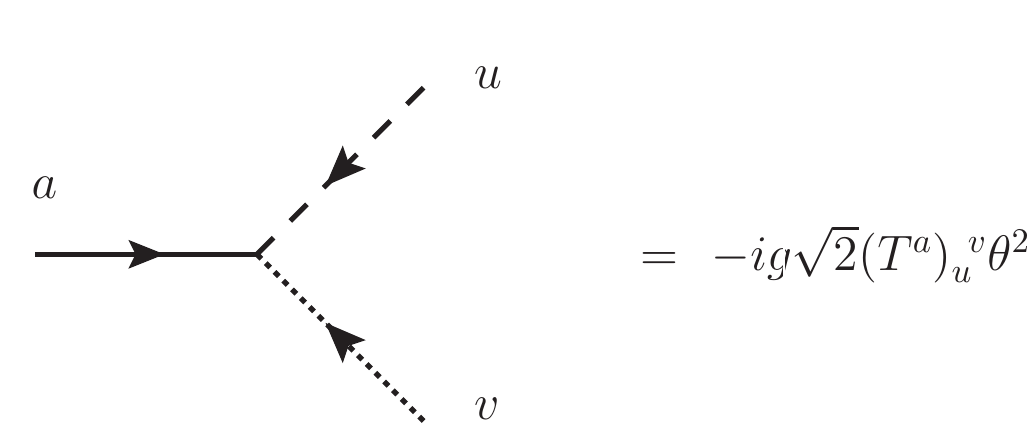} \\
	\includegraphics[scale=0.6]{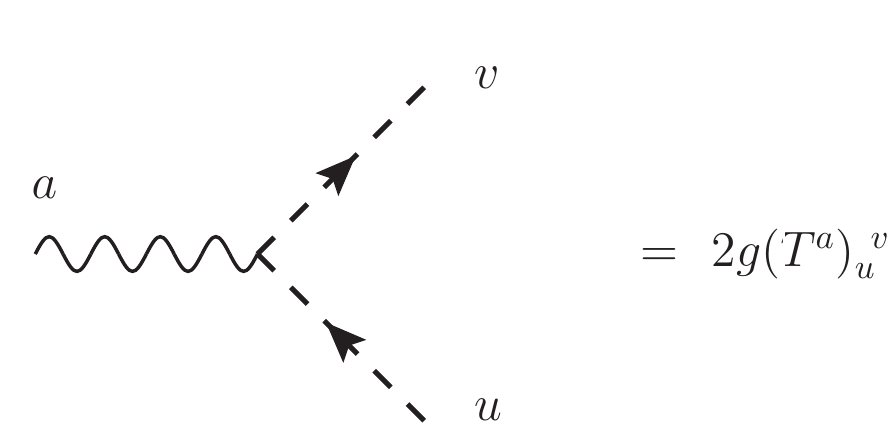}\hspace{2cm}
	\includegraphics[scale=0.6]{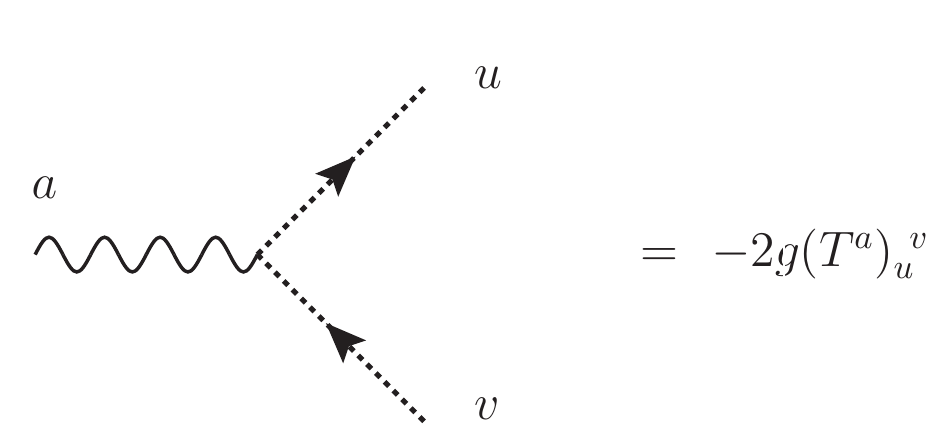} \\ \vspace{0.2cm}
	\includegraphics[scale=0.6]{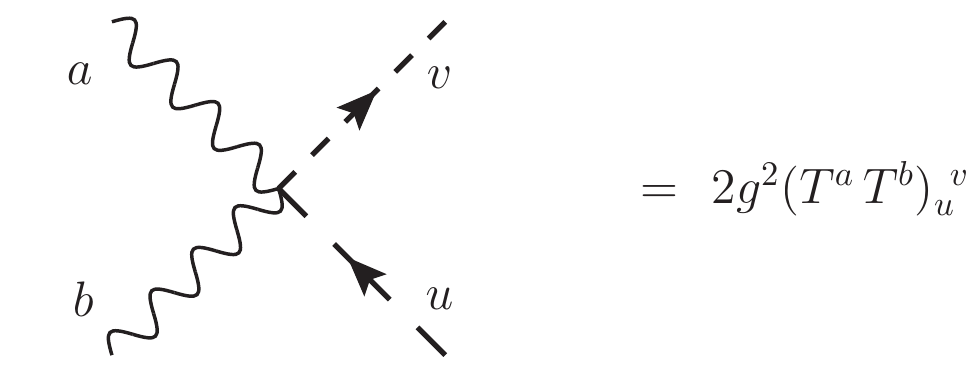}\hspace{1.5cm}
	\includegraphics[scale=0.6]{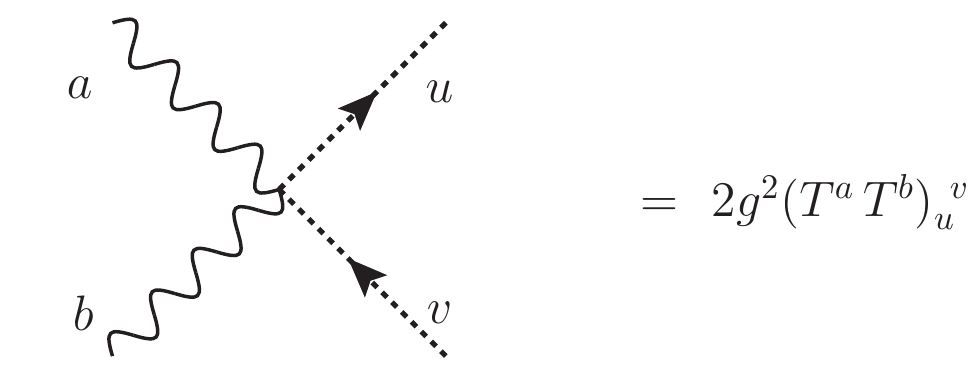}
	\caption{Feynman rules involving the matter superfields that are relevant
	for our calculations.}
	\label{fig:Feynmatter}
\end{center}
\end{figure}

\noindent
We write the action for the matter part:
\begin{align}\label{Smatter1}
S_{\mathrm{matter}} =& \!\int\!d^4x\,d^2\theta\,d^2\bar{\theta}\, \left( Q^\dagger e^{2gV} Q + \widetilde{Q} e^{-2gV}\widetilde{Q}^\dagger\right)+ \left(\ii \sqrt{2}g\!\int\!d^4x\,d^2\theta\,\widetilde{Q} \Phi Q +\mathrm{h.c.}\right)
\end{align}
We use a compact notation to specify the representation: $Q$ and $\tilde{Q}$ have  an index $u=1,\dots ,\mathrm{dim}_\cR$, which includes also the case $\cR$ is reducible, in particular when it contains several copies of an irreducible representation\footnote{For instance, if $\cR$ is the direct sum of $N_F$ fundamental representations, here $u=1,\dots ,N_F N$}. Expanding at the second order in $V$:
\begin{align}
	\label{Smatter}
		S_{\mathrm{matter}}& =
		\!\int\!d^4x\,d^2\theta\,d^2\bar{\theta}\,\Big(Q^{\dagger\,u} Q_{u}
		+ 2g\,Q^{\dagger\,u} V^a (T^a)_{u}^{\,v}\,Q_{v}
		+  2g^2\,Q^{\dagger\,u}
		V^a\,V^b(T^a \,T^b)_{u}^{\,v} \, Q_{v}\notag\\
		&\qquad\qquad+ \widetilde{Q}^{u}\,\widetilde{Q}^\dagger_{u} 
		- 2g\,\widetilde{Q}^{u} \,V^a (T^a)_{u}^{\,v}\,\widetilde{Q}^\dagger_{v}
		+2g^2\,\widetilde{Q}^{u} V^a \,V^b (T^a\,T^b)_{u}^{\,v}
		\,\widetilde{Q}^\dagger_{v}+\cdots\notag\\
		&\qquad\qquad
		+ \ii\sqrt{2}g\,\widetilde{Q}^{u}\Phi^a (T^a)_{u}^{\, v} Q_{v}\,\bar{\theta}^2
		- \ii\sqrt{2}g\,Q^{\dagger\,u}\Phi^{\dagger\,a}(T^a)_{u}^{\,v} 
		\widetilde{Q}^\dagger_{v }\,\theta^2
		\Big)
\end{align}
The Feynman rules derived from this action are illustrated in Figure \ref{fig:Feynmatter}.

By expanding the superfields in terms of the superspace variables $\theta,\bar{\theta}$ and integrating over them one can recover the standard $\cN=2$ action in components.

\subsection{Difference theory}\label{subsec::difference}
We discuss a convenient diagrammatic method to produce perturbative computations for theories with extended supersymmetry.
This method has been applied in several contexts (see for example \cite{Andree:2010na,Pomoni:2011jj,Pomoni:2013poa,Fiol:2015spa,Mitev:2015oty,Billo:2017glv,Gomez:2018usu}). The idea is that there exists a specific set of observables (like those we are considering in this thesis) which are in common with the $\cN=4$ theory, and that can be studied in the \emph{difference} between $\cN=2$ and $\cN=4$. This procedure allows to isolate the "pure" $\cN=2$ correction and thus significantly reduces the number of Feynman diagrams to be studied. We see the $\cN=4$ theory as a pure YM $\cN=2$ theory with the addition of a hypermultiplet in the adjoint representation of the gauge group. Indeed the unique $\cN=4$ vector multiplet can be decomposed as:
\begin{align}\label{N4fields}
\mathrm{Vector}_{(\cN=4)} = \begin{cases}
\mathrm{Vector}_{(\cN=2)} &= \big(V, \Phi \big) ~~\mathrm{adj~of~SU}(N)  \\
\mathrm{Hyper}_{(\cN=2)} &= \big(H, \widetilde{H} \big)~~\mathrm{adj~of~SU}(N)~.
\end{cases}
\end{align}
The hypermultiplet components have an adjoint index $H_a$, $\tilde{H}_a$, $a= 1,\ldots N^2-1$,
and their action $S_H$ has the same structure as $S_{\mathrm{matter}}$  with $Q_u$ and $\widetilde Q^u$ replaced by $H_a$ and $\widetilde H_a$ and the generator components $(T_a)_u^{\, v}$ by the structure constants $\ii f_{abc}$.
Thus we can write
\begin{equation}
	\label{S4}
		S_{\cN=4}= S_{\mathrm{YM}} + S_H~.
\end{equation}
Doing the same substitutions on the Feynman rules of Figure \ref{fig:Feynmatter} yields the Feynman rules for the $H$ and $\widetilde{H}$ superfields.

{From} (\ref{Stot}) and (\ref{S4}) it is easy to realize that the total action of our $\cN=2$ theory can be written as
\begin{equation}
S_{\cN=2} = S_{\cN=4} - S_H + S_{\mathrm{matter}}~.
\end{equation}
Actually, given any observable $\cA$ of the $\cN=2$ theory, which also exists in the
$\cN=4$ theory, we can write
\begin{equation}
\label{difference}
\Delta\cA= \mathcal{A}_{\cN=2} - \mathcal{A}_{\cN=4} = \mathcal{A}_{\mathrm{matter}} - \mathcal{A}_H~.
\end{equation}
Thus, if we just compute the difference with respect to the $\cN=4$ result, we have to consider only 
diagrams where the hypermultiplet fields, either of the $Q$, $\widetilde Q$ type or of the $H$, $\widetilde{H}$ type, propagate in the internal lines, and then consider the difference between the $(Q,\widetilde{Q})$ and the $(H,\widetilde{H})$ diagrams. Since the difference between the actions $ S_{\mathrm{matter}}$ and $S_H$ lies in the representation of the gauge group only, we expect the $Q$ and $H$ diagrams to have the same spacetime contributions. As we will see performing some specific computations this is precisely what happens in the conformal case. We show how this method works with a simple example.

\subsubsection*{Example: one-loop scalar propagator}
The tree level propagator for the adjoint scalar field $\varphi$ of
the vector multiplet can be extracted from the propagator of the superfield $\Phi$ given
in the first line of Figure~\ref{fig:Feyngauge} by imposing $\theta_1 = \theta_2 =0$:
\begin{equation}
	\label{prop-tree}
		\Delta^{bc}_{(0)}(q)= \frac{\delta^{bc}}{q^2}~.
\end{equation}
Since we consider conformal  $\cN=2$ theories, the quantum corrected propagator will 
depend on the momentum only through the factor $1/q^2$, 
and by gauge symmetry it can only be proportional to $\delta^{bc}$. So we will have
\begin{equation}
	\label{propPi}
		\Delta^{bc}(q)= \frac{\delta^{bc}}{q^2}\,\big(1 + \Pi\big)
\end{equation}
where $\Pi$ is a $g$-dependent constant describing the effect of the perturbative 
corrections. 
 
\begin{figure}[ht]
	\begin{center}
		\includegraphics[scale=0.45]{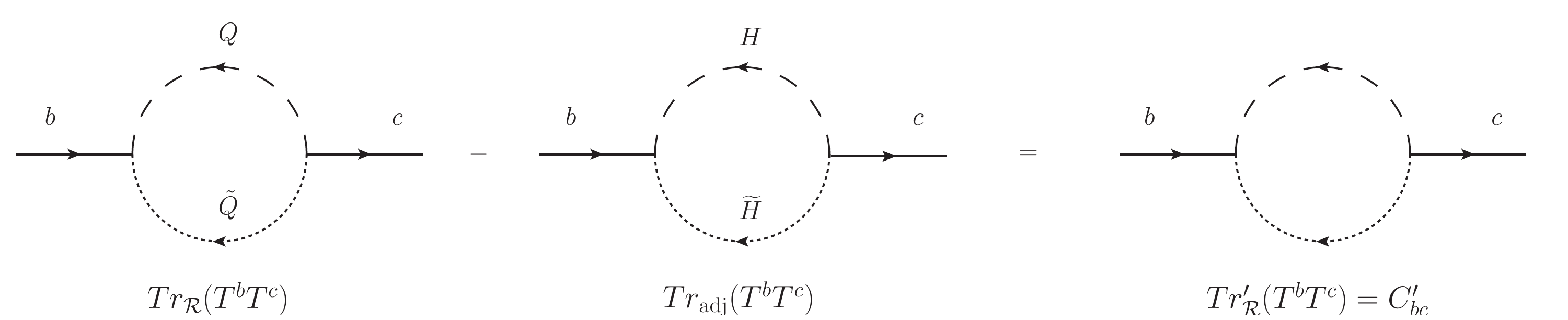}
	\end{center}
	\caption{One-loop correction to $\Phi$ propagator in the difference theory. }
	\label{fig:1loop}
\end{figure}

At order $g^2$ the first diagram we have to consider is
\begin{equation}
	\label{1loopQ}
		\parbox[c]{.4\textwidth}{\includegraphics[width 
		= .4\textwidth]{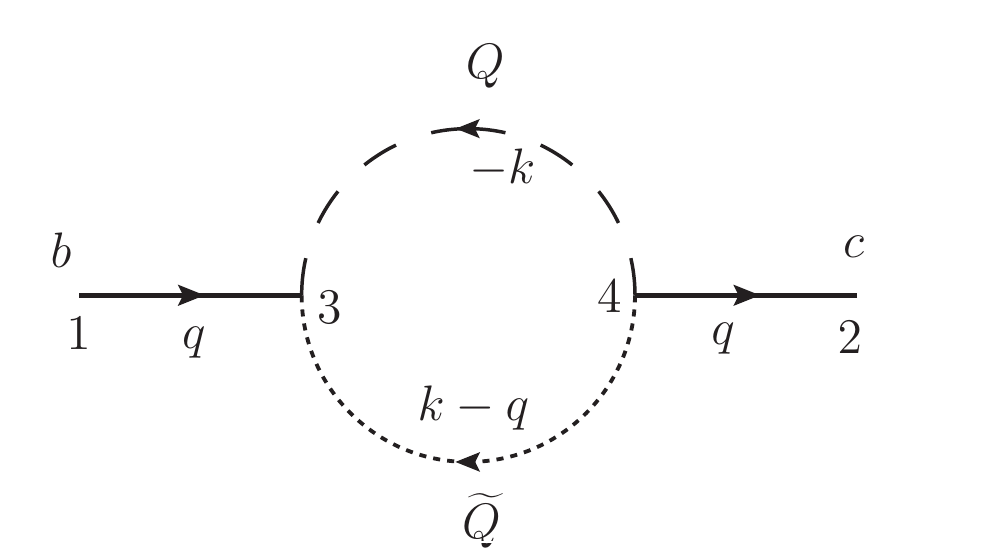}}
		\hspace{-0.6cm}= 2 g^2 \times \Tr_\cR(T^b T^c)\, \times
		\int \frac{d^dk}{(2\pi)^d} \frac{1}{(q^2)^2}\frac{1}{k^2(k-q)^2}\, \cZ(k,q)~.
\end{equation}
Here, and in all following diagrams, we adopt the notation explained in detail in 
Appendix~\ref{app:diagrams} (see in particular (\ref{gen-diag}) and the following 
sentences): we write the diagram as the product of three pieces:
\begin{itemize}
\item
a normalization factor, $2 g^2$ 
in this case, which takes into account the combinatorical factor and the strength 
of the vertices;
\item
the color factor;
\item
integral over the internal momenta, where the factor $\cZ(k,q)$ is the result of the integration over the Grassmann variables at each 
internal vertex\,%
\footnote{The Grassmann variables in the external points 1 and 2 are set to zero to pick up 
the lowest component $\varphi$ of the superfield, namely we have 
$\theta_1=\bar\theta_2=0$. 
Note that if we do not do this and consider the propagator of the full superfield $\Phi$ the 
color factor remains the same.} and, according to the rules in Figures \ref{fig:Feyngauge} 
and \ref{fig:Feynmatter} reads
\begin{align}
	\label{Z1loopQ}
		\cZ(k,q) = \int d^4\theta_3 \,d^4\theta_4\, (\theta_3)^2 ({\bar\theta}_4)^2\,
		\exp{\big(\!-2 \,\theta_4 \,q\, \bar{\theta}_3\big)} = - q^2~.
\end{align}
\end{itemize}
The momentum integral in (\ref{1loopQ}) is divergent for $d\to 4$; however in the 
difference theory we have to subtract an identical diagram in which the adjoint superfields 
$H$ and $\widetilde H$ run in the loop instead of $Q$ and $\widetilde Q$. This diagram 
has the same expression except for the color factor which is now given by 
$\Tr_\mathrm{adj}(T^b T^c)$. The difference of the two diagrams is therefore proportional 
to 
\begin{align}
	\label{Trprbc}
		\Tr_\cR(T^b T^c) - \Tr_\mathrm{adj}(T^b T^c) = 	\Tr_\cR^\prime(T^b T^c) 
		= C^\prime_{bc}~. 
\end{align}
{From} now on, we will use the graphical notation introduced in Figure~\ref{fig:1loop}, 
according to which a hypermultiplet loop stands for the difference between 
the $(Q,\widetilde Q)$ and the $(H,\widetilde H)$ diagrams, with a color factor 
that is directly given by a primed trace.

In particular, if we evaluate the color factor \eqref{Trprbc} using the conventions of Appendix \ref{app:group}:
\begin{equation}
	\label{C2is} 
		C^\prime_{b_1 b_2} = \left(i_{\cR} - i_{\mathrm{adj}}\right) \delta_{b_1 b_2} = 
		\left(i_{\cR} - N\right) \delta_{b_1 b_2} = - \frac{\beta_0}{2}\, \delta_{b_1 b_2}~,
\end{equation}
where $i_{\cR}$ is the index of the representation $\cR$ and $\beta_0$ the one-loop coefficient of the $\beta$-function of the corresponding $\cN=2$ gauge theory. 
In superconformal theories, one has $\beta_0=0$.  
Thus the constant $\Pi$ in (\ref{propPi}) starts at order $g^4$ and all diagrams including 
the one-loop correction to the $\Phi$ propagator as a sub-diagram vanish.  

\section{Supersymmetric theories on curved spaces}\label{sec2:susycurved}

We consider the problem of placing supersymmetric field theories on a non-trivial manifold. The relevance of this problem has old origins \cite{Witten:1988ze}, and in recent years has encountered a renowned appeal, since it leads to the application of supersymmetric localization approaches. We briefly review the work of Festuccia-Seiberg \cite{Festuccia:2011ws}, who built a very general procedure. Then we see how this procedure has been successfully applied to two specific cases, the four-dimensional sphere and ellipsoid, which we will use in the following.

\subsection{Overview of the formalism}
We consider a generic supersymmetric theory on a flat-space and our intention is to uplift it to a supersymmetric theory on a curved space.
Placing supersymmetric field theories on a manifold  with a curved metric generally breaks all flat-space supercharge. If we simply minimally couple the theory to the metric $g_{\m\n}$, we get a curved-space supercharge $Q$ for each covariantly constant spinor $\zeta_\a$:
\begin{equation}\label{constzeta}
\nabla_\m \zeta_\a=0~,
\end{equation}
where $\z$ is the SUSY parameter associated to the supercharge $Q$.\\
This equation is very restrictive, there are not so many compact four-manifolds that admit covariantly constant spinors. However, the condition \eqref{constzeta} can be relaxed explicitly using the presence of supersymmetry. \\
In general, starting from the flat space $g_{\m\n}=\delta_{\m\n}$ and defining a metric deformation $\Delta g_{\m\n}$, we add to the Lagrangian a term coupled to $\Delta g_{\m\n}$ through the stress tensor:
\begin{equation}\label{Lcoupled}
g_{\m\n}=\delta_{\m\n}+\Delta g_{\m\n}~, \hspace{1.cm} \Delta\cL=-\frac{1}{2}\Delta g^{\m\n}T_{\m\n} + O(\Delta g^2)~.
\end{equation}
The additional term $\Delta \cL$ does not preserve supersymmetry because the stress tensor is not a BPS operator, but in a supersymmetric theory $T_{\m\n}$ is part of a supermultiplet, which contains other bosonic and fermionic operators, generically denoted as $\cJ^i_B$ and $\cJ^i_F$. At the same time, $g_{\m\n}$ resides in a supergravity multiplet as well, together with bosonic and fermionic fields $\cB^i_B$ and $\cB^i_F$. As explained in  \cite{Festuccia:2011ws}, the condition to be imposed is that the gravity multiplet must be set off-shell and acts as a \emph{non-dynamical background}. At this point the bosonic stress tensor superpartners $\cJ^i_B$ can be coupled to corresponding bosonic background fields $\cB^i_B$ and added to the Lagrangian \eqref{Lcoupled}:
\begin{equation}
\Delta\cL=-\frac{1}{2}\Delta g^{\m\n}T_{\m\n} + \sum_i \cB^i_B\cJ^i_B ~.
\end{equation}
This construction is viewed as a rigid limit of dynamical off-shell supergravity. The requirement that $\Delta\cL$ should preserve a supercharge $Q$ corresponds to the equation:
\begin{equation}\label{deltaQ}
\delta_Q \cB^i_F =0
\end{equation} 
These equations are composed by some non-trivial bosonic expressions involving $g_{\m\n}$, $\cB^i_B$ and the spinor $\zeta_\a$ corresponding to the supercharge $Q$. They determine the allowed configurations for the bosonic background fields and the spinor parameter $\z$.  The described procedure is still pretty general (we will provide some explicit examples in the following), but already at this level it allows some useful considerations. 
\begin{itemize}
\item
First of all, the fermionic sources $\cB^i_F$ always include the gravitino $\Psi_{\m\a}$ (also in a minimal $\cN=1$ theory), whose supersymmetry variation takes the form
\begin{equation}\label{KSE}
\delta_Q \Psi_{\m\a} = \nabla_\m \zeta_\a +\dots
\end{equation}
which therefore generalizes \eqref{constzeta}, involving (in the $\dots$ part) specific constraints on the bosonic part of the supergravity multiplet. For this reason the equations \eqref{KSE} are also known as generalized Killing spinor equations.
\item
Secondly, a rigid supersymmetric invariant theory is characterized by a full set of bosonic background fields, namely specifying the metric does not determine the background in a unique way. This fact can be seen by some arbitrariness in the solutions of the Killing spinor equations (we will see some specific examples about this fact).
\item
Finally, we stress that all this construction does not depend on the specific content of the theory. If we restrict to cases with a Lagrangian description in terms of fields, the transformation rules for the fields directly follow from the corresponding off-shell supergravity rules.
\end{itemize}

\subsection{$\cN=2$ superconformal symmetry on Euclidean manifolds}
We now specify to the case of theories with extended supersymmetry which are also conformally invariant and can be studied by coupling the theory to conformal supergravity. The approach follows the general procedure we described before, with some generalization which we try to stress. The equations to be satisfied in this case are the conformal Killing spinor equations \cite{Klare:2012gn}, which are a generalization of \eqref{deltaQ}. Then, the higher level of supersymmetry increases the number of constraints. In particular in the gravity supermultiplet we find another dynamical fermion (the dilatino), in addition to the gravitino. Its supersymmetry variation leads to additional differential equations involving the various background fields.
We outline the construction of a generic $\cN=2$ superconformal invariant theory on a curved space with Euclidean signature, following \cite{Klare:2013dka} and \cite{Hama:2012bg}.

The fields of the conformal supergravity multiplet, also called Weyl multiplet, are (see for example \cite{Freedman:2012zz})
\begin{equation}
	\label{offsm}
		g_{\mu\nu}~,\quad
		\psi^{\cI}_\mu~,\quad
		\sT_{\mu\nu}~,\quad 
		\bsT_{\mu\nu}~,\quad
		\widetilde{M}~,\quad
		\eta^{\cI}~,\quad
		V_{\mu}^0~,\quad
		(V_{\mu})^{\cI}_{\cJ}~,
\end{equation}
where $g_{\mu\nu}$ is the metric, $\psi^{\cI}_\mu$ (with $\cI=1,2$  SU$(2)$ R symmetry index) is the gravitino, 
$\sT_{\mu\nu}$ and $\bsT_{\mu\nu}$ are, respectively, real self-dual and anti self-dual tensors\,%
\footnote{Do not confuse $\sT_{\mu\nu}$, written in an upright font, with the stress-energy tensor $T_{\mu\nu}$.}, $\widetilde{M}$ is a scalar field, $\eta^{\cI}$ is the dilatino, and finally 
$V_{\mu}^0$ and $(V_{\mu})^{\cI}_{\cJ}$ 
are the gauge fields of the  SO$(1,1)_R\,\times\,$SU$(2)_R$ R-symmetry. 

Superconformal symmetry is characterized by two pairs of Killing spinors $(\zeta_{\a\cI},\bar{\zeta}^{\dot{\a}}_{\cI})$ and $(\zeta'_{\a\cI},\bar{\zeta'}^{\dot{\a}}_{\cI})$ associated to $Q$ and $S$ fermionic operators respectively. On a Euclidean manifold, they are required to satisfy a reality condition, \textit{i.e.} they are symplectic Majorana-Weyl spinors:
\begin{equation}
(\zeta_{\a\cI})\dagger= \epsilon^{\a\b}\epsilon^{\cI\cJ}\zeta_{\b\cJ} = \zeta^{\a\cI}~, \hspace{1.5cm}(\bar{\zeta}_{\dot{\a}\cI})\dagger= \epsilon^{\dot{\a}\dot{\b}}\epsilon^{\cI\cJ}\bar{\zeta}_{\dot{\b}\cJ} = \bar{\zeta}^{\dot{\a}\cI}~,
\end{equation}
and the same for $\zeta'_{\cI}$.\\
The condition for preserving supersymmetry is then
\begin{equation}\label{susyinv}
\delta\, \psi^{\cI}_\mu = 0~, \hspace{1.5cm} \delta\, \eta^{\cI}=0~,
\end{equation}
which translates into two sets of equations, associated to the variation of the gravitino and the dilatino. The gravitino equations read (we drop the spinor indices):
\begin{align}\label{gravitino}
\nabla_\mu^{(V)} \zeta_{\cI}+\sT^{\rho\lambda}\,\sigma_{\rho\lambda}\,\sigma_\m \, \bar{\zeta}_{\cI}&= -\ii \sigma_\m\, \bar{\zeta}'_{\cI}~,  \notag \\
\nabla_\mu^{(V)} \bar{\zeta}_{\cI}+\bsT^{\rho\lambda\,}\bs_{\rho\l}\,\bs_\m \, \zeta_{\cI} &= -\ii \bs_\m\, \zeta'_{\cI}~,
\end{align}
while the dilatino equations:
\begin{align}\label{dilatino}
\s^\m\,\bs^\n \nabla_\mu^{(V)}\nabla_\nu^{(V)}\zeta_{\cI} + 4 \nabla_\rho^{(V)} \sT_{\m\n}\,\s^{\m\n}\,\s^\rho \,\bar{\zeta}_{\cI}&=\left(\widetilde{M}-\frac{R}{3}\right)\,\zeta_{\cI}~, \notag \\
\bs^\m\,\s^\n \nabla_\mu^{(V)}\nabla_\nu^{(V)}\bar{\zeta}_{\cI} + 4 \nabla_\rho^{(V)} \bsT_{\m\n}\,\bs^{\m\n}\,\bs^\rho \,\zeta_{\cI}&=\left(\widetilde{M}-\frac{R}{3}\right)\,\bar{\zeta}_{\cI}~.
\end{align}
where $R$ is the Ricci scalar.\\
The covariant derivatives here are improved with R-symmetry gauge fields and read:
\begin{align}
\nabla_\mu^{(V)} \zeta_{\cI} &= \partial_\m \zeta_{\cI}+\frac{1}{4} \Omega^{mn}_\m \s_{mn}\, \zeta_{\cI} - \ii\,\zeta_{\cI}V_\m^0 + \ii\,\zeta_{\cJ}V_{\m~\cI}^{\cJ}~, \notag \\
\nabla_\mu^{(V)} \bar{\zeta}_{\cI} &=\partial_\m \bar{\zeta}_{\cI}+\frac{1}{4} \Omega^{mn}_\m \bs_{mn}\,\bar{\zeta}_{\cI} -\ii\, \bar{\zeta}_{\cI}V_\m^0 + \ii\,\bar{\zeta}_{\cJ}V_{\m~\cI}^{\cJ}~.
\end{align}
Solving this equations in terms of the background fields of the gravity supermultiplet allows to find the conditions to define a superconformal theory on a generic curved space. This procedure has been pursued in \cite{Klare:2013dka} following a general method, which limits the possible choices of $g_{\m\n}$. The only constraints on the metric $g_{\m\n}$ come from the gravitino equations and turn out to be the existence of a conformal Killing vector. The dilatino equations do not impose conditions on the geometry, but fix the background values of the supersymmetric partners, up to some arbitrariness. Such arbitrariness will be an important point of discussion in the following. 

At this point we have all the ingredients to discuss the construction of $\cN=2$ SCFTs on two specific backgrounds, the four dimensional sphere and ellipsoid, that we are going to use in the following.

\section{Superconformal theories on ellipsoids}\label{sec2:ellips}
We build $\mathcal{N}=2$ SYM theories on four-dimensional ellipsoids preserving rigid supersymmetry. We follow the analysis of \cite{Hama:2012bg,Hosomichi:2015jta}, whose conventions we largely adopt.
First we briefly review the four sphere subcase, in order to ease the generalization to the ellipsoid.

\subsection{Four Sphere}
The problem of putting a four-dimensional theory with extended supersymmetry was addressed by Pestun in a seminal paper \cite{Pestun:2007rz}. The goal was to build a Lagrangian on a compact manifold in order to perform supersymmetric localization. We will briefly review Pestun's result in the next Section. \\
The sphere case is particularly simple, since the Killing spinor equations have an explicit solution without turning on  any of the background fields of the supergravity multiplet. Therefore here the only main modification for the Lagrangian with respect to the flat case is the coupling to the non-trivial metric. Given the metric of a four sphere  $S^4$ with radius $r$, described by polar coordinates $\xi^\m=(\rho,\theta,\varphi,\chi)$ and specified by the following vielbein one-forms \footnote{The index $m=1,\dots,4$ runs over the flat directions.} $E^m$:
\begin{equation}
E^1=r \sin \rho \cos\theta d\phi~, \quad E^2= r \sin\rho\sin\theta d\chi~, \quad E^3=r \sin \rho d\theta~, \quad E^4=r d\rho  ~,
\end{equation}
where $\rho\in [0,\pi]$, $\theta\in [0,\pi/2]$, $\varphi\in [0,2\pi]$ and $\chi\in [0,2\pi]$.
Since we want the theory to preserve conformal invariance, the only variation with respect to the flat space case is the addition of the
$R\,\bar{\phi}\phi$-term to the scalar kinetic term
\begin{equation}\label{confcouple}
D_{\mu}\bar \phi D^{\mu} \phi \longrightarrow \Big(D_{\mu}\bar \phi D^{\mu} \phi+ \frac{R}{6}\,\bar\phi \phi\Big)~,
\end{equation}
in such a way that the scalar fields of the vector multiplet are conformally coupled to the sphere metric. The same happens for the scalar fields of hypermultiplets.\\
The problem of solving the Killing spinor equations has been addressed by \cite{Hama:2012bg}. After switching all $\sT_{\mu\nu}~, \bsT_{\mu\nu}~,\widetilde{M}~,V_{\mu}^0~,(V_{\mu})^{\cI}_{\cJ}$ off in equations \eqref{gravitino} and \eqref{dilatino}, it is possible to explicitly solve the four dimensional Killing spinor equations by relating them to the three dimensional case \cite{Hama:2011ea}. We report here their result:
\begin{align}\label{KSsolutions}
\zeta_{\cI}&= (\zeta_1,\zeta_2)= \left( \sin \frac{\rho}{2}\cdot \kappa_+, \sin \frac{\rho}{2}\cdot \kappa_-  \right)    \notag \\
\bar{\zeta}_{\cI}&= (\bar\zeta_1,\bar\zeta_2) = \ii \left( \cos \frac{\rho}{2}\cdot \kappa_+,- \cos \frac{\rho}{2}\cdot \kappa_-  \right)     ~,
\end{align}
where $\kappa_\pm$ are the three-dim solutions.
\begin{align}
\kappa_\pm = \frac{1}{2} \begin{pmatrix}
~~e^{\frac{i}{2}(\pm \varphi\pm\chi-\theta)}\\
\mp e^{\frac{i}{2}(\pm \varphi\pm\chi+\theta)}
\end{pmatrix}
\end{align}
The square of the supersymmetry transformation from these Killing spinor solutions gives rise to the Killing vector:
\begin{align}
v^\m\partial_\m = 2 \bar{\zeta}^\cI \bs^\m \zeta_\cI \partial_\m = \frac{1}{r} (\partial_\varphi+\partial_\chi)
\end{align} 
This solution is important since the strategy of \cite{Hama:2012bg} to perform a similar analysis for the ellipsoid is to impose that the Killing spinors \eqref{KSsolutions} are still Killing spinors on the ellipsoid.

\subsection{The ellipsoid geometry}
\label{subsecn:ellips}

A four-dimensional ellipsoid can be defined as the surface in $\mathbb{R}^5$ described 
by the equation
\begin{equation}
	\label{defellipsoid}
		\frac{x_1^2+x_2^2}{\ell^2}+\frac{x_3^2+x_4^2}{\widetilde{\ell}^{\,2}}+\frac{x_5^2}{r^2}=1~.
\end{equation}
When $\ell=\widetilde\ell=r\equiv \mathsf{r}$, the ellipsoid becomes a round sphere $S^4$ of radius $\mathsf{r}$.
It is convenient to introduce the squashing parameter
\begin{equation}
b=\sqrt{\frac{\,\ell\,}{\,\widetilde{\ell}\,}}~,
\label{b}
\end{equation}
\iffalse
and use the following parametrization 
\begin{equation}
\label{parametrization}
\ell=l(b)\,b~,\quad \widetilde{\ell}=\frac{l(b)}{b}~,\quad
r=r(b)~,
\end{equation}
where $l(b)$ and $r(b)$ are such that $l(1)=r(1)=\mathsf{r}$. In this way, 
the limit $b\to 1$ corresponds to the sphere limit.
\fi
Again we adopt polar coordinates like in the sphere case, such that
\begin{equation}
\begin{aligned}
	\label{coords}
		x_1&=\ell\, \sin \rho \cos \theta \cos \varphi~,\\
		x_2&=\ell\, \sin \rho \cos \theta \sin \varphi~,\\
		x_3&=\widetilde{\ell}\,\sin \rho \sin \theta \cos \chi~,\\
		x_4&=\widetilde{\ell}\, \sin \rho \sin \theta \sin \chi~,\\
		x_5&=r\, \cos \rho~,
\end{aligned}
\end{equation}
 We denote the polar coordinates as $\xi^\mu$, to distinguish them from the $\mathbb{R}^5$ coordinates $x_M$. 

The ellipsoid metric $g_{\mu\nu}$ is simply given by the pullback of the flat Euclidean metric of the embedding space $\mathbb{R}^5$, namely
\begin{equation}
g_{\mu\nu}= \frac{\partial x_M}{\partial \xi^\mu}\,\frac{\partial x_N}{\partial \xi^\nu}\,\delta^{MN}
\label{metric}
\end{equation}
In our coordinate system, this metric is not diagonal and the corresponding vierbein are 
\begin{align}
	\label{vierbein}
		E^1 &= \ell\, \sin \rho \cos \theta\, d\varphi~,\quad\quad
		E^2 = \widetilde{\ell}\,\sin \rho \sin \theta\, d\chi~,\notag \\ 
		E^3 &= f_1 \,\sin \rho\, d\theta+f_3\, d\rho~,\quad\quad
		E^4 = f_2\, d\rho~,
\end{align}
where we defined three functions \cite{Hama:2012bg}
\begin{align}
	\label{fhg}
		f_1 &=\sqrt{\ell^2\, \sin^2 \theta+\widetilde{\ell}^{\,2}\, \cos^2 \theta}~,\quad
		f_2 = \sqrt{r^2\,\sin^2 \rho+ \frac{\ell^2\,\widetilde{\ell}^{\,2}}{f^2}\, \cos^2 \rho}~,\notag \\
		f_3 &= \frac{\widetilde{\ell}^{\,2}-\ell^2}{f}\,\cos \rho\, \sin \theta\, \cos \theta~.
\end{align}
It is easy to see that $f_1\to \mathsf{r}$, $f_2\to \mathsf{r}$ and $f_3\to 0$ when $b\to 1$. Notice that since the polar coordinates $\xi^\mu$ are dimensionless, the metric $g_{\mu\nu}$ carries 
dimensions of $\mathrm{(length)}^2$; however, for the conformal invariant 
theories which we will consider, these dimensions can always be scaled away.

\subsection{Supersymmetric Lagrangians}
\label{subsecn:susy}
We explicitly show how to couple the $\cN=2$ Lagrangian to the non-dynamical supergravity multiplet., following the procedure described before.

The action for a $\mathcal{N}=2$ SYM theory on an ellipsoid with squashing parameter $b$ has been
derived in \cite{Hama:2012bg} and is given by
\begin{equation}
\label{Sb}
S_b= \frac{1}{g^2}\,\int \!d^4\xi \,\sqrt{\det g}\,\left(L_{\tmb{YM}}+L_{\text{matter}}\right)
\end{equation}
The first term, ${L}_{\tmb{YM}}$, accounts for the couplings of the gauge vector multiplet, which comprises the gauge connection $A_\mu$, the gaugino $\lambda_\cI$ and its conjugate
$\bar\lambda_\cI$, the scalar fields
$\phi$ and $\bar{\phi}$, and the auxiliary field $\cD_{\cI\cJ}$ -- all in the adjoint of the 
gauge group $G$. 
The explicit expression of $L_{\tmb{YM}}$ is
\begin{equation}
\begin{aligned}
	\label{LYM}
 		L_{\tmb{YM}}=\tr\bigg[&\frac12 F^{\mu\nu} F_{\mu\nu}+ 16 F_{\mu\nu}(\bar \phi \mathsf{T}^{\mu\nu}+\phi \bar{\mathsf{T}}^{\mu\nu})+ 64\, \bar\phi^2 \mathsf{T}^{\mu\nu} \mathsf{T}_{\mu\nu}+ 64\, \phi^2 \bar{\mathsf{T}}^{\mu\nu} \bar{\mathsf{T}}_{\mu\nu}\\
 		&-4 D_{\mu}\bar \phi D^{\mu} \phi
 		+2 \Big(\widetilde{M}-\frac{R}{3}\Big) \bar \phi \phi -2 \ii \lambda^{\mathcal{I}}\s^{\mu} D_{\mu} \bar \lambda_{\mathcal{I}}- 2 \lambda^{\mathcal{I}} [\bar \phi,\lambda_{\mathcal{I}}]\\&+ 2 \bar \lambda^{\mathcal{I}}[\phi,\bar \lambda_{\mathcal{I}}]+4 [\phi, \bar \phi]^2
 		-\frac{1}{2}\cD^{\cI\cJ}\cD_{\cI\cJ}\bigg]
\end{aligned}
\end{equation}
where $R$ is the Ricci scalar associated to the ellipsoid metric $g_{\mu\nu}$. 
Our conventions for the traces and the spinors are explained in Appendix~\ref{app:Notations}. Here we simply recall that the sum over repeated indices $\cI$ involves an $\epsilon$-tensor. For
example
\begin{equation}
\lambda^\cI \lambda_\cI = \epsilon^{\cI\cJ}\lambda_\cJ\lambda_\cI
\end{equation}
with $\epsilon^{12}=1$.

A few comments are in order. Following \cite{Hosomichi:2016flq}, we have written the coefficient of the $\bar\phi \phi$-term as twice $\big(\widetilde{M}-\frac{R}{3}\big)$. This combination is equivalent to the field $M$ used in \cite{Hama:2012bg}, but for our purposes it is more convenient to distinguish the contribution due the background field $\widetilde{M}$ from the one due to the curvature. 
Indeed, if we add the
$R\,\bar{\phi}\phi$-term to the scalar kinetic term, we obtain 
\begin{equation}
-4\tr\Big(D_{\mu}\bar \phi D^{\mu} \phi+ \frac{R}{6}\,\bar\phi \phi\Big)~,
\end{equation}
namely the same combination as \eqref{confcouple}, in order to have conformally coupled scalars.
The coefficient of $1/6$ in front of the curvature shows that the scalar fields of the vector multiplet are conformally coupled to the ellipsoid metric. We also note that the $\mathrm{SU}(2)_R$ connection 
$(V_{\mu})^{\cI}_{\cJ}$ does not appear explicitly in the Lagrangian, but only 
through the covariant derivative of the gaugino, which is defined as 
\begin{align}
	\label{gaugcovder}
		D_{\mu} \bar \lambda^{\dot \a}_{\mathcal{I}} 
		=\pa_{\mu} \bar \lambda^{\dot \a}_{\mathcal{I}}-\ii [A_{\mu}, \bar \lambda^{\dot \a}_{\mathcal{I}}]+ \frac14 \omega_{\mu}^{mn} (\bar \s_{mn})^{\dot \a}{}_{\dot \b} \bar \lambda^{\dot \b}_{\mathcal{I}}+ \ii \bar \lambda^{\dot \a}_{ \mathcal{J}} (V_{\mu})^{\mathcal{J}}{}_{\mathcal{I}}~,		
\end{align}
where $\omega_{\mu}^{mn}$ is the spin-connection, and similarly for the left-handed components.
Note that the gauge field $V_{\mu}^0$ has been set to zero, as in \cite{Hama:2012bg}. We discuss this choice at the end of this Subsection.

The matter part of the Lagrangian, $L_{\text{matter}}$, accounts for the couplings of $\cN=2$ 
hypermultiplets transforming in a (generically reducible) representation $\mathcal{R}$ of the gauge group. The number of these hypermultiplets is clearly equal to the dimension of $\mathcal{R}$, which we denote simply by $d_r=\mathrm{dim} \cR$. If the index $i_{\mathcal{R}}$ of the matter representation equals that of the adjoint, then the resulting $\cN=2$ SYM theory is conformal, see \eqref{C2is}.
If we denote the scalar fields of the hypermultiplets by $q_{\cI\cA}$ and their fermionic
partners by $\psi_{\cA}$ and $\bar\psi_\cA$, with $\cA=1,\ldots,2d_r$ being an index of Sp($d_r$),
the matter Lagrangian takes the form \footnote{Following \cite{Hama:2012bg} we are using a compact notation, which is different from Section \ref{sec:FTactions}. The map is the following: the hypermultiplet scalars $(q, \bar q, \tilde q,\tilde{ \bar q})$ with an index $u=1,\dots,d_r$ are mapped in the present Section to $q_{\cI\cA}$, $\cI=1,2$ and $\cA=1,\ldots,2d_r$.}
\begin{equation}
\begin{aligned}
	\label{Lhm} 	 
	 	L_{\text{matter}}\!&=\frac12 D_{\mu} q^\cI D^{\mu} q_\cI- q^\cI\{\phi, \bar \phi\} q_\cI
	 	-\frac18 q^\cI q_\cI q_\cJ q^\cJ+\frac18 \Big(\widetilde{M}-\frac{2}{3}R\Big)q^\cI q_\cI -\frac{\ii}{2} \bar \psi \bar \sigma^{\mu} D_{\mu} \psi\\
 		&~~ -\frac12 \psi \phi \psi +\frac12 \bar \psi \bar \phi \bar \psi+\frac{\ii}{2}\psi \s^{\mu\nu} \mathsf{T}_{\mu\nu} \psi -\frac{\ii}{2}
 		\bar \psi \bar \s^{\mu\nu} \bar{\mathsf{T}}_{\mu\nu} \bar \psi- q^\cI \lambda_\cI \psi+ \bar \psi \bar \lambda_\cI q^\cI ~.
\end{aligned}
\end{equation}
Here the sum over the Sp($r$) indices has been understood. If one wants to write it explicitly, one
has for example
\begin{align}
 q^{\mathcal{I}}  q_{\mathcal{I}}= \Omega^{\mathcal{A}\mathcal{B}}\,
 q^{\mathcal{I}}_{~\mathcal{B}}\,q_{\mathcal{I}\mathcal{A}} ~,
\end{align}
where $\Omega^{\mathcal{A}\mathcal{B}}$ is the real anti-symmetric invariant tensor of Sp($d_r$).
Notice that the matter fields are coupled to the vector multiplet through an embedding of the gauge group into $\mathrm{Sp}(d_r)$ and that, as before, 
the $\mathrm{SU}(2)_R$ connection appears only in the covariant derivatives defined by
\begin{align}
	\label{covderpsi}
		D_{\mu} q_{\mathcal{I} \mathcal{A}}&=\pa_{\mu} q_{\mathcal{I} \mathcal{A}}-\ii (A_{\mu})_{\mathcal{A}}{}^{\mathcal{B}} q_{\mathcal{I} \mathcal{B}}+ \ii q_{\mathcal{J}\mathcal{A}} (V_{\mu})^{\mathcal{J}}{}_{\mathcal{I}}~.
\end{align}
Again, in the Lagrangian (\ref{Lhm}) we have replaced the scalar $M$ appearing in \cite{Hama:2012bg} with $\big(\widetilde{M}-\frac{R}{3}\big)$ in order to disentangle the contribution due to
the curvature from that due to the scalar field of the supergravity multiplet. And again, combining the
$R\,q^\cI q_\cI$-term with the kinetic terms we obtain
\begin{equation}
\frac12 \Big(D_{\mu} q^\cI D^{\mu} q_\cI+\frac{R}{6}\,q^\cI q_\cI\Big)
\end{equation}
which shows that also the scalar fields of the matter hypermultiplets are conformally coupled to
the curvature of the ellipsoid.

The action $S_b$ in (\ref{Sb}) is invariant under the $\cN=2$ supersymmetry transformations 
of the gauge and matter fields given in Appendix \ref{app:SUSYtransf} provided the supergravity
background is carefully chosen. In particular, the metric $g_{\mu\nu}$ must be that of the ellipsoid
as in (\ref{metric}), while $\sT_{\mu\nu}$, $\bar{\sT}_{\mu\nu}$, $\widetilde{M}$ 
and $(V_{\mu})^{\cI}_{\cJ}$ must assume background values determined by solving 
the Killing spinor equations that ensure the vanishing of the supersymmetry transformations of the gravitino and dilatino. The expressions for the ellipsoid Killing spinors are the same as the sphere case \eqref{KSsolutions} (this is an assumption of \cite{Hama:2012bg}). 
Under this assumption the Killing vector on the ellipsoid becomes:
\begin{align}
v^\m\partial_\m = 2 \bar{\zeta}^\cI \bs^\m \zeta_\cI \partial_\m = \frac{1}{\ell}\partial_\varphi+\frac{1}{\widetilde{\ell}}\partial_\chi~.
\end{align} 
The explicit background values of the supergravity multiplet depend on the geometric properties of the ellipsoid, and in particular on the squashing parameter $b$. We recall them in the following Subsection.
As already mentioned, the SO$(1,1)_R$ connection $V_{\mu}^0$ can be consistently set to zero, since the Killing spinor equations determine the background geometry up to some residual degrees of freedom. This choice pursued in \cite{Hama:2012bg} is justified also by the necessity of reproducing the so-called $\Omega$-background \cite{Nekrasov:2002qd} at the North and South poles of the ellipsoid and is allowed by a residual symmetry from the 
supersymmetry conditions, as widely explained in \cite{Klare:2013dka}. 

\subsection{Supergravity background}
\label{subsecn:background}
The Killing spinor equations provide specific geometric constraints that allow to fix the 
profile of the background fields, although not uniquely. In \cite{Hama:2012bg}  it was found
that these fields are given by\footnote{To be precise \cite{Hama:2012bg} contains the explicit expression of $M$, not $\widetilde{M}$. To obtain the latter, one can simply use the relation $\widetilde{M}=M+\frac{R}{3}$ and the Ricci curvature associated to the metric \eqref{metric}, $R=3\left(\frac{1}{(f_2)^2}+\frac{r^2}{(f_1)^2(f_2)^2}\right)$.\\ A second remark is that here $\zeta_{\a\cI}$ can be seen as a $2\times 2$ matrix: when it acts on the left it saturates the SU$(2)_R$ indices, when it acts on the right is saturates the spinor indices.}
\begingroup
\allowdisplaybreaks
\begin{subequations}
\label{HHsol}
\begin{align}	
	\widetilde M & =\frac{1}{f_1^2}+\frac{f_3^2+r^2}{f_1^2f_2^2}-\frac{4}{f_1f_2}
		+ \Delta \widetilde M~,\label{Msol}\\[2mm]
		\mathsf{T}_{\a}{}^{\b} & = \frac14\Big( \frac{1}{f_1}-\frac{1}{f_2}\Big) (\t_{\theta}^1)_{\a}{}^{\b} 
		+\frac{f_3}{4 f_1f_2} (\t_{\theta}^2)_{\a}{}^{\b} + \Delta \mathsf{T}_{\a}{}^{\b}~,\label{Tsol}\\[2mm]
		\bar{\mathsf{T}}^{\dot \a}{}_{\dot \b} & = \frac14\Big( \frac{1}{f_1}-\frac{1}{f_2}\Big) 
		(\t_{\theta}^1)^{\dot \a}{}_{\dot \b} -\frac{f_3}{4 f_1 f_2} (\t_{\theta}^2)^{\dot \a}{}_{\dot \b} 
		+ \Delta \bar{\mathsf{T}}^{\dot \a}{}_{\dot \b}~,\label{barTsol}\\[2mm]
(\zeta \cdot \tilde{V}_1)_{\a \cI} &= \left\{\frac{\cos\theta}{2\sin\rho} \bigg(\frac{1}{f_1}-\frac{1}{f_2}\bigg)-\frac{\sin\theta \cos\rho}{2\sin\rho}\frac{f_3}{f_1f_2}   \right\}(\tau^1_\theta \cdot \zeta)_{\a \cI}\notag \\&\quad+\frac{\sin\theta\cos\rho}{2f_1\sin\rho}\bigg(1-\frac{\widetilde{\ell}^2}{f_1f_2}\bigg)(\tau^2_\theta \cdot \zeta)_{\a \cI}+(\zeta \cdot \Delta\tilde{V}_1)_{\a \cI}~,\label{V1sol}\\[2mm]
(\zeta \cdot \tilde{V}_2)_{\a \cI} &= \left\{\frac{\sin\theta}{2\sin\rho} \bigg(\frac{1}{f_1}-\frac{1}{f_2}\bigg)+\frac{\cos\theta \cos\rho}{2\sin\rho}\frac{f_3}{f_1f_2}   \right\}(\tau^1_\theta \cdot \zeta)_{\a \cI}\notag \\&\quad-\frac{\cos\theta\cos\rho}{2f_1\sin\rho}\bigg(1-\frac{\ell^2}{f_1f_2}\bigg)(\tau^2_\theta \cdot \zeta)_{\a \cI}+(\zeta \cdot \Delta\tilde{V}_2)_{\a \cI}~,\label{V2sol}\\[2mm]
(\zeta \cdot \tilde{V}_3)_{\a \cI} &= -\frac{\cos\rho}{2f_1\sin\rho} \bigg(1-\frac{\ell^2\widetilde\ell^2}{f_1^3f_2}\bigg)(\tau^3\cdot \zeta)_{\a \cI}+(\zeta \cdot \Delta\tilde{V}_3)_{\a \cI}~,\label{V3sol}\\[2mm]
(\zeta \cdot \tilde{V}_4)_{\a \cI} &= \frac{f_3\cos\rho}{2f_1f_2\sin\rho} \bigg(1-\frac{\ell^2\widetilde\ell^2}{f_1^3f_2}\bigg)(\tau^3\cdot \zeta)_{\a \cI}+(\zeta \cdot \Delta\tilde{V}_4)_{\a \cI}~,\label{V4sol}
		\end{align}
\end{subequations}
\endgroup
where the functions $f_1$, $f_2$ and $f_3$ are defined in (\ref{fhg}), while the matrices $\t^i_\theta$ are
\begin{align}
	\label{tthetamat}
		\t^i_\theta = \t^i \,\begin{pmatrix} \rme^{+\ii \theta} &0  \\ 0& \rme^{-\ii \theta} 
		\end{pmatrix} ~,
\end{align}
with $\t^i$ being the usual Pauli matrices. Moreover we introduced the combination
\begin{align}
(\tilde{V}_m)^\cI_\cJ= E^\m_m(V_\m)^\cI_\cJ + V^{[3]}\,(\tau^3)^\cI_\cJ~, 
\end{align}
where $V^{[3]}$ is the three-dimensional solution \cite{Hama:2011ea}
\begin{align}
V^{[3]}=\frac{1}{2}\bigg(1-	\frac{\ell}{f_1}\bigg)d\varphi +\frac{1}{2}\bigg(1-\frac{\widetilde\ell}{f_1}\bigg)d\chi~.
\end{align}
The self-dual and anti self-dual
tensors $\mathsf{T}_{\mu\nu}$ and $\bar{\mathsf{T}}_{\mu\nu}$ are related to the matrices 
$\mathsf{T}_{\a}{}^{\b} $ and $\bar{\mathsf{T}}^{\dot \a}{}_{\dot \b} $ in 
(\ref{Tsol}) and (\ref{barTsol}) according to
\begin{align}
	\label{Tab}
		\mathsf{T}_{\a}{}^{\b} = -\ii\left(\sigma^{\mu\nu}\right)_{\a}{}^{\b}\, \mathsf{T}_{\mu\nu}~,~~~
		\bar{\mathsf{T}}^{\dot \a}{}_{\dot \b} =-\ii \left(\bar{\sigma}^{\mu\nu}\right)^{\dot \a}{}_{\dot \b} \,\bar{\mathsf{T}}_{\mu\nu}~.
\end{align}
Finally, in each line of (\ref{HHsol}) the last contribution, indicated with a $\Delta$, depends 
on three arbitrary functions $c_1$, $c_2$ and $c_3$, which parameterize the ambiguity of the background solution. In fact we have \cite{Hama:2012bg}
\begin{equation}
\begin{aligned}
	\label{DeltaM}
		\D \widetilde{M}&= 8\,\Big(\frac{1}{f_2} \pa_{\rho}- \frac{f_3}{f_1f_2 \sin \rho} \pa_{\theta} + \frac{\ell^2 \tilde \ell^2 \cos \rho}{(f_1)^4 f_2 \sin \rho}+ \frac{(\ell^2+\tilde{\ell}^2-(f_1)^2)\cos \rho}{f_2 (f_1)^2 \sin \rho}- \frac{\cos \rho}{f_1 \sin \rho} \Big) c_1\\[1mm]
		&~~+8\,\Big(\frac{1}{f_1\sin \rho}\pa_\theta+ \frac{\ell^2 \tilde{\ell}^2 f_3\,\cos \rho}{(f_2)^2 (f_1)^4 \sin \rho} + \frac{2 \cot 2\theta}{f_1 \sin \rho}- \frac{f_3 \cos \rho}{f_1 f_2 \sin \rho} \Big) c_2 -16 \sum_i c_i^2~,
\end{aligned}
\end{equation}
and
\begingroup
\allowdisplaybreaks
\begin{subequations}\label{Deltas}
\begin{align}
		\D  \mathsf{T}_{\a}{}^{\b} &= \tan \frac{\rho}{2} \,\Big(c_1  (\t_{\theta}^1)_{\a}{}^{\b}+c_2  (\t_{\theta}^2)_{\a}{}^{\b}+c_3  (\t^3)_{\a}{}^{\b} \Big)~,\\[2mm]
		\D  \bar{\mathsf{T}}^{\dot \a}{}_{\dot \b} &= \cot \frac{\rho}{2} \,\Big(\!-c_1  (\t_{\theta}^1)^{\dot \a}{}_{\dot \b}+c_2  (\t_{\theta}^2)^{\dot \a}{}_{\dot \b}+c_3  (\t^3)^{\dot \a}{}_{\dot \b}\Big)~,\\[2mm]
		(\zeta \cdot \Delta\tilde{V}_1)_{\a \cI}&= -2\sin\theta\, \Big(c_2 (\t^1_\theta\cdot\zeta)_{\a\cI}-c_1 (\t^2_\theta\cdot\zeta)_{\a\cI}\Big)~,\\[2mm]
		(\zeta \cdot \Delta\tilde{V}_2)_{\a \cI}&= 2\cos\theta\, \Big(c_2 (\t^1_\theta\cdot\zeta)_{\a\cI}-c_2 (\t^2_\theta\cdot\zeta)_{\a\cI}\Big)~,\\[2mm]
		(\zeta \cdot \Delta\tilde{V}_3)_{\a \cI}&= -2 c_1 (\t^3_\theta\cdot\zeta)_{\a\cI}+2c_3 (\t^1_\theta\cdot\zeta)_{\a\cI}~,\\[2mm]
		(\zeta \cdot \Delta\tilde{V}_4)_{\a \cI}&= 2 c_2 (\t^3_\theta\cdot\zeta)_{\a\cI}-2c_3 (\t^2_\theta\cdot\zeta)_{\a\cI}~.
\end{align}
\end{subequations}
\endgroup
It is easy to check that in the sphere limit when $b\to 1$, all non $\Delta$-terms in (\ref{HHsol})
vanish. Therefore, since on the sphere the only surviving background field
is the metric, we must require that also $\Delta\widetilde{M}$, $\Delta \mathsf{T}$,
$\Delta \bar{\mathsf{T}}$ and $\zeta \cdot \Delta\tilde{V}_m$ vanish when $b=1$. In turn this requirement implies the $c_i$'s must be
zero at $b=1$, {\it{i.e.}} they must have the following form
\begin{align}
c_i= c'_i (b-1) + {O}\big((b-1)^2\big)~.
\label{ci}
\end{align}
The three arbitrary functions $c_i$ will not affect the computation of physical quantities, as we will show in Chapter \ref{chap:6}.

\section{Supersymmetric localization}\label{sec2:localization}
Supersymmetric localization has become a really powerful tool in recent years to compute exact quantities in QFT. Indeed the presence of extended supersymmetry is used to prove that the path integral only receives contribution from the locus of certain fixed points. This technique represents the supersymmetric extension of the equivariant localization formula for ordinary integrals with bosonic symmetries \cite{Duistermaat:1982vw,Witten:1988ze,Blau:1995rs}. We briefly introduce this technique, for more detailed reviews see for example \cite{Marino:2011nm, Cremonesi:2014dva} and the collection of recent achievements in localization techniques \cite{Pestun:2016zxk}.

Consider a Lagrangian theory in a Euclidean spacetime described by an action $S$, in presence of a global fermionic symmetry $\cQ$, such that its square either vanishes or yields a bosonic symmetry $\delta_B$ of the action. We deform the path integral in the following way:
\begin{equation}\label{Sloc}
Z(t) = \int [\cD \Phi] e^{-S[\Phi]-t \,\cQ V[\Phi]}~,
\end{equation}
where $t$ is a real parameter and $V$ is some functional such that $\delta_B V=0$. If the measure is $\cQ$-invariant, then $Z$ does not depend on $t$:
\begin{equation}
\frac{\partial Z}{\partial t} = -  \int [\cD \Phi]~\cQ V~ e^{-S-t \,\cQ V}= -\int [\cD \Phi]~\cQ~\Big( V e^{-S-t \,\cQ V}\Big) = 0~.
\end{equation}
So $\cQ$ acts as a total derivative. The same argument can be repeated after the insertion of $\cQ$ invariant operators $\cO[\Phi]$ inside the path integral \eqref{Sloc}. Therefore any correlation function inside \eqref{Sloc} does not depend on the parameter $t$, but only on the $\cQ$-cohomology class of the operator. \\
If we take a functional $V$ such that the bosonic part of $\cQ V \geq 0$, then the limit $t \to \infty$ selects only the field configurations for which $\cQ V $ is suppressed. Then the path integral \emph{localizes} to the bosonic zeroes $\Phi_0$ of $\cQ V $. If we parametrize the fields around $\Phi_0$ as
\begin{equation}
\Phi = \Phi_0+t^{-1/2} \delta \Phi~,
\end{equation} 
we can expand the action around $\Phi_0$:
\begin{equation}
S+t \,\cQ V = S[\Phi_0]+ \frac{1}{2}\frac{\delta^2 (\cQ V)}{\delta\Phi^2}\Bigg|_{\Phi_0} \delta\Phi^2+(t^{-1/2})~.
\end{equation}
This result is ``one-loop exact", since higher orders in the functional Taylor expansions are weighted by negative powers of $t$.\\
After a Gaussian integration over the fluctuations $\delta\Phi$, the resulting partition function becomes an integral over the localization locus $\cM$, defined by field configurations $\cQ V=0$:
\begin{equation}
Z= \int_M \cD \Phi_0 e^{-S[\Phi_0]}Z_{1-loop}[\Phi_0]~,
\end{equation}
where the $Z_{1-loop}$ term is the ratio of the determinants of the operators appearing at quadratic order in the bosonic and fermionic fluctuations:
\begin{equation}
Z_{1-loop}[\Phi_0]= \frac{1}{\mathrm{SDet}\left[\frac{\delta^2 (\cQ V)}{\delta\Phi^2}\Big|_{\Phi_0} \right]}
\end{equation}
This is an exact formula, and if the space $\cM$ is finite-dimensional we have been able to reduce an ordinary path integral to an ordinary integral. In particular, a canonical choice for the functional $V$ is:
\begin{equation}
V= \sum_{\mathrm{fermions \Psi}} (\cQ \Psi)^\dagger \Psi+\Psi^\dagger (\cQ\Psi^\dagger)^\dagger~,
\end{equation}
so that the bosonic part of $\cQ V$ is a sum of squares of supersymmetry variations and it satisfies the previous conditions. The localization locus is then defined by
\begin{equation}\label{locallocus}
\cQ \Psi =\cQ \Psi^\dagger=0~.
\end{equation}
The crucial point of this procedure is the computation of the 1-loop determinant, but this is doable thanks to cancellations due to supersymmetry.

\subsection{Matrix model on the four sphere}\label{subsec2:MMsphere}
Localization techniques have been exploited to compute exactly certain observables in $\cN=2$ SYM theories. In particular, in a seminal paper \cite{Pestun:2007rz}, Pestun was able to evaluate the partition function on a 4-sphere $S^4$ and the vacuum expectation value of BPS Wilson loops. We are going to review the final result after the localization procedure.\\
We consider $\cN=2$ SYM theories with gauge group SU($N$) and matter hypermultiplets transforming in a generic representation $\cR$. According to the localization principle, the only non-vanishing contributions to the path integral arise from the localization locus \eqref{locallocus}, which in this case is defined by the following saddle points:
\begin{align}
\label{saddlesphere}
	A_\mu=0~, ~~~ \varphi=\bar \varphi=\,\frac{a}{\sqrt{2}}~, 
\end{align}
where $a$ is a $N\times N$ constant Hermitean matrix taking values in the $\mathfrak{su}(N)$ Lie Algebra.

The partition function on a 4-sphere $S^4$ with unit radius\,\footnote{The dependence on 
the radius $R$ can be trivially recovered by replacing $a$ with $R\,a$.}, can be expressed as follows:
\begin{equation}
	\cZ_{S^4}=\int \prod_{u=1}^N da_u \,\,\Delta(a) \,
	\big| Z(\ii a,g)\big|^2\,\delta\Big(\sum_{u=1}^N a_u\Big)
	\label{ZS4}
\end{equation}
where $a_u$ ($u=1,\ldots,N$) are the real eigenvalues of $a$, $\Delta$ is the
Vandermonde determinant
\begin{equation}
	\Delta(a)=\prod_{u<v=1}^N (a_u-a_v)^2~,
	\label{vandermonde}
\end{equation}
and $Z(\ii a,g)$ is the partition function for a gauge theory with coupling $g$ defined 
on $\mathbb{R}^4$ with $a$ parametrizing the Coulomb branch. 

Before considering $Z(\ii a,g)$ in more detail, let us remark that the integration
over the eigenvalues $a_u$ in (\ref{ZS4}) can be rewritten simply as the integral over all 
components of the Hermitean traceless matrix $a$, namely
\begin{equation}
	\label{intda}
		\cZ_{S^4} = \int da\,\big| Z(\ii a,g)\big|^2~. 
\end{equation}
The matrix $a$ can be decomposed over a basis of generators $t_a$ of $\mathfrak{su}(N)$:
\begin{equation}
	\label{aont}
	a = a^b \,t_b~,~~~b = 1,\ldots, N^2-1~;
\end{equation}
we will normalize these generators so that the index of the fundamental representation equals $1/2$:
\begin{equation}
	\label{normtatb}
	\tr \,t_a t_b = \frac 12\, \delta_{ab}~.
\end{equation}
In Appendix \ref{app:group} we collect our group theory conventions and other useful 
formulas. The integration measure is then simply proportional to ${\prod_b da^b}$.\\
The $\mathbb{R}^4$ partition function $Z(\ii a,g)$ can be written as
\begin{equation}
	Z= Z_{\mathrm{tree}}\,Z_{\mathrm{1-loop}}\,
	Z_{\mathrm{inst}}~.
\end{equation}
$Z_{\mathrm{inst}}$ is the Nekrasov's instanton partition function \cite{Nekrasov:2002qd,Nekrasov:2003rj} of the gauge theory in the $\Omega$ deformation of $\mathbb{R}^4$.
In perturbation theory, we can neglect the instanton contributions and put 
$Z_{\mathrm{inst}}=1$. Throughout the present thesis we will not care about the instanton part.\\ The tree-level term is given by
\begin{equation}
\big|Z_{\mathrm{tree}}\big|^2 = \rme^{-\frac{8\pi^2}{g^2}
	\,{\mathrm{tr}\,a^2}}~,
\end{equation} 
providing a free matrix model with a Gaussian term. 
The 1-loop part contains interaction terms, which we write as follows: 
\begin{equation}
	\label{Z1ltoS}
	\big|Z_{\mathrm{1-loop}}\big|^2 \equiv \rme^{-\widehat{S}(a)}~.
\end{equation}
The matrix model corresponding to the $\cN=4$ SYM theory has $\widehat{S}(a)=0$ and is 
purely Gaussian.
For $\cN=2$ SYM theories, instead, there are interaction terms.
In general, let us denote by $\mathbf{a}$ the $N$-dimensional vector of 
components $a_u$, and by $W(\cR)$ the set of the weights $\mathbf{w}$ of the 
representation $\cR$ and by
$W(\mathrm{adj})$ is the set of weights of the adjoint representation. Then, 
\begin{equation}
	\big|Z_{\mathrm{1-loop}}\big|^2 =
	\frac{\prod_{\mathbf{w}\in W(\mathrm{adj}) } H(\ii \mathbf{w}\cdot\mathbf{a})}
	{\prod_{\mathbf{w}\in W(\mathcal{R})} H(\ii \mathbf{w}\cdot\mathbf{a})})
	~,
	\label{Z1loop}
\end{equation} 
where
\begin{equation}
H(x)=G(1+x)\,G(1-x)
\end{equation}
and $G$ is the Barnes $G$-function. 

\subsubsection{The interaction action}
\label{subsec:intSa}
Let us now consider the interaction action $\widehat{S}(a)$. {From} (\ref{Z1ltoS}) it follows that 
\begin{align}
	\label{StologH}
	\widehat{S}(a) & =\!  \sum_{\mathbf{w}\in W(\mathcal{R})}\!\! \log H(\ii \mathbf{w}\cdot
	\mathbf{a})
	~~- \sum_{\mathbf{w}\in W(\mathrm{adj})}\!\! \log H(\ii \mathbf{w}\cdot\mathbf{a}) 
	\notag\\[1mm]
	&= \Tr_{\cR} \log H(\ii a) - \Tr_{\mathrm{adj}} \log H(\ii a)  \,=\, \Tr_{\cR}^\prime \log 
	H(\ii a)~, 
\end{align}
where in the last step  we introduced the notation   
\begin{align}
	\label{deftrp}
	\Tr_{\cR}^\prime \bullet \,= \,\Tr_{\cR} \bullet - \Tr_{\mathrm{adj}} \bullet~.
\end{align}
This indeed vanishes for the $\cN=4$ SYM theory, where the representation $\cR$
of the hypermultiplets is the adjoint. For $\cN=2$ models, this combination of 
traces is non-vanishing and precisely accounts for the matter content of the ``difference 
theory'' which we reviewed in Subsection \ref{subsec::difference} and is often used in field theory computations. The power of the form \eqref{StologH} of the matrix model  is precisely that it can be used as a guideline for the field theory side.

Using the properties of the Barnes $G$-function, one can prove that
\begin{equation}
	\label{logHexp}
		\log H(x)=-(1+\gamma_{\mathrm{E}})\,x^2-\sum_{n=1}^\infty
		\frac{\zeta(2n+1)}{n+1}\,
		x^{2n+2}
\end{equation}
where $\zeta(n)$ are the Riemann $\zeta$-values. Then, we can rewrite (\ref{StologH}) 
as follows 
\begin{equation}
	\label{Saexp}
		\widehat{S}(a) = (1+\gamma_{\mathrm{E}})\trp a^2 + \sum_{n=2}^\infty(-1)^{n}
		\frac{\zeta(2n-1)}{n} \trp a^{2n}~.
\end{equation} 
With the rescaling 
\begin{equation}
	\label{rescaling}
		a\to \sqrt{\frac{g^2}{8\pi^2}}~a~,
\end{equation}
we bring the partition function on $S^4$ to the form 
\begin{equation}
	\label{ZS4resc}
		\cZ_{S^4} =  \Big(\frac{g^2}{8\pi^2}\Big)^{\frac{N^2-1}{2}}
		\int da~\rme^{-\mathrm{tr}\, a^2 - S(a)}~,
\end{equation}
where
\begin{align}
	\label{Sint}
		S(a)&=\trp \log H\Big(\ii\,\sqrt{\frac{g^2}{8\pi^2}}\,a\Big)\nonumber\\
		&=\frac{g^2}{8\pi^2}\,(1+\gamma_{\mathrm{E}})\,\trp a^{2}
		-\left(\frac{g^2}{8\pi^2}\right)^2\frac{\zeta(3)}{2}\,\trp a^{4}
		+\left(\frac{g^2}{8\pi^2}\right)^3\frac{\zeta(5)}{3}\,\trp a^{6}
		+\ldots
\end{align}
The overall $g$-dependent pre-factor in (\ref{ZS4resc}) is irrelevant in computing 
matrix model correlators, and thus can be discarded. Using the expansion (\ref{aont}),
the traces appearing in $S(a)$ can be expressed as 
\begin{equation}
	\label{tra2ntoC}
		\trp a^{2k} = C^\prime_{(b_1\ldots b_k)} \, a^{b_1}\ldots a^{b_k}~,	
\end{equation}
where
\begin{equation}
	\label{defC}
		C^\prime_{b_1\ldots b_n} = \trp T_{b_1}\ldots T_{b_n}~.
\end{equation}
These tensors are cyclic by definition. Note that we have already encountered the combination \eqref{defC} in the computation of the color factor of the one-loop correction of the scalar propagator (see Subsection \ref{subsec::difference}). This represents a clear example of the deep relation between matrix model and the field theory computations. And since we found $C^\prime_{b_1 b_2} \propto \beta_0\, \delta_{b_1 b_2}$, this implies that $\trp a^{2}=0$ in superconformal models,
so the interaction action $S(a)$ starts at order $g^4$, {\it{i.e.}} at two loops.

Therefore, localization technique allows to map a $\cN=2$ theory on to a matrix model, which has the explicit form \eqref{ZS4resc}. We will see in the following how this partition function can be used to compute BPS observables of the theory.

\subsection{Localization result on the ellipsoid}\label{subsec2:MMellips}
It is possible to extend Pestun's result to a similar computation for superconformal field theories on the ellipsoid. The result is a new matrix model where new kinds of observables can be evaluated. Again we simply want to report the explicit form of this matrix model, the reader should refer to Section 4 of \cite{Hama:2012bg} for further details on the derivation. 

Similarly as the sphere case, according to the localization principle, the only non-vanishing contributions to the path integral (\ref{Sb}) arise from the following saddle point values of the fields \footnote{The normalization for the vector multiplet scalars in \eqref{LYM} induces a different value for $\phi$ and $\bar\phi$ at the saddle point locus with respect to the sphere case \eqref{saddlesphere}. This is simply a matter of conventions, in this case we want to be consistent with \cite{Hama:2012bg}}
\begin{align}
\label{saddle}
	A_\mu=0~, ~~~ \phi=\bar{\phi}=-\frac{\ii}{2}\,a_0~, ~~~\cD_{\cI\cJ}=-\ii \,w_{\cI\cJ}\,a_0~,
\end{align}
where again $a_0$ is a $N\times N$ matrix taking values in the $\mathfrak{su}(N)$ Lie Algebra. The explicit expression of $w_{\cI\cJ}$ can be found in \cite{Hama:2012bg}.
The classical action (\ref{Sb}) coming from the pure Yang -Mills part \eqref{LYM} on the saddle point becomes 
\begin{align}
	\label{clsad}
		\cS_b=\frac{8\pi^2}{g^2} \ell \tilde\ell \,\tr a_0^2~,
\end{align} 
The path integral measure appearing in the partition function and in any other expectation value, reduces to the integration over the matrix $a_0$.
Besides the Gaussian factor arising from $\rme^{-\cS_b}$, the integrand comprises also 
a one-loop determinant, that accounts for the fluctuations around the saddle point, 
and a non-perturbative instanton part. Both of these terms turn out to depend only 
on the ellipsoid scales $\ell$ and $\tilde\ell$ appearing in (\ref{defellipsoid}) and not on $r$. Moreover, the product $\ell\tilde{\ell}$ and the matrix $a_0$ always occur together in the combination 
\begin{equation}
\hat{a_0}=\sqrt{\ell\tilde\ell}\, a_0~,
\end{equation}
as can see also from the explicit form of the 1-loop determinant:
\begin{equation}\label{1loopellips}
\big|Z_{\mathrm{1-loop}}(\hat a_0)\big|^2 =
	\prod_{\mathbf{w}\in W(\mathrm{adj})} \Upsilon(\ii \mathbf{w}\cdot\mathbf{\hat a_0})\Upsilon(-\ii \mathbf{w}\cdot\mathbf{\hat a_0})
	\prod_{\mathbf{w}\in W(\mathcal{R})} \Upsilon\left(\ii \mathbf{w}\cdot\mathbf{\hat a_0}+\frac{b+b^{-1}}{2}\right)^{-1}~.
\end{equation}
Here the $\Upsilon$ are special functions were introduced by \cite{Zamolodchikov:1995aa} (see equation (3.10) of that paper for the explicit integral representation and special properties). The most important property of the $\Upsilon$ functions is that they are invariant under $b\rightarrow b^{-1}$. The non-perturbative term $\big|Z_{\mathrm{inst}}\big|^2$ preserves this property, since it accounts for the instanton contribution from north pole and the anti-instanton from south pole.

One can thus eliminate entirely the dependence on the product $\ell\tilde\ell$ by changing the integration variable from $a_0$ to the matrix\,%
\footnote{
Note that the overall minus sign in (\ref{defa}) is irrelevant.} 
\begin{align}
	\label{defa}
		a = - \sqrt{\ell\tilde{\ell}} \,\sqrt{\frac{8\pi^2}{g^2}}\, a_0~.
\end{align}
We prefer to rescale $a_0$ also with a factor of $\sqrt{{8\pi^2}/{g^2}}$ so that the classical action $\cS_b$
becomes simply $\tr a^2$. This leads to a Gaussian term $\exp(-\tr a^2)$ in the matrix model integrand, while the one-loop determinant and the instanton factor get organized, respectively,
into a perturbative and a non-perturbative expansion in $g$, in the same way we did in the sphere case.
The overall constant factors arising from the Jacobian for this change of variable cancel out in all properly normalized expectation values between the integral in the numerator and the partition function in the denominator. Therefore we can write the ellipsoid partition function as:
\begin{align}
	\label{partb}
		\cZ_b = \int \! da~\rme^{-\tr a^2} \,\big|\cZ^\text{1-loop}_b\big|^2
		\, \big|\cZ^\text{inst}_b\big|^2
\end{align}
When written in terms of the matrix $a$, both the one-loop determinant and the instanton terms
only depend on the squashing parameter $b = \sqrt{\ell/\tilde \ell}$, and for $b=1$ they 
reduce to the expressions obtained on the sphere in \cite{Pestun:2007rz}.
Moreover, exploiting the properties outlined before, \eqref{partb} is symmetric in 
the exchange $b\rightarrow b^{-1}$, namely the partition function does not depend on $b$ at first order:
\begin{align}
	\label{derZb1}
		\partial_b \cZ_b\,\Big|_{b=1} = 0~.
\end{align}
This property will be crucial for the computation of special observables in this matrix model, as we will see in Chapter \ref{chap:6}.

\chapter{Wilson loop vev in $\mathcal{N}=2$ SCFTs}\label{chap:3}
We introduced the supersymmetric Wilson loop in Section \ref{sec1:WL} as a fundamental observable in the AdS/CFT correspondence. The high degree of symmetry generates many simplifications already at a perturbative level, such that the $\cN=4$ 1/2 BPS Wilson loop vacuum expectation value can be computed exactly in terms of a Gaussian matrix model \cite{Erickson:2000af, Drukker:2000rr}. Such matrix model description arises naturally using supersymmetric localization techniques \cite{Pestun:2007rz}, as we reviewed in Section \ref{sec2:localization}. Localization method is valid in $\cN=2$ theories also, in which case
the resulting matrix model is not Gaussian any longer
but contains interaction terms.
This has been very useful in the study of the AdS/CFT duality in 
the $\mathcal{N}=2$ setting \cite{Rey:2010ry,Passerini:2011fe,Russo:2012ay,Buchel:2013id,Russo:2013qaa,Russo:2013kea,Russo:2013sba,Fiol:2015mrp}, 
since the interacting matrix model allows one to study the large-$N$ limit in an efficient way, 
also in the strong coupling regime. 

In the present Chapter we introduce the Wilson loop vev computation starting from a review of the $\cN=4$ case, then in Section \ref{sec:mmvev} we describe the technical procedure to compute expectation values in the $\cN=2$ matrix model introduced in Section \ref{sec2:localization}. Using this machinery, in Section \ref{sec:propWilson} we first compute the quantum correction to the ``propagator'' of  the interacting matrix model up to three loops, and then use it to obtain the leading terms of the vacuum expectation value of the 1/2 BPS circular Wilson loop in the fundamental representation. We also derive the exact expressions in $g$ and $N$ for the corrections proportional to $\zeta(3)$ and $\zeta(5)$ in this vacuum expectation value, and exploit them to study the large-$N$ limit. In Section~\ref{secn:fieldtheory} we perform a perturbative field-theory computation in the $\cN=2$ superconformal theories at order $g^8$ using the $\cN=1$ superfield formalism, following the convention of Section \ref{sec:FTactions}. By computing (super) Feynman diagrams in the ``difference theory", we show the perfect agreement with the matrix model results.

\section{From the $\mathcal{N}=4$ to the $\cN=2$ Wilson loop}\label{sec3:N4WL}
We start by reviewing the perturbative computation of a 1/2 BPS Wilson loop vev in a $\mathcal{N}=4$ theory. From its relation with the localization computation we can report some remarkable exact result that will be important in the following.

We consider a 1/2 BPS circular Wilson loop, 
placed on a circle $C$ of radius $R$ and defined as
\begin{equation}
\label{defWR}
W_{\mathcal{R}}(C)=\frac{1}{N}\Tr_{\mathcal{R}} \: \mathcal{P}
		\exp \left\{g \oint_C d\tau \Big[\ii \,A_{\mu}(x)\,\dot{x}^{\mu}(\tau)
+R\,n^{u}(\tau)\phi_u(x)\Big]\right\}
\end{equation}

For definiteness, from now on we will take the representation $\mathcal{R}$ to be the fundamental representation of SU($N$) (denoted simply by ``$\tr$'') and denote the corresponding Wilson loop simply as $W(C)$.\\
We take $n^{u}(\tau)= \delta^{u1}$, in order to induce a scalar coupling which holds for the $\cN=2$ case also \footnote{As we stressed in Section \ref{sec:FTactions}, we will perform all the field theory computations in the difference between $\cN=2$ and $\cN=4$, so it is convenient to choose the $\cN=4$ scalar coupling which remains unchanged in the $\cN=2$ case.}, and introduce the chiral and anti-chiral combinations which sits in the chiral field $\Phi$ which belongs to the $\cN=2$ vector multiplet (see \eqref{N2fields}):
\begin{equation}\label{fibarfi}
\varphi=\frac{1}{\sqrt{2}}\big(\phi_1+\ii\,\phi_2\big)~,~~
\bar\varphi=\frac{1}{\sqrt{2}}\big(\phi_1-\ii\,\phi_2\big)~,
\end{equation}
so that \eqref{defWR} becomes
\begin{equation}
	\label{WLdef}
		W(C)=\frac{1}{N}\tr \: \mathcal{P}
		\exp \left\{g \oint_C d\tau \Big[\ii \,A_{\mu}(x)\,\dot{x}^{\mu}(\tau)
		+\frac{R}{\sqrt{2}}\big(\varphi(x) + \bar\varphi(x)\big)\Big]\right\}
\end{equation}
We parametrize the loop as:
	\begin{equation}\label{circle}
		x^\mu(\tau)=R\,\big(\cos\tau,\sin\tau,0,0\,\big)
\end{equation}
with $\tau\in[\,0,2\pi\,]$. The tree-level propagators of the gauge field and of the adjoint 
scalar can be extracted from their superspace realization (see Figure \ref{fig:Feyngauge}), and in configuration space read
\begin{equation}
	\label{propsAphi}
		\big\langle \bar\varphi^a (x_1)\, \varphi^b (x_2)\big\rangle_{\mathrm{tree}}
		= \frac{\delta^{ab}}{4\pi^2 x_{12}^2}~,~~~ \big\langle
		A_{\mu}^{a} (x_1)\, A_{\nu}^{b} (x_2)\big\rangle_{\mathrm{tree}}
		= \frac{\delta^{ab} \delta_{\mu\nu}}{4\pi^2 x_{12}^2}~.
\end{equation}
They are identical, a part from the different space-time indices, since they are part of the same supermultiplet. We will widely use this property, which will lead to many simplifications in the following, starting from the notation: we will denote the sum of a scalar and a gluon propagator with a straight/wiggly line.

\begin{figure}[htb]
	\begin{center}
		\includegraphics[scale=0.4]{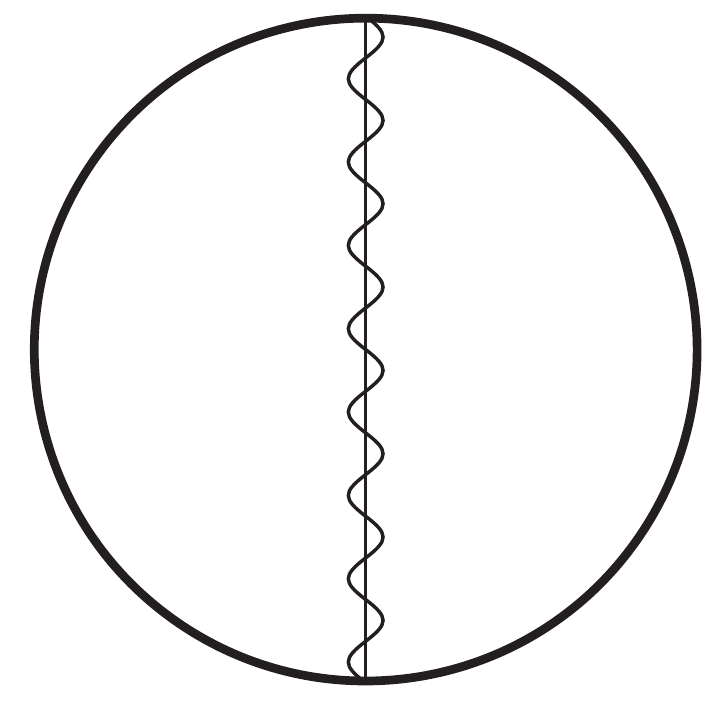} \\
		\includegraphics[scale=0.7]{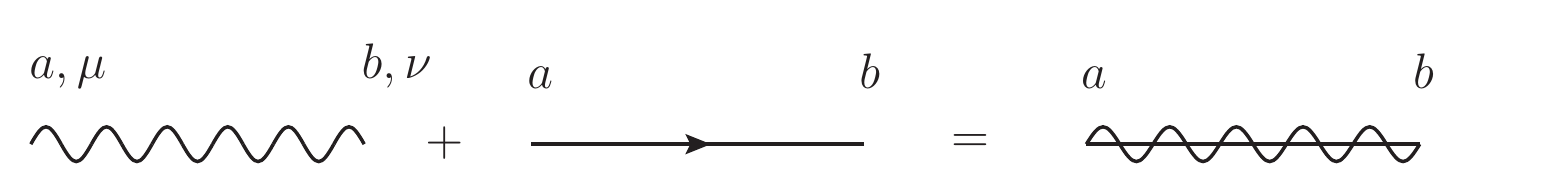}
	\end{center}
	\caption{The graphical representation of the $g^2$-correction to $\big\langle
	W(C)\big\rangle$. The wiggled/straight line stands for $V$ or $\Phi$ propagators, as explained in the second row 
	of the figure.}
	\label{fig:WLtree}
\end{figure}

Expanding \eqref{WLdef} at order $g^2$, one gets an integral over $C$ of the sum of 
the tree-level propagators of the gluon and of the scalar fields between the points 
$x(\tau_1)$ and $x(\tau_2)$. This contribution is represented in Figure \ref{fig:WLtree}.

Using (\ref{propsAphi}), one finds  
\begin{equation}
	\label{WLg2}
		\big\langle W(C)\big\rangle = 1+\frac{g^2 (N^2-1)}{4N} \oint \frac{d\tau_1 d\tau_2}{4\pi^2}~ \frac{R^2 -\dot{x}(\tau_1)\cdot \dot{x}(\tau_1)}{|x(\tau_1)-x(\tau_2)|^2} + \cO (g^4)~.
\end{equation}
Exploiting the parametrization \eqref{circle}, one can easily show that the integrand 
is $\tau$-independent; indeed
\begin{equation}
	\label{propwl}
		\frac{R^2 -\dot{x}(\tau_1)\cdot \dot{x}(\tau_1)}{4\pi^2 |x(\tau_1)-x(\tau_2)|^2}
		= \frac 12~.	
\end{equation}
Inserting this (\ref{WLg2}), one finally obtains
\begin{equation}
	\label{WLg2res}
		\big\langle W(C)\big\rangle = 1+\frac{g^2 (N^2-1)}{8N}+ \cO (g^4)~.
\end{equation}

\begin{figure}[htb]
	\begin{center}
		\includegraphics[scale=0.6]{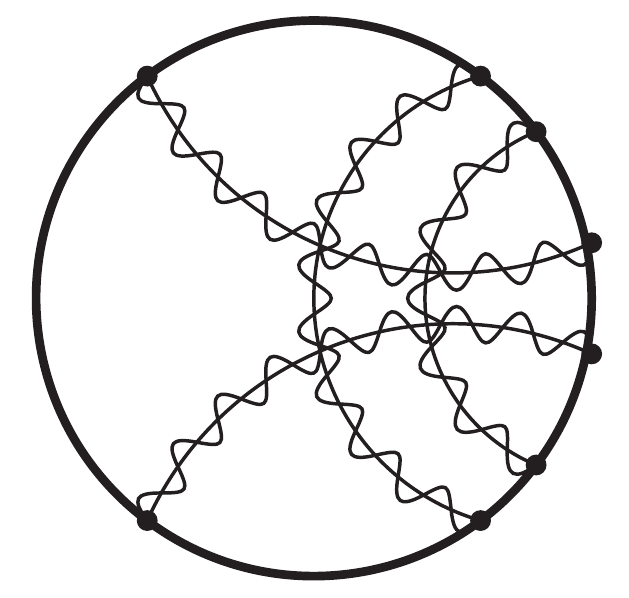}
	\end{center}
	\caption{Example of ladder diagrams without internal vertices contributing to the Wilson loop vev in $\cN=4$}
	\label{fig:WLladder}
\end{figure}

We obtain a space-time independent quantity, where the UV divergences (usually present in a non supersymmetric computation) are mutually canceled between gauge and scalar field contributions. This highly non-trivial behavior is preserved at next order, where all the diagrams with internal vertices vanishes, as shown by \cite{Erickson:2000af}. It turns out that the only possible contributions at each perturbative order come from the so called ``rainbow'' diagrams, see Figure \ref{fig:WLladder}.

The computation of the Wilson loop vev is reduced to a combinatorial analysis of the color factors coming from the propagator insertions. Therefore it was conjectured \cite{Drukker:2000rr} the existence of a Gaussian matrix model which could provide the full result. This matrix model was then proven by Pestun's localization computation, which we reviewed in Subsection \ref{subsec2:MMsphere}. In particular for the $\cN=4$ case we have neither instanton nor 1-loop determinant contributions and the matrix model in purely Gaussian. The Wilson loop operator \eqref{WLdef} evaluated on the localization locus \eqref{saddlesphere} for $R=1$ becomes \footnote{Note that to obtain the form \eqref{WLa} the further rescaling \eqref{rescaling} needs to be performed.}
\begin{equation}
	\label{WLa}
		\mathcal{W}(a)=\frac{1}{N}\,\mathrm{tr}\exp\Big(\frac{g }{\sqrt{2}}\,a\Big)~,
\end{equation}
so that its expectation value in the matrix model reads:
\begin{align}
	\label{defWN4}
	\big\langle \cW(a)\big\rangle_0 = \frac{1}{\cZ} \int da~ e^{-tr a^2} \frac{1}{N} e^{\frac{g}{\sqrt{2}}\,a} ~,
\end{align}	
where the subscript $0$ stands for Gaussian matrix model.
This expression can be resummed to obtain \cite{Erickson:2000af,Drukker:2000rr}:	
\begin{align}
	\label{WN4exact}
		W(g) =
		\frac{1}{N}\,L_{N-1}^1\left(\!-\frac{g^2}{4}\right)\,
		\exp\left[\frac{g^2}{8}\Big(1-\frac{1}{N}\Big)\right]~,
\end{align}
where $L_{n}^m(x)$ is the generalized Laguerre polynomial of degree $n$. This formula represents a remarkable example of exact result in $\cN=4$, since \eqref{WN4exact} holds for any value of $g$ and $N$. It will represent a fundamental benchmark throughout the present thesis.

\subsubsection{Introduction to $\cN=2$ computations}

The natural question is to ask how the full story goes in the $\cN=2$ case, and the computation of the Wilson loop vev in a $\cN=2$ theory is the main topic of the rest of this Chapter. In particular we consider the vev of the fundamental 1/2 BPS 
circular loop in conformal $\cN=2$ SU($N$) theories with matter transforming in a generic representation with the 
the requirement that the $\beta$-function vanishes. The approach is to exploit the localized matrix model in $S^4$ (which is no longer Gaussian, see Subsection \ref{subsec2:MMsphere}), to obtain information about the field theory structure on the flat space $\mathbb{R}^4$, using the ``difference theory", namely computing only the diagrammatic difference with respect to $\cN=4$ SYM, as we explained in Subsection \ref{subsec::difference}. This technique is powerful enough to push the perturbative analysis to $g^8\,\zeta(5)$ at four loops. The motivations and the outcomes are several.
\begin{itemize}
\item
Considering theories with a generic matter content, as we see in the matrix model description, makes evident that  the matrix model itself naturally organizes its outcomes in terms of the ``difference theory''.
\item
The matrix model also suggests that the lowest-order contributions to the circular Wilson loop vev proportional to 
a given Riemann $\zeta$-value, namely the terms of the type $g^{2n+2}\,\zeta(2n-1)$, are 
entirely due to the $n$-th loop correction to a single propagator inserted in the 
Wilson loop in all possible ways.
\item
Appendix~\ref{app:grass-super} describes a 
method to carry out the Grassmann integrations appearing in $\cN=1$ superdiagrams with chiral/anti-chiral multiplet and vector multiplet lines. 
We have found this method, which follows a different route from the use of the $D$-algebra proposed long ago in \cite{Grisaru:1979wc}, quite efficient in dealing with the type of diagrams involved in our computation. 
\item
Being able to treat generic conformal $\cN=2$ theories allows us to select special cases that exhibit a particular behavior in the large-$N$ limit. In particular we find that for for two specific theories (\textbf{D} and \textbf{E} in Table \ref{tab:scft}) the Wilson loop vacuum expectation value is equivalent to the $\cN=4$ case at leading order in the large-$N$ limit. These two classes of theories were shown to have a holographic dual \cite{Ennes:2000fu} of the type $\mathrm{AdS}_5\times S^5/\mathbb{Z}$ for an appropriate discrete group 
$\mathbb{Z}$, which is a simple modification of the $\mathrm{AdS}_5\times S^5$ geometry
corresponding to the $\cN=4$ SYM theory.

\end{itemize}

We begin with a description of the matrix model techniques, starting from the matrix model description of Subsection \ref{subsec2:MMsphere}, then we apply them to the Wilson loop vev case. Section \ref{secn:fieldtheory} is devoted to the field theory computations, using the Lagrangian formalism developed in Section \ref{sec:FTactions}.

\section{Interacting matrix model techniques}
\label{sec:mmvev}
The $\cN=2$ matrix model in the zero-instanton sector can be written as (we drop the overall factor of \eqref{ZS4resc}):
\begin{equation}
\cZ_{S^4} = \int da~\rme^{-\mathrm{tr}\, a^2 - S(a)}~,
\end{equation}
where the interacting action is perturbatively given by \eqref{Sint}.
The basic observation which is crucial for $\cN=2$ matrix model computations is that for each perturbative order we deal with a Gaussian matrix model. For any observable represented by a 
function $f(a)$ in the matrix model, its vacuum expectation value is  
\begin{align}
	\label{vevmat}
		\big\langle f(a) \big\rangle\, 
		= \,\frac{\displaystyle{ \int \!da ~\rme^{-\tr a^2-S(a)}\,f(a)}}
		{ \displaystyle{\int \!da~\rme^{-\tr a^2-S(a)}} }\,	= \,\frac{\big\langle\,
		\rme^{- S(a)}\,f(a)\,\big\rangle_0\phantom{\Big|}}
		{\big\langle\,\rme^{- S(a)}\,\big\rangle_0
		\phantom{\Big|}}~,
\end{align} 
where the subscript 0 in the right-hand-side 
indicates that the vacuum expectation value is evaluated 
in the free Gaussian model describing the $\mathcal{N}=4$ theory. 
These free vacuum expectation values can be computed in a straightforward way 
via Wick's theorem in terms of the propagator\,%
\footnote{We  normalize the flat measure as 
$da = \prod_{b} \left(da^b/\sqrt{2\pi}\right)$,
so that  $\int da\, \rme^{-\tr a^2}=1$.
In this way the contraction (\ref{wickabc}) immediately follows.}
\begin{equation}
	\label{wickabc}
		\big\langle a^b \,a^c\big\rangle_0 \,=\, \delta^{bc}~.
\end{equation}
As discussed in \cite{Billo:2017glv,Billo:2018oog,Billo:2019job}, 
%using the basic contraction (\ref{wickabc}) and the so-called fusion/fission relations for traces in the fundamental representation of SU($N$), 
it is possible to recursively evaluate the quantities
\begin{equation}
	\label{tn}
		t_{k_1,k_2,\cdots} = \big\langle \tr a^{k_1}\,\tr a^{k_2} \cdots\big\rangle_0
\end{equation}
and obtain explicit expressions for generic values of $k_1,k_2,\ldots$. Indeed using \eqref{wickabc} and
\begin{equation}
\tr \big(T^bT^c\big)=\frac{1}{2}\,\delta^{bc}~,~~~\tr T^b=0 \quad\mbox{with}~~b,c=1,\cdots,N^2-1~,
\label{normT}
\end{equation}
we evidently have 
\begin{equation}
\label{rrrecursion-2}
\begin{aligned}
t_0 =\big\langle\, \tr 1\,\big\rangle_0 =N~,~~~
t_1 = \big\langle\, \tr a\,\big\rangle_0=0~,~~~
t_2 =  \big\langle\, \tr a^2\,\big\rangle_0 =\frac{N^2-1}{2}~.
\end{aligned}
\end{equation}
Higher order traces can be computed performing consecutive Wick contractions with (\ref{wickabc})
and using the fusion/fission identities
\begin{equation}
\begin{aligned}
\tr \big(T^b B \,T^b C\big)&=\frac{1}{2}\,\tr B~\tr C-\frac{1}{2N}\,\tr \big(B\, C\big)~,\\
\tr \big(T^b C\big)~\tr \big(T^b C\big)&=\frac{1}{2}\,\tr \big(B\, C\big)-\frac{1}{2N}\,\tr B~\tr C~,
\end{aligned}
\label{fusionfission}
\end{equation}
which hold for any two matrices $B$ and $C$.
In this way we can build recursion relations and, for example, get:
\begin{align}
 t_{n}    &=   \frac12 \sum_{m=0}^{n-2}  \Big( t_{m,n-m-2}
  -\frac{1}{N}\, t_{n-2}  \Big)  ~,\notag\\
 t_{n,n_1} &=  \frac12 \sum_{m=0}^{n-2}  \Big( t_{m,n-m-2,n_1}
  -\frac{1}{N}\,   t_{n-2,n_1}  \Big)  
  + \frac{n_1}{2} \,\Big(  t_{n+n_1-2 } -\frac{1}{N} \,t_{n-1,n_1-1} \Big)~,\label{rrecursion}\\
  t_{n,n_1,n_2}  &=\frac12 \sum_{m=0}^{n-2}  \Big( t_{m,n-m-2,n_1,n_2}
  -\frac{1}{N} \,  t_{n-2,n_1,n_2}  \Big)  + \frac{n_1}{2} \Big(t_{n+n_1-2,n_2}
  -\frac{1}{N} \,  t_{n-1,n_1-1,n_2}\Big)  \notag\\
   &~~~~~~~~~~~+ \frac{n_2}{2} \Big( t_{n+n_2-2,n_1 }
   -\frac{1}{N}  t_{n-1,n_1,n_2-1} \Big) ~,\notag
\end{align} 
and so on. 
These  relations, together with the initial conditions (\ref{rrrecursion-2}), give an efficient way to obtain
multi-trace vacuum expectation values in the Gaussian model and will be the basic ingredients for the 
computations of the correlators in the $\cN=2$ superconformal theory.

To compute perturbatively the vacuum expectation value $\big\langle f(a)\big\rangle$ 
in the interacting theory, one starts from the right-hand-side of (\ref{vevmat}) 
and expands the action $S(a)$ as in (\ref{Sint}). Proceeding in this way, for conformal theories where the $g^2$-term vanishes, one gets
\begin{align}
	\label{espvev}
		\big\langle f(a)\big\rangle
		& =   \big\langle f(a)\big\rangle_0 
		+ \left(\frac{g^2}{8\pi^2}\right)^2 \frac{\zeta(3)}{2}\,
		\big\langle f(a)\, \trp a^4\big\rangle_{0,\mathrm{c}}
		- \left(\frac{g^2}{8\pi^2}\right)^3 \frac{\zeta(5)}{3}\, \big\langle
		f(a)\, \trp a^6\big\rangle_{0,\mathrm{c}}\nonumber\\[2mm]
		&~~ + \ldots~,
\end{align}
where the notation $\langle ~\rangle_{0,\mathrm{c}}$ stands for
the connected part of a free correlator, namely
\begin{align}
	\label{defccor}
		\big\langle f(a)\, g(a)\big\rangle_{0,\mathrm{c}}  \,=\,  
		\big\langle f(a) \, g(a)\big\rangle_0 \,-\, \big\langle f(a)\big\rangle_0 
		\,\,\big\langle g(a)\big\rangle_0~.
\end{align}
We may regard (\ref{espvev}) as an expansion in ``trascendentality'', in the sense that each term 
in the sum has a given power of Riemann $\zeta$-values since it comes from the expansion of the
exponential of the interaction action (\ref{Sint}). 
For example the second term is the only one proportional to $\zeta(3)$, the third term is the only one 
proportional to $\zeta(5)$, while the ellipses stand for terms proportional to $\zeta(7)$, 
$\zeta(3)^2$ and so on.

Often $f(a)$ is  a ``gauge-invariant'' quantity, expressed in terms of traces of powers 
of $a$ in some representations. Also the quantities $\trp a^{2k}$ are traces of this type.
As shown in Appendix \ref{app:group}, relying on the Frobenius theorem it is possible 
to express such traces in terms of traces in the fundamental representation. 
At this point, the vacuum expectation value (\ref{espvev}) is reduced to a 
combinations of the quantities $t_{k_1,k_2,\ldots}$ introduced in (\ref{tn}) and which can be computed using \eqref{rrecursion}. 
This is the computational strategy we adopt in the following Sections.

\subsection{A class of conformal $\cN=2$ theories}
\label{subsec:CN2class}
Let us consider a class of theories with $N_F$ matter hypermultiplets transforming in the fundamental 
representation, $N_S$ in the symmetric and $N_A$ in the anti-symmetric representation of order 2. 
This corresponds to taking 
\begin{equation}
	\label{RNFNASNA}
		\cR = N_F\, \Yfund \oplus N_S\, \Ysymm \oplus N_A\, \Yasymm~.
\end{equation}
The traces $\trp a^{2k}$ appearing in the interaction action $S(a)$ 
can be re-expressed in terms of traces in the fundamental representation, as discussed in 
Appendix~\ref{app:group}. 

For example, for $k=1$ one has
\begin{equation}
	\label{trpa2}
		\trp a^2 = 2 \left(i_{\cR} - i_{\mathrm{adj}}\right) \tr a^2 
		= - \beta_0 \tr a^2~, 
\end{equation}
with
\begin{equation}
	\label{b0is}
		\beta_0=2N-N_F-N_S(N+2)-N_A(N-2)~.
\end{equation}
Superconformal theories must have $\beta_0=0$. It is easy to see that imposing 
this condition leads to five families of $\mathcal{N}=2$ superconformal field theories 
with gauge group SU($N$), and matter in the fundamental, symmetric or anti-symmetric 
representations. They were identified long ago in \cite{Koh:1983ir} and recently reconsidered in
\cite{Fiol:2015mrp,Bourget:2018fhe}. They are displayed in table \ref{tab:scft}.
\renewcommand{\arraystretch}{1}
\begin{table}[ht]
	\begin{center}
		{\small
		\begin{tabular}{c|c|c|c}
			\hline
			\hline
			\,\,theory \phantom{\bigg|}& $N_F$ & $N_S$ & $N_A$ \\
			\hline
			$\mathbf{A}\phantom{\Big|}$& $~~2N~~$ & $~~0~~$ & $~~0~~$  \\
			$\mathbf{B}\phantom{\Big|}$& $~~N-2~~$ & $~~1~~$ & $~~0~~$  \\
			$\mathbf{C}\phantom{\Big|}$& $~~N+2~~$ & $~~0~~$ & $~~1~~$  \\
			$\mathbf{D}\phantom{\Big|}$& $~~4~~$ & $~~0~~$ & $~~2~~$  \\
			$\mathbf{E}\phantom{\Big|}$& $~~0~~$ & $~~1~~$ & $~~1~~$  \\
			\hline
			\hline
			
		\end{tabular}
		}
	\end{center}
	\caption{The five families of $\cN=2$ superconformal theories with SU($N$) gauge 
	group and matter in fundamental, symmetric and anti-symmetric representations.}
	\label{tab:scft}
\end{table}

Theory $\mathbf{A}$ is the $\cN=2$ conformal SQCD which is often considered as
the prototypical example of a $\cN=2$ superconformal theory.
On the other hand, theories $\mathbf{D}$ and $\mathbf{E}$ are quite interesting:
for these superconformal models a holographic dual of the form 
$\mathrm{AdS}_5\times S^5/\mathbb{Z}$ with an appropriate discrete group $\mathbb{Z}$ 
has been identified \cite{Ennes:2000fu}. We will discuss some properties 
of these theories in the following.

For higher traces with $k>1$, one finds (see again Appendix \ref{app:group} for details)
\begin{align} 
	\label{S2n}
		\trp a^{2k} = &
		\frac{1}{2}\sum_{\ell=2}^{2k-2}
		\binom{2k}{\ell}\left(N_S+N_A-2\,(-1)^\ell\right)\,
		\tr a^\ell\,\tr a^{2k-\ell}\nonumber\\
		& + \left(\big(2^{k-1}-2\big)\, (N_S- N_A)-\beta_0\right)\, \tr a^{2k}~.
\end{align} 
Inserting this into the expansion (\ref{Sint}) we can express the interaction action 
in terms of traces in the fundamental representation. For the superconformal theories of table \ref{tab:scft} we find the results displayed in table \ref{tab:S4S6conf}.
\begin{table}[ht]
	\begin{center}
		{\small
		\begin{tabular}{c|c|c}
			\hline
			\hline
			\,\,theory \phantom{\bigg|}& $\trp a^{4}$ & 
			$\trp a^{6}$ 
			\\
			\hline
			$\mathbf{A}\phantom{\Big|}$& $~~6\,\big(\mathrm{tr} \,a^2\big)^2~~\phantom{\bigg|}$ & $~~10\,\Big[2\big(\mathrm{tr}\,a^3\big)^2-3\,\mathrm{tr}\,a^4\,\mathrm{tr}\,a^2
			\Big]~~$   \\
			$\mathbf{B}\phantom{\Big|}$& $~~3\,\Big[\big(\mathrm{tr}\,a^2\big)^2-2\,\mathrm{tr}\,a^4
			\Big]~~\phantom{\bigg|}$ & $~~15\,\Big[2\big(\mathrm{tr}\,a^3\big)^2
			-\mathrm{tr}\,a^4\,\mathrm{tr}\,a^2+2\,\mathrm{tr}\,a^6
			\Big]~~$  \\
			$\mathbf{C}\phantom{\Big|}$& $~~3\,\Big[\big(\mathrm{tr}\,a^2\big)^2+2\,\mathrm{tr}\,a^4
			\Big]~~\phantom{\bigg|}$ & $~~15\,\Big[2\big(\mathrm{tr}\,a^3\big)^2
			-\mathrm{tr}\,a^4\,\mathrm{tr}\,a^2-2\,\mathrm{tr}\,a^6
			\Big]~~$   \\
			$\mathbf{D}\phantom{\Big|}$& $~~12\,\mathrm{tr}\,a^4~~\phantom{\bigg|}$ & $~~20\,\Big[2\big(\mathrm{tr}\,a^3\big)^2-3\,\mathrm{tr}\,a^6
			\Big]~~$     \\
			$\mathbf{E}\phantom{\Big|}$& $~~0~~\phantom{\bigg|}$ & $~~40\,\big(\mathrm{tr}\,a^3\big)^2~~$   \\
			\hline
			\hline
	\end{tabular}
	}
\end{center}
\caption{The quartic and sextic interaction terms in the action $S(a)$
 for the five families of conformal 
theories defined in table \ref{tab:scft}.}
\label{tab:S4S6conf}
\end{table}

Notice that for theory $\mathbf{E}$ the quartic term vanishes and thus in this case the
effects of the interactions appear for the first time at order $g^6$, {\it{i.e.}} at three loops, 
and are proportional to $\zeta(5)$. This feature, which has been recently pointed out  
also in \cite{Bourget:2018fhe}, is a simple consequence of the properties of the quartic
trace in a representation $\cR$ formed by one symmetric and one anti-symmetric 
representation. Altogether, the matter hypermultiplets fill a generic $N\times N$ matrix; this is to
be compared with the $\cN=4$ case in the hypermultiplets are in the 
adjoint representation, which is equivalent to $N\times \overline{N}$ minus one singlet. The strong similarity of the two representations explains why theory
$\mathbf{E}$ is the $\cN=2$ model which is more closely related to the $\cN=4$ SYM theory.
For theory $\mathbf{D}$, instead, the quartic term is a single fundamental 
trace and thus is simpler than in the other theories. In the following we will see that these
features of theories $\mathbf{D}$ and $\mathbf{E}$ have a bearing on their large-$N$ behavior.

\section{Propagator and Wilson loops in superconformal matrix models}
\label{sec:propWilson}
We now discuss in detail two specific applications of the formula (\ref{espvev}): first 
the ``propagator'' $\vev{a^b\, a^c}$ and later the 1/2 BPS circular Wilson loops $\cW(a)$
in the fundamental representation.

\subsection{The propagator}
\label{subsec:propagator}
If in (\ref{espvev}) we take $f(a)=a^b\,a^c$, we get 
\begin{align}
	\label{corrbc}
		\big\langle a^b\, a^c\big\rangle & = \big\langle a^b\, a^c\big\rangle_0 
		+ \left(\frac{g^2}{8\pi^2}\right)^2 \frac{\zeta(3)}{2}\, C^\prime_{(d_1 d_2 d_3 d_4)}
		\vev{a^b\, a^c\, a^{d_1}\, a^{d_2}\, a^{d_3}\, a^{d_4}}_{0,\mathrm{c}}
		\nonumber\\
		&~~ 
		- \left(\frac{g^2}{8\pi^2}\right)^3 \frac{\zeta(5)}{3}\, C^\prime_{(d_1 d_2 d_3 d_4 d_5 d_6)}
		\vev{a^b\, a^c\, a^{d_1}\, a^{d_2}\, a^{d_3}\, a^{d_4}\, a^{d_5}\, a^{d_6}}_{0,\mathrm{c}} +\ldots~,
\end{align}
where inside each connected correlator we cannot contract $a^b$ with $a^c$. Doing all legitimate contractions we obtain
\begin{align}
	\label{corrbcres}
		\vev{a^b\, a^c} & = \delta^{bc} 
		+ \left(\frac{g^2}{8\pi^2}\right)^2\zeta(3)\times 6\, C^\prime_{(bcdd)}
		- \left(\frac{g^2}{8\pi^2}\right)^3 \zeta(5)\times 30\, C^\prime_{(bcddee)}
		+ \ldots~. 
\end{align}
The above contracted tensors are proportional to $\delta^{bc}$, and thus if define
\begin{align}
	\label{c4c6delta}
		6 \,C^\prime_{(bcdd)} & = \cC^\prime_4\, \delta^{bc}~,~~~
		30 \,C^\prime_{(bcddee)}  = \cC^\prime_6\, \delta^{bc}~,	
\end{align}
we can rewrite (\ref{corrbcres}) as
\begin{align}
	\label{corrbcres1}
		\big\langle a^b\, a^c\big\rangle& = \delta^{bc} \,\big(1 + \Pi\big)
\end{align}
with
\begin{align}
	\label{Piis}		 
		\Pi & = \left(\frac{g^2}{8\pi^2}\right)^2\zeta(3)\,\cC^\prime_4
		- \left(\frac{g^2}{8\pi^2}\right)^3 \zeta(5)\,\cC^\prime_6
		+ \ldots~. 
\end{align}
Using the expressions of the tensors $C^\prime$ for the five families of superconformal SU(N)
theories that can be obtained from the formul\ae~ in Appendix \ref{app:group} 
with the help of Form Tracer \cite{Cyrol:2016zqb}, one finds
\begin{equation}
	\begin{aligned}
		\cC^\prime_4 &= \frac{6\,C^\prime_{(ccdd)}}{N^2-1} = 3
		\Big[(N_S+N_A-2)\,\frac{N^2+1}{2}+(N_S-N_A)\,\frac{2N^2-3}{N}\Big]~,\\[2mm]
		\cC^\prime_6&=\frac{30\,C^\prime_{(ccddee)}}{N^2-1}=15\Big[
		(N_S+N_A-2)\,\frac{2N^4+5N^2-17}{4N}\\[1mm]
		&~~~\qquad\qquad\qquad\qquad+
		(N_S-N_A)\,\frac{5(N^4-3N^2+3)}{2N^2}+\frac{2(N^2-4)}{N}\Big]~.
		\end{aligned}
	\label{C4C6prime}
\end{equation}
These coefficients are tabulated in table
\ref{tab:C4C6conf}. 
\begin{table}[ht]
	\begin{center}
		{\small
		\begin{tabular}{c|c|c}
			\hline
			\hline
			\,\,theory \phantom{\bigg|}& $\cC^\prime_4 $ & 
			$\cC^\prime_6 $ 
			\\
			\hline
			$\mathbf{A}\phantom{\Big|}$& $~~-3(N^2+1)~~\phantom{\bigg|}$
			& $~~-\frac{15(N^2+1)(2N^2-1)}{2 N}~~\phantom{\bigg|}$   \\
			$\mathbf{B}\phantom{\Big|}$& $~~-\frac{3(N+1)(N-2)(N-3)}{2N}~~\phantom{\bigg|}$
			& $~~-\frac{15 (N-2) (2N^4-6N^3-15N^2+15)}{4 N^2}~~\phantom{\bigg|}$ \\
			$\mathbf{C}\phantom{\Big|}$& $~~-\frac{3(N-1)(N+2)(N+3)}{2N}~~\phantom{\bigg|}$
			& $~~-\frac{15 (N+2) (2N^4+6N^3-15N^2+15)}{4 N^2}	~~\phantom{\bigg|}$ \\
			$\mathbf{D}\phantom{\Big|}$&$~~-\frac{6(2N^2-3)}{N}~~\phantom{\bigg|}$
			& $~~-\frac{15(5N^4-2N^3-15N^2+8N+15)}{N^2}~~\phantom{\bigg|}$ \\
			$\mathbf{E}\phantom{\Big|}$&$~~0~~\phantom{\bigg|}$
			& $~~\frac{30 (N^2-4)}{N}~~\phantom{\bigg|}$\\
			\hline
			\hline
		\end{tabular}
		}
	\end{center}
	\caption{The coefficients $\cC^\prime_4 $ and $\cC^\prime_6$ 
	for the five families of conformal theories defined in table \ref{tab:scft}.}
	\label{tab:C4C6conf}
\end{table}

For the comparison with perturbative field theory calculations presented 
in Section \ref{secn:fieldtheory}, it is useful to make explicit the symmetrization of the $C^\prime$-tensors appearing in (\ref{corrbcres}).  For the 4-index tensor, we have
\begin{align}
	\label{symmC4is}
		6\, C^\prime_{(bcdd)} & = 2\, \big(
		C^\prime_{bcdd} + C^\prime_{bdcd} + C^\prime_{bddc}\big) ~.
\end{align}
Indeed, due to the cyclic property and the fact that two indices are identified, a subgroup 
$\mathbb{Z}_4\times \mathbb{Z}_2$ of permutations leaves $C^\prime_{bcdd}$ invariant and one has to average only over the $4!/8=3$ permutations in the coset with respect to this stability subgroup. In a similar way, for the 6-index tensor we have
\begin{align}
	\label{symmC6is}
		30\, C^\prime_{(bcddee)} & = 2\, \big(C^\prime_{bcddee} + C^\prime_{bcdede} + C^   
		\prime_{bcdeed} + C^\prime_{bdcdee} + C^\prime_{bdcede} \notag\\[1mm]
		&~~~~~+ C^\prime_{bdceed} +\, C^\prime_{bddcee} + C^\prime_{bdecde} 
		+ C^\prime_{bdeced} + C^\prime_{bddece}\notag\\[1mm]
		&~~~~~+ C^\prime_{bdedce} + C^\prime_{bdeecd}+ C^\prime_{bddeec} 
		+ C^\prime_{bdedec} + C^\prime_{bdeedc} \big)~.		
\end{align}     
In this case, the stability subgroup is $\mathbb{Z}_6\times \mathbb{Z}_2\times 
\mathbb{Z}_2\times \mathbb{Z}_2$ and the coset has $6!/48 = 15$ elements. 

We would like to remark that even if we have considered theories with SU$(N)$ gauge group
and matter in the fundamental, symmetric and anti-symmetric representations, the color tensors
$C^\prime_{b_1\ldots b_n}$ in (\ref{defC}) and the corresponding coefficients $\cC^\prime_n$
can be defined also for other representations of SU($N$) (or U$(N)$) using the Frobenius theorem,
as indicated in Appendix~\ref{secn:frob}, and also for other gauge groups. Thus, the structure
of the propagator corrections in (\ref{corrbcres1}) is very general.

\subsection{Wilson loops}
\label{subsec:wilson}
We consider the 1/2 BPS circular Wilson loop in the fundamental representation.
If this operator is inserted on the equator of $S^4$, in the matrix model we can represent it 
by the operator \eqref{WLa}, which can be expanded as:
\begin{equation}
	\label{WLaexpanded}
		\mathcal{W}(a)=\frac{1}{N}\,\mathrm{tr}\exp\Big(\frac{g}{\sqrt{2}}\,a\Big)
		=\frac{1}{N}\,\sum_{k=0}^\infty\frac{1}{k!}\,\frac{g^k}{2^{\frac{k}{2}}}\,\mathrm{tr}\,a^k~.
\end{equation}
Its vacuum expectation value is computed starting from (\ref{espvev}), following the strategy outlined in Section \ref{sec:mmvev}. We write
\begin{align}
	\label{Wexpchis}
		\Delta\cW\,\equiv\, \big\langle\cW(a)\big\rangle - \big\langle\cW(a)\big\rangle_0 = \cX_3 + \cX_5 + \ldots~,
\end{align}
where
\begin{align}
\cX_3&=\left(\frac{g^2}{8\pi^2}\right)^2 \frac{\zeta(3)}{2}\,
		\big\langle \cW(a)\, \trp a^4\big\rangle_{0,\mathrm{c}}~,\label{chi3is}\\
\cX_5&=- \left(\frac{g^2}{8\pi^2}\right)^3 \frac{\zeta(5)}{3}\, \big\langle
		\cW(a)\, \trp a^6\big\rangle_{0,\mathrm{c}}~,\label{chi5is}
\end{align}
and so on. {From} these expressions it is easy to realize that for each Riemann $\zeta$-value
(or product thereof) the term with the lowest power of $g$ in $\Delta\cW$ arises from 
the quadratic term in the expansion \eqref{WLaexpanded} of the Wilson loop operator. Indeed, we have
\begin{align}
\cX_3&=\left(\frac{g^2}{8\pi^2}\right)^2\frac{\zeta(3)}{2}\,\frac{g^2}{4N}\,
		\big\langle \mathrm{tr}\,a^2\, \trp a^4\big\rangle_{0,\mathrm{c}}+O(g^8)\notag\\[1mm]
		&=\frac{g^6\,\zeta(3)}{512\pi^4	}\,\frac{N^2-1}{N}\,\cC_4^\prime+O(g^8)
		\label{chi3expg}
\end{align} 
where $\cC_4^\prime$ is the coefficient of the two-loop correction of the ``propagator'' of the
matrix model defined in (\ref{C4C6prime}). This result is valid for any superconformal theory, and in particular for the five families introduced in 
Subsection~\ref{subsec:CN2class}. Clearly, for theory $\mathbf{E}$ the correction is zero; actually the whole $\cX_3$ vanishes in this case. 
In a similar way we find
\begin{align}
\cX_5&=-\left(\frac{g^2}{8\pi^2}\right)^3\frac{\zeta(5)}{3}\,\frac{g^2}{4N}\,
		\big\langle \mathrm{tr}\,a^2\, \trp a^6\big\rangle_{0,\mathrm{c}}+O(g^{10})\notag\\[1mm]
		&=-\frac{g^8\,\zeta(5)}{4096\pi^6}\,\frac{N^2-1}{N}\,\cC_6^\prime+O(g^{10})
		\label{chi5expg}
\end{align}
where $\cC_6^\prime$ is the three-loop correction of the matrix model ``propagator''.
Combining (\ref{chi3expg}) and (\ref{chi5expg}) we see that at the lowest orders in $g$ 
the difference of the vacuum expectation value of the Wilson loop with respect to the 
$\cN=4$ expression is given by
\begin{equation}
\Delta\cW= \frac{N^2-1}{8N}\,g^2\,\Pi+\ldots
\label{DeltaWPi}
\end{equation}
where $\Pi$ is the quantum correction to the ``propagator" given in (\ref{Piis}).
In the following Sections we will prove that these results are in perfect agreement with perturbative
field theory calculations using ordinary (super) Feynman diagrams.

Actually, as explained in \cite{Billo:2018oog}, within the matrix model it is possible to evaluate 
$\cX_3$, $\cX_5$ and so on, without making any expansion in $g$. To obtain these 
{\emph{exact}} results, one has to write the traces $\trp a^{2k}$ in terms of the traces
in the fundamental representation by means of (\ref{S2n}). In this way everything is reduced 
to combinations of the quantities $t_{k_1,k_2,\ldots}$ defined in (\ref{tn}), 
which in turn can be evaluated in an algorithmic way using the fusion/fission identities
\eqref{fusionfission}. In the end, this procedure allows one to express the result 
in terms of derivatives of the exact vacuum expectation value of the Wilson loop $W(g)$ in the $\mathcal{N}=4$ theory, namely the result \eqref{WN4exact}.\\
Applying this procedure to the five families of superconformal theories introduced in Section~\ref{subsec:CN2class}, we find
\begin{align}
	\label{X3exact}
		\mathcal{X}_3& =	\left(\frac{g^2}{8\pi^2}\right)^2\frac{3\,\zeta(3)}{16N^2}
		\bigg[2\big(N_S+N_A-2\big)N^2
		\Big(\big(2N^2+1\big)\,g\,\partial_g W(g)+g^2\,\partial_g^2 W(g)\Big)
		\nonumber\\[1mm]
	& ~~
	\qquad\quad\qquad\quad
		+\big(N_S-N_A\big)\Big(\big(N^2-1\big)\,g^2\,W(g)+
		\big(g^2+8N^3 - 12N\big)\,g\,\partial_g W(g)
\nonumber \\[1mm]
		& \qquad\qquad\quad\qquad\quad\qquad\quad
		-4N g^2\,\partial_g^2 W(g)+16N^2\,g\,\partial_g^3 W(g)\Big)\bigg]
		~.
\end{align}
Expanding in $g$, it is easy to check the validity of (\ref{chi3expg}).
The case of theory $\mathbf{A}$, namely $N_S=N_A=0$, 
was already described in \cite{Billo:2018oog}.  For theory $\mathbf{E}$, as we have already remarked, $\cX_3=0$ since $\trp a^4=0$. Therefore, in this case the first correction with respect to 
the $\cN=4$ result for the Wilson loop is $\cX_5$, which turns
out to be
\begin{align}
	\label{X5exact}
		\mathcal{X}_5\,\big|_{\mathbf{E}}& =	
		-\left(\frac{g^2}{8\pi^2}\right)^3\frac{5\,\zeta(5)}{12N^2}
		\bigg[\big(N^4+5N^2-6\big)\,g^2\,W(g)\nonumber\\
		&\qquad\qquad~~\qquad
		+\big(2g^2N^2+6g^2-8N^3-48N\big)\,g\,\partial_g W(g)\nonumber\\[1mm]
		&\qquad\qquad~~\qquad+\big(g^2-8N^3-48N\big)\,g^2\,\partial_g^2W(g)\nonumber
		\\
		&\qquad\qquad~~\qquad
-8N\big(g^2-10N\big)\,g\,\partial_g^3W(g)+16N^2\,g^2\,\partial_g^4W(g)\bigg]~.
\end{align}
Similar formul\ae~ can be easily obtained for the other families of superconformal theories. We have derived them but we do not report their explicit expressions since for theories $\mathbf{A}$,
$\mathbf{B}$, $\mathbf{C}$, and $\mathbf{D}$ the leading term in the difference with respect 
to the $\cN=4$ result is given by $\cX_3$.

We stress once more that this procedure allows one to obtain in an algorithmic way the {\emph{exact}} expression in $g$ and $N$ of any term of the vacuum expectation value 
of the circular Wilson loop with a fixed structure of Riemann $\zeta$-values. This fact will now
be used to study the behavior of the matrix model in the large-$N$ limit.

\subsection{The large-$N$ limit}
\label{subsec:largeN}
The large-$N$ limit is defined by taking $N\to\infty$ and keeping the 't Hooft coupling
\begin{equation}
\lambda=g^2\,N
\label{lambda}
\end{equation}
fixed. In this limit the perturbative correction $\Pi$ to the ``propagator'' given in (\ref{Piis}) becomes
\begin{align}
\Pi&=\big(N_S+N_A-2\big)\left(\frac{3\zeta(3)\,\lambda^2}{128\pi^4}-\frac{15\zeta(5)\,\lambda^3}{1024\pi^6}+O(\lambda^4)\right)\notag\\[1mm]
&~~~+\big(N_S-N_A\big)\left(\frac{3\zeta(3)\,\lambda^2}{32\pi^4}-\frac{75\zeta(5)\,\lambda^3}{1024\pi^6}+O(\lambda^4)\right)\,\frac{1}{N}\label{PilargeN}\\[1mm]
&~~~+\bigg[
\big(N_S+N_A-2\big)\left(\frac{3\zeta(3)\,\lambda^2}{128\pi^4}-\frac{75\zeta(5)\,\lambda^3}{2048\pi^6}\right)-\frac{15\zeta(5)\,\lambda^3}{256\pi^6}
+O(\lambda^4)\bigg]\,\frac{1}{N^2}+ O\left(\frac{1}{N^3}\right)\notag~.
\end{align}
{From} this expression we easily see that in the planar limit $\Pi$ is non-zero for 
theories $\mathbf{A}$, $\mathbf{B}$ and $\mathbf{C}$, whereas it 
vanishes for theories $\mathbf{D}$ and $\mathbf{E}$:
\begin{equation}
\lim_{N\to\infty} \Pi\,\big|_{\mathbf{D}~\mathrm{or}~\mathbf{E}} = 0~.
\label{PiplanarDE}
\end{equation}
In particular for theory $\mathbf{D}$ the correction to the ``propagator'' goes
like $1/N$, whereas for theory $\mathbf{E}$ it goes like $1/N^2$:
\begin{align}
\Pi\,\big|_{\mathbf{D}}&
=-\left(\frac{3\zeta(3)\,\lambda^2}{16\pi^4}-\frac{75\zeta(5)\,\lambda^3}{512\pi^6}+O(\lambda^4)\right)\,\frac{1}{N}+O\left(\frac{1}{N^2}\right)~,\label{PiD}\\[1mm]
\Pi\,\big|_{\mathbf{E}}&=-\left(\frac{15\zeta(5)\,\lambda^3}{256\pi^6}+O(\lambda^4)\right)\,\frac{1}{N^2}+O\left(\frac{1}{N^3}\right)~.\label{PiE}
\end{align}
Therefore, in the planar limit, the ``propagator'' of the matrix model for these two families is
identical to that of the free matrix model describing the $\cN=4$ SYM theory.

Let us now consider the vacuum expectation value of the circular Wilson loop. Taking the 
large-$N$ limit in the $\cN=4$ expression (\ref{WN4exact}) one obtains 
\cite{Erickson:2000af}
\begin{equation}
\lim_{N\to\infty} W\big(\sqrt{\lambda/N}\big)\,
=\,\frac{2}{\sqrt{\lambda}}\,I_1\left(\sqrt{\lambda}\right)
\label{Wlambda}
\end{equation}
where $I_\ell$ is the modified Bessel function of the first kind.

Using this result in the $\zeta(3)$-correction (\ref{X3exact}), we get
\begin{equation}
\cX_3 = \big(N_S+N_A-2\big)\,\frac{3\zeta(3)\,\lambda^2}{128\pi^4}\,
I_2\left(\sqrt{\lambda}\right)+ O\left(\frac{1}{N}\right)~.
\label{X3largeN}
\end{equation}
This is a generalization of the formula obtained in \cite{Billo:2018oog} for the SQCD 
theory. With the same procedure we have also derived the planar limit of the $\zeta(5)$-
correction, finding
\begin{equation}
\cX_5 = -\big(N_S+N_A-2\big)\,\frac{5\zeta(5)\,\lambda^3}{1024\pi^6}\,\Big(
3I_2\big(\sqrt{\lambda}\big)+I_4\big(\sqrt{\lambda}\big)\Big)+ O\left(\frac{1}{N}\right)~.
\label{X5largeN}
\end{equation}
These results indicate that for theories $\mathbf{A}$, $\mathbf{B}$ and $\mathbf{C}$ the 
vacuum expectation value of the circular Wilson loop in the planar limit is different from 
the one of the $\cN=4$ SYM theory. On the other hand, for theories $\mathbf{D}$ 
and $\mathbf{E}$ this difference vanishes, namely
\begin{equation}
\lim_{N\to\infty} \Delta\cW \,\big|_{\mathbf{D}~\mathrm{or}~\mathbf{E}} = 0
\label{DeltaDE}
\end{equation}
in analogy with the ``propagator'' result (\ref{PiplanarDE}).
Working out the details at the next-to-leading order for theory $\mathbf{D}$, we find
\begin{align}
\Delta\cW \,\big|_{\mathbf{D}} &= -\bigg[
\frac{3\zeta(3)\,\lambda^2}{32\pi^4}\,\Big(
2I_2\big(\sqrt{\lambda}\big)+I_4\big(\sqrt{\lambda}\big)\Big)\label{DeltaWD}\\[1mm]
&\quad\qquad-\frac{15\zeta(5)\,\lambda^3}{256\pi^6}\,\Big(
5I_2\big(\sqrt{\lambda}\big)+4I_4\big(\sqrt{\lambda}\big)+
I_6\big(\sqrt{\lambda}\big)\Big)+\ldots
\bigg]\,\frac{1}{N}+O\left(\frac{1}{N^2}\right)\notag
\end{align}
where the ellipses stand for terms with higher Riemann $\zeta$-values (or product 
thereof). Similarly, at the next-to-next-to-leading order for theory $\mathbf{E}$, we find
\begin{align}
\Delta\cW \,\big|_{\mathbf{E}} &= -\bigg(
\frac{15\zeta(5)\,\lambda^{7/2}}{1024\pi^6}\,
I_1\big(\sqrt{\lambda}\big)+\ldots\bigg)\,\frac{1}{N^2}+O\left(\frac{1}{N^3}\right)~.
\label{DeltaWE}
\end{align}

Our findings have been obtained with a weak-coupling analysis at small $\lambda$. They 
are, however, in agreement with the strong-coupling results at large $\lambda$ presented 
in \cite{Fiol:2015mrp}, in the sense that also at strong coupling the vacuum expectation
value of the circular Wilson loop in the planar limit is different from that of the 
$\cN=4$ SYM theory for theories $\mathbf{A}$, $\mathbf{B}$ and $\mathbf{C}$, while
it is the same for theories $\mathbf{D}$ and $\mathbf{E}$. This observation suggests that
also the interpolating function between weak and strong coupling shares the same features
for the various theories.
The fact that for theories $\mathbf{D}$ and $\mathbf{E}$ the vacuum expectation value
of the circular Wilson loop is identical to that of the $\cN=4$ SYM theory in the planar 
limit is also in agreement with the fact that the holographic dual of 
theories $\mathbf{D}$ and $\mathbf{E}$ is of the form $\mathrm{AdS}_5\times S^5/\mathbb{Z}$ 
with an appropriate discrete group $\mathbb{Z}$ \cite{Ennes:2000fu}. Indeed, 
for the 1/2 BPS circular Wilson loop, the relevant part of the geometry is the 
Anti-de Sitter factor $\mathrm{AdS}_5$,
which is the same one that appears in the famous $\mathrm{AdS}_5\times S^5$ holographic dual of 
the $\cN=4$ SYM theory \cite{Maldacena:1997re}. It would be interesting 
to identify other observables that have this property in the planar limit and check 
the holographic correspondence, and also to find which observables 
of the theories $\mathbf{D}$ and $\mathbf{E}$ instead 
feel the difference with the $\cN=4$ SYM theory in the planar limit. Investigating
which sectors of our $\cN=2$ theories are planar equivalent to those  of the
$\cN=4$ SYM theory would be useful to better clarify the relations among the
various models and also to understand to which extent the arguments discussed
for example in \cite{Armoni:2004uu} in the so-called orientifold models 
can be applied to our case. We leave however this issue for future work.

We conclude by observing that the coefficients $\big(N_S+N_A-2\big)$ and
$\big(N_S-N_A\big)$ appearing in the planar limit results (see, for example, 
(\ref{PilargeN}), (\ref{X3largeN}) and
(\ref{X5largeN})) have an interesting meaning in terms of the central charges of 
the $\cN=2$ superconformal gauge theories corresponding to the matrix model. 
Indeed, taking into account the matter content corresponding to the representation
(\ref{RNFNASNA}) and using the formul\ae ~for the $c$ and $a$ central charges derived in
\cite{Anselmi:1997ys}, we find
\begin{equation}
\begin{aligned}
c&=-\frac{1}{24}\Big(\big(N_S+N_A-8\big)N^2+3\big(N_S-N_A\big)N+4\Big)~,\\
a&=-\frac{1}{48}\Big(\big(N_S+N_A-14\big)N^2+3\big(N_S-N_A\big)N+10\Big)~,
\end{aligned}
\label{ca}
\end{equation}
implying that 
\begin{align}
\frac{48(a-c)}{N^2}=\big(N_S+N_A-2\big)+\frac{3\big(N_S-N_A)}{N}-\frac{2}{N^2}
\label{c-a}
\end{align}
Using this, we can rewrite our results for the Wilson loop in the following way
\begin{equation}
\Delta\cW=\frac{a-c}{N^2}\bigg[
\frac{9\zeta(3)\,\lambda^2}{8\pi^4}\,I_2\big(\sqrt{\lambda}\big)
-\frac{15\zeta(5)\,\lambda^3}{64\pi^6}\,\Big(
3I_2\big(\sqrt{\lambda}\big)+I_4\big(\sqrt{\lambda}\big)
\Big)+\ldots\bigg]\,+\,O\left(\frac{1}{N}\right)~.
\label{DeltaWc-a}
\end{equation}
It would be nice to have an interpretation of this formula, and in particular of its prefactor, based on general principles.

\section{$\cN=2$ Wilson loop from field theory computations}
\label{secn:fieldtheory}

In this Section we consider the field-theoretic counterpart of the computations we performed in Section \ref{sec:propWilson} using the matrix model.

\subsection{The scalar propagator}
\label{subsec:scalprop}
In Subsection \ref{subsec::difference} we computed the one-loop correction to the scalar propagator, as an example of a computation in the difference theory. Now, starting from the form \eqref{propPi}, which we rewrite here for convenience:
\begin{equation}
	\label{propPi2}
		\Delta^{bc}(q)= \frac{\delta^{bc}}{q^2}\,\big(1 + \Pi\big)~,
\end{equation}
we look at higher perturbative orders.\\
Since conformal invariance imposes that the momentum dependence lies in the $1/q^2$ term, the $\Pi$ factor  should be captured by the matrix model and thus should be the same as the 
quantity $\Pi$ defined in (\ref{corrbcres1}). We will now check explicitly that this is indeed 
the case, up to the three-loop order corrections proportional to $\zeta(5)$. 
We already showed that the one-loop term vanishes for conformal theories, we now consider higher order diagrams in the difference theory.

\subsubsection*{Building blocks for higher order diagrams} 
Similarly to what happens at one-loop as shown in Figure~\ref{fig:1loop}, the 
contributions of the $(Q,\widetilde Q)$ and $(H,\widetilde H)$ hypermultiplets always 
have a color factor that contains a ``primed'' trace of generators, 
{\it{i.e.}} they contain the tensor $C^\prime_{b_1 \ldots b_n}$ defined in (\ref{defC}). 
We will use the symbol $C^\prime_{(n)}$ to denote such a tensor 
when we do not need to specify explicitly its $n$ indices. Notice that, according to the 
Feynman rules, each insertion of a generator on the hypermultiplet loop carries a factor 
of $g$, so that the color factor $C^\prime_{(n)}$ is always accompanied by a factor of 
$g^n$.

\begin{figure}[ht]
	\begin{center}
		\includegraphics[scale=0.55]{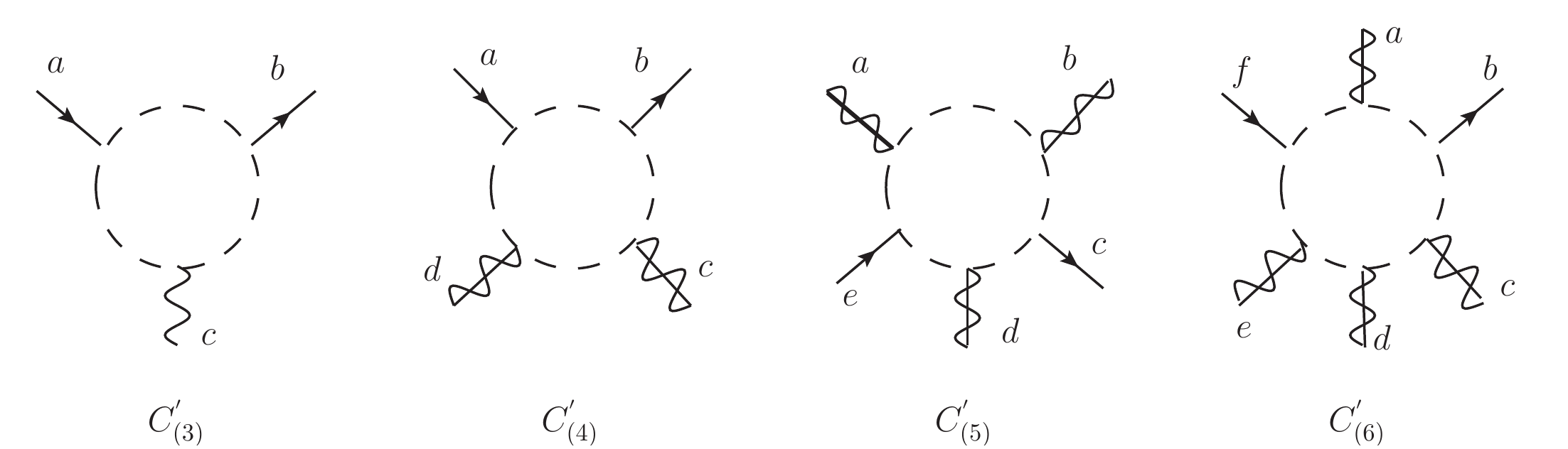} 
	\end{center}
	\caption{Each building block is accompanied by its color coefficient of the type $C^\prime_{(n)}$.
	Here we used a generic dashed lines for hypermultiplets loops. In reality some part of 
	the loop should be dashed and some dotted, in accordance with the Feynman rules.}
	\label{tracesdraw}
\end{figure}

In the difference theory all diagrams up to order $g^6$ can be formed using the building 
blocks depicted in Figure~\ref{tracesdraw}, and suitably contracting the adjoint lines, 
corresponding to $V$ or $\Phi$ propagators, inserted in the loops. 

As a matter of fact, we can also have quartic vertices with two gluon lines inserted in the 
same point along the hypermultiplet loop, each of which comes with a factor of $g^2$ and 
two generators. 
However, for the purpose of identifying the color factors, these contributions do not 
substantially differ from those produced by two separate insertions. Therefore, the possible 
color structures that occur up to the order $g^6$ can all be derived from the diagrams in 
Figure~\ref{tracesdraw}.
Organizing the Feynman diagrams according to their color 
coefficients $C'_{(n)}$ in the way we have outlined facilitates the comparison with the 
matrix model. 

\begin{figure}[htb]
	\begin{center}
		\includegraphics[scale=0.4]{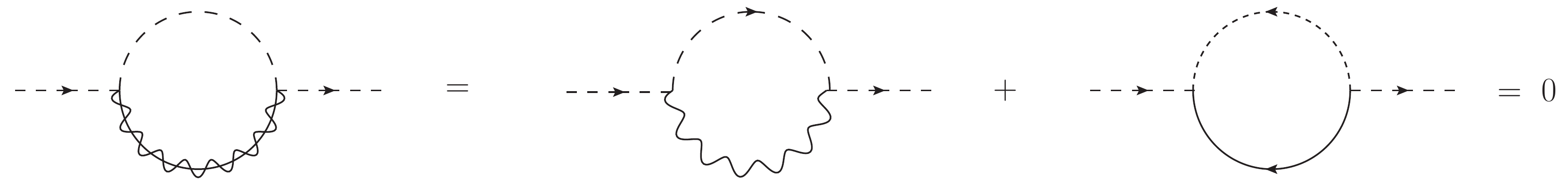}
	\end{center}
	\caption{The one-loop correction to hypermultiplet propagator vanishes.}
	\label{fig:hyper1loop}
\end{figure}

In constructing higher order diagrams we exploit a further simplification: in $\cN=2$ 
theories the one-loop correction to any hypermultiplet propagator vanishes. This is 
illustrated in Figure~\ref{fig:hyper1loop}. Such one-loop corrections cannot therefore 
appear as sub-diagrams of higher loop diagrams. 

\subsubsection*{Two loops} 
\begin{figure}[htb]
	\begin{center}
		\includegraphics[scale=0.6]{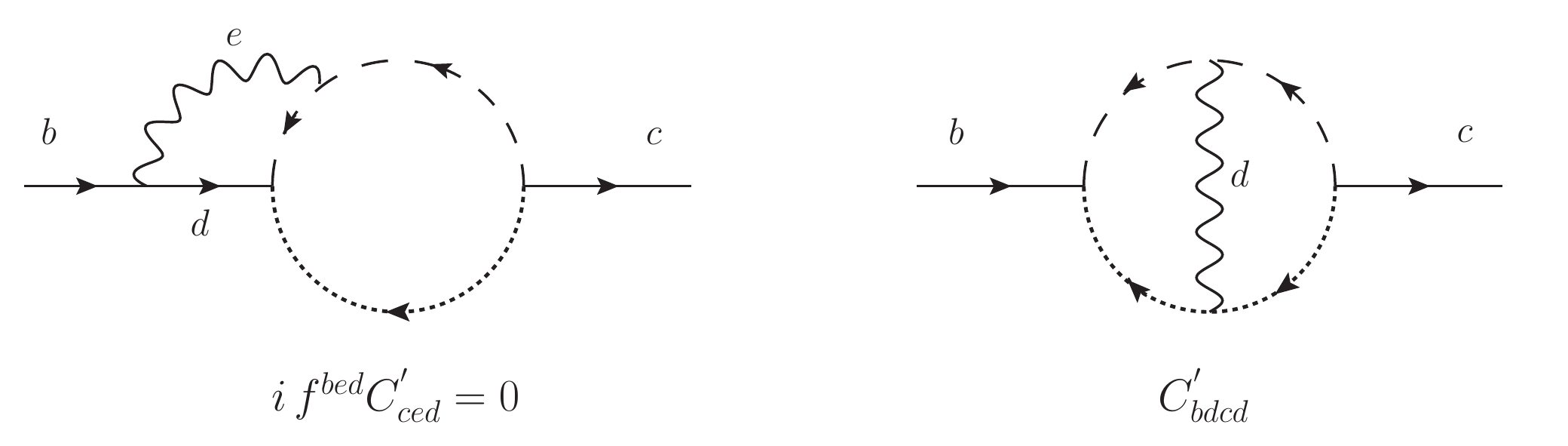}
	\end{center}
	\caption{Two-loop diagrams and their color factors}
	\label{fig:2loopp}
\end{figure}

At order $g^4$ there are two classes of diagrams that may contribute, whose color 
coefficients are proportional to $C^\prime_{(3)}$ or to $C^\prime_{(4)}$. 
The diagrams proportional to $g^3\,C^\prime_{(3)}$ always contain also an 
adjoint vertex proportional to $g$ with which they are contracted. 
This is the case represented on the left in  Figure~\ref{fig:2loopp}. However, due the 
symmetry properties of the tensor $C^\prime_{(3)}$ 
(see  (\ref{C3confsym})), they vanish and one is left only with the diagrams with four 
adjoint insertions in the hypermultiplet loop.

Let us now consider these diagrams. As remarked before, a building block with four adjoint 
lines inserted on the hypermultiplet loop is proportional to $g^4\,C^\prime_{(4)}$, so at
this order we cannot add any other vertices to it. Moreover, there is a unique contraction 
allowed, since each hypermultiplet field has a vanishing one-loop propagator. Thus, the 
only diagram at this order is the one represented on the right in Figure~\ref{fig:2loopp}. 
This has already been computed in \cite{Andree:2010na} (see also \cite{Billo:2017glv}). 
Performing the Grassmann algebra and the momentum integral, we obtain 
a finite result proportional to $\zeta(3)$, which explicitly reads
\begin{equation}
	\label{2loopprop}
		\parbox[c]{.33\textwidth}{\includegraphics[width 
		= .33\textwidth]{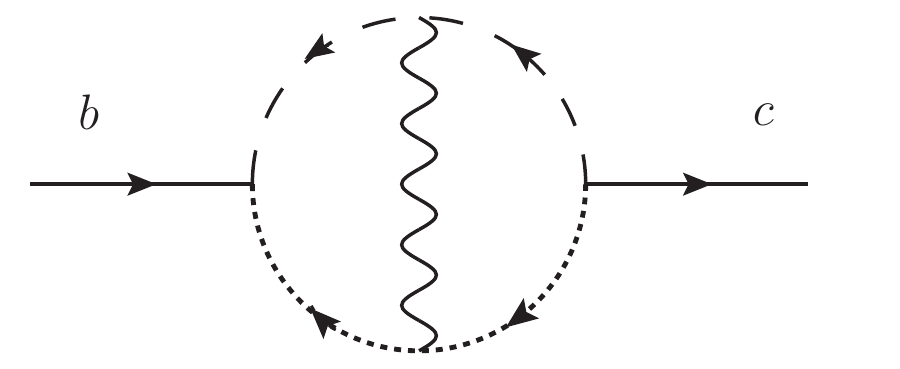}}
		\hspace{-0.4cm}= \frac{1}{q^2} \left(\frac{g^2}{8\pi^2}\right)^2\zeta(3)
		\times  6\, C^{\prime}_{bdcd}~.
\end{equation}
Using the properties of the $C^\prime$-tensors - see in particular (\ref{C4sw1}) and 
(\ref{C4sw12}) - we have
\begin{equation}
6 \,C^{\prime}_{bdcd} = 6 \,C^{\prime}_{(bcdd)}= \cC^{\prime}_{4}\,\delta^{bc}~.
\end{equation} 
Since this is the only correction to the propagator at this order, from (\ref{propPi2}) we find 
\begin{align}
	\label{Pig2}
		\Pi = \left(\frac{g^2}{8\pi^2}\right)^2\zeta(3) \, \cC^{\prime}_{4} + \cO(g^6)~,
\end{align}
in perfect agreement with  the matrix model result reported in (\ref{c4c6delta}) and 
(\ref{Piis}). This is an extension to a generic $\cN=2$ SYM theory of the check 
originally performed in \cite{Andree:2010na} for conformal SQCD.

\subsubsection*{Three loops}
At order $g^6$ many diagrams survive even in the difference theory. Moreover, some
of them can be divergent in $d=4$. However, since we are dealing with conformal
field theories, all divergences cancel when one sums all contributing diagrams. 
Therefore, we can concentrate on extracting the finite part, which the matrix
model result (\ref{corrbcres}) suggests to be proportional to $\zeta(5)$.
Thus we only look for diagrams which provide $\zeta(5)$ contributions, and we check 
that their sum reproduces exactly the matrix model result.

To scan all the possible $\zeta(5)$-contributions we use the same approach we 
applied above. We start from the building blocks in Figure~\ref{tracesdraw} and 
contract their adjoint lines in all the possible ways, introducing new vertices 
when necessary. It is quite simple to realize that many of the diagrams 
that are created in this way have a vanishing color factor.
For example, the diagrams proportional to $C'_{(3)}$ vanish for the same reason
we discussed before. As far as the diagrams with $C'_{(4)}$ are concerned, 
we can discard those containing as a sub-diagram the two-loop contribution 
on the right of Figure \ref{fig:2loopp} since this latter is proportional to $\zeta(3)$, 
and no $\zeta(5)$-contribution can arise from this kind of diagrams. 
All other possible diagrams that one can construct using as building block a sub-diagram
with $C'_{(4)}$ vanish by manipulations of their color factors. 

We are left with diagrams whose color factor is proportional either to $C'_{(5)}$ 
or to $C'_{(6)}$. In the first case, the building block is proportional to $g^5$ and thus we
have insert a further cubic vertex to obtain the desired power of $g$; in the second case,
instead, the building block is already of order $g^6$, and so we can only contract its 
adjoint lines among themselves. 
We have made a systematic search of all diagrams that can be obtained in this way. 
Many of them vanish either because of their color factor or because of 
the $\theta$-algebra, while in other cases the momentum integral does not 
produce any $\zeta(5)$-contribution. 
In the following we list all of the diagrams that \emph{do} yield a 
$\zeta(5)$-term. There are seven such diagrams, named $\cW_{bc}^{(I)}(q)$
with $I=1,\ldots 7$, which are explicitly computed in Appendix~\ref{app:diagrams}.  
Here we simply report the result in a schematic way, writing each of them in the
form
\begin{align}
	\label{resWI}
		\cW_{bc}^{(I)}(q) = 
		-\frac{1}{q^2} \left(\frac{g^2}{8\pi^2}\right)^3 \zeta(5)
		\times x^{(I)}\,\cT_{bc}^{(I)}
\end{align}  
where $\cT_{bc}^{(I)}$ is the color factor, which is in fact proportional to $\delta_{bc}$, 
and $x^{(I)}$ is a numerical coefficient determined by the explicit evaluation of the
 integrals over the loop momenta.
In detail, we have
\begingroup
\allowdisplaybreaks
\begin{align}
	\label{resz5-1}
		\cW_{bc}^{(1)} (q) = 
		\parbox[c]{.28\textwidth}{\includegraphics[width 
		= .28\textwidth]{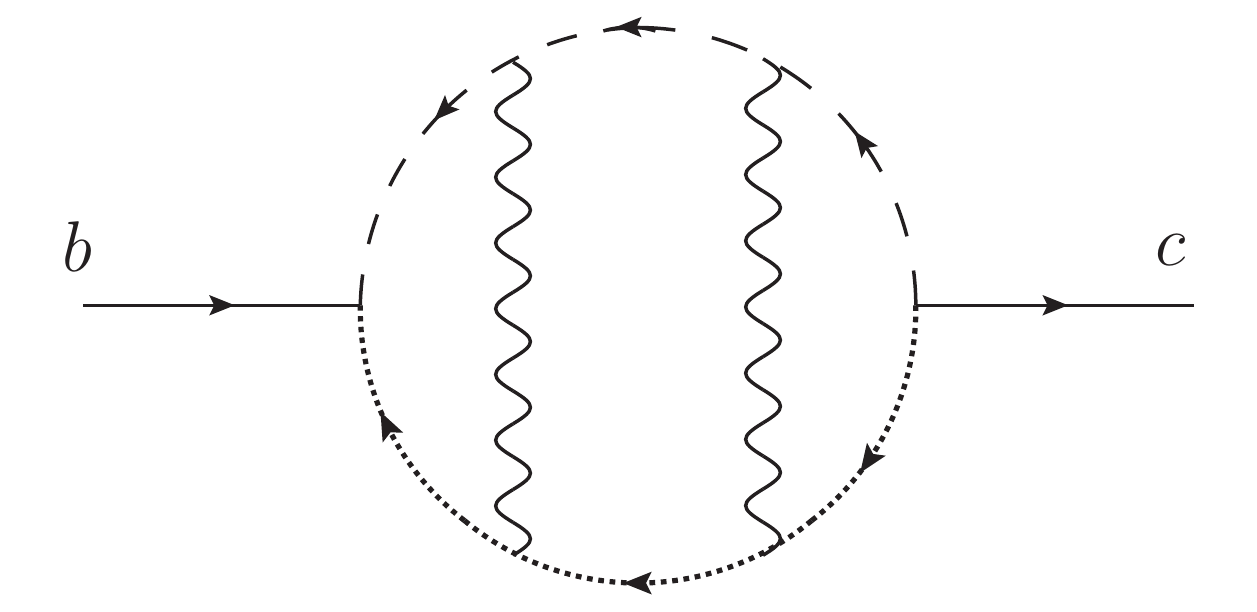}}
		~ \rightarrow ~~ 
		x^{(I)}\,\cT_{bc}^{(1)} & = 
		20\, C^\prime_{bdeced}~,\\
		\cW_{bc}^{(2)} (q) = 		
		\parbox[c]{.28\textwidth}{\includegraphics[width 
		= .28\textwidth]{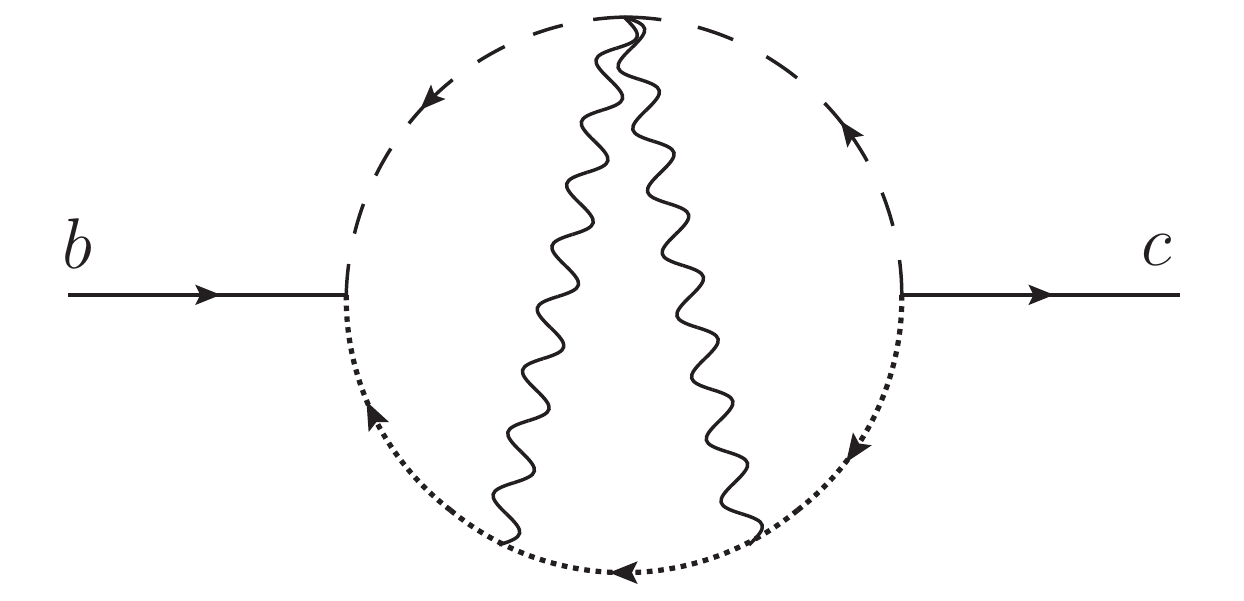}}
		~ \rightarrow ~~  
		x^{(2)}\,\cT_{bc}^{(2)}
		& =  -20\,C^\prime_{bdeced}-20\,C^\prime_{bdecde}~,\\
		\cW_{bc}^{(3)} (q) = 		
		\parbox[c]{.28\textwidth}{\includegraphics[width 
		= .28\textwidth]{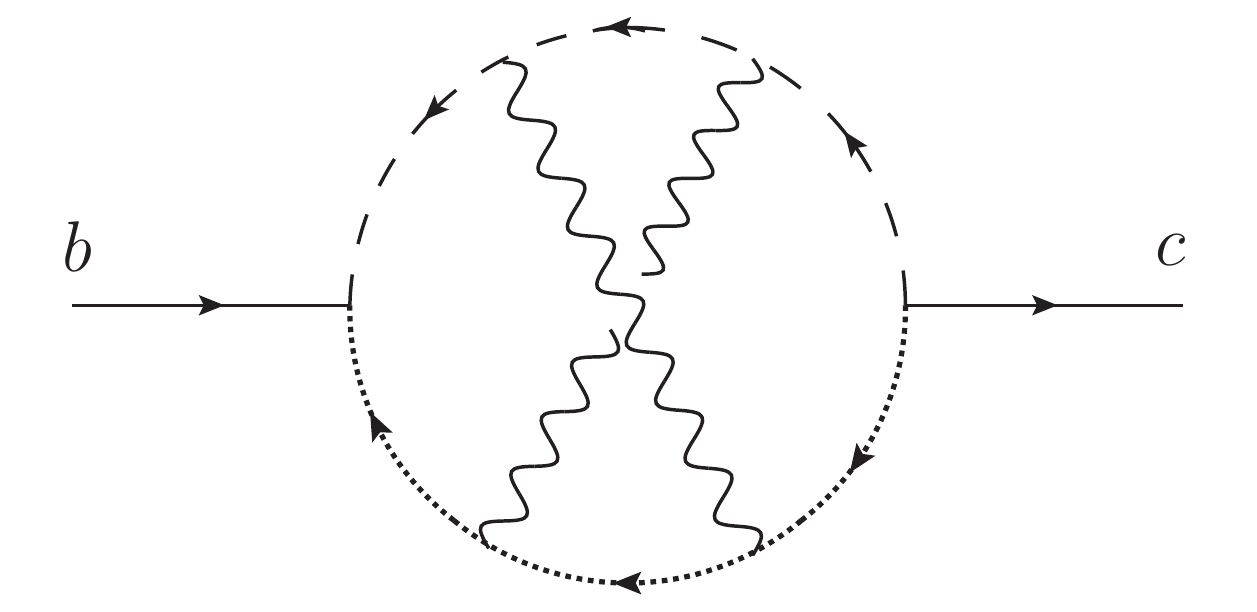}}
		~ \rightarrow ~~  
		x^{(3)}\,\cT_{bc}^{(3)}
		& =  10\, C^\prime_{bdecde}~,\\
		\cW_{bc}^{(4)} (q) = 		
		\parbox[c]{.28\textwidth}{\includegraphics[width 
		= .28\textwidth]{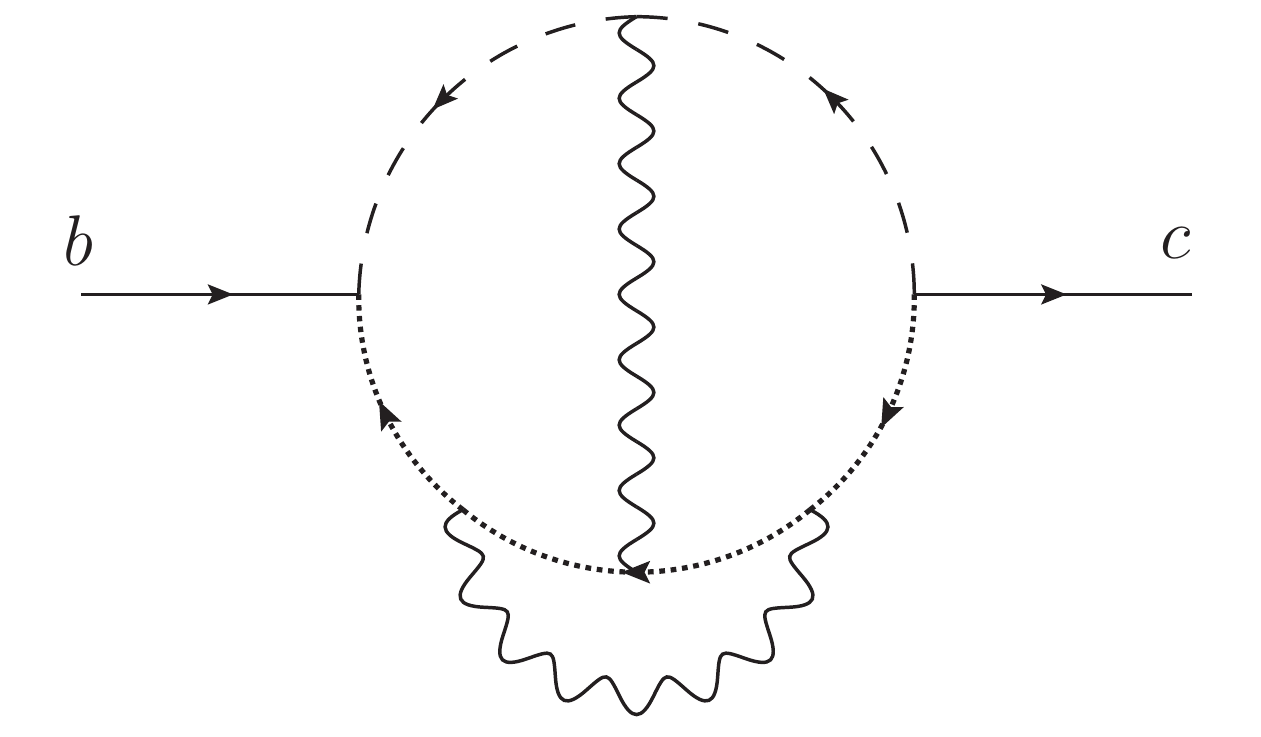}}
		~ \rightarrow ~~  
		x^{(4)}\,\cT_{bc}^{(4)}
		& =  20\,C^\prime_{bdcede}+20\,C^\prime_{bedecd}~,\\
		\cW_{bc}^{(5)} (q) = 		
		\parbox[c]{.28\textwidth}{\includegraphics[width 
		= .28\textwidth]{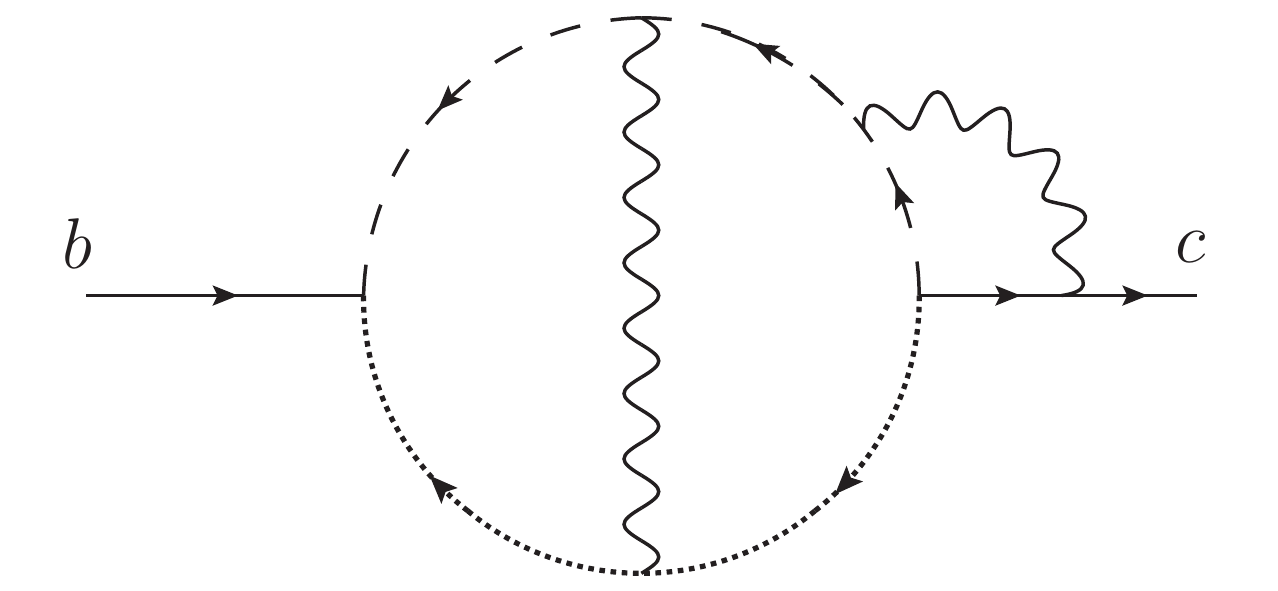}}
		~ \rightarrow ~~  
		x^{(5)}\,\cT_{bc}^{(5)}
		& =  -40\,\ii f_{cef} C^\prime_{bdefd} -40\, \ii f_{bef} C^\prime_{cdefd}~,\\
		\cW_{bc}^{(6)} (q) = 		
		\parbox[c]{.28\textwidth}{\includegraphics[width 
		= .28\textwidth]{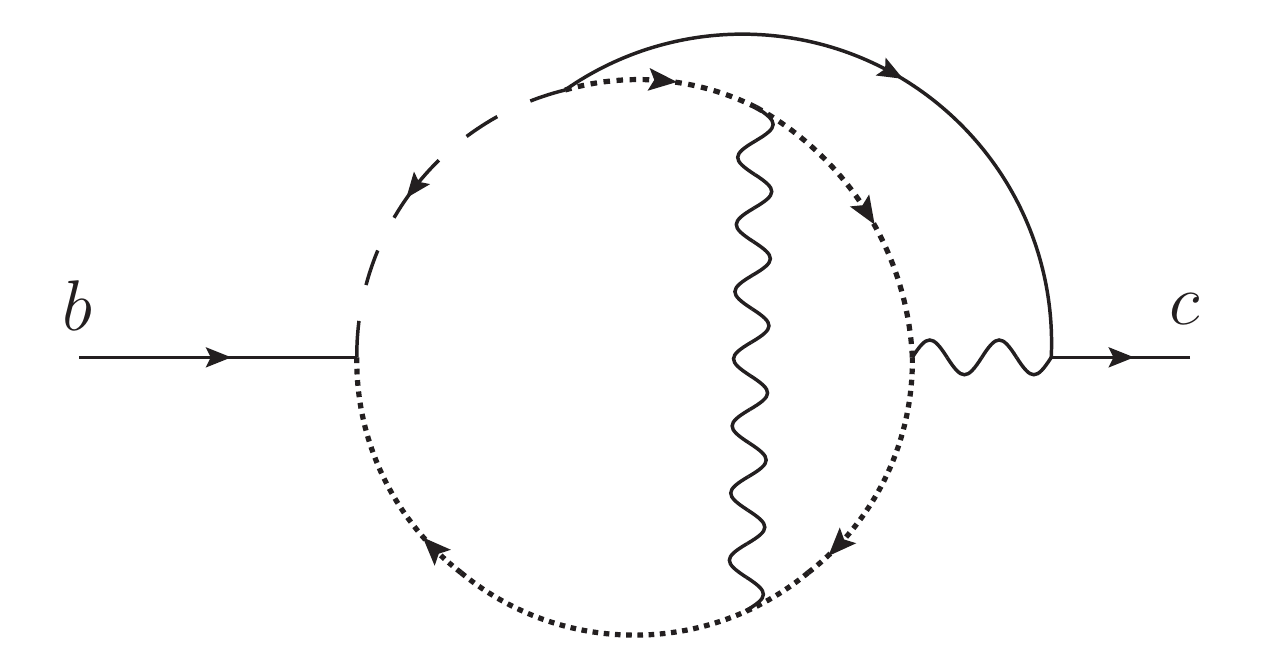}}
		~ \rightarrow ~~
		x^{(6)}\,\cT_{bc}^{(6)}
		& = -20\,\ii f_{ced} C^\prime_{bfdfe} +20\, \ii f_{ced} C^\prime_{befdf}
		\nonumber\\[-5mm]
		&~~~-20\, \ii f_{bed} C^\prime_{cfdfe} +20\, \ii f_{bed} C^\prime_{cefdf}~,\\
		\cW_{bc}^{(7)} (q) = 		
		\parbox[c]{.28\textwidth}{\includegraphics[width 
		= .28\textwidth]{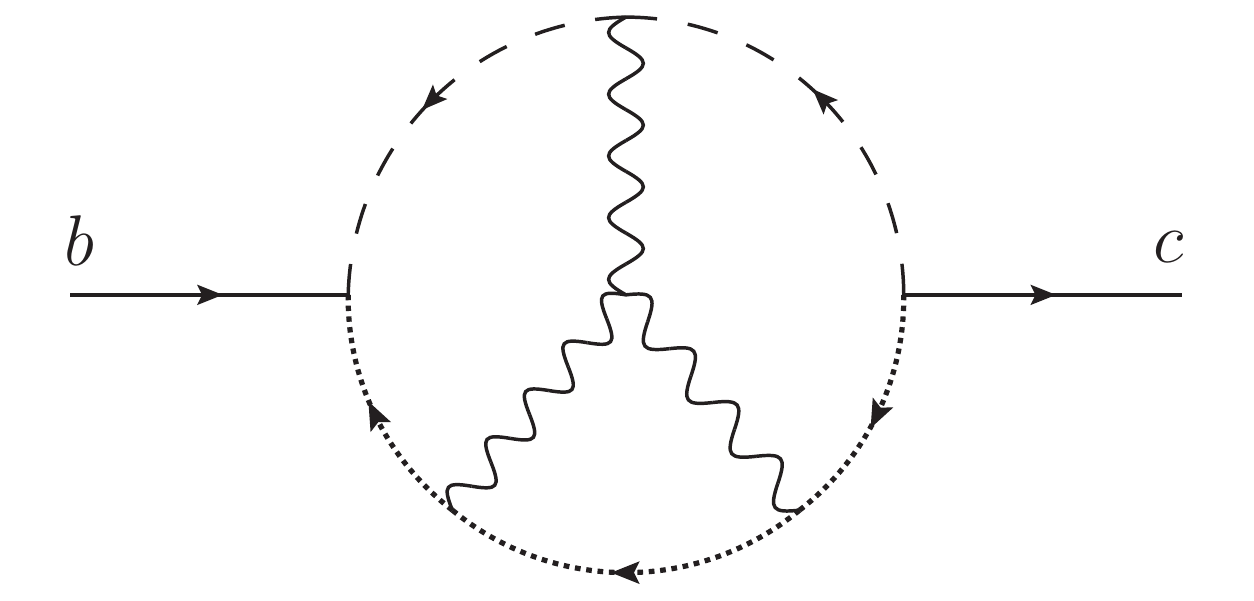}}
		~ \rightarrow ~~  
		x^{(7)}\,\cT_{bc}^{(7)}
		& = 10\, \ii f_{def} C^\prime_{bfecd} + 10\,\ii f_{def} C^\prime_{cfebd}~.
\end{align}
\endgroup
Since each color factor is proportional to $\delta^{bc}$, we can identify terms that are equal up to an exchange of $b$ and $c$. In this way we get
\begin{equation}
	\label{sumTI}
		\sum_{I=1}^7 x^{(I)}\,\cT_{bc}^{(I)} = 
		- 80\, \ii f_{ced} C^\prime_{bfdfe} + 80\, \ii f_{ced} C^\prime_{bfdef} 
		- 10\, C^\prime_{bdecde} 
		+ 40 \,C^\prime_{bdcede} 
		+ 20\, \ii f_{def} C^\prime_{bfecd}~.
\end{equation}
Using the relation (\ref{switchC}), it is easy to see that the first two terms actually 
cancel, and that the remaining ones can be written as follows:
\begin{equation}
	\label{sumTIbis}
		\sum_{I=1}^7 x^{(I)}\,\cT_{bc}^{(I)} = 
		30\, C^\prime_{bdcede} -10 \ii f_{ced} C^\prime_{bfdfe} 
		+ 20\, \ii f_{def} C^\prime_{bfecd}~.
\end{equation}
This expression is apparently different from the color tensor in the $g^6$-term of the matrix model result (\ref{corrbcres}). In fact, the latter contains the totally symmetric 
combination $30 C^\prime_{(bdcede)}$ and does not contain any $C^\prime$ with five indices. However, using again (\ref{switchC}) and the properties of the $C^\prime$ tensors described in Appendix~\ref{app:group}, it is possible to show that the last two terms in (\ref{sumTIbis}) 
precisely symmetrize the first term. 
The total three-loop contribution is therefore
\begin{align}
	\label{sumWI}
		\sum_{I=1}^7 \cW_{bc}^{(I)}(q) &
		=	-\frac{1}{q^2} \left(\frac{g^2}{8\pi^2}\right)^3 \zeta(5)
		\times 30\, C^\prime_{(bdcede)}\notag\\
		& = -\frac{1}{q^2} \left(\frac{g^2}{8\pi^2}\right)^3 \zeta(5)
		\times \cC^\prime_6\,\delta_{bc}~,
\end{align}
where in the last step we used (\ref{c4c6delta}). 
Altogether, adding the two-loop term (\ref{Pig2}), the quantum corrections
of the scalar propagator proportional to $g^4\,\zeta(3)$ and $g^6\,\zeta(5)$ are
\begin{align}
	\label{Pig3}
		\Pi = \zeta(3) \left(\frac{g^2}{8\pi^2}\right)^2 \cC^{\prime}_{4}
		- \zeta(5) \left(\frac{g^2}{8\pi^2}\right)^3 \cC^{\prime}_{6} + \cO(g^8)~. 
\end{align}
This result fully agrees with the matrix model prediction given in (\ref{Piis}).

As already mentioned at the end of Subsection~{\ref{subsec:propagator}}, we observe that
the color tensors $C^\prime_{b_1\ldots b_n}$ and the coefficients $\cC^\prime_n$ can be 
defined for any representation of SU($N$) (or U($N$)). Moreover, the steps that we performed above
to show the agreement with the matrix model predictions only rely on the symmetry/anti-symmetry
properties of these tensors and their group-theory properties, and not on their specific expressions
for the SU($N$) theories with matter in the fundamental, symmetric or anti-symmetric representations. For this reason we believe that the same match could be proved and realized 
also in more general superconformal theories with other gauge groups and matter representations.

\subsection{Supersymmetric Wilson loop}
\label{subsec:swl}
We apply the results for the scalar propagator on the Wilson loop vev, following the same logic as the matrix model case.

We already fix our convention for the Wilson loop shape, by taking the same as the $\cN=4$ case, see eq \eqref{WLdef}. Therefore, the leading order in $g$ is of course the same in $\cN=2$ and $\cN=4$, and there is no $g^2$- contribution to the vacuum expectation value of $W(C)$
in the difference theory. Also at order $g^4$ there are no
contributions in the difference, since the only possible sources for such contributions are
the one-loop corrections to the scalar and gluon propagators, which however vanish for
superconformal theories in the Fermi-Feynman gauge \cite{Grisaru:1979wc,Kovacs:1999rd}, 
see Figure \ref{fig:1loop}. One begins to see a difference between the $\cN=4$ and the conformal
$\cN=2$ results at order $g^6$. Indeed, as we have seen in the previous Section, in a generic
conformal $\cN=2$ theory the propagator of the adjoint scalar 
gets corrected by loop effect starting at order $g^4$.
Due to supersymmetry, also the gluon propagator in the Fermi-Feynman gauge
gets corrected in the same way and thus (\ref{propsAphi}) can be replaced by
\begin{equation}
	\label{propsAphicorr}
		\big\langle \bar\varphi^a (x_1) \varphi^b (x_2) \big\rangle
		= \frac{\delta^{ab}}{4\pi^2 x_{12}^2}\,
		 \big(1+ \Pi\big)~,~~~ 
		\big\langle A_{\mu}^{a} (x_1) A_{\nu}^{b} (x_2)\big\rangle
		= \frac{\delta^{ab} \delta_{\mu\nu}}{4\pi^2 x_{12}^2}\,\big(1+\Pi\big)~,
\end{equation}
where $\Pi$ is the quantity introduced in (\ref{propPi}).

\begin{figure}[htb]
	\begin{center}
		\includegraphics[scale=0.38]{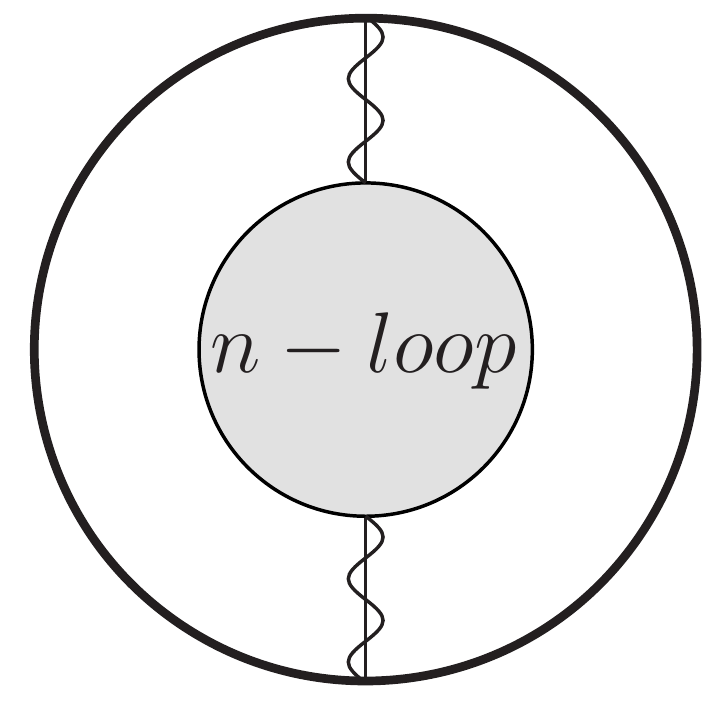}
	\end{center}
	\caption{The graphical representation of the contribution to $\big\langle
	W(C)\big\rangle$ arising from the $n$-loop correction of the gluon and scalar propagators.}
	\label{fig:WLnloops}
\end{figure}

Exploiting this fact, and repeating the same steps as before, we can easily compute the contribution
to the vacuum expectation value of $W(C)$ corresponding to the diagram in Figure~\ref{fig:WLnloops}, which yields a term proportional to $g^{2n+2}\,\zeta(2n-1)$.

Using (\ref{Pig3}), for $n=2$ this calculation yields
\begin{align}
	\label{WC3}
		\frac{g^2 (N^2-1)}{8N} \left(\frac{g^2}{8\pi^2}\right)^2 \zeta(3)\, \cC^{\prime}_4~,
\end{align}
while for $n=3$ it gives
\begin{align}
	\label{WC5}
		-\frac{g^2 (N^2-1)}{8N} \left(\frac{g^2}{8\pi^2}\right)^3 \zeta(5)\,\cC^{\prime}_6~.
\end{align}
Comparing with (\ref{chi3expg}) and (\ref{chi5expg}), we find a perfect agreement with
the matrix model predictions for the lowest order terms in the $g$-expansion of
$\cX_3$ and $\cX_5$. The precise match with the matrix model results
suggests that in the vacuum expectation value of $W(C)$ the terms proportional 
to a given Riemann $\zeta$-value with the lowest power of $g$, namely the terms 
proportional to $g^{2n+2}\,\zeta(2n-1)$, are \emph{entirely} captured by the $n$-th loop
correction of a single gluon or scalar propagator inserted in the Wilson loop. Moreover,
the agreement with the matrix model also suggests that all diagrams contributing to $\big\langle
W(C)\big\rangle$ have an \emph{even} number of legs attached to the Wilson loop.
We shall now check that this is indeed true, at the first relevant orders. 

\subsubsection*{Absence of other contributions} 
\begin{figure}[htb]
	\begin{center}
		\includegraphics[scale=0.38]{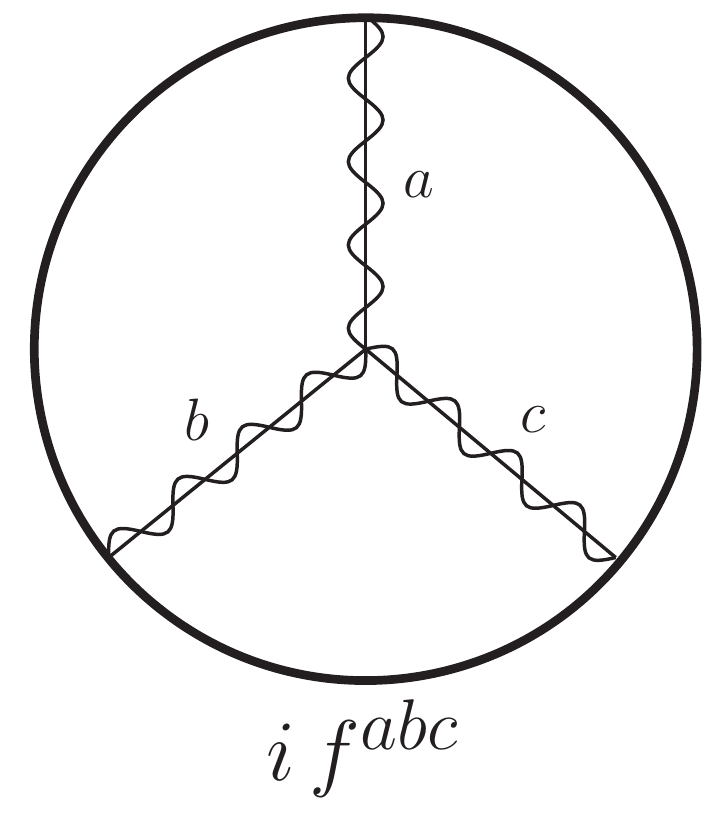}
	\end{center}
	\caption{The vertex correction to $\big\langle W(C)\big\rangle$ 
	in the $\cN=4$ theory at order $g^4$.}
	\label{fig:WLvertex0}
\end{figure}

Let us consider diagrams with three insertions on the Wilson loop contour. 
In the $\cN=4$ theory there is such a diagram already at order $g^4$ which is shown in Figure~\ref{fig:WLvertex0}. Here the internal vertex can be with three gluons or with two scalars and
one gluon. In both cases it carries a color factor proportional to $f_{abc}$.
This contribution has been proven to vanish long ago \cite{Erickson:2000af,Bassetto:2008yf}. The 
cancellation is justified by symmetry properties of the (finite) integral over the insertion points 
along the circular loop.

\begin{figure}[htb]
	\begin{center}
		\includegraphics[scale=0.38]{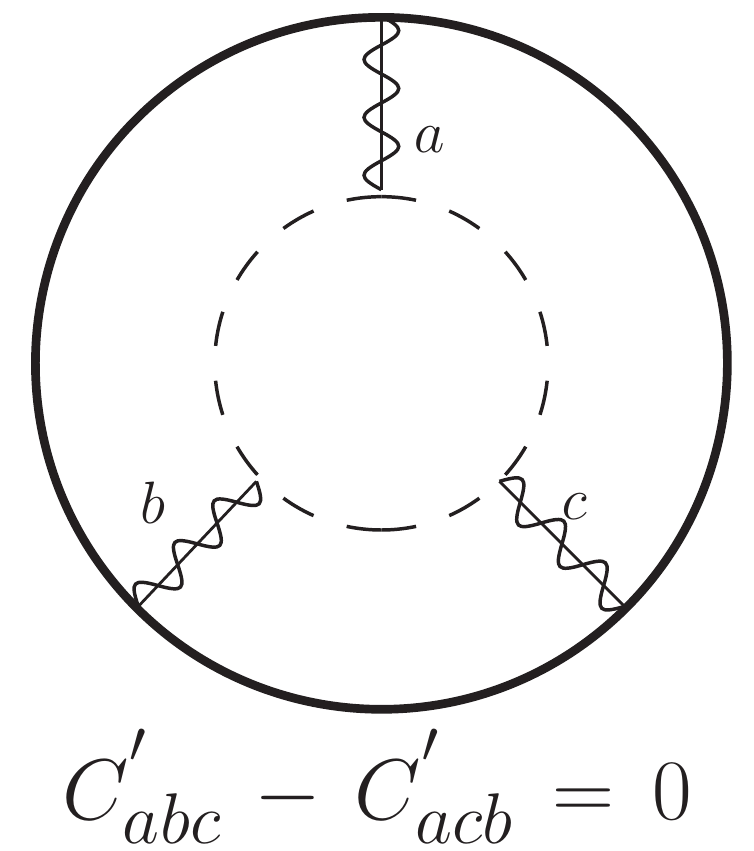}
	\end{center}
	\caption{The one-loop vertex corrections to $\big\langle W(C)\big\rangle$ at $g^6$ 
	order in the difference theory is vanishing.}
	\label{fig:WLvertex1}
\end{figure}

In the difference theory, instead, the first three-leg diagram appears at order $g^6$ 
and is depicted in Figure~\ref{fig:WLvertex1}.
This contribution, however, has a vanishing color factor (see also \cite{Gomez:2018usu}). 
This is due to the different roles of the $Q$ or $H$ superfields, transforming in the 
representation $\cR$, and of the $\widetilde Q$ or $\widetilde H$ ones, transforming 
in the representation $\bar\cR$.
This implies that the color factor is
\begin{align}
\Tr_{\cR}^\prime T^aT^bT^c +\Tr_{\bar\cR}^\prime T^aT^bT^c = C^\prime_{abc} - C^\prime_{acb}~,
\label{Cabc}
\end{align}
which is automatically zero due to the complete symmetry of $C^\prime_{(3)}$ as shown in (\ref{C3confsym}).

\begin{figure}[h]
\begin{center}
\includegraphics[scale=0.38]{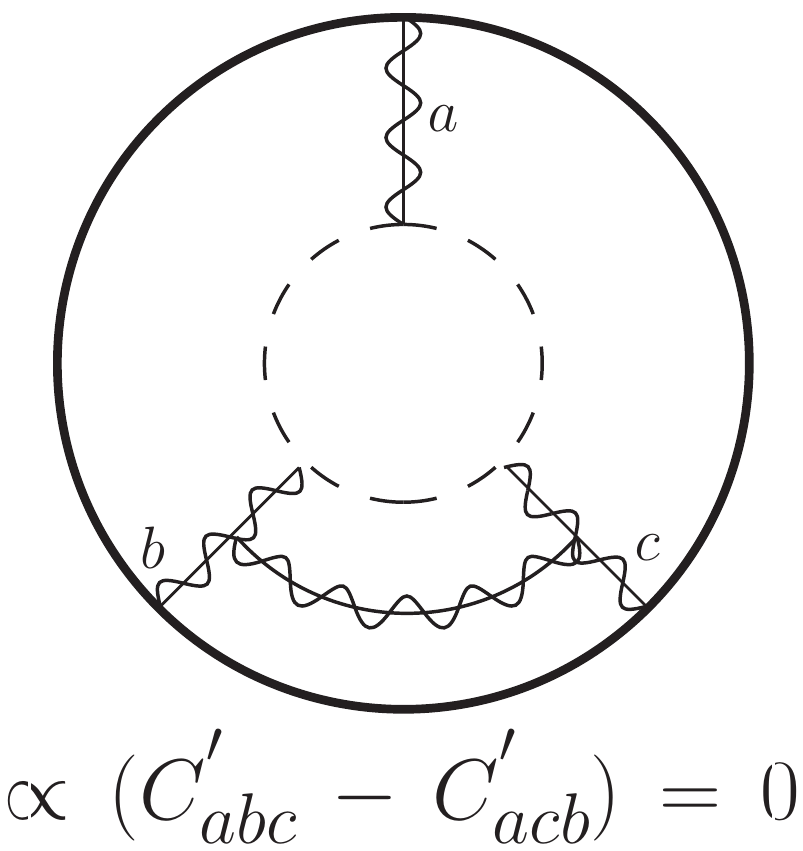}
\hspace{0.5cm}
\includegraphics[scale=0.38]{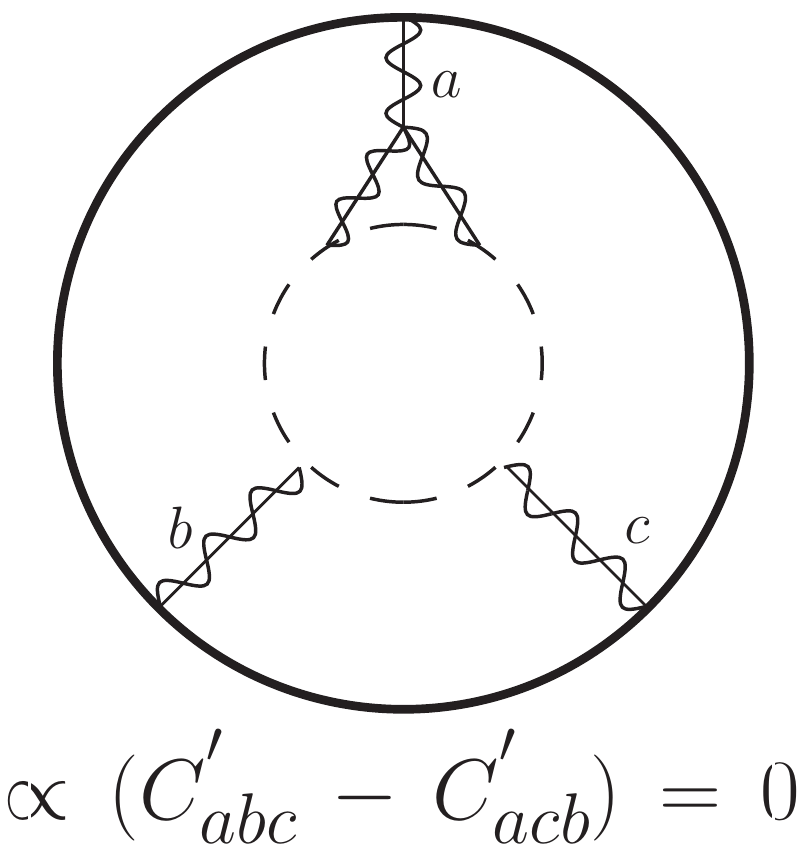}
\hspace{0.5cm}
\includegraphics[scale=0.38]{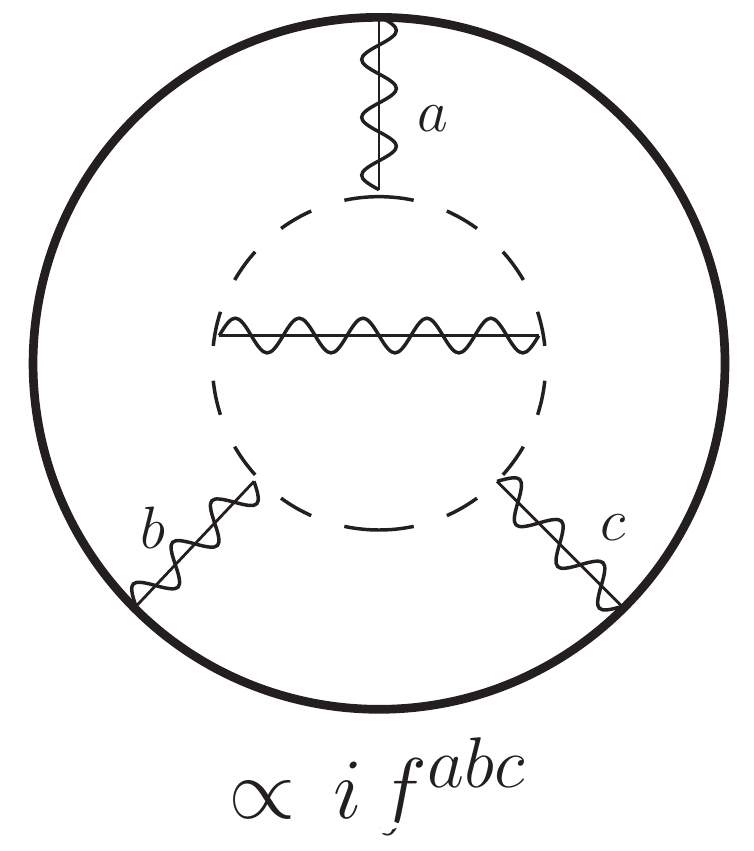}
\hspace{0.5cm}
\caption{Possible two-loop vertex corrections contributing to 
$\big\langle W(C)\big\rangle$ at order $g^8$ together with their color factors.}
\label{fig:WLvertex2}
\end{center}
\end{figure}

At order $g^8$ there are several possible three-leg diagrams. Again, if we classify them in terms 
of their color factor, we can distinguish three classes, 
represented in Figure \ref{fig:WLvertex2}. The first two have again a color factor proportional to the combination
(\ref{Cabc}) which vanishes, while the last type has a color factor proportional to $f_{abc}$. 
We have not performed a detailed calculation of this class of diagrams, but it is natural 
to expect that they cancel by a mechanism analogous to the one at work 
in the $g^4$ diagrams of the $\cN=4$ theory represented in Figure \ref{fig:WLvertex0}, since they have the same color structure and symmetry properties.

This concludes our analysis on the check of the agreement between the matrix model prediction and
the field theory results of $\big\langle W(C)\big\rangle$ at order $g^8$.

%%%%%%%%%%%%%%%%%%%%%%%%%%%%%%%%%%%%%%%%%%
%%%%%%%%%%%%%%%%%%%%%%%%%%%%%%%%%%%%%
%%%%%%%%%%%%%%%%%%%%%%%%%%%%%%%%

\part{Wilson loop observables}
\chapter{Defect Conformal field theories}\label{chap:4}
In this Chapter we study modifications of a CFT which preserve a large subgroup of the conformal group (see for instance \cite{Cardy:1984bb,McAvity:1995zd, Kapustin:2005py, Billo:2016cpy}). We introduce the notion of \textit{conformal defect}, which can be seen as an extended operator inserted into the theory; the physical meaning of this operation can be found in checking the behavior of a theory in presence of an external probe. Indeed the goal of these Chapter is to outline some techniques, in order to consider the insertion of a Wilson loop as a conformal defect and therefore to compute some specific sets of observables in presence of such a probe.

The structure of this Chapter is the following. In Section \ref{X13} we describe the residual conformal symmetry after the defect insertion. In Section \ref{sec::4.2} we outline the methods to fix the correlation functions in this case, and we provide explicit examples of correlation functions that we will use in the following Chapters. In Section \ref{X14} we discuss one of the key points of the insertion of a defect in the vacuum, namely the energy-momentum conservation. Finally in Section \ref{sec4:WLdefect} we introduce the Wilson operator as a specific example of conformal defect, following the approach that we will keep in the next Chapters.

\section{Restricted conformal group}\label{X13}
We consider a $d$-dimensional conformal field theory defined on a flat Euclidean space $ \mathbb{R}^{d} $ described by coordinates $ x^{\mu} = (x^{a},x^{i})$, where $a=1, \dots, p $ and $ i=1,\dots, d-p $, with the insertion of a $ p $-dimensional hyperplane fixed in $ x^{i} = 0 $. This means that $ x^{i} $ is the perpendicular distance of the defect from a general point $ x^{\mu} $ belonging to the bulk. The quantity $ d-p \equiv q $ is known as the \textit{codimension}.

The original conformal symmetry SO$(1,d+1) $ is broken, then we have to restrict the conformal group to those transformation leaving $ x^{i} = 0 $ invariant. This subgroup is given by SO$(p+1,1) \times $SO$(q) \subset $ SO$(1,d+1) $, namely a conformal group along the defect plus rotations of the transverse directions with respect to the defect. The generators of this subgroup are then:
\begin{equation} \label{d1}
 \begin{split}
\textrm{Parallel translations} \hspace{0.5 cm} &P_{a} = -i\partial_{a} \\
\textrm{Parallel rotations} \hspace{0.5 cm} &M_{a b} = i(x_{a}\partial_{b}-x_{b}\partial_{a}) \\
\textrm{Dilatations} \hspace{0.5 cm} &D = -ix^{\mu}\partial_{\mu} \\
\textrm{SCT} \hspace{0.5 cm} &K_{a} = -i(2x_{a}x^{\nu}\partial_{\nu}-x^{2}\partial_{a}) \\
\textrm{Transverse rotations} \hspace{0.5 cm} &M_{i j} = i(x_{i}\partial_{j}-x_{j}\partial_{i}) \\
\end{split}
 \end{equation}
where the special conformal transformations (SCT) are simply the ordinary SCT restricted to the values preserving the defect, namely $ b^{\mu} = (b^{a},0) $. The algebra associated to (\ref{d1}) is given by:
\begin{equation}
\begin{split}
&[K_{a}, P_{b}] = 2i(\delta_{a b}D-L_{a b})\\
&[P_{c},M_{a b}]=i(\delta_{c a}P_{b}- \delta_{c b}P_{a})\\
&[K_{c},M_{a b}]= i(\delta_{c a}K_{b}- \delta_{c b}K_{a})\\
&[M_{a b},M_{c d}] = i(\delta_{a d} M_{b c} + \delta_{b c} M_{a d} - \delta_{a c} M_{b d} - \delta_{b d} M_{a c}) \\
&[D,P_{a}]=iP_{a} \\
&[D,K_{a}]=-iK_{a} \\
&[P_{a}, P_{b}]=[K_{a}, K_{b}]=[D,M_{a b}]=0 \\
&[M_{i j},M_{k l}] = i(\delta_{i l} M_{j k} + \delta_{j k} M_{i l} - \delta_{i k} M_{j l} - \delta_{j l} M_{i k}) \\
&[P_{a}, M_{i j}]=[K_{a}, M_{i j}]=[D,M_{i j}]=0 
\end{split}
\end{equation}

The main feature of this algebra rules is that the original rotation group SO$(d) $ is broken to SO$(p) ~\times $ SO$(q) $, and we can consider SO$(q) $ as an "internal symmetry" group on the defect: thus from the defect point of view we remain with a $ p $-dimensional conformal field theory (CFT$_{p}$) with a SO$(q) $ flavor group. However, note that there is an important difference between this theory and a general conformal field theory with an internal symmetry group: in such CFTs generally there exists a stress-tensor, while no such defect stress-tensor is available in the spectrum of defect operators, as we will see in the following. 

\subsubsection*{Bulk and defect operators}
Apart from the symmetry breaking pattern, a DCFT is defined by the CFT data that specify the correlation functions of local operators. In this context, we have to distinguish between the insertion of operators in the bulk or on the defect. \\
Bulk operators  depend on the bulk coordinates $ x^{\mu} $ and have the properties we described in Section \ref{X3}. In general they are rank-L tensors of SO$(d)$, of the form $ \mathcal{O}_{\Delta}^{\mu_{1}, \dots, \mu_{L}}(x^{\mu}) $.
In this Chapter we denote defect operators as $ \hat{\mathcal{O}} $. These excitations live on the defect ($ x^{i}=0 $) and only admit parallel dependence $ x_{a} $. Moreover, they have "parallel" and "transverse" components, since defect operators are rank-l tensors of SO$(p) $ and rank-s tensors of their flavor symmetry group SO$(q) $. Hence in general they take the form $ \hat{\mathcal{O}}_{\hat{\Delta}}^{a_{1}, \dots, a_{l}i_{1}, \dots, i_{s}}(x^{a}) $.  

And now we have to reckon with this important distinction to discuss the form of the Operator Product Expansion in a DCFT. Clearly the fusion of primary operators in the bulk is a local property and then is unaffected by the presence of the defect. Hence we still have the usual bulk OPE defined in (\ref{c22}):
\begin{equation}
 \mathcal{O}_{i}(x_{i})\mathcal{O}_{j}(x_{j}) \stackrel{x_{i} \rightarrow x_{j}}{=} \sum_{\mathcal{O}} c_{ij\mathcal{O}}\frac{1}{x_{ij}^{\Delta_{i}+\Delta_{j}-\Delta_{\mathcal{O}}}} C_{\mathcal{O}} (x_{i}-x_{j}, \partial_{j}) \mathcal{O}(x_{j}).
\end{equation}
However, as we saw the defect also possesses local excitations $ \hat{\mathcal{O}} $ not related by symmetry to the bulk ones and when a bulk excitation is brought close to the defect it becomes indistinguishable from a defect excitation. This process is captured by a new expansion, the \textit{bulk-to-defect OPE}, which will be singular when the bulk operator approaches to the defect:
\begin{equation}\label{d2}
\mathcal{O}(x^{a}, x^{i}) = b_{\mathcal{O}\hat{\mathcal{O}}} \frac{1}{\norm{x^{i}}^{\Delta_{\mathcal{O}}-\hat{\Delta}_{\hat{\mathcal{O}}}}} C_{\hat{\mathcal{O}}} (|x^{i}|,\partial^{2}_{a})\hat{\mathcal{O}}(x^{a}). 
\end{equation}
where again the differential operator $ C_{\hat{\Delta}} $ creates the whole conformal family belonging to the primary defect operator $ \hat{\mathcal{O}} $. Note that among the $b_{\mathcal{O}\hat{\mathcal{O}}}$ coefficients we find $b_{ \mathcal{O}\hat{1}}$ which plays a special role, since it allows bulk operators to acquire a non trivial one point function:
\begin{equation}
\langle O(x)= \frac{A_O}{\norm{x^i}^\Delta}~.
\end{equation}

\section{DCFT correlators}\label{sec::4.2}
Considering the p-dimensional defect as an extended operator $ \mathcal{D} $ placed in the vacuum of a CFT, the correlation functions of the theory are intended to be measured in presence of this extended operator, whose expectation value is divided out. In general we can define a correlator with bulk and defect insertions as follows:
\begin{equation} \label{d3}
 \langle \mathcal{O}_{1} (x_{1}) \dots \hat{\mathcal{O}}_{1} (x_{1}^{a})\dots \rangle_{\mathcal{D}} \equiv
 \frac{1}{\langle \mathcal{D}\rangle_{0}} \langle \mathcal{O}_{1} (x_{1}) \dots \hat{\mathcal{O}}_{1} (x_{1}^{a})\dots 
 \mathcal{D} \rangle_{0}~,
 \end{equation}
where the subscript 0 denotes expectation values taken in the conformal invariant vacuum. \\

\subsection{Preliminary: tensors as polynomials on the light-cone}
We review a method to deal with correlation functions of operators with spin, using the embedding space formalism we described in Section \ref{X4}. Let us start with  a set up without any defect. We consider operators that are symmetric, traceless tensors, and can be encoded in polynomials by introducing an auxiliary vector $z^\mu$:
\begin{equation}\label{4.2.2}
F_{\m_1,\dots \m_J} (x)\rightarrow F_J(x,z) =z^{\m_1}\dots z^{\m_J}F_{\m_1,\dots \m_J} (x)~, 
\end{equation}
with $z^2=0$ to enforce tracelesness of $F$. The index structure can be recovered by using the Todorov differential operator:
\begin{align}
\label{todorov}
\cD_{\mu}=\Big(\frac{d-2}{2}+z\cdot \frac{\pa}{\partial z} \Big)\frac{\partial}{\partial z^{\mu}}-\frac12 z_{\mu} \frac{\partial^2}{\partial z \cdot \partial z}~.
\end{align}
If we apply this operator $J$ times we can free all the indices:
\begin{equation}
F_{\m_1,\dots \m_J} = \frac{\cD_{\m_1} \dots \cD_{\m_J} F_J(x,z)}{J! \left( \frac{d-2}{2}\right)_J}
\end{equation}
We move to the embedding space, where any field $F_{\m_1,\dots \m_J}(x)$ can be obtained from a field $F_{\cM_1,\dots \cM_J}(P)$ by restricting it to the Poincar\`e section. We also want $F_{\cM_1,\dots \cM_J}(P)$ to be
\begin{itemize}
\item
homogeneous of degree $-\Delta$, \textit{i.e.} $F_{\cM_1,\dots \cM_J}(\lambda P)=\lambda^{-\Delta}F_{\cM_1,\dots \cM_J}(P)$, $\lambda>0$;
\item
transverse, \textit{i.e.} $P_{\cM_1}F^{\cM_1,\dots \cM_J}(P)=0$,
\end{itemize} 
which guarantees $F_J$ to project to a primary operator in physical space. We can encode symmetric traceless tensors in embedding space in a polynomial, as before:
\begin{equation}\label{4.2.5}
F_{\cM_1,\dots \cM_J}(P)\rightarrow F_J(P,Z) \equiv Z^{\cM_1}\dots Z^{\cM_J}F_{\cM_1,\dots \cM_J}(P)~, \hspace{1cm} \begin{cases} Z^2=0 \\ P\cdot Z =0 \end{cases}~.
\end{equation}
where the two conditions preserve tracelessness and transversality. the two conditions \eqref{4.2.5} and \eqref{4.2.2} agree if
\begin{equation}
Z=(0,2x\cdot z, x^\m)~.
\end{equation}
In general it is convenient to write polynomials in the variable $Z$ and constrain the coefficients such that the polynomials satisfy the required properties. We can rephrase the transversality condition as:
\begin{equation}\label{4.2.7}
F_J(P,Z+\alpha P)= F_J(P,Z)~, \hspace{0.5cm} (\forall\alpha)
\end{equation}
 In particular we define the transverse tensors that can be used as building blocks. If no defect is present, there is only one tensor with these features:
\begin{equation}\label{4.2.6}
C_{\cM\cN} = Z_\cM P_\cN-Z_\cN P_\cM~.
\end{equation}
Finally, let us point out some considerations about conserved tensor. The conservation condition in physical space can be written using the Todorov operator \eqref{todorov}:
\begin{equation}
\partial^\m \cD_\m T(x,z)=0~.
\end{equation}
This condition is consistent if the dimension of $T$ is $\Delta=d-2+J$. This conservation law has consequances on the correlators, so it is convenient to rephrase it on the light cone, where it reads:
\begin{equation}
\partial^\cM \cD_\cM \hat{T}(P,Z)=0~.
\end{equation}
$\cD_\cM$ has the same expression as \eqref{todorov} with $z$ replaced by $Z$, while $\hat{T}(P,Z)$ is obtained from $T(x,z)$ with the conditions $Z^2=Z\cdot P=0$.

\subsubsection*{Generalization to mixed symmetry case}
We consider the case of a tensor in a mixed symmetry representation of the orthogonal group SO$(d)$. We briefly outline the main idea (for a complete and general derivation, see \cite{Lauria:2018klo}), but we especially concentrate on the  case of a rank 2 antisymmetric operator that we will encounter in Section \ref{sec:6.4}. \\
A tensor $t_\ell$ in an irreducible representation $\ell$ is described by a Young tableau, with indices in each box. The indices in the rows are symmetrized, those in the columns are anti-symmetrized. To make these operation manifest we contract all the indices of the $i$-th row with the same polarization vector $z^{(i)}$:
\begin{align}\label{YoungGeneral}
t_\ell (z) =t_\ell \left(\begin{footnotesize}
\begin{ytableau}z^{(1)} &\dots&\dots&\dots&\dots&\dots& z^{(1)} \cr z^{(2)} & \dots & \dots & \dots & z^{(2)} \cr \vdots & \vdots & \vdots & \vdots \cr z^{(k)} & \dots & z^{(k)} \cr \end{ytableau}\end{footnotesize}\right)= t_\ell \left(\begin{footnotesize}
\begin{ytableau}\m_1^1 &\dots&\dots&\dots&\dots&\dots&\m_{l_1}^1 \cr \m_1^2 & \dots & \dots & \dots & \m_{l_2}^1 \cr \vdots & \vdots & \vdots & \vdots \cr \m_1^k & \dots & \m_{l_k}^k \cr \end{ytableau}\end{footnotesize}\right)~~\prod_{i=1}^{k} z^{(i)}_{\m^i_1}\cdots z^{(i)}_{\m^i_{\ell_i}}
\end{align}
The result is a polynomial $t_\ell (z)$ which satisfies the following properties:
\begin{itemize}
\item
$t_\ell (z)$ is defined on the subspace $(z^{(i)}\cdot z^{(j)})=0$
\item
$t_\ell (z)$ has homogeneity $\ell_i$ for all the $z^{(i)}$, so ti satisfies:
\begin{equation}
z^{(i)}\cdot \partial_{z^{(i)}}\, t_\ell (z) = \ell_i\, t_\ell (z)~.
\end{equation}
\item
$t_\ell (z)$ is transverse, so it satisfies:
\begin{equation}\label{4.2.12}
z^{(i)}\cdot\partial_{z^{(j)}}\, t_\ell (z) =0~, \hspace{0.5cm} \forall j>i~.
\end{equation}
\item
To recover the original tensor one needs to apply the Todorov operator \eqref{todorov} according to the tableau structure \eqref{YoungGeneral}.
\end{itemize}  

The way to lift mixed symmetry operators to the embedding space follows in a straightforward way. We encode the operator in a polynomial:
\begin{align}
O(P,Z^{(i)}) \equiv O(P) \left(\begin{footnotesize}
\begin{ytableau}\cM_1^1 &\dots&\dots&\dots&\cM_{l_1}^1  \cr \vdots & \vdots & \vdots & \vdots \cr \cM_1^k & \dots & \cM_{l_k}^k \cr \end{ytableau}\end{footnotesize}\right)~~\prod_{i=1}^{k} Z^{(i)}_{\cM^i_1}\cdots Z^{(i)}_{\cM^i_{\ell_i}}~,
\end{align}
where the auxiliary vectors satisfy:
\begin{equation}
P\cdot P=0~, \hspace{0.5cm} P\cdot Z^{(i)}=0~, \hspace{0.5cm} Z^{(i)}\cdot Z^{(j)}=0~,
\end{equation}
and the operators satisfy the following transversality and scaling conditions:
\begin{align}
&P\cdot\partial_{Z^{(j)}}\, O(P,Z^{(i)})=0~, \hspace{1.2cm} Z^{(k)}\cdot\partial_{Z^{(j)}}\, O(P,Z^{(i)}) =0~, \hspace{0.3cm} \forall k>j~,\notag\\
&O(\alpha P,\beta_i z^{(i)})=O(P,Z^{(i)})\, \alpha^{-\Delta} \prod_{i=1}^{[d/2]}\beta_i^{\ell_i}~.
\end{align}
Thanks to the linear action of SO$(1,d+1)$, we can think the operator $O(P,Z^{(i)})$ as contracted with antisymmetric tensors of the form
\begin{equation}
C_{P~~\cM_1,\dots,\cM_m}^{(m)} \equiv P^{\phantom{x}}_{[\cM_1}~Z^{(1)}_{\cM_2}\cdots~Z^{(m-1)}_{\cM_m]}~, \hspace{0.5cm} m=1,\dots,\left[\frac{d}{2}\right]+1~.
\end{equation}
which is the antisymmetric analogous of \eqref{4.2.6}.

We do not want to provide a detailed analysis of this construction, we simply extract the cases we are interested in.
First of all note that the totally symmetric tensor defined in \eqref{4.2.2} corresponds to a \eqref{YoungGeneral} with a single row Young diagram: its polynomial has been defined in terms of a single auxiliary vector $z^{(1)}_\mu\equiv z_\m$.\\
We provide a more interesting example, namely a totally antisymmetric rank-2 tensor $H_{\m\n}$. In this case the Young diagram is simply $\Yasymm$, and following \eqref{YoungGeneral} we need two auxiliary vectors:
\begin{equation}\label{4.2.18}
H_{\m\n}(x)\rightarrow H(x,z^{(1)},z^{(2)})= z^{(1)~[\m}z^{(2)~\n]} H_{\m\n}(x)~.
\end{equation}
When we uplift this operator in the embedding space we get
\begin{equation}\label{4.2.19}
H(P,Z^{(1)},Z^{(2)}) =Z^{(1)\,[\cM_1}Z^{(2)\,\cM_2]} H_{\cM_1,\cM_2}(P)
\end{equation} 
and we constrain its correlation functions using the structure associated to the rank-2 antisymmetric tensor \eqref{4.2.18}, which is simply:
\begin{equation}\label{4.2.20}
C^{\phantom{x}}_{\cM_1,\cM_2,\cM_3} \equiv P^{\phantom{x}}_{[\cM_1}~Z^{(1)}_{\cM_2}~Z^{(2)}_{\cM_3]}~.
\end{equation}

\subsection{Defect tensorial structures}
In presence of a defect, defect operators carry both SO$(p)$ and SO$(q)$ quantum numbers, corresponding to transverse ($s$) and parallel ($j$) spins. We can still encode spinning operators into polynomials, using two auxiliary variables ($w^i$ and $z^a$) associated to transverse and parallel spin respectively. Again we impose conditions $w^iw_i=z^az_a=0$ to have symmetric traceless representations of both SO$(p)$ and SO$(q)$. To remove the polarization vectors we use two kinds of Todorov operators:
\begin{align}
\label{todorovzw}
\cD_{a}&=\Big(\frac{p-2}{2}+z^b \frac{\pa}{\partial z^b} \Big)\frac{\partial}{\partial z^{a}}-\frac12 z_{a} \frac{\partial^2}{\partial z^b \partial z_b}~,\notag \\
\cD_{i}&=\Big(\frac{q-2}{2}+w^j \frac{\pa}{\partial w^j} \Big)\frac{\partial}{\partial w^i}-\frac12 w_i \frac{\partial^2}{\partial w^j \partial w_j}~.
\end{align}
In the embedding space, we split the coordinates in two sets, distinguishing the ``parallel" directions denoted by $A,B,\dots$ indices, and ``orthogonal" directions denoted by $I,J,\dots$. SO$(1,p+1)$ and SO$(q)$ act on these sets respectively:
\begin{equation}\label{4.2.22}
\cM=(A,I)~, \hspace{1.5cm} A=0,1,\dots p+1~, ~~~I=p+2,\dots ,d+1~.
\end{equation}
The symmetry is still linearly realized in embedding space, we simply have to build two scalar products:
\begin{equation}\label{4.2.23}
P\bullet Q=P^A\eta_{AB}Q^B \hspace{1cm} P\circ Q= P^I\delta_{IJ} Q^J
\end{equation}
For the symmetric tensors case these are the only structures possible. Since bulk insertions still satisfy the conditions $P^2=Z^2=Z\cdot P=0$, only a subset of the scalar products \eqref{4.2.23} is independent
\begin{equation}
P\bullet P =- P\circ P~, \hspace{0.5cm} Z\bullet Z=-Z\circ Z~, \hspace{0.5cm} Z\bullet P = -Z\circ P~.
\end{equation}
Many correlation functions in the embedding space which satisfy transversality can be written in terms of the broken transverse tensor \eqref{4.2.6}. Among the three possible structures $C^{AB},~C^{IJ},~C^{AI}$, only the last one
\begin{equation}\label{4.2.14}
C^{AI}= P^AZ^I-P^IZ^A
\end{equation}
will be necessary for bulk correlation functions with a non trivial tensorial structure. Instead, $C^{AB}$ will define the transversality rule for defect operators, which are simply defined as $p+1$-dimensional light cone extensions of operators defining a CFT$_p$. 

\subsubsection*{Mixed symmetry case}
The set of all the structures arising for a generic bulk or defect operator can be implemented using parallel or orthogonal projection. the general procedure is quite involved and is thoroughly described in Section 3 of \cite{Lauria:2018klo}. Here we simply extract the case we are interested in, namely the rank-2 antisymmetric tensor whose expression in the embedding space is \eqref{4.2.19}. The structures that can enter in the game (in particular in its one-point function in presence of a defect) correspond to the projection of \eqref{4.2.20} along the parallel and orthogonal directions:
\begin{equation}\label{4.2.26}
C^{\phantom{x}}_{A_1,A_2,A_3} = P^{\phantom{x}}_{[A_1}~Z^{(1)}_{A_2}~Z^{(2)}_{A_3]}~, \hspace{0.5cm}C^{\phantom{x}}_{I_1,I_2,I_3} = P^{\phantom{x}}_{[I_1}~Z^{(1)}_{I_2}~Z^{(2)}_{I_3]}~,
\end{equation}
where the indices run as in \eqref{4.2.22}.

\subsection{Examples of correlation functions}\label{subsec::DCFTexamples}
We provide some examples of the correlation functions that can be computed and constrained using the tools of previous Subsection. In particular we will explicitly write down the results that we are going to use in the following. For a more detailed discussion see \cite{Billo:2016cpy, Lauria:2018klo}.
\subsubsection*{Defect channel}
As already remarked, from the defect point of view, this is just a CFT$ _{p} $ with a SO$(q) $ flavour symmetry; thus all the correlation functions containing defect operators will maintain the same form as we found in Section \ref{X3}. The only difference is that they can also carry some irreducible representations of SO$(q) $. \\
Given two defect operators with the same dimension $ \hat{\Delta} $, scalar under SO$(p) $, they belong to two representations of SO$(q)$. Therefore they will need auxiliary transverse vectors $W^I$ and will be denoted on the light-cone as $\hat{O}_{\hat{\Delta},0,s}(P_i,W_i)$. Then the two-point function of defect scalar primaries reads:
\begin{equation} \label{d4}
\langle \hat{O}_{\hat{\Delta},0,s}(P_1,W_1)  \hat{O}_{\hat{\Delta},0,s}(P_2,W_2) \rangle_{\mathcal{D}} =C_O\, \frac{(W_1\circ W_2)^s}{(2P_1\circ P_2)^\Delta},
\end{equation}
where $C_O$ is the two-point coefficient.\\
This expression can be projected down to the physical space, then it is possible to free the indices associated to the SO$(q) $ global symmetry using the orthogonal Todorov operator \eqref{todorovzw}. \\
In the following we will only need correlation functions for defect primaries which are scalar under SO$(q)$. In such a case the final result is precisely equal to a common two-point function. We only need to specify the section we are projecting to. In particular, in the following we will consider a physical space with a (1-dimensional) spherical defect with radius $R$. In this case it is convenient to split $x^\m$ into $p+1$ directions (denoted by a \textasciitilde) in which the $p$-sphere is embedded:
\begin{equation}\label{4.2.28}
x^\m = (x^{\tilde{a}}, x^i)~, \hspace{0.5cm} \tilde{a}=1,\dots , p+1~, \hspace{0.5cm} i=p+2,\dots ,d~.
\end{equation}
Using this notation the defect is placed in $r=R$, where $r=\norm{x^{\tilde{a}}}$. After defined the operators in the $p+2$-dimensional space the Poincar\'e section will be now defined as:
\begin{equation}
P^\cM=(P^0,P^1,\dots,P^d,P^{d+1}) = \left(\frac{1+x^2}{2}, x^\m, \frac{1-x^2}{2}\right)~,
\end{equation}
where the coordinates will be splitted into parallel and orthogonal directions consistently with \eqref{4.2.22}:
\begin{equation}
P^A=(P^0,P^1,\dots ,P^{p+1})~, \hspace{1cm} P^I = (P^{p+2}, \dots, P^{d+1})~.
\end{equation}
A point on the defect is defined by coordinates $\tau^a$,  and in general will have a parametrization $x^{\tilde{a}}(\tau)$.\\
Therefore starting from \eqref{d4} we project down to physical space and we obtain the generic two point function of scalar primaries in presence of a spherical defect $\cD$ takes the form:
\begin{equation} \label{4.2.31}
\langle \hat{O}_{\hat{\Delta},0,0}(\tau_1^a)  \hat{O}_{\hat{\Delta},0,0}(\tau_2^a) \rangle_{\mathcal{D}} = \frac{C_O}{x_{12}^{2\hat{\Delta}}}~.
\end{equation}
where $ x_{12}=x(\tau_1)-x(\tau_2) $ is fixed by the parametrization of the defect and $C_O$ is the two-point coefficient which in general depends on the couplings of the theory. It can also have an important physical meaning, as we will see in the next Chapters.

\subsubsection*{Bulk channel: symmetric tensor}
As we already noticed in Section \ref{X13}, bulk operators in presence of a defect acquire a non-trivial one point function, due to the broken translational invariance. The structure of the one-point function of a primary is constructed using the machinery introduced before. Scale invariance implies the form:
\begin{equation}\label{4.2.32}
\vev{O_{\Delta,J}(P,Z)}_\cD = A_O \frac{Q_J(P,Z)}{(P\circ P)^{\Delta/2}}~,
\end{equation}
where $A_O$ is the one-point coefficient.
The structure of this one-point function will be our guideline in the following.\\
$Q_J(P,Z)$ is a homogeneous polynomial of degree $J$ in $Z$. It must also have degree zero in $P$ and be transverse. The unique function with these properties can be built out of the transverse tensor \eqref{4.2.14}:
\begin{equation}\label{4.2.33}
Q_J = \left(\frac{C^{AI}C_{AI}}{2P\circ P}\right)^{J/2} = \left(\frac{(P\circ Z)^2}{P\circ P}-Z\circ Z\right)^{J/2}~.
\end{equation}
Since $Z$ appears with even powers only, only \emph{even spin} operators can acquire an expectation value in a parity preserving theory. Indeed it is impossible to act once with the Todorov operator \eqref{todorov} and obtain a finite result. This will be a crucial point in the following. \\
We provide as an example the exact expression of the one-point function of the stress tensor, which is a totally symmetric $ L=2 $ bulk primary with conformal dimension $\Delta=d$. This result is obtained from \eqref{4.2.32} and \eqref{4.2.33}, using a Poincar\`e section for a flat defect case:
\begin{align}
 P^A = (1,x^2,x^a)~,~~~~P^I=0~,~~~~  Z^A=(0,2x^az_a,z^a)~,
\end{align}
and the indices are then opened using \eqref{todorovzw}. This procedure yields:
\begin{equation} \label{d7}
\langle T^{\mu \nu} (x) \rangle = \left\{ \begin{array}{ll}
\langle T^{a b} (x) \rangle =\frac{A_{T}}{|x^{i}|^{d}} \frac{d-p-1}{d}\delta^{a b}\\[0.2cm]
\langle T^{a i} (x) \rangle = 0\\[0.2cm]
\langle T^{i j} (x) \rangle =-\frac{A_{T}}{|x^{i}|^{d}} \left( \frac{p+1}{d}\delta^{i j} - \frac{x^{i}}{|x^{i}|}\frac{x^{j}}{|x^{j}|} \right)\\
\end{array} \right.
\end{equation}

Finally we notice that the structure of the one-point function is compatible with conservation. The condition
\begin{equation}
\partial_\cM \cD^\cM \frac{Q_J(P,Z)}{(P\circ P)^{\Delta/2}} = J(q+J-3)(d-\Delta+J-2) \frac{P\circ Z \,Q_J}{2(P\circ P)^{\frac{\Delta+2}{2}}}=0
\end{equation}
is satisfied when $\Delta=d-2+J$ (then it holds for the stress tensor, as expected).

\subsubsection*{Bulk channel: rank-2 antisymmetric tensor for $d=4$, $p=1$} 
We also treat the case of the one-point function for a rank-2 antisymmetric tensor $H(P,Z^{(1)},Z^{(2)})$ introduced in \eqref{4.2.19}.\\
Also in this case the structure is fixed by the building blocks introduced before: we found 2 possible tensors \eqref{4.2.26} preserving transversality and antisymmetry in presence of a defect. For a single operator insertion, the result should be a linear combination of these tensors opportunely saturated.\\
It turns out that there is a specific situation where we can obtain a non-zero result. If we consider $d=4$, $p=1$ (a line defect in a four dimensional theory)  we can saturate the antisymmetrized indices for each structure in \eqref{4.2.26} with parallel and orthogonal epsilon tensors
\begin{equation}
\epsilon_{A_1A_2A_3}~,~~~ A_a=0,1,2~, \hspace{1cm}\epsilon_{I_1I_2I_3}~,~~~ I_i=3,4,5~.
\end{equation}
Therefore in this case the structure of the one-point function is the following:
\begin{align}\label{4.2.38}
\vev{H(P,Z^{(1)},Z^{(2)})}_\cD = \frac{1}{(P\circ P)^{\Delta/2}}\left(\frac{k_1\epsilon_{I_1I_2I_3}P_{I_1}~Z^{(1)}_{I_2}~Z^{(2)}_{I_3}+k_2\epsilon_{A_1A_2A_3}P_{A_1}~Z^{(1)}_{A_2}~Z^{(2)}_{A_3}}{P\circ P}\right)
\end{align}
where $k_1$ and $k_2$ are the one-point coefficients.

\section{Energy-momentum conservation and displacement operator}\label{X14}
This Section is devoted to discuss the energy-momentum conservation in a DCFT. We do not want to propose a complete analysis (see Section 5 of \cite{Billo:2016cpy} for further details), but rather to try to outline the main steps, focusing on the most relevant physical features. 

We consider a DCFT defined on a manifold $ \mathcal{M} $ described by a metric $ g_{\m\n} $, in presence of a defect sub-manifold $ \mathcal{D} $ parametrized by the coordinates $ \tau^{a} $, $ a=1,\dots,p $, and whose embedding is defined as $x^{\mu} = X^{\mu}(\sigma^{a})$.
\iffalse
Considering an action functional $ S[g,X,\phi,\partial\phi, \dots] $, we can define the stress tensor as:
\begin{equation}\label{d12}
T^{\mu \nu} = -\frac{2}{\sqrt{g}}\frac{\delta S}{\delta g_{\mu \nu}}.
\end{equation}
\fi
We want to determine how conformal invariance get modified due to the presence of the defect. We write the conformal Ward identities that follows from the fact that conformal transformations are a subgroup of diffeomorphisms $ \times $ Weyl transformations. Thus we impose the theory to be invariant under diffeomorphisms $ x'^{\mu} = x^{\mu}+ \xi^{\mu}(x) $ on $ \mathcal{M} $ and under Weyl rescaling $ \delta_{\sigma} g_{\mu \nu} = 2 \sigma(x) g_{\mu \nu} $. A conformal killing vector $ \tilde{\xi} $ is defined as
\begin{equation}
 \nabla_{(\mu}\tilde{\xi}_{\nu )} = -\frac{\nabla_{\rho}\tilde{\xi}^{\rho}}{d} g_{\mu \nu} \equiv - \tilde{\sigma} g_{\mu \nu}~.
 \end{equation} 
This equation means that effecting a diffeomorphism of parameter $ \tilde{\xi}^{\mu} $ and compensating with a Weyl rescaling of parameter $ \tilde{\sigma} $ corresponds to a conformal transformation that leaves the metric invariant. However, no Weyl transformation can compensate the action of diffeomorphisms on the embedding functions $X^\m$, then the variation of the partition function is non-vanishing, due to the presence of the defect.\\
We see this effect through the action of a conformal transformation on a correlation function of bulk local operators $\cX\equiv \mathcal{O}_{1}(x_{1}), \dots, \mathcal{O}_{n}(x_{n}) $. In particular here we provide the explicit expression of the conformal Ward identity in a flat space in presence of a flat defect, since it has a simple shape  but presents a strong physical meaning:
\begin{equation}\label{confWard}
 (\delta_{\tilde{\xi}} +\delta_{\tilde{\sigma}}) \vev{\cX} = \int_{\mathcal{D}} \tilde{\xi}^{i} \vev{\mathbb D_{i}~ \cX}~.
 \end{equation}
Here we note the presence of a new primary operator $\mathbb D_{i}$ defined on the defect: this is the \textit{displacement operator} and encodes the effects of conformal transformations on correlation functions in the presence of a defect.
The full physical meaning of the displacement operator appears in the stress tensor conservation laws.
We have already mentioned that, unless there is a decoupled sector on the defect, there is no conserved stress tensor on the defect, since energy is expected to be exchanged with the bulk, so that only the global stress tensor, which we denote here as $ T^{\mu \nu}_{tot} $, is conserved. In term of $ T^{\mu \nu}_{tot} $ the conservation laws read:
\begin{equation} \label{d8}
\partial_{\mu} T^{\mu a}_{tot} = 0~, \hspace{1 cm} \partial_{\mu} T^{\mu i}_{tot} = - \delta_{\mathcal{D}}(x)\mathbb D^{i}+\mathrm{total~derivatives}~,  
\end{equation}
and obviously $ (T_{tot})^{\mu}_{\mu} = 0 $ to keep scale invariance.\\
We observe that the breaking of translational invariance due to the defect induces the displacement operator to appear as a delta function contribution to the divergence of the stress tensor, \textit{i.e.} it determines its discontinuity across the defect. As a proper defect excitation, the displacement operator has a well-defined two-point function, so its form is fixed as in equation \eqref{4.2.31}:
\begin{equation}\label{d13}
 \langle \mathbb D_{i}(x)\mathbb D_{j}(0) \rangle = C_{D} \frac{\delta_{ij}}{x^{2 \Delta_{D}}},
 \end{equation} 
equation (\ref{d8}) assures that the scale dimension $ \Delta_{\mathbb D} $  of the displacement operator is fixed in terms of the stress tensor, namely $ \Delta_{\mathbb D} = p+1 $. Its normalization $ C_{D} $ is part of CFT data, and it represents an important physical observable, as we will see in the following.
%Such relationship has an important physical meaning, and will be the key-point of section 6 of the present thesis.
\subsection{Relation between $ A_{T} $ and $C_D$}
Equation \eqref{d8} suggests an interesting physical connection between the stress tensor and the displacement operator.
We analyze how conformal symmetry relates the displacement coefficient $ C_{D} $ to the one-point coefficient of the stress tensor $ A_{T} $, defined in \eqref{d7}.\\
Choosing $\cX=T_{\m\n}$ in the conformal Ward identities \eqref{confWard}, we can relate the one-point function $\vev{T_{\m\n}}$ with the two-point function $\vev{\mathbb D^i\,T_{\m\n}}$. The latter can be evaluated exactly using the bulk-to-defect OPE \eqref{d2} in terms of three coefficients $b_{TD}^{1,2,3}$ (see Subsection 5.2 of \cite{Billo:2016cpy} for further details), so the conformal Ward identities yield two simple conditions:
\begin{align}\label{4.3.5}
b_{TD}^{2} &= \frac{1}{p+1}\left(2b_{TD}^{3}-b_{TD}^{1}\right)~,\notag \\
b_{TD}^{3} &= 2^{p+2}\pi^{-\frac{p+1}{2}}\Gamma\left(\frac{p+3}{2}\right)\,A_T~.
\end{align}
Then we exploit the conservation law \eqref{d8}, and we evaluate it inside a proper correlation function
\begin{equation}
\vev{\partial_\m T^{\m i}(x)\mathbb D^j(0)}=-\delta_\cD \vev{\mathbb D^i(x^a)\mathbb D^j(0)}~,
\end{equation}
which returns a further relation, this time involving $b_{TD}^{1,2,3}$ and $C_D$:
\begin{equation}\label{4.3.7}
b_{TD}^{2}=\frac{1}{p}\left((d+q-2)\frac{b_{TD}^{3}}{2}+q\,\frac{C_d}{\Omega_{q-1}}\right)~,
\end{equation}
where $\Omega_{q-1}$ is the volume of a sphere $S_{q-1}$. We see that combining \eqref{4.3.5} and \eqref{4.3.7} we fix the bulk-to-defect coefficients $b_{TD}^{1,2,3}$ in terms of $(A_T,C_D)$ quantities, but there is \emph{no universal relation between $A_T$ and $C_D$} only using conformal constraints. In Chapter \ref{chap:6} (and in particular Section \ref{sec6:BhWCD}) we will clarify this statement, since we will analyze the further constraints coming from extended supersymmetry.

\section{Wilson loop as a line defect}\label{sec4:WLdefect}
We discuss here a specific example which represents the central point of the present thesis, \textit{i.e.} the Wilson loop operator as a $p=1$ defect in conformal gauge theories. In particular we concentrate on a $\mathcal{N}=2$ SYM theory on $\mathbb{R}^4$ with gauge group SU($N$) with a superconformal matter content, of the same class as Chapter \ref{chap:3}. In the following Chapter we will need to evaluate specific observables of the theory, associated to one-point function of local operators in presence of a supersymmetric Wilson loop. Here we fix the set up from a DCFT point of view.

We consider a $1/2$ BPS Wilson loop in the fundamental representation of SU($N$), the same as \eqref{WLdef}:
\begin{equation}
\label{WLdefsection4}
W(C)=\frac{1}{N}\tr\, \mathcal{P}
\exp \left\{g \oint_C d\tau\, \Big[\ii \,A_{\mu}(y)\,\dot{y}^{\mu}(\tau)
+\frac{R}{\sqrt{2}}\big(\varphi(y) + \bar\varphi(y)\big)\Big]\right\}~.
\end{equation}
Without any loss of generality, we can place the
circle $C$ in the plane $(x^1,x^2)\subset\mathbb{R}^4$. The loop $C$ is parameterized as in \eqref{circle}, so $y^\mu(\tau)=R\,\big(\cos\tau,\sin\tau,0,0\,\big)$, with $\tau\in[\,0,2\pi\,]$. 

We are going to evaluate specific observables in presence of the Wilson operator. The fact that \eqref{WLdefsection4} has an explicit form in terms of fundamental fields allows to find very explicit formulas for these observables: conformal symmetry fixes the kinematic factors, up to a coefficient which depends on the couplings of the theory. This coefficient can be then computed using both supersymmetric localization and perturbative computations in field theory. This will be the general goal of Chapters \ref{chap:5} and \ref{chap:6}.

\subsection{One-point functions}\label{subsec4:WLonepoint}

The first quantity of interest is the one-point function
\begin{equation}
\big\langle\, W(C)\,O_{\Delta,J}(x)\,\big\rangle~,
\label{WLO}
\end{equation}
where $O_{\Delta,J}(x)$ is a local operator inserted in the bulk. We discuss the generic form of this one-point function following the results obtained in Subsection \ref{subsec::DCFTexamples}: we can explicitly write the spacetime structures fixed by conformal symmetry.\\

\begin{figure}[ht]
\begin{center}
\begingroup%
  \makeatletter%
  \providecommand\color[2][]{%
    \errmessage{(Inkscape) Color is used for the text in Inkscape, but the package 'color.sty' is not loaded}%
    \renewcommand\color[2][]{}%
  }%
  \providecommand\transparent[1]{%
    \errmessage{(Inkscape) Transparency is used (non-zero) for the text in Inkscape, but the package 'transparent.sty' is not loaded}%
    \renewcommand\transparent[1]{}%
  }%
  \providecommand\rotatebox[2]{#2}%
  \ifx\svgwidth\undefined%
    \setlength{\unitlength}{240bp}%
    \ifx\svgscale\undefined%
      \relax%
    \else%
      \setlength{\unitlength}{\unitlength * \real{\svgscale}}%
    \fi%
  \else%
    \setlength{\unitlength}{\svgwidth}%
  \fi%
  \global\let\svgwidth\undefined%
  \global\let\svgscale\undefined%
  \makeatother%
  \begin{picture}(1,0.47275008)%
    \put(0,0){\includegraphics[width=\unitlength,page=1]{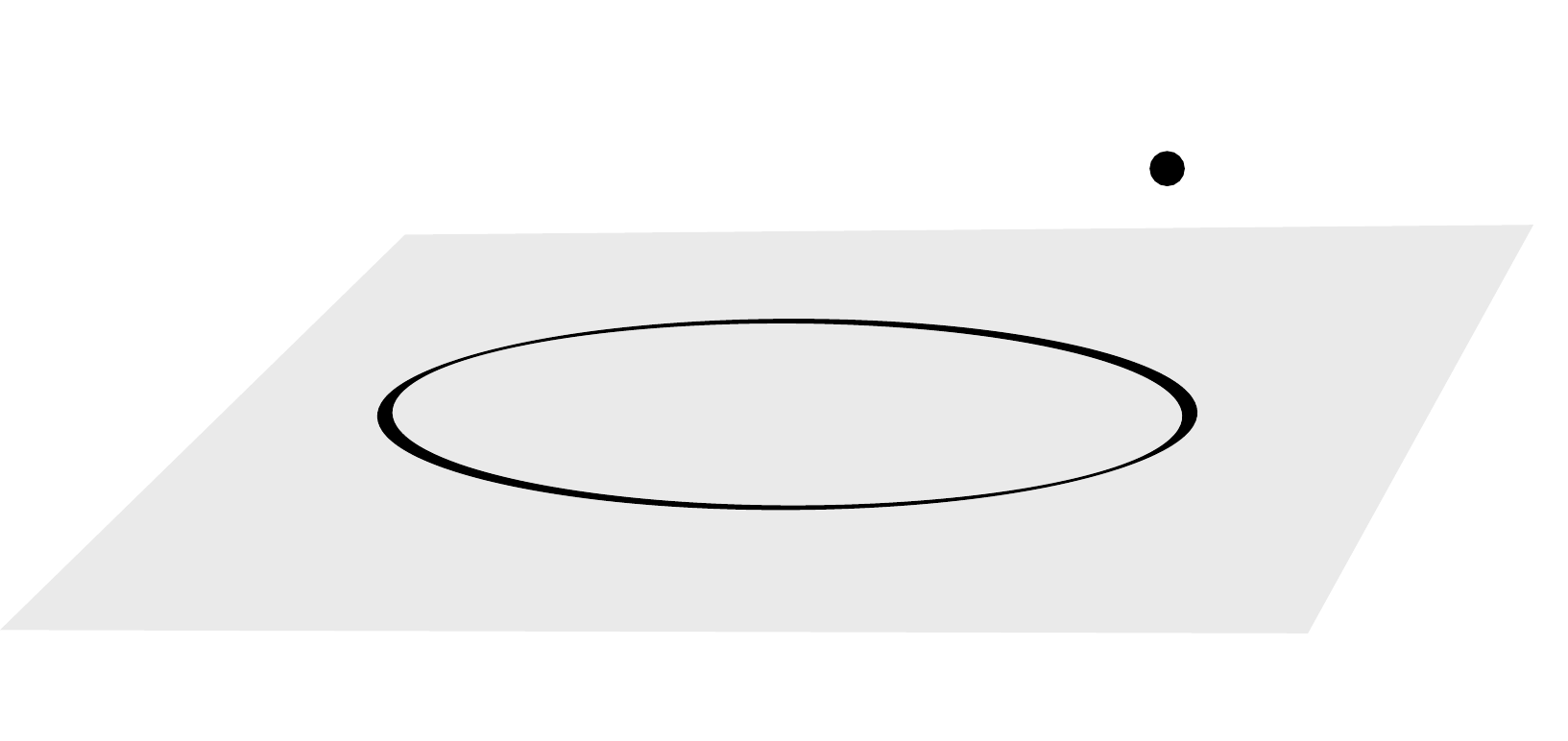}}%
    \put(0.7625012,0.37810715){\color[rgb]{0,0,0}\makebox(0,0)[lb]{\smash{$x$}}}%
    \put(0.1530294,0.12147429){\color[rgb]{0,0,0}\makebox(0,0)[lb]{\smash{$W(C)$}}}%
    \put(0,0){\includegraphics[width=\unitlength,page=2]{WOngeom.pdf}}%
    \put(0.92552756,0.16817663){\color[rgb]{0,0,0}\makebox(0,0)[lb]{\smash{$x_2$}}}%
    \put(0.34953029,0.0060181){\color[rgb]{0,0,0}\makebox(0,0)[lb]{\smash{$x_1$}}}%
    \put(0.50767365,0.45336258){\color[rgb]{0,0,0}\makebox(0,0)[lb]{\smash{$x_3,x_4$}}}%
    \put(0,0){\includegraphics[width=\unitlength,page=3]{WOngeom.pdf}}%
    \put(0.60292707,0.11029535){\color[rgb]{0,0,0}\makebox(0,0)[lb]{\smash{$r$}}}%
    \put(0.75362858,0.25823033){\color[rgb]{0,0,0}\makebox(0,0)[lb]{\smash{$L$}}}%
    \put(0,0){\includegraphics[width=\unitlength,page=4]{WOngeom.pdf}}%
    \put(0.35439669,0.21746727){\color[rgb]{0,0,0}\makebox(0,0)[lb]{\smash{$R$}}}%
  \end{picture}%
\endgroup%
\end{center}
\caption{The geometric set-up for the configuration we consider.}
\label{fig:WOngeom}
\end{figure}

We start by considering a generic scalar operator $O_{\Delta,0}(x)$ to obtain the simplest example for the one-point function \eqref{WLO}. We fix the denominator of the one-point function defined in \eqref{4.2.32} projecting to a flat space Poincar\`e section, in this specific case of a circular defect $W(C)$ of radius $R$. 
Using the embedding formalism, a point $x\in\mathbb{R}^4$ is associated to a null section $P$ of the form
\begin{equation}
\label{Pdef}
P = \Big(\frac{R^2+x^2}{2R},x^\mu,\frac{R^2-x^2}{2R}\Big)~,
\end{equation} 
which satisfies $P^2 \equiv P^T\eta \,P = 0$ with $\eta=\mathrm{diag}(-1,1,1,1,1,1)$.
In presence of the Wilson loop, we can split the spacetime coordinates into ``parallel'' and ``transverse'' ones: 
\begin{equation}
x^\mu \to (x^a,x^i)~,~~ \mathrm{where}~~ a=1,2~, ~~~~ i=3,4~.
\end{equation}
We will denote $x^a x_a = r^2$ and $x^i x_i = L^2$, so that $x^2 = r^2 + L^2$.
The symmetry is reduced according to the pattern
\begin{equation}
\label{sbp}
\mathrm{SO}(1,5)\to \mathrm{SO}(1,2)\times \mathrm{SO}(3)~,
\end{equation} 
with SO$(1,2)$ and SO$(3)$ linearly acting, respectively, on 
\begin{equation}
\label{SOs}
P_\parallel = \Big(\frac{R^2+x^2}{2R},x^a\Big)~~~\text{and}~~~
P_\perp = \Big(x^i,\frac{R^2-x^2}{2R}\Big)~.
\end{equation} 
There are two scalar products invariant with respect to the two symmetry factors, 
which  we denote as
\begin{equation}
\label{scpr}
P\bullet P \equiv P_\parallel^T\eta \,P_\parallel~~\text{with}~
\eta=\mathrm{diag}(-1,1,1)~~~~~\mbox{and}~~~~~
P\circ P \equiv P_\perp^T \, P_\perp~. 
\end{equation}
We know they are not independent, since $P\bullet P + P\circ P = P^2 = 0$. Therefore, we
can take as the single independent invariant the quantity
\begin{equation}
\label{PcPis}
\|x\|_C \equiv 2 \sqrt{P\circ P} = \frac{\sqrt{(R^2 - x^2)^2 - 4 R^2 L^2}}{R}~.
\end{equation}
This is the ``average distance'' between $x$ and $C$, which is invariant under 
the $\mathrm{SO}(1,2)\times \mathrm{SO}(3)$ subgroup of the conformal symmetry that is preserved by the Wilson 
loop, see Figure \ref{fig:WOngeom}

The one-point function $\big\langle \,W(C) O_{\Delta,0}(x)\,\big\rangle 
= \big\langle\, W(C)\, O_{\Delta,0}(P)\, \big\rangle$ must depend on 
$\|x\|_C$, and must be homogeneous of degree $\Delta$ in it; thus it must necessarily be of the form
\begin{equation}
\label{opW}
\big\langle\, W(C) \, O_{\Delta,0}(P) \,\big\rangle = \frac{A_O}{(2\pi \|x\|_C)^\Delta}~.
\end{equation} 
The $2\pi$ factor is inserted for convenience and the constant 
$A_O$ is the one-point coefficient, related to the value of the correlator at $x=0$, {\it{i.e.}} at 
$P = P_0 = (\frac R2,\frac R2, \vec 0)$ where $\|x\|_C \to R$, so that
\begin{equation}
\label{opW0}
\big\langle\, W(C) \, O_{\Delta,0}(P_0)\, \big\rangle = \frac{A_O}{(2\pi R)^\Delta}~.
\end{equation} 
The explicit computation of $A_O$, both using localization and perturbative computations, for $O_{\Delta,0}$ being a \emph{chiral primary operator} will be the main topic discussed in Chapter \ref{chap:5}.

The other important observable to be discussed in the following involves the stress energy tensor, which in this formalism is associated to an operator $O_{\Delta,2}$. We already discussed how residual conformal symmetry constrains its one-point function in presence of a defect in Subsection \ref{subsec::DCFTexamples}. In that case we obtain the one-point function $\vev{T_{\m\n}}_W$ in the flat space, see equation \eqref{d7}. In Chapter \ref{chap:6} we will need the equivalent of \eqref{d7} in a curved space, in order to discuss the computation of the one point coefficient $a_T$ in supersymmetric theories defined on ellipsoids.

\subsection{Two-point function of the displacement operator}\label{subsec4:WLdispl}
Another relevant physical quantity that we want to introduce in presence of a Wilson loop is an example of defect correlation function, \textit{i.e.} the two-point function of the displacement operator. This special defect operator has been introduced in Section \ref{X14} from the conformal Ward identities in presence of a defect, see \eqref{confWard}. For a Wilson loop $W=\tr\cP \exp\Big[\ii \int \cA\Big]$ it can be written as:
\begin{equation}
\delta\vev{W} = \tr\cP\left\{  \int d\t\, \delta x^i(\t)~ e^{\ii \int_{-\infty}^\t \cA}\, \mathbb D_i(\t) \, e^{\ii \int_{\t}^{\infty} \cA}\right\}
\end{equation}
Since the Wilson loop \eqref{WLdefsection4} has an explicit expression in terms of elementary fields, the same can be done for the displacement operator.
In this case we find (see \cite{Bianchi:2018zpb} for a consistent description in terms of conservation laws):
\begin{equation}\label{Displacementexplicit}
\mathbb D_i(\t) = \ii g \dot x^\m F_{\m i}(\t) + g \big( D_i \varphi+D_i\bar \varphi\big)(\t)~.
\end{equation}
Notice that the form \eqref{Displacementexplicit} follows from (super)conformal invariance only. \\
The two-point function of the displacement operator is fixed by DCFT: following the prescriptions of Section \ref{sec::4.2} and in particular the form \eqref{d13} we find:
\begin{equation} \label{displ2point4}
\vev{\mathbb D_i(\t) \mathbb D_j (0)}_W = \frac{C_D\, \delta_{ij}}{\t^4}~.
\end{equation}
The two-point function coefficient $C_D$ also determines the expectation value of a Wilson loop with a wavy line profile (see \cite{Semenoff:2004qr}). The physical interpretation is quite clear: the displacement operator describes small deviation from the original defect shape, and its normalization \eqref{displ2point} represents an important piece of CFT data, as we will see in Chapter \ref{chap:6}.

\chapter{Correlators between Wilson loop and chiral operators}\label{chap:5}
We introduced the class of chiral primary operators (CPOs) in Chapter \ref{chap:1} in the context of representations of superconformal symmetry. They are protected observables, so have always played an important role in $\cN=4$ theories \cite{Minwalla:1997ka}, and their correlation functions have been explored in the AdS/CFT context \cite{Witten:1998qj,Gubser:1998bc,Lee:1998bxa} and in perturbative computations \cite{Penati:1999ba,Penati:2000zv,Bianchi:1999ge}. Correlators for CPOs in presence of Wilson loops have also been considered \cite{Semenoff:2001xp,Pestun:2002mr,Semenoff:2006am}; in particular, correlators of a 1/8-BPS circular loop and chiral primaries in $\mathcal{N}=4$ SYM theory have been computed \cite{Giombi:2009ds,Giombi:2012ep,Bassetto:2009rt,Bassetto:2009ms,Bonini:2014vta}, mapping them to multi-matrix models. Also correlators with local chiral operators and Wilson loops in higher representations have been discussed \cite{Giombi:2006de,Gomis:2008qa}.

A crucial progress has been made also in $\cN=2$ SYM theories, mainly thanks to the localization computation \cite{Pestun:2007rz} of the partition function and the Wilson loop vev, which are reduced to a matrix model on a four-sphere, as seen in Chapter \ref{chap:3}. The computation of observables in such a matrix model is allowed when the $\cN=2$ theory is also conformal: it has been shown that the matrix model for the partition function on $S^4$ also contains information about correlators of chiral operators on $\mathbb{R}^4$ \cite{Baggio:2014ioa,Baggio:2014sna,Gerchkovitz:2014gta,Baggio:2015vxa,Baggio:2016skg}, provided one disentangles 
the operator mixing induced by the map from $S^4$ to $\mathbb{R}^4$ \cite{Gerchkovitz:2016gxx, Rodriguez-Gomez:2016ijh,Rodriguez-Gomez:2016cem}. 
In \cite{Billo:2017glv} this disentangling of operators has been realized as a
normal-ordering procedure and the relation between field theory and matrix model correlators has been
shown to hold also in non-conformal situations for a very special class of operators. \\
It is natural to conjecture that, as it is the case in the $\cN=4$ theory, also in superconformal $\cN=2$ theories the matrix model for the circular Wilson loop on $S^4$ may contain information on correlators of chiral operators in the presence of a circular loop in $\mathbb{R}^4$.
In particular, from Section \ref{sec4:WLdefect} we know that the functional form of the one-point function in presence of a Wilson loop is completely fixed up to a coefficient $A_O(g)$, see eq. \eqref{opW}; this coefficient can be encoded in the Pestun matrix model. 

This Chapter is structured as follows: in Section \ref{sec:chiral} we identify the matrix model counterparts of chiral operators in the field theory through a normal-ordering prescription, whereas in Section \ref{secn:mmcwl} we compute the one-point functions of such operators in the matrix model. Then in Section \ref{sec:pert} we compare them with the corresponding field theory one-point functions in presence of the Wilson loop computed in standard perturbation theory up to two loops for finite $N$, and to all orders in perturbation theory in planar limit for the $\zeta(3)$ dependent part. 
Again we consider the diagrammatic difference between  
$\mathcal{N}=4$ and $\mathcal{N}=2$, following Subsection \ref{subsec::difference}. We find complete agreement between the matrix model and field theory results.

These achievements are valid for any $\cN=2$ Lagrangian SCFTs. In particular, we verified them for any SCFTs with SU$(N)$ gauge group, which correspond to the 5 theories described in Table \ref{tab:scft}. For brevity, in the present Chapter we will show only the results related to the most famous $\cN=2$ SCFT, namely the Superconformal QCD, corresponding to the \textbf{A} theory in Table \ref{tab:scft}.

\section{Chiral operators in $\cN=2$ SCFTs}\label{sec:chiral}
\subsection{Flat space CPOs}
Chiral primaries in $\cN=2$ SCFTs were introduced in Section \ref{sec1:SCFT}. They transform as scalars of SU$(2)_R$ and their conformal dimension is completely determined by their U$(1)_R$ charge. Moreover they parametrize the Coulomb branch of vacua, thus in Lagrangian theories they are written as combinations of the scalar field $\varphi$ ($\bar\varphi$ for the anti-chiral operators) of the $\cN=2$ vector multiplet in a gauge invariant way. Therefore they can be labeled by a vector of integers $\vec{n}=(n_1,n_2,\cdots,n_\ell)$ and take 
a multi-trace expression of the form
\begin{equation}
O_{\vec{n}}(x)=\tr \varphi^{n_1}(x)\,\tr \varphi^{n_2}(x)\cdots\tr \varphi^{n_\ell}(x)~,~~~~~~~~\sum_{k=1}^\ell n_k=n~,
\label{On}
\end{equation}
where $n$ is the conformal dimension. Equivalently, by expanding $\varphi(x) = \varphi^b(x)\,T^b$, where $T^b$
are the generators of SU($N$) in the fundamental representation canonically normalized as in \eqref{normT} we can write 
\begin{equation}
\label{OnR}
O_{\vec{n}}(x) = R_{\vec{n}}^{\,b_1\dots b_n}\, \varphi^{b_1}(x)\ldots \varphi^{b_n}(x)
\end{equation}
where $R_{\vec{n}}^{\,b_1\dots b_n}$ is a totally symmetric 
$n$-index tensor whose expression is encoded%
\footnote{Explicitly,
\begin{equation*}
R_{\vec{n}}^{\,b_1\dots b_n} 
= \tr \big(T^{(b_1}\cdots T^{b_{n_1}}\big)~
\tr \big(T^{b_{n_1+1}}\cdots T^{b_{n_1+n_2}}\big)\ldots
\tr \big(T^{b_{n_1 + \ldots + n_{\ell-1}+1}}\cdots T^{b_n)}\big)
\end{equation*}
where the indices are symmetrized with strength 1. \label{footnote:R}}
in (\ref{On}). 
The quantity of interest is the one-point function
\begin{equation}
\big\langle\, W(C)\,O_{\vec{n}}(x)\,\big\rangle~,
\label{WLOsec5}
\end{equation}
where the definition of the Wilson loop corresponds to eq. \eqref{WLdefsection4} and is reported here for convenience:
\begin{equation}
\label{WLdefchapter5}
W(C)=\frac{1}{N}\tr\, \mathcal{P}
\exp \left\{g \oint_C d\tau\, \Big[\ii \,A_{\mu}(y)\,\dot{y}^{\mu}(\tau)
+\frac{R}{\sqrt{2}}\big(\varphi(y) + \bar\varphi(y)\big)\Big]\right\}~.
\end{equation}

\noindent Following the analysis of Section \ref{sec4:WLdefect}, the correlator (\ref{WLOsec5}) takes the form
\begin{equation}
\big\langle\,W(C)\,O_{\vec{n}}(x)\,\big\rangle= \frac{A_{\vec{n}}}{\,\big(2\pi\|x\|_C\big)^n\phantom{\Big|}}
\label{WLO1}
\end{equation}
where $A_{\vec{n}}$ is a $g$-dependent constant, which corresponds 
to the one-point function evaluated in the origin:
\begin{equation}
A_{\vec{n}}(g)= (2\pi R)^n\,\big\langle\, W(C)\,O_{\vec{n}}(0)\,\big\rangle~.
\label{Ang}
\end{equation}
In the next Sections we will compute this function in two different ways: one by using the matrix model 
approach of Section \ref{sec:mmvev}, and the other by using standard perturbative field theory methods, following Section \ref{sec:FTactions}.

\subsection{Chiral operators in the matrix model}\label{subsec:chiralMM}
As shown in Section \ref{sec2:localization}, evaluating the physical object in the matrix model, corresponds to reduce the fields on the saddle point configuration, see \eqref{saddlesphere}, where $A_\m=0$ and $\varphi=\bar \varphi =  a/\sqrt{2}$. Therefore the Wilson loop (\ref{WLdefchapter5})
in the fundamental representation and on a circle of radius $R=1$ is given by the following operator in the matrix model (already introduced in \eqref{WLa}):
\begin{equation}
\cW(a)=\frac{1}{N}\,\tr~\exp\frac{g}{\sqrt{2}}\,a=\frac{1}{N}
\sum_{k=0}^\infty \frac{g^k}{2^{\frac{k}{2}}\,k!}\,\tr a^k ~.
\label{WLmm}
\end{equation}
On the other hand, it would seem natural to associate to any multi-trace chiral operator $O_{\vec{n}}(x)$ of the SYM theory defined in (\ref{On}) a matrix operator 
$O_{\vec{n}}(a)$ with precisely the same expression but with the field $\varphi(x)$ replaced by the matrix $a$ (up to an overall numerical factor), namely
\begin{equation}
O_{\vec{n}}(a)=\tr a^{n_1}\,\tr a^{n_2}\cdots\tr a^{n_\ell}\,=\,R_{\vec{n}}^{\,b_1\dots b_n}\,
\,a^{b_1}\,a^{b_2}\cdots a^{b_n}~.
\label{Ona}
\end{equation}
However, since the field theory propagator only connects $\varphi$ with $\bar\varphi$, 
all operators $O_{\vec{n}}(x)$ have no self-contractions, whereas the operators 
$O_{\vec{n}}(a)$ defined above do not share this property. This means that the dictionary
between the SYM theory and the matrix model is more subtle. Indeed, we have to subtract from
$O_{\vec{n}}(a)$ all its self-contractions by making it orthogonal to all the lower dimensional operators, or
equivalently by making it normal-ordered. As discussed in \cite{Billo:2017glv}, given any operator 
$O(a)$ we can define its normal-ordered version $\cO(a)$ as follows. Let be $\Delta$ the dimension of
$O(a)$ and $\big\{O_p(a)\big\}$ a basis of in the finite-dimensional space of matrix operators with dimension smaller than $\Delta$. Denoting by $C_\Delta$ the (finite-dimensional) matrix of correlators
\begin{equation}
\big(C_\Delta\big)_{pq} = \big\langle\,O_p(a)\,O_q(a)\,\big\rangle~,
\label{Cpq}
\end{equation}
 we define the normal-ordered operator 
\begin{equation}
\cO(a) =\,\, \nordg{O(a)}\,\,
=\, O(a) - \sum_{p,q} \big\langle\,O(a)\, O_{p}(a)\,\big\rangle\, (C_\Delta^{-1})^{pq}\, O_q(a)~.
\label{normalo}
\end{equation}
As emphasized by the notation, the normal-ordered operators are $g$-dependent, since the 
correlators in the right hand side of (\ref{normalo}) are computed in the interacting $\cN=2$ matrix model.

Using these definitions, the correspondence between field theory and matrix model operators takes
the following simple form
\begin{equation}
\label{corrOcO}
O_{\vec{n}}(x) \,\to\, \cO_{\vec{n}}(a) =\,\, \nordg{O_{\vec{n}}(a)}~.
\end{equation} 
This is a prescription to compute chiral correlation function in a $\cN=2$ matrix model. \\
Let us make a recall of the matrix model shape:
\begin{align}
	\label{MMchapter5}
		\cZ_{S^4} = \int da~\rme^{-\mathrm{tr}\, a^2 - S_{\mathrm{int}}(a)}~,~~~~~	
		S_{\mathrm{int}}(a)= \sum_{n=2}(-1)^{n}\left(\frac{g^2}{8\pi^2}\right)^{n}\,\frac{\zeta(2n-1)}{n}\trp a^{2n}~,
\end{align}
and a generic observable is computed in this matrix model according to (\ref{vevmat}), namely:
\begin{align}
	\label{vevmatChapter5}
		\big\langle f(a) \big\rangle\, 
		= \,\frac{1}{\cZ_{S^4}} \int \!da ~\rme^{-\tr a^2-S_{\mathrm{int}}(a)}\,f(a)
			= \,\frac{\big\langle\,
		\rme^{- S_{\mathrm{int}}(a)}\,f(a)\,\big\rangle_0\phantom{\Big|}}
		{\big\langle\,\rme^{- S_{\mathrm{int}}(a)}\,\big\rangle_0
		\phantom{\Big|}}~,
\end{align} 
where the subscript $0$ stands for Gaussian matrix model.
Throughout the present Chapter, we will make explicit references to Section \ref{sec:mmvev} for all the details about the matrix model techniques. Since in the present Chapter we concentrate on the SCQCD case (Theory \textbf{A} in Table \ref{tab:scft}), all the computations are written in the specialized $\cN=2$ SCQCD matrix model defined by the interactive action (see first row of Table \ref{tab:S4S6conf}):
\begin{equation}
\label{Sintresc}
S_{\mathrm{int}}(a) =\frac{3\,\zeta(3)\,g^4}{(8\pi^2)^2} \,\left(\tr a^2\right)^2  -
\frac{10\,\zeta(5)\,g^6}{3(8\pi^2)^3}\, 
\Big[3\, \tr a^4 \, \tr a^2 -2\,\left(\tr a^3\right)^2 \Big]+ \cdots ~.
\end{equation}

We now provide some explicit examples by considering the first few low-dimensional operators. At level $n=2$
we have just one operator:
\begin{equation}
\label{noO2} 
\cO_{(2)}(a) = \,\,\nordg{\tr a^2}\,\,= \tr a^2-\frac{N^2-1}{2}
+\frac{3\,\zeta(3)\,g^4}{(8\pi^2)^2}\,\frac{(N^2-1)(N^2+1)}{2}+O(g^6)~.
\end{equation} 
Similarly, at level $n=3$ we have one operator, which in the SU($N$) theory does not receive any correction:
\begin{equation}
\label{noO3}
\cO_{(3)}(a) = \,\,\nordg{\tr a^3}\,\,= \tr a^3~.
\end{equation}
At level $n=4$, we have instead two independent operators corresponding to 
$\vec{n}=(4)$ and $\vec{n}=(2,2)$. Their normal-ordered expressions are given, respectively, by
\begin{align}
\cO_{(4)}&(a) = \,\,\nordg{\tr a^4}\nonumber\\
&= \tr a^4 
-\frac{2N^2-3}{N}\,\tr a^2+\frac{(N^2-1)(2N^2-3)}{4N}
\phantom{\Big|}\label{noO4}\\
&+\frac{3\,\zeta(3)\,g^4}{(8\pi^2)^2}
\Big[\frac{(2N^2-3)(N^2+5)}{N}\:\tr a^2-\frac{2(N^2-1)(N^2+4)(2N^2-3)}{4N}\Big] 
+O(g^6)~,\phantom{\Big|}\nonumber
\end{align}
and
\begin{align}
\cO_{(2,2)}(a) &= \,\,\nordg{\left(\tr a^2\right)^2}\nonumber \\
&=\left(\tr a^2\right)^2-(N^2-1)\,\tr a^2+\frac{N^4-1}{4}
\phantom{\Big|}\label{noO22}\\
&~~+\frac{3\,\zeta(3)\,g^4}{(8\pi^2))^2}\Big[
(N^2-1)(N^2+5)\,\tr a^2-\frac{(N^4-1)(N^2+4)}{2}\Big]
+\mathcal{O}(g^6)~.\phantom{\Big|}\nonumber
\end{align}
 Up to the order $g^6$ we have considered, it is easy to check that these operators satisfy
\begin{equation}
\begin{aligned}
\big\langle\,\cO_{\vec{n}}(a)\,\big\rangle&=0~,\\
\big\langle\,\cO_{\vec{n}}(a)\,\cO_{\vec{m}}(a)\,\big\rangle&=0~,\\
\end{aligned}
\end{equation}
for $n\neq m$. 
Normal-ordered operators of higher dimension can be constructed without any problem along these same
lines.

We observe that the $g$-independent parts of the above expressions correspond to the normal-ordered 
operators in the Gaussian model, {\it{i.e.}} in the $\cN=4$ theory. Since we will often compare our
$\cN=2$ results with those of the $\cN=4$ theory, we find convenient to introduce a specific notation
for the $g\to 0$ limit of the normal ordering and write
\begin{equation}
\widehat{\cO}_{\vec{n}}(a) \equiv \lim_{g\to 0} \cO_{\vec{n}}(a) =\,\,\, \nord{O_{\vec{n}}(a)}~,
\label{hatOn}
\end{equation}
so that most of the formulas will look simpler.

In the following Section we will explicitly compute the one-point functions between the Wilson loop and
the chiral operators in the $\cN=2$ matrix model, namely
\begin{equation}
\cA_{\vec{n}}=\big\langle\,\cW(a)\,\cO_{\vec{n}}(a)\,\big\rangle
\label{Angdef}
\end{equation}
which will later compare with the field theory amplitudes defined in (\ref{Ang}).

\section{Matrix model correlators in presence of a Wilson loop}
\label{secn:mmcwl}

Our main goal here is the computation of $\cA_{\vec{n}}$ 
in the interacting matrix model described above. 
As a warming-up, but also for later applications, we begin by presenting the results in the 
Gaussian matrix model, {\it{i.e.}} in the $\cN=4$ theory.

\subsection{The $\cN=4$ theory}
\label{subsubsec:N4mmc}
In this case we should consider the operators $\widehat{\cO}_{\vec{n}}(a)$ defined in (\ref{hatOn}) and
compute
\begin{equation}
\widehat{\cA}_{\vec{n}}=\big\langle\,\cW(a)\,\widehat{\cO}_{\vec{n}}(a)\,\big\rangle_0
\label{hatAn}
\end{equation}
where $\vev{f(a)}_0= \int da\, e^{-\tr a^2} f(a)$.

The simplest example is the amplitude with the identity ($\vec{n}=(0)$), 
which yields the vacuum expectation value of the Wilson loop operator (\ref{WLmm}):
\begin{equation}
\label{WvevN4}
\widehat{\cA}_{(0)}=
\big\langle\,\cW(a)\,\big\rangle_0=\frac{1}{N}\,\sum_{k=0}^\infty
\frac{g^k}{2^{\frac{k}{2}}\,k!}\,t_k 
\end{equation}
with $t_k$ defined in (\ref{tn}). We already review the computation of this quantity in Chapter \ref{chap:3}, it yields the well known $\cN=4$ exact result for the Wilson loop vev \eqref{WN4exact}, which we report here for convenience: 
\begin{align}
\widehat{\cA}_{(0)}
=\frac{1}{N}\,L_{N-1}^{1}\Big(-\frac{g^2}{4}\Big)\,\exp\Big[\frac{g^2}{8}\Big(1-\frac{1}{N}\Big)\Big]
\label{WvevN4a}
\end{align}
where $L_n^m$ is the generalized Laguerre polynomial of degree $n$. 

Next we consider the amplitude between the Wilson loop and the operator $\widehat{\cO}_{(2)}(a)$ at level
2. This is given by
\begin{equation}
\widehat{\cA}_{(2)}=\big\langle\,\cW(a)\,\nord{\tr a^2}\big\rangle_0
=\frac{1}{N}\,\sum_{k=0}^\infty
\frac{g^k}{2^{\frac{k}{2}}\,k!}\,\Big(t_{k,2}-\frac{N^2-1}{2}\,t_k\Big) ~.
\label{A20}
\end{equation}
The recursion relations (\ref{rrecursion}) imply 
\begin{equation}
t_{k,2}=\Big(\frac{k}{2}+\frac{N^2-1}{2}\Big)t_k~,
\end{equation}
and thus the amplitude (\ref{A20}) becomes
\begin{equation}
\widehat{\cA}_{(2)}=\frac{1}{N}\,\sum_{k=0}^\infty\frac{k}{2}\,
\frac{g^k}{2^{\frac{k}{2}}\,k!}\,t_k =\frac{g}{2}\,\partial_g\widehat{\cA}_{(0)}~.
\label{A21}
\end{equation}
Expanding for small $g$, we get
\begin{equation}
\widehat{\cA}_{(2)}=g^2\,\frac{N^2-1}{8N}+g^4\,\frac{(N^2-1)(2N^2-3)}{192N^2}
+g^6\,\frac{(N^2-1)(N^4-3N^2+3)}{3072N^3}+\cdots~.
\label{A21g}
\end{equation}
This same procedure can be used to compute the amplitudes $\widehat{\cA}_{\vec{n}}$ for
any $\vec{n}$. The remarkable fact is that, thanks to the recursion relations (\ref{rrecursion}),
it is always possible to obtain compact expressions in terms of $\widehat{\cA}_{(0)}$ and its
derivatives that are exact, {\it{i.e.}} valid for any $N$ and any $g$. For example, at level $n=3$ we find
\begin{equation}
\widehat{\cA}_{(3)}=\frac{g}{\sqrt{2}}\,\partial_g^2\widehat{\cA}_{(0)}-\frac{g^2}{4\sqrt{2}N}
\,\partial_g\widehat{\cA}_{(0)}-\frac{g(N^2-1)}{4\sqrt{2}N}
\,\widehat{\cA}_{(0)}~,
\label{A3}
\end{equation}
while at level $n=4$ we have
\begin{equation}
\widehat{\cA}_{(4)}=g\,\partial_g^3\widehat{\cA}_{(0)}+\frac{g^2}{4N}\,\partial_g^2\widehat{\cA}_{(0)}
+\frac{g^3-4gN(2N^2-3)}{16N^2}\,\partial_g\widehat{\cA}_{(0)}+
\frac{g^2(N^2-1)}{16N^2}\,\widehat{\cA}_{(0)}~,
\label{A4}
\end{equation}
and
\begin{equation}
\widehat{\cA}_{(2,2)}=\frac{g^2}{4}\,\partial_g^2\widehat{\cA}_{(0)}
-\frac{g}{4}\,\partial_g\widehat{\cA}_{(0)}~.
\end{equation}
We have performed similar calculations for higher dimensional operators, but we do not report the
results since they would not add much to what we have already exhibited. Instead, we point out that
the lowest order term in the small $g$ expansion of $\widehat{\cA}_{\vec{n}}$, 
which we call ``tree-level term'', can be compactly written as
\begin{equation}
\begin{aligned}
\widehat{\cA}_{\vec{n}}\Big|_{\mathrm{tree-level}}\,&=\frac{g^n}{N\,2^{\frac{n}{2}}\,n!}\,
R_{\vec{n}}^{\,b_1\dots b_n}\,
\big\langle\,\tr a^n\,\nord{a^{b_1}\dots \,a^{b_n}}\big\rangle_0\\
&=\frac{g^n}{N\,2^{\frac{n}{2}}}\,R_{\vec{n}}^{\,b_1\dots b_n}\,\tr \big(T^{b_1}\dots T^{b_n}\big)
\end{aligned}
\label{Antree}
\end{equation}
where $R_{\vec{n}}^{\,b_1\dots b_n}$ is the symmetric tensor 
associated to the operator $O_{\vec{n}}(a)$ according to (\ref{Ona}).
For later convenience, in Tab.~\ref{tab1} we collect the explicit expressions of 
$\widehat{\cA}_{\vec{n}}\big|_{\mathrm{tree-level}}$ for all operators up to level $n=4$.
\begin{table}[ht]
\begin{center}
{
%\footnotesize
\begin{tabular}{|c|c|}
\hline
$\vec{n}\phantom{\bigg|}$&$\widehat{\cA}_{\vec{n}}\big|_{\mathrm{tree-level}}$\\
      \hline\hline
$(2)\phantom{\bigg|}$ & $g^2\frac{N^2-1}{8N\phantom{\big|}}$ \\
\hline
$(3)\phantom{\bigg|}$ & $g^3\frac{(N^2-1)(N^2-4)}{32\sqrt{2}N^2\phantom{\big|}}$ \\
\hline
$(4)\phantom{\bigg|}$ & $g^4\frac{(N^2-1)(N^4-6N^2+18)}{384 N^3\phantom{\big|}}$ \\
\hline
$(2,2)\phantom{\bigg|}$ & $g^4\frac{(N^2-1)(2N^2-3)}{192N^2\phantom{\big|}}$ \\
\hline
\end{tabular}
}
\end{center}
\caption{The tree-level contribution to $\widehat{\cA}_{\vec{n}}$ for operators up to order $n=4$.}
\label{tab1}
\end{table}

\subsection{The $\mathcal{N}=2$ SCQCD theory}

Let us now return to our main goal, namely the computation of the one-point amplitudes in 
the interacting matrix model that describes the $\cN=2$ SCQCD theory.
Comparing $\cA_{\vec{n}}$ with the $\cN=4$ amplitudes $\widehat{\cA}_{\vec{n}}$, we see
two main differences:
\begin{enumerate}
\item the normal-ordered operators $\cO_{\vec{n}}$ explicitly contain $g$-dependent terms;
\item the vacuum expectation value is computed in a $g$-dependent matrix model.
\end{enumerate}
Both effects arise from the interaction terms of $S_{\mathrm{int}}(a)$ given in (\ref{Sintresc}); 
thus we can write
\begin{equation}
\cA_{\vec{n}}=\widehat{\cA}_{\vec{n}}+\delta\cA_{\vec{n}}
\label{diffAn}
\end{equation}
with
\begin{equation}
\delta\cA_{\vec{n}}=\frac{3\,\zeta(3)\,g^4}{(8\pi^2)^2}\,\cX_{\vec{n}}
-\frac{10\,\zeta(5)\,g^6}{3(8\pi^2)^3}\,\cY_{\vec{n}}+\,\cdots
\label{deltaAnexp}
\end{equation}
where the ellipses stand for terms of higher transcendentality,
proportional to  $\zeta(7)$, $\zeta(3)^2$ and so on. 
The quantities $\cX_{\vec{n}}$, $\cY_{\vec{n}}$ and the analogous ones at higher transcendentality
depend on the coupling constant $g$ and can be expressed using vacuum expectation values in the 
Gaussian model and, eventually, $\widehat{\cA}_{(0)}$ and its derivatives in a compact way.
Since $\delta\cA_{\vec{n}}$ starts at order $g^4$, {\it{i.e.}} at two loops, we clearly have
\begin{equation}
\delta\cA_{\vec{n}}\Big|_{\mathrm{tree-level}}=0\qquad\mbox{and}\qquad
\delta\cA_{\vec{n}}\Big|_{\mathrm{1-loop}}=0\label{deltaAntree1loop}
\end{equation}
for any $\vec{n}$.
In the following we will restrict our analysis to the first correction $\cX_{\vec{n}}$ for which we will 
provide explicit formulas in several examples.

Let us start with the Wilson loop, {\it{i.e.}} with the identity operator ($n=0$). In this case there 
is no normal-ordering to do and thus the only contribution to $\cX_{(0)}$ comes from the
interactions in the matrix model. Focusing on the $\zeta(3)$-term which is proportional 
to $\left(\tr a^2\right)^2$, after some straightforward algebra we get
\begin{equation}
\begin{aligned}
\cX_{(0)}&=-\,\big\langle\, \cW(a)\left(\tr a^2\right)^2\big\rangle_0+
\big\langle\, \cW(a)\,\big\rangle_0~\big\langle\, \left(\tr a^2\right)^2\big\rangle_0~.
\end{aligned}
\label{X0}
\end{equation}
Evaluating the vacuum expectation values by means of 
the recursion relations (\ref{rrecursion}) and expressing the results in terms of the $\cN=4$ Wilson loop, 
we can rewrite the above expression as
\begin{equation}
\cX_{(0)}=-\frac{g^2}{4}\,\partial_g^2\widehat{\cA}_{(0)}-\frac{g(2N^2+1)}{4}\,\partial_g\widehat{\cA}_{(0)}~.
\end{equation}
Using (\ref{WvevN4a}) and expanding for small $g$, we easily get
\begin{equation}
\begin{aligned}
\cX_{(0)}&=-\,g^2\,\frac{(N^2-1)(N^2+1)}{8N}-g^4\,\frac{(N^2-1)(2N^2-3)(N^2+2)}{192N^2}\\
&~~~\,-g^6\,\frac{(N^2-1)(N^4-3N^2+3)(N^2+3)}{8N}
+\cdots~.
\end{aligned}
\end{equation}
Therefore, in the difference $\delta\cA_{(0)}$ the leading term, which is a 2-loop effect
induced by the $g^4$-part of $S_{\mathrm{int}}(a)$ proportional to $\zeta(3)$, 
turns out to be
\begin{equation}
\delta\cA_{(0)}\Big|_{\mathrm{2-loop}}= -g^6\,\frac{\zeta(3)}{(8\pi^2)^2}\,
\frac{3(N^2-1)(N^2+1)}{8N}~.
\label{deltaA02loop}
\end{equation}
This expression has been successfully checked in \cite{Andree:2010na} against an explicit perturbative
2-loop calculation in field theory.

Let us now consider the operator $\cO_{(2)}$ at level $n=2$. In this case we have
\begin{align}
\cX_{(2)}&=
-\,\big\langle\, \cW(a)\,\widehat{\cO}_{(2)}(a)\left(\tr a^2\right)^2\big\rangle_0+
\big\langle\, \cW(a)\,\widehat{\cO}_{(2)}(a)\,\big\rangle_0~\big\langle \!
\left(\tr a^2\right)^2\big\rangle_0\notag\phantom{\Big|}\\
&~~~\,+\frac{(N^2-1)(N^2+1)}{2}\,\big\langle\,\cW(a)\,\big\rangle_0
\label{X2}
\end{align}
where the last term is due to the normal-ordering procedure in the interacting theory which indeed
yields a part proportional to $(N^2-1)(N^2+1)/2$ (see (\ref{noO2})).
Evaluating the vacuum expectation values, this expression becomes
\begin{equation}
\cX_{(2)}=
-\frac{g^3}{8}\,\partial_g^3\widehat{\cA}_{(0)}
-\frac{g^2(2N^2+7)}{8}\,\partial_g^2\widehat{\cA}_{(0)}
-\frac{5g(2N^2+1)}{8}\,\partial_g\widehat{\cA}_{(0)}~,
\label{X2a}
\end{equation}
while its perturbative expansion is
\begin{equation}
\begin{aligned}
\cX_{(2)}&=-g^2\,\frac{3(N^2-1)(N^2+1)}{8N}-g^4\,\frac{(N^2-1)(2N^2-3)(N^2+2)}{48N^2}\\
&~~~\,-g^6 \,\frac{5(N^2-1)(N^4-3 N^2+3)(N^2+3)}{3072 N^3}+\cdots~.
\end{aligned}
\label{X2b}
\end{equation}
The leading term tells us that the 2-loop correction to the $\cN=2$ amplitude $\cA_{(2)}$ is
\begin{equation}
\delta\cA_{(2)}\Big|_{\mathrm{2-loop}}= -g^6\,\frac{\zeta(3)}{(8\pi^2)^2}\,
\frac{9(N^2-1)(N^2+1)}{8N}~.
\label{deltaA22loop}
\end{equation}

This procedure can be easily applied to operators of higher dimensions. 
For example, skipping the intermediate steps, at level $n=3$
we find
\begin{equation}
\begin{aligned}
\cX_{(3)}
&=-\,g^3\,\frac{3(N^2-1)(N^2-4)(N^2+3)}{32\sqrt{2}N^2}
-g^5\,\frac{(N^2-1)(N^2-4)(N^4+2N^2-8)}{128\sqrt{2}N^3}\\
&~~~\,-g^7\,\frac{(N^2-1)(N^2-4)(3 N^6+5 N^4-35 N^2+75)}{12288
   \sqrt{2} N^4}+\cdots~,
\end{aligned}
\label{X3exp}
\end{equation}
while at level $n=4$ we get
\begin{equation}
\begin{aligned}
\cX_{(4)}&=-\,g^4\,\frac{(N^2-1)(N^6+2N^4-18N^2+81)}{96N^3}\\
&~~~\,-g^6
\frac{(N^2-1)(2 N^8+5 N^6-41 N^4+270 N^2-486)}{3072 N^4}\\
&~~~\,-g^8\,\frac{(N^2-1)(2 N^{10}+9 N^8-53 N^6+270 N^4-960 N^2+1710)}{122880 N^5}+\cdots~,
\end{aligned}
\label{X4exp}
\end{equation}
and
\begin{align}
\cX_{(2,2)}&=-\,g^4\,\frac{(N^2-1)(2N^2-3)(N^2+3)}{32 N^2}
-g^6\frac{(N^2-1)(7N^2+27)(N^4-3 N^2+3)}{1536 N^3}\notag
\\
&~~~\,-g^8\,\frac{(N^2-1)(4 N^2+19)(2 N^6-8 N^4+15 N^2-15)}{61440 N^4}+\cdots~.
\label{X22exp}
\end{align}
Multiplying the leading terms in these expansions by $\frac{3\,\zeta(3)\,g^4}{(8\pi^2)^2}$,
we obtain the 2-loop corrections to the amplitudes
$\cA_{\vec{n}}$, whose explicit expressions are collected in Tab.~\ref{tab2} for all operators
up to dimension $n=4$.

\begin{table}[ht]
\begin{center}
{
\begin{tabular}{|c|c|}
\hline
$\vec{n}\phantom{\bigg|}$&$\delta \cA_{\vec{n}}\big|_{\mathrm{2-loop}}$\\
      \hline\hline
$(2)\phantom{\bigg|}$ & $-g^6\frac{\zeta(3)}{(8\pi^2)^2\phantom{\big|}}
\frac{9(N^2-1)(N^2+1)}{8N\phantom{\big|}}$ \\
\hline
$(3)\phantom{\bigg|}$ & $-g^7\frac{\zeta(3)}{(8\pi^2)^2\phantom{\big|}}\frac{9(N^2-1)(N^2-4)(N^2+3)}{32\sqrt{2}N^2\phantom{\big|}}$ \\
\hline
$(4)\phantom{\bigg|}$ & $-g^8\frac{\zeta(3)}{(8\pi^2)^2\phantom{\big|}}
\frac{(N^2-1)(N^6+2N^4-18N^2+81)}{32N^3\phantom{\big|}} $ \\
\hline
$(2,2)\phantom{\bigg|}$ & $-g^8\frac{\zeta(3)}{(8\pi^2)^2\phantom{\big|}}
\frac{3(N^2-1)(2N^2-3)(N^2+3)}{32 N^2\phantom{\big|}}$ \\
\hline
\end{tabular}
}
\end{center}
\caption{The 2-loop contribution to the difference $\delta\cA_{\vec{n}}$ between the $\cN=2$
and the $\cN=4$ amplitudes for operators up to order $n=4$.}
\label{tab2}
\end{table}

It should be clear by now that this procedure can be used to find $\cX_{\vec{n}}$ for any $\vec{n}$, 
and also that it can be straightforwardly generalized to obtain the exact 
expressions of the  corrections with higher transcendentality, like for example $\cY_{\vec{n}}$ 
in (\ref{deltaAnexp}).
Of course, the resulting formulas become longer and longer when one goes higher and higher 
in $n$ or in transcendentality; however this approach, which is essentially based on the use of the 
recursion relations (\ref{rrecursion}), provides a systematic way to obtain exact expressions to any desired
order.

\subsection{The large-$N$ limit}
\label{secn:largeN}

We now study the behavior of the matrix model amplitudes in the planar limit $N\to \infty$
with the 't~Hooft coupling
\begin{equation}
\lambda=g^2 N
\end{equation}
kept fixed.
We begin with the $\cN=4$ theory and later turn to the superconformal $\cN=2$ model.

\subsubsection*{The $\cN=4$ theory}
Taking the planar limit of the expectation value of the Wilson loop, from (\ref{WvevN4a}) we get
\begin{equation}
\widehat{\cA}_{(0)}\Big|_{\mathrm{planar}} = 1+\frac{\lambda}{8}+\frac{\lambda^2}{192}
+\frac{\lambda^3}{9216}+\cdots =\frac{2}{\sqrt{\lambda}}\,I_1\big(\sqrt{\lambda}\big)
\label{hatA0planar}
\end{equation}
where $I_n$ is the modified Bessel function of the first kind.
This is a well-known and established result \cite{Erickson:2000af}.

Next, let us consider the amplitude between the Wilson loop and the operator at level $n=2$ given in
(\ref{A21g}). In the planar limit it becomes
\begin{equation}
\widehat{\cA}_{(2)}\Big|_{\mathrm{planar}} = \frac{\lambda}{8}+\frac{\lambda^2}{96}
+\frac{\lambda^3}{3072}+\cdots =I_2\big(\sqrt{\lambda}\big)~.
\label{hatA2planar}
\end{equation}
Also this is a known result \cite{Semenoff:2001xp}.

Proceeding systematically in this way and using the explicit results in the Gaussian matrix model, 
it is not difficult to find the weak-coupling
expansion of the amplitude $\widehat{\cA}_{\vec{n}}$ in the planar limit for a generic operator, and also
to obtain its exact resummation in terms of Bessel functions. Indeed, for 
a generic vector $\vec{n}$ one can show that
\begin{equation}
g^{n-2\ell}\,\widehat{\cA}_{\vec{n}}\Big|_{\mathrm{planar}} 
= \frac{\big(\sqrt{\lambda}\big)^{n-\ell-1}}{2^{\frac{n}{2}+\ell-1}\phantom{\big|}}
\,I_{n-\ell+1}\big(\sqrt{\lambda}\big)\,\prod_{i=1}^\ell n_i
\label{hatAnplanar}
\end{equation}
where $n$ is, as usual, the sum of the components of $\vec{n}$ (see (\ref{On})), 
while $\ell$ is the number of these components, namely the number of traces that appear in the
corresponding operator. We have verified the validity of this formula by explicitly computing 
the planar limit of the amplitudes between the Wilson loop and all operators up to dimension $n=7$. 
In Tab.~\ref{tab3} we collect our results up to level $n=4$. We point out that
for $\ell=1$, {\it{i.e.}} for the single trace operators, our formula (\ref{hatAnplanar}) agrees 
with the findings of \cite{Semenoff:2001xp}.

\begin{table}[ht]
\begin{center}
{
\begin{tabular}{|c|c|c|}
\hline
\multirow{2}{*}{$\vec{n}$} & Expansion of
& ~~Exact expression of ~~\\
      & ~~$g^{n-2\ell}\,\widehat{\cA}_{\vec{n}}\big|_{\mathrm{planar}}
\phantom{\Big|}$ &~~$g^{n-2\ell}\,\widehat{\cA}_{\vec{n}}\big|_{\mathrm{planar}}
\phantom{\Big|}$\\
      \hline\hline
$(2)$ & $
\frac{\lambda}{8}
+ \frac{\lambda^2}{96}
+ \frac{\lambda^3}{3072} 
+\cdots\phantom{\bigg|}$ 
& $I_2\big(\sqrt{\lambda}\big)$
\\ \hline
$(3)$ & $
\!\frac{\lambda^2}{32\sqrt{2}}+\frac{\lambda^3}{512\sqrt{2}}
+\frac{\lambda^4}{20480\sqrt{2}}
+\cdots\phantom{\bigg|}$ 
& $\frac{3\sqrt{\lambda}}{2\sqrt{2}}\,I_3\big(\sqrt{\lambda}\big)$
\\ \hline
$(4)$ & $
\frac{\lambda^3}{384}+\frac{\lambda^4}{7680}
+\frac{\lambda^5}{368640}
+\cdots\phantom{\bigg|}$ 
& $\lambda\,I_4\big(\sqrt{\lambda}\big)$
\\ \hline
$(2,2)$ & $
\frac{\lambda^2}{96}+ \frac{\lambda^3}{1536} +\frac{\lambda^4}{61440}
+\cdots\phantom{\bigg|}$ 
&$\frac{\sqrt{\lambda}}{2}\,I_3\big(\sqrt{\lambda}\big)$
\\ \hline
\end{tabular}
}
\end{center}
\caption{Results for the $\mathcal{N}=4$ matrix model in the planar limit. As explained in the text, $n$ is the 
sum of the components of $\vec{n}$ while $\ell$ is the number of these components.}
\label{tab3}
\end{table}

\subsubsection*{The $\cN=2$ SCQCD theory}
Multiplying (\ref{X0}) by $\frac{3\,\zeta(3)\,g^4}{(8\pi^2)^2}$ and then taking the large $N$ limit, it is
straightforward to obtain
\begin{equation}
\delta{\cA}_{(0)}\Big|_{\mathrm{planar}}\!\!=-\frac{3\,\zeta(3)\,\lambda^2}{(8\pi^2)^2}\Big(
\frac{\lambda}{8}+\frac{2\lambda^2}{192}
+\frac{3\lambda^3}{9216}+\cdots\!\Big)\!+\!\cdots=-\frac{3\,\zeta(3)\,\lambda^2}{(8\pi^2)^2}\,I_2\big(
\sqrt{\lambda}\big)+\cdots
\label{deltaA0planar}
\end{equation}
where the last ellipses stand for terms of higher transcendentality.

In a similar way, from (\ref{X2b}) we easily get
\begin{equation}
\delta{\cA}_{(2)}\Big|_{\mathrm{planar}}=-\frac{3\,\zeta(3)\,\lambda^2}{(8\pi^2)^2}\Big(
\frac{3\lambda}{8}+\frac{4\lambda^2}{96}
+\frac{5\lambda^3}{3072}+\cdots\Big)+\cdots~.
\label{deltaA2planar}
\end{equation}
It is interesting to observe that if one compares this expression with the expansion of
the planar limit of the $\cN=4$ amplitude $\widehat{\cA}_{(2)}$ given in
(\ref{hatA2planar}), one sees that each term of the latter proportional to $\lambda^k$
gets multiplied by
\begin{equation}
-\frac{3\,\zeta(3)\,\lambda^2}{(8\pi^2)^2}\,(k+2)~.
\label{shift2}
\end{equation}
As we will see in Section~\ref{sec:pert}, this fact has a simple and nice diagrammatic interpretation. The
expansion (\ref{deltaA2planar}) can be resummed in terms of modified Bessel functions as
follows
\begin{equation}
\delta{\cA}_{(2)}\Big|_{\mathrm{planar}}=
-\frac{3\,\zeta(3)\,(\sqrt{\lambda})^5}{2(8\pi^2)^2\phantom{\big|}}\Big(I_1\big(\sqrt{\lambda}\big)
+\frac{2}{\sqrt{\lambda}} I_2\big(\sqrt{\lambda}\big)\Big)+\cdots~.
\label{deltaA2planar1}
\end{equation}
Taking into account the different normalization of the operator $\cO_{(2)}(a)$ we have used, 
our result agrees with \cite{Rodriguez-Gomez:2016cem}.

Proceeding in this way and using (\ref{X3exp})--(\ref{X22exp}), 
it is not difficult to obtain the weak-coupling expansions of $\delta\cA_{(3)}$, $\delta\cA_{(4)}$ and 
$\delta\cA_{(2,2)}$ in the planar limit, and eventually their exact expressions.
In Tab.~\ref{tab4} we have collected our findings for the terms proportional 
to $\zeta(3)$ in $\delta\cA_{\vec{n}}$ for all operators up to dimension $n=4$.

\begin{table}[ht]
\begin{center}
{
\begin{tabular}{|c|c|c|}
\hline
\multirow{2}{*}{$\vec{n}$} & Expansion of the $\zeta(3)$-term of
& ~Resummation of the $\zeta(3)$-term of ~\\
      & ~~$g^{n-2\ell}\,\delta{\cA}_{\vec{n}}\big|_{\mathrm{planar}}
\phantom{\Big|}$ &~~$g^{n-2\ell}\,\delta{\cA}_{\vec{n}}\big|_{\mathrm{planar}}
\phantom{\Big|}$\\
      \hline\hline
$(2)$ & $-\frac{3\,\zeta(3)\,\lambda^2}{(8\pi^2)^2\phantom{\big|}}\Big(
\frac{3\lambda}{8}
+ \frac{4\lambda^2}{96}
+ \frac{5\lambda^3}{3072} +\cdots\!\Big)\!
\!\!\phantom{\bigg|}$ 
& $-\frac{3\,\zeta(3)\,(\sqrt{\lambda})^5}{2(8\pi^2)^2\phantom{\big|}}\Big(I_1\big(\sqrt{\lambda}\big)
+\frac{2}{\sqrt{\lambda}} I_2\big(\sqrt{\lambda}\big)\Big)$
\\ \hline
$(3)$ & $\!\!
-\frac{3\,\zeta(3)\,\lambda^2}{(8\pi^2)^2\phantom{\big|}}\Big(\frac{3\lambda^2}{32\sqrt{2}}\!
+\!\frac{4\lambda^3}{512\sqrt{2}}
+\frac{5\lambda^4}{20480\sqrt{2}}\!+\cdots\!\Big)\!\!
\!\!\phantom{\bigg|}$ 
& $-\frac{9\,\zeta(3)\,\lambda^3}{4\sqrt{2}(8\pi^2)^2\phantom{\big|}}\,I_2\big(\sqrt{\lambda}\big)$
\\ \hline
$(4)$ & $
-\frac{3\,\zeta(3)\,\lambda^2}{(8\pi^2)^2\phantom{\big|}}\Big(\frac{4\lambda^3}{384}+\frac{5\lambda^4}{7680}
+\frac{6\lambda^5}{368640}
+\cdots\!\Big)\phantom{\bigg|}$ 
& $-\frac{3\,\zeta(3)\,(\sqrt{\lambda})^7}{2(8\pi^2)^2\phantom{\big|}}\,I_3\big(\sqrt{\lambda}\big)$
\\ \hline
$(2,2)$ & $
-\frac{3\,\zeta(3)\,\lambda^2}{(8\pi^2)^2\phantom{\big|}}\Big(\frac{6\lambda^2}{96}+ \frac{7\lambda^3}{1536} +\frac{8\lambda^4}{61440}
+\cdots\!\Big)\phantom{\bigg|}$ 
&$-\frac{3\,\zeta(3)\,\lambda^3}{4(8\pi^2)^2\phantom{\big|}}\Big(I_2\big(\sqrt{\lambda}\big)
+\frac{6}{\sqrt{\lambda}} I_3\big(\sqrt{\lambda}\big)\Big)$
\\ \hline
\end{tabular}
}
\end{center}
\caption{Results for the $\mathcal{N}=2$ superconformal matrix model in the planar limit. As before, $n$ is 
the sum of the components of $\vec{n}$ while $\ell$ is their number.}
\label{tab4}
\end{table}

{From} these explicit results it is possible to infer the following general formula
\begin{align}
g^{n-2\ell}\,\delta{\cA}_{\vec{n}}\Big|_{\mathrm{planar}}
&=-\frac{3\,\zeta(3)}{(8\pi^2)^2}\,
\frac{\big(\sqrt{\lambda}\big)^{n-\ell+4}}{2^{\frac{n}{2}+\ell}\phantom{\big|}}\left\{\bigg[
I_{n-\ell}\big(\sqrt{\lambda}\big)
+\frac{2(\ell-1)}{\sqrt{\lambda}}\,I_{n-\ell+1}\big(\sqrt{\lambda}\big)\bigg]\prod_{k=1}^\ell n_k
\right.
\notag\\
&\left.~~~+\bigg(\sum_{i=1}^\ell \delta_{n_i,2}\bigg)\frac{2}{\sqrt{\lambda}}\,
I_{n-\ell+1}\big(\sqrt{\lambda}\big)\prod_{k=1}^\ell n_k\right\}+\cdots
\label{deltaAnexact}
\end{align}
which we have verified in all cases up to $n=7$. We observe that there is a contribution, represented by 
the second line above, which occurs only when the operator $\cO_{\vec{n}}(a)$ contains at least
a factor of the type $\tr a^2$. This fact has a precise diagrammatic counterpart, as we will see
in the next Section.

Comparing the two exact expressions (\ref{deltaAnexact}) and (\ref{hatAnplanar}) and using the 
properties of the modified Bessel functions, it is not difficult to realize that
\begin{equation}
g^{n-2\ell}\,\delta\cA_{\vec{n}}\Big|_{\mathrm{planar}}= -\frac{3\,\zeta(3)\,\lambda^2}{(8\pi^2)^2}\,
\bigg(\lambda\,\frac{\partial}{\partial\lambda}+\ell+
\sum_{i=1}^\ell\delta_{n_i,2}\bigg)\bigg(g^{n-2\ell}\,\widehat{\cA}_{\vec{n}}\Big|_{\mathrm{planar}}\bigg)+\cdots
\label{deltaAnhatAn}
\end{equation}
where, as usual, the ellipses stand for terms of higher transcendentality. Such a relation implies
that if we multiply each term $\lambda^k$ in the weak-coupling expansion
of $g^{n-2\ell}\,\widehat{\cA}_{\vec{n}}\big|_{\mathrm{planar}}$ by
\begin{equation}
-\frac{3\,\zeta(3)\,\lambda^2}{(8\pi^2)^2}\,\Big(k+\ell+\sum_{i=1}^\ell \delta_{n_i,2}\Big)~,
\label{shiftn}
\end{equation}
then we obtain the expansion of the $\zeta(3)$-correction of the corresponding $\cN=2$ planar amplitude
$g^{n-2\ell}\,\delta\cA_{\vec{n}}\big|_{\mathrm{planar}}$. Also this formula, which generalizes 
(\ref{shift2}) to any $\vec{n}$, has a simple and nice interpretation in terms of field theory diagrams, 
as we will see in the next Section. 

\section{Perturbative checks in field theory}
\label{sec:pert}
We now consider the direct field theory computation of the expectation values of chiral operators 
with a circular BPS Wilson loop in a superconformal $\mathcal{N}=2$ theory defined 
on $\mathbb{R}^4$.

As explained in Section~\ref{sec4:WLdefect}, conformal invariance implies that all information about these 
expectation values is contained in the amplitudes $A_{\vec{n}}$ defined in (\ref{Ang}). 
The conjecture we want to test is that these amplitudes match the corresponding ones $\cA_{\vec{n}}$
in the matrix model that we introduced in (\ref{Angdef}), namely we want to show that
\begin{equation}
\label{AisCA}
A_{\vec n} \,=\, \cA_{\vec n}~.
\end{equation}
The diagrammatic evaluation in field theory of the correlators $A_{\vec n}$ beyond tree-level is 
in general quite complicated. However, it becomes tractable if one only computes 
the difference between the $\mathcal{N}=2$ result and the one we would have in the $\mathcal{N}=4$ theory. So we exploit the same techniques introduced in Section \ref{sec:FTactions}, and we shall check, up to two loops,  the following equality:
\begin{equation}
\label{checkN2N4}
\delta A_{\vec n} = \delta \cA_{\vec n}~, 
\end{equation}
where $\delta \cA_{\vec n}$ is the difference between the 
$\cN=2$ and $\cN=4$ matrix model results introduced in (\ref{diffAn}). We stress again that, even though we concentrate on the SCQCD theory, these results are valid for any $\cN=2$ SCFT with SU$(N)$ gauge group.

\subsection{Tree-level}
\label{subsec:tree}
\noindent At the lowest order in $g$ the $\cN=2$ and $\cN=4$ amplitudes coincide:
\begin{equation}
\label{Atree}
A_{\vec n}\Big|_{\mathrm{tree-level}} = 
\widehat A_{\vec n}\Big|_{\mathrm{tree-level}}~;
\end{equation}
in other words, 
\begin{equation}
\delta A_{\vec n}\Big|_{\mathrm{tree-level}} = 0~.
\end{equation}
Also in the matrix model this difference vanishes at the lowest order, see (\ref{deltaAntree1loop}).
Thus, the equality (\ref{checkN2N4}) is satisfied at tree level. 

\begin{figure}[ht]
\begin{center}
\begingroup%
  \makeatletter%
  \providecommand\color[2][]{%
    \errmessage{(Inkscape) Color is used for the text in Inkscape, but the package 'color.sty' is not loaded}%
    \renewcommand\color[2][]{}%
  }%
  \providecommand\transparent[1]{%
    \errmessage{(Inkscape) Transparency is used (non-zero) for the text in Inkscape, but the package 'transparent.sty' is not loaded}%
    \renewcommand\transparent[1]{}%
  }%
  \providecommand\rotatebox[2]{#2}%
  \ifx\svgwidth\undefined%
    \setlength{\unitlength}{200bp}%
    \ifx\svgscale\undefined%
      \relax%
    \else%
      \setlength{\unitlength}{\unitlength * \real{\svgscale}}%
    \fi%
  \else%
    \setlength{\unitlength}{\svgwidth}%
  \fi%
  \global\let\svgwidth\undefined%
  \global\let\svgscale\undefined%
  \makeatother%
  \begin{picture}(1,0.63419123)%
    \put(0,0){\includegraphics[width=\unitlength,page=1]{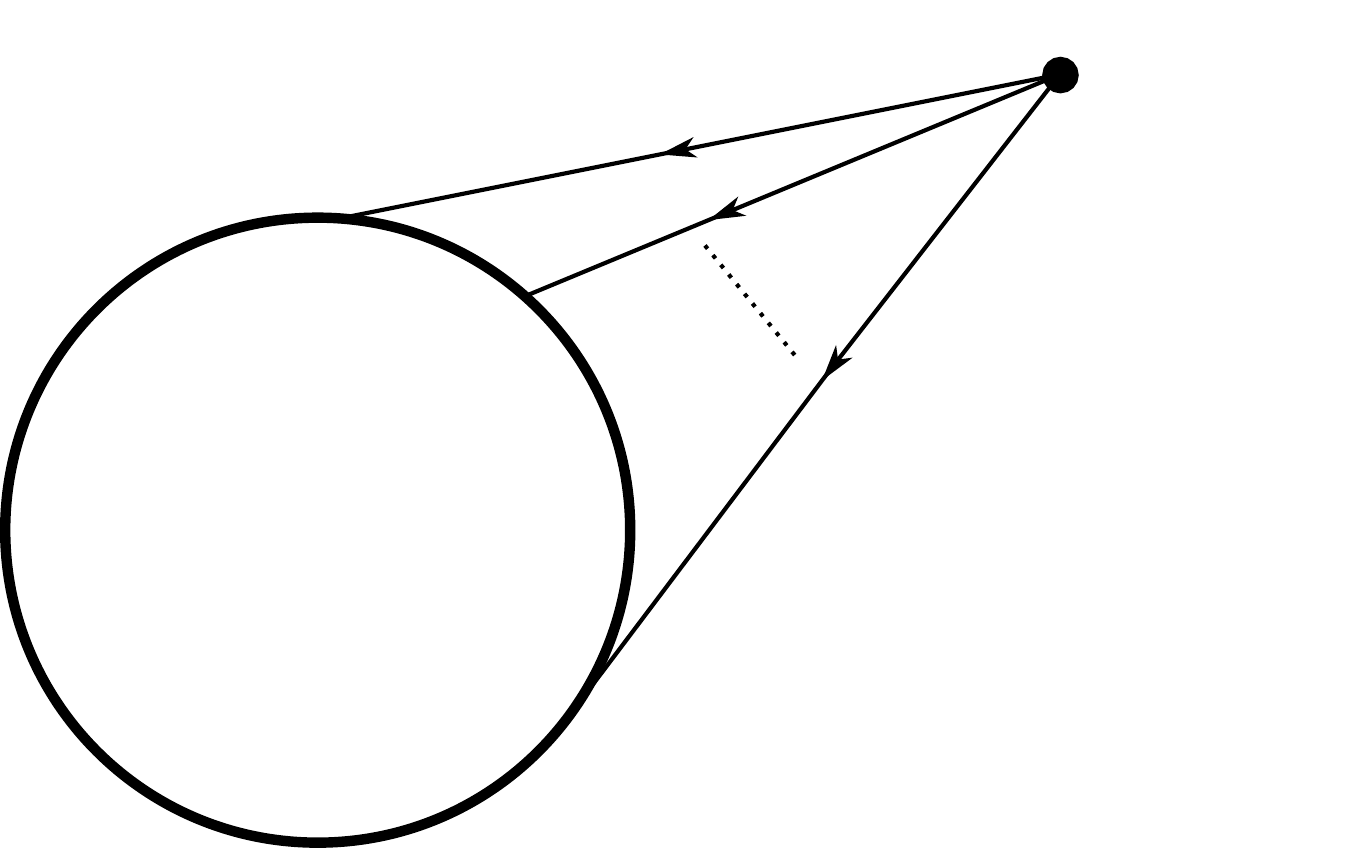}}%
    \put(0.71542891,0.61051217){\color[rgb]{0,0,0}\makebox(0,0)[lb]{\smash{$O_{\vec{n}}(x)$}}}%
    \put(0.05320224,0.12273485){\color[rgb]{0,0,0}\makebox(0,0)[lb]{\smash{$W(C)$}}}%
  \end{picture}%
\endgroup%
\end{center}
\caption{At the lowest perturbative order, the operator $O_{\vec n}(x)$ is connected to the Wilson loop by $n$ scalar propagators. Exploiting conformal invariance, we can place the operator in the origin, {\it{i.e.}}
in the center of the Wilson loop. Nevertheless, in this and in the following pictures we will continue to 
place it outside the loop to avoid graphical clutter.}
\label{fig:WOntree}
\end{figure}

Actually, in this case it is easy (and also convenient for later purposes) 
to check directly the validity of (\ref{AisCA}). 
Performing this check is helpful also to establish some facts that will be useful at higher orders; 
in particular, the way the path-ordered integration over the Wilson loop simplifies in the tree-level case 
will be exploited also in the two-loop order computations. Thus, for later convenience
we briefly show some details.
At the lowest order in $g$, the $n$ chiral fields $\varphi$ appearing
in the operator $O_{\vec n}$ must be contracted with the $n$ antichiral fields
present in the term of order $n$ in the expansion on the Wilson loop operator (\ref{WLdefsection4}).
This is represented by the diagram in Figure~\ref{fig:WOntree}. 

Thus, we have
\begin{equation}
\label{cosW2}
\big\langle\, W(C)\,O_{\vec n}(0) \,\big\rangle\Big|_{\mathrm{tree-level}} =\frac{1}{N}\,\frac{g^{n}}{n!}\,
\Big\langle\, \mathcal{P}\,\tr\Big({\prod_{i=1}^n} \oint_{C}d\tau_{i}\,
\frac{R}{\sqrt{2}}\,\bar{\varphi}(x_i)\Big) O_{\vec n}(0) \,\Big\rangle 
\end{equation}
where we have denoted by $x_i = x(\tau_i)$ the positions along the Wilson loop $C$.
Using (\ref{OnR}), we rewrite this expression as
\begin{align}
\label{cosW3}
\big\langle\,W(C)\,O_{\vec n}(0)\,\big\rangle\Big|_{\mathrm{tree-level}} &= 
\frac{1}{N}\,\frac{g^n \,R^n}{2^{\frac n2}\,n!}\,
\cP\, {\prod_{i=1}^n} \oint_{C}d\tau_{i}\,
\tr \big(T^{a_1} \cdots  T^{a_n} \big) \,R^{b_1\ldots b_n}_{\vec{n}}\nonumber\\
& ~~~\times \big\langle\,\bar{\varphi}^{a_1}(x_1)\cdots \bar{\varphi}^{a_n}(x_n)
\,\varphi^{b_1}(0) \cdots \varphi^{b_n}(0)\,\big\rangle~.
\end{align}
The vacuum expectation value in the second line above is computed using the free scalar propagator
\begin{equation}
\label{scalprop}
\big\langle\,\bar\varphi^{a}(x_i)\,\varphi^{b}(0)\,\big\rangle 
= \frac{\delta^{ab}}{4\pi^2\,x_i^2}= \frac{\delta^{ab}}{4\pi^2 R^2}
\end{equation} 
where we have exploited the fact that $x_i = x(\tau_i)$ belongs to the circle $C$ of radius $R$ and thus can
be parameterized as in (\ref{circle}). In view of this, 
when we apply Wick's theorem in (\ref{cosW3}) we obtain an integrand that  
does not depend on the variables $\tau_i$. 
The path ordering becomes therefore irrelevant and, from the integration over $\tau_i$, we 
simply get a factor of $(2\pi)^n$. Moreover the $n!$ different contraction patterns all yield the same expression, due to the symmetry of the tensor $R_{\vec n}$. 
Taking all this into account, we get
\begin{equation}
\label{cosW4}
\big\langle\,W(C)\,O_{\vec n}(0)\,\big\rangle\Big|_{\mathrm{tree-level}} 
= \frac{1}{N}\,\frac{g^n}{2^{\frac n2}}\,
\frac{1}{(2\pi R)^n}\,R_{\vec{n}}^{\,b_1\dots b_n}\,\tr \big(T_{b^1}\dots T^{b_n}\big)~,
\end{equation}
which implies that
\begin{equation}
\label{Antree0}
A_{\vec n}\Big|_{\mathrm{tree-level}} \,=\, 
\widehat A_{\vec n}\Big|_{\mathrm{tree-level}} 
=\frac{g^n}{N\,2^{\frac{n}{2}}}\,R_{\vec{n}}^{\,b_1\dots b_n}\,\tr \big(T^{b_1}\dots T^{b_n}\big)~,
\end{equation}
in full agreement with the matrix model result (\ref{Antree}).

\subsection{Loop corrections}
\label{subsec:2loop} 
\begin{figure}[ht]
	\begin{center}
		\includegraphics[scale=0.5]{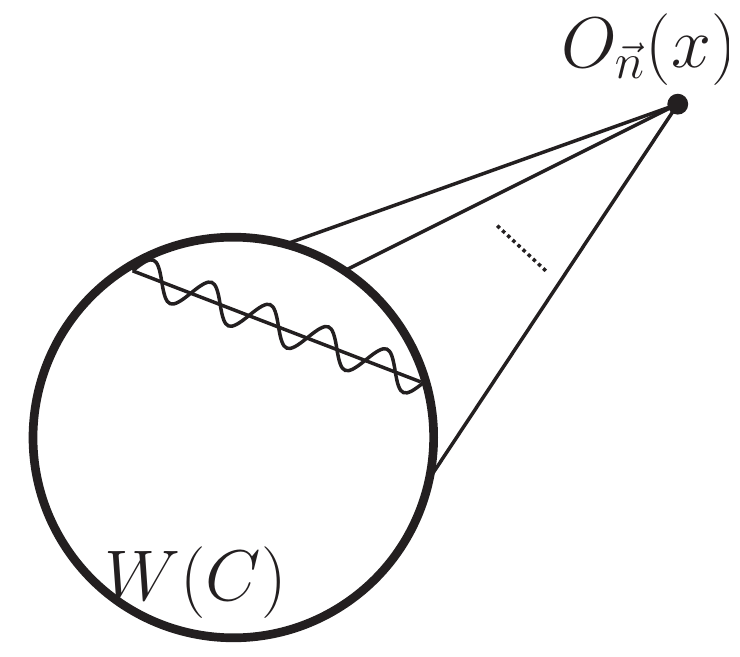}\hspace{1cm}
		\includegraphics[scale=0.5]{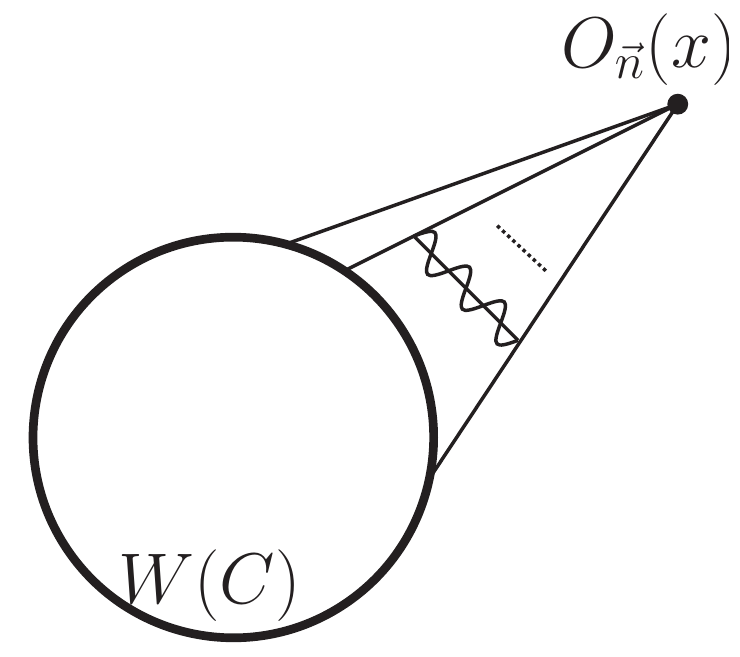}
		\hspace{1cm}
		\includegraphics[scale=0.5]{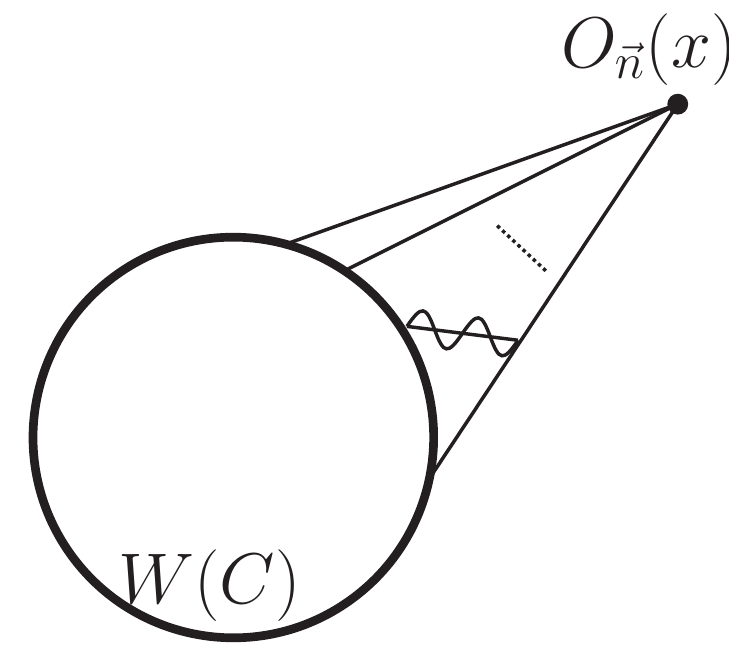}
	\end{center}
	\caption{Examples of diagrams which appear at order $g^2$ with respect to the tree-level amplitude and which do not contain hypermultiplets and therefore vanish in the difference between the $\cN=2$ and the $\cN=4$ theory. We have used the straight/wiggle line to denote the sum of the gluon and the scalar propagator, see Figure \ref{fig:WLtree}}
	\label{fig:WOnN4}
\end{figure}

At higher orders in $g$ we concentrate on the difference $\delta A_{\vec{n}}$. As we already
pointed out, the number of diagrams which contribute to this difference
is massively reduced. For example, all diagrams represented in Figure~\ref{fig:WOnN4} yield a $g^2$
correction with respect to the tree-level amplitude $A_{\vec{n}}$ but they should not be considered
in the computation of $\delta A_{\vec{n}}$ since they do not contain internal lines with $H$ or $Q$ hypermultiplets.

\subsubsection*{One loop}
\begin{figure}[ht]
\begin{center}
\begingroup%
  \makeatletter%
  \providecommand\color[2][]{%
    \errmessage{(Inkscape) Color is used for the text in Inkscape, but the package 'color.sty' is not loaded}%
    \renewcommand\color[2][]{}%
  }%
  \providecommand\transparent[1]{%
    \errmessage{(Inkscape) Transparency is used (non-zero) for the text in Inkscape, but the package 'transparent.sty' is not loaded}%
    \renewcommand\transparent[1]{}%
  }%
  \providecommand\rotatebox[2]{#2}%
  \ifx\svgwidth\undefined%
    \setlength{\unitlength}{200bp}%
    \ifx\svgscale\undefined%
      \relax%
    \else%
      \setlength{\unitlength}{\unitlength * \real{\svgscale}}%
    \fi%
  \else%
    \setlength{\unitlength}{\svgwidth}%
  \fi%
  \global\let\svgwidth\undefined%
  \global\let\svgscale\undefined%
  \makeatother%
  \begin{picture}(1,0.63419123)%
    \put(0,0){\includegraphics[width=\unitlength,page=1]{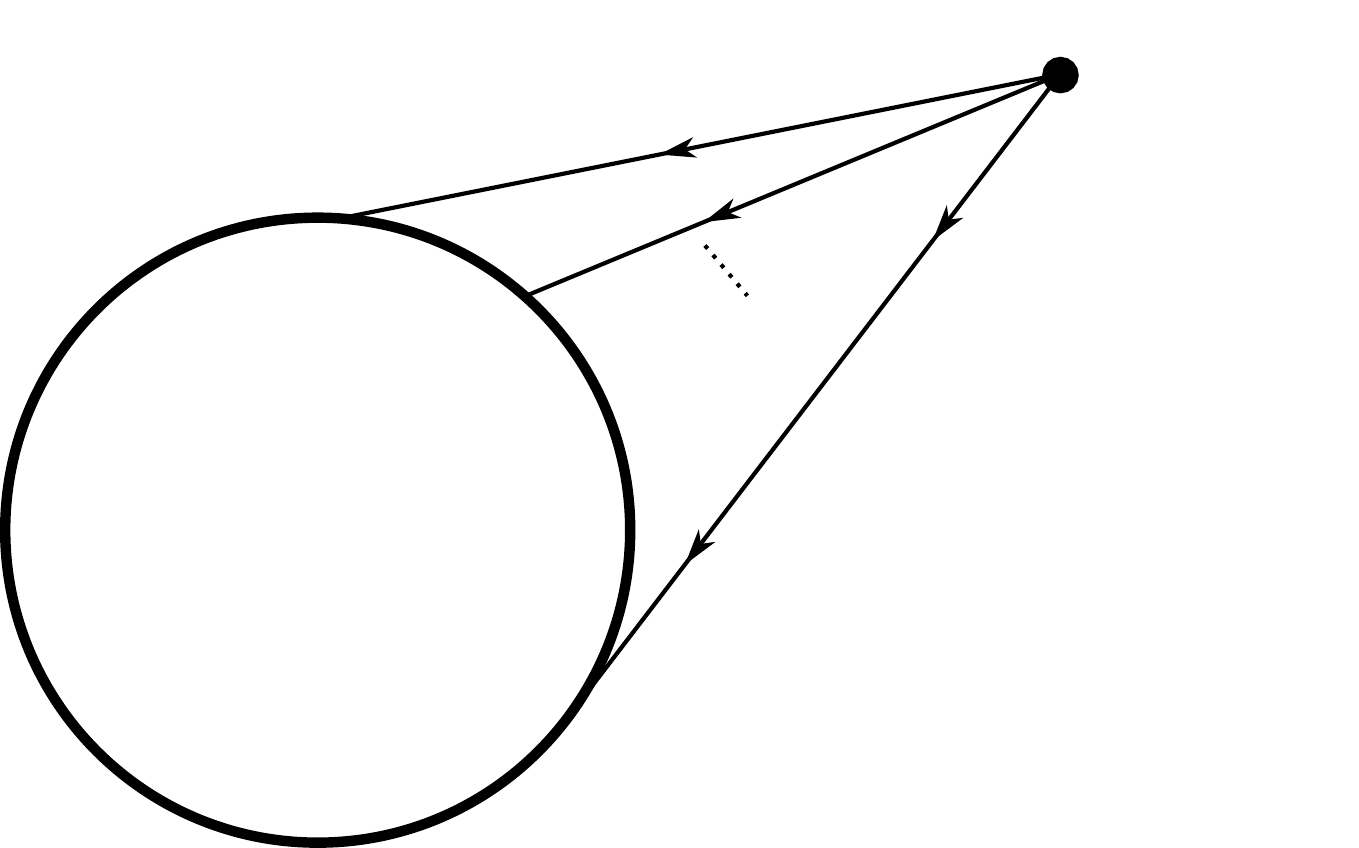}}%
    \put(0.70542891,0.60051217){\color[rgb]{0,0,0}\makebox(0,0)[lb]{\smash{$O_{\vec{n}}(x)$}}}%
    \put(0.05320224,0.12273485){\color[rgb]{0,0,0}\makebox(0,0)[lb]{\smash{$W(C)$}}}%
    \put(0,0){\includegraphics[width=\unitlength,page=2]{WOn1loop.pdf}}%
    \put(0.59002566,0.32527747){\color[rgb]{0,0,0}\makebox(0,0)[lb]{\smash{$1$}}}%
  \end{picture}%
\endgroup%
\end{center}
\caption{The only diagrams that yield a $g^2$ correction to the tree-level amplitude $A_{\vec{n}}$
and contain $Q$ and $H$ hypermultiplets arise from the one-loop correction of the external 
scalar propagators.}
\label{fig:WOn1loop}
\end{figure}

It is easy to see that in any $\cN=2$ superconformal theory there are no corrections of 
order $g^2$ with respect to the tree-level result. 
In fact, at this order the only possible diagrams containing $H$ and $Q$ hypermultiplets arise 
from the one-loop correction of the external scalar propagators as shown in Figure~\ref{fig:WOn1loop}. 
This one-loop correction vanishes, as we showed in Subsection \ref{subsec::difference}, due to the vanishing of the one-loop coefficient of the $\beta$ function in any conformal theory. Therefore, in any superconformal $\cN=2$ theory we have
\begin{equation}
\label{delta1loop}
\delta A_{\vec n}\Big|_{\rm 1-loop} = 0~,
\end{equation}
in full agreement with the matrix model result (see (\ref{deltaAntree1loop})).

\subsubsection*{Two loops}
\begin{figure}[ht]
	\begin{center}
		\includegraphics[scale=0.5]{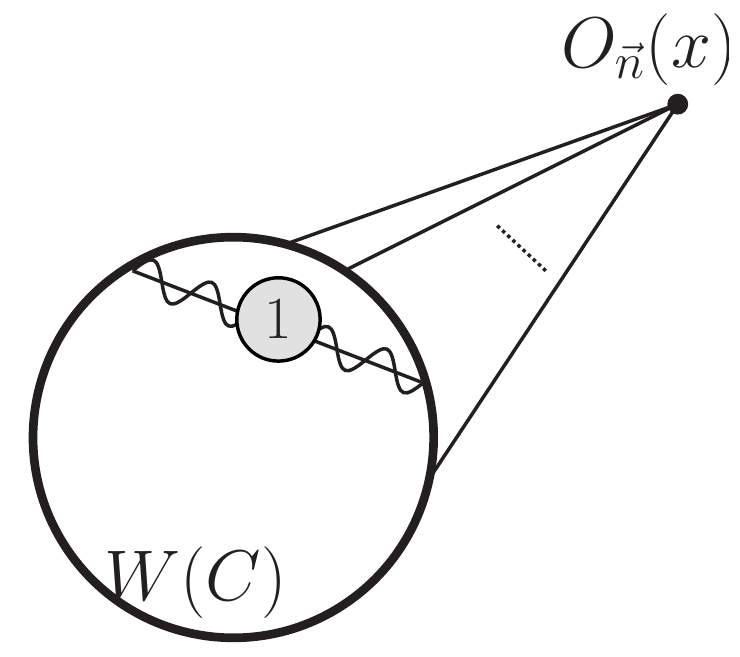}\hspace{1cm}
		\includegraphics[scale=0.5]{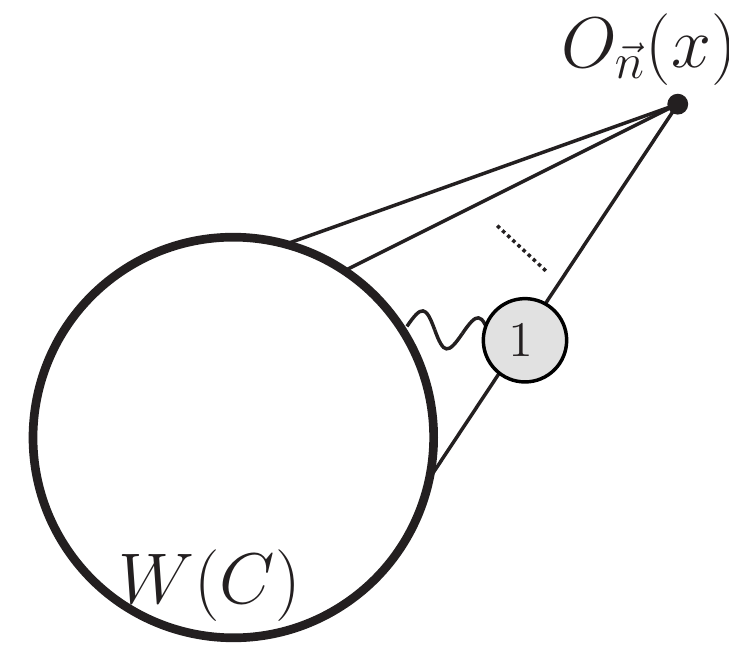}
	\end{center}
	\caption{Some examples of diagrams contributing to $\delta A_{\vec{n}}$ at two loops. 
The one-loop correction of the 
gluon propagator vanishes like the scalar case, see Subsection \ref{subsec:swl} for a full discussion. The one-loop 
correction to the 3-point vertex vanishes in the superconformal theory, see Figure \ref{fig:2loopp}.}
\label{fig:WOn1loopg4}
\end{figure}

Let us now consider the two-loop corrections, {\it{i.e.}} those at order $g^4$ with respect to the
tree-level amplitudes, and focus on the difference $\delta A_{\vec{n}}$.
The $H$ or $Q$ diagrams which contribute at this order can be divided into two classes. The first one
is formed by those diagrams which contain a sub-diagram with the one-loop correction to the scalar
propagator, or to the gluon propagator or to the 3-point vertex. Some examples of such diagrams are shown in Figure~\ref{fig:WOn1loopg4}.

\begin{figure}[ht]
\vspace{0.3cm}
\begin{center}
\begingroup%
  \makeatletter%
  \providecommand\color[2][]{%
    \errmessage{(Inkscape) Color is used for the text in Inkscape, but the package 'color.sty' is not loaded}%
    \renewcommand\color[2][]{}%
  }%
  \providecommand\transparent[1]{%
    \errmessage{(Inkscape) Transparency is used (non-zero) for the text in Inkscape, but the package 'transparent.sty' is not loaded}%
    \renewcommand\transparent[1]{}%
  }%
  \providecommand\rotatebox[2]{#2}%
  \ifx\svgwidth\undefined%
    \setlength{\unitlength}{305bp}%
    \ifx\svgscale\undefined%
      \relax%
    \else%
      \setlength{\unitlength}{\unitlength * \real{\svgscale}}%
    \fi%
  \else%
    \setlength{\unitlength}{\svgwidth}%
  \fi%
  \global\let\svgwidth\undefined%
  \global\let\svgscale\undefined%
  \makeatother%
  \begin{picture}(1,0.31558777)%
    \put(0,0){\includegraphics[width=\unitlength,page=1]{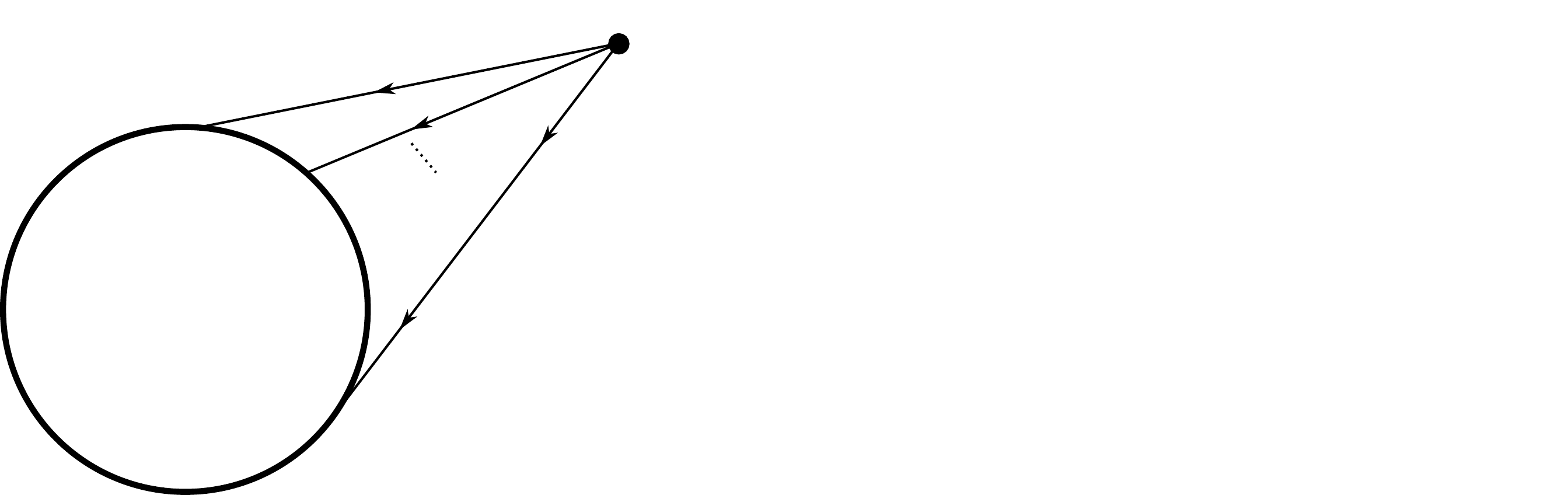}}%
    \put(0.33890409,0.30380454){\color[rgb]{0,0,0}\makebox(0,0)[lb]{\smash{$O_{\vec n}(x)$}}}%
    \put(0.05647463,0.04107561){\color[rgb]{0,0,0}\makebox(0,0)[lb]{\smash{$W(C)$}}}%
\put(0.27647463,0.04107561){\color[rgb]{0,0,0}\makebox(0,0)[lb]{\smash{\textbf{(i)}}}}%
    \put(0,0){\includegraphics[width=\unitlength,page=2]{WOn2loop.pdf}}%
    \put(0.29708383,0.16184158){\color[rgb]{0,0,0}\makebox(0,0)[lb]{\smash{$2$}}}%
    \put(0,0){\includegraphics[width=\unitlength,page=3]{WOn2loop.pdf}}%
    \put(0.83325455,0.30380454){\color[rgb]{0,0,0}\makebox(0,0)[lb]{\smash{$O_{\vec n}(x)$}}}%
    \put(0.56082511,0.04107561){\color[rgb]{0,0,0}\makebox(0,0)[lb]{\smash{$W(C)$}}}%
    \put(0,0){\includegraphics[width=\unitlength,page=4]{WOn2loop.pdf}}%
    \put(0.73047399,0.23086561){\color[rgb]{0,0,0}\makebox(0,0)[lb]{\smash{$2$}}}%
\put(0.78582511,0.04107561){\color[rgb]{0,0,0}\makebox(0,0)[lb]{\smash{\textbf{(j)}}}}%
  \end{picture}%
\endgroup%
\end{center}
\caption{Diagrams that contribute to $\delta A_{\vec{n}}$ at two loops in the $\cN=2$ superconformal
theory. 
Diagram $\mathbf{(i)}$ on the left contains the irreducible two-loop correction of the scalar propagator, while diagram $\mathbf{(j)}$ on the right contains the 
two-loop effective vertex}
\label{fig:WOn2loop}
\end{figure}

\begin{figure}[ht]
	\begin{center}
		\includegraphics[scale=0.7]{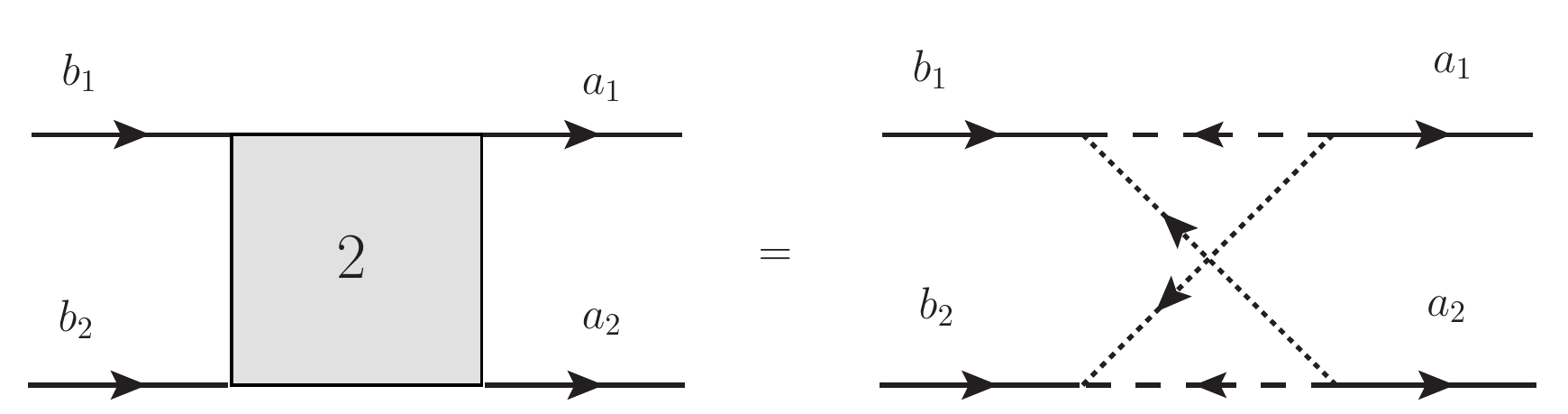}
	\end{center}
	\caption{The two-loop effective vertex that can contribute to the amplitude $A_{\vec{n}}$
in the difference theory.}
\label{fig:scalbox2l}
\end{figure}

All these diagrams vanish in the $\cN=2$ superconformal theory. 
The only class of diagrams that can contribute to $\delta A_{\vec{n}}$ at two loops in the
superconformal theory are those of the type displayed in Figure~\ref{fig:WOn2loop}.
They contain either the irreducible two-loop correction of the scalar propagator that we already computed in Subsection \ref{subsec:scalprop} (see Figure \ref{fig:2loopp}), or the two-loop effective vertex represented 
in Figure~\ref{fig:scalbox2l}. Thus, we can write
\begin{equation}
\delta A_{\vec{n}}\Big|_{\mathrm{2-loop}} = \,I_{\vec{n}}+ J_{\vec{n}}
\label{deltaAn2loop}
\end{equation}
where $I_{\vec{n}}$ and $J_{\vec{n}}$ correspond, respectively, to the diagrams of type $\mathbf{(i)}$
and $\mathbf{(j)}$.

Let us first consider the irreducible two-loop correction
of the scalar propagator. We already computed it in momentum space, see equation \eqref{2loopprop}.
Moving to the configuration space, we find that the two-loop correction of the scalar propagator is
\begin{equation}
\parbox[c]{.35\textwidth}{\includegraphics[width = .35\textwidth]{2loops-nonzero.pdf}}
=
g^4\,\frac{\zeta(3)}{(8\pi^2)^2} \cC_4^\prime\Big[\frac{\delta^{bc}}{4\pi^2(x_1-x_2)^2}\Big]
\label{scalprop2la}
\end{equation}
where $\cC_4^\prime$ for the SCQCD theory is explicitly given in the first row of Table \ref{tab:C4C6conf}, and the expression in square brackets is the tree-level propagator. Therefore, when we compute
the amplitude $I_{\vec{n}}$ corresponding to the diagram $\mathbf{(i)}$ of Figure~\ref{fig:WOn2loop}, 
we simply obtain an expression which is proportional to the tree-level result (\ref{Antree0}). Indeed we 
get for the SCQCD theory:
\begin{equation}
I_{\vec{n}}= -n\,g^4\,\frac{3\,\zeta(3)}{(8\pi^2)^2}\Big[\frac{g^n}{N\,2^{\frac{n}{2}}}
\,R_{\vec{n}}^{\,b_1\dots b_n}\,\tr \big(T^{b_1}\dots T^{b_n}\big)\Big](N^2+1)
\label{In}
\end{equation}
where the overall factor of $n$ is due to the fact that the two-loop correction (\ref{scalprop2la})
can be inserted in any of the $n$ external propagators.

Let us now consider the two-loop diagram $\mathbf{(j)}$ of Figure~\ref{fig:WOn2loop}. 
To compute the corresponding amplitude $J_{\vec{n}}$, we have to perform all contractions as in 
the tree-level diagram but with two scalar propagators replaced by the sub-structure corresponding
to the two-loop effective vertex of Figure~\ref{fig:scalbox2l}. The latter has been analyzed
in \cite{Billo:2017glv} to which we refer again for details. Considering that the two external legs with
color indices $b_1$ and $b_2$ are inserted at the point $x$ where the operator $O_{\vec{n}}$ is located,
and the other two external legs with indices $a_1$ and $a_2$ are inserted at two points $x_1$ and $x_2$
on the circular Wilson loop, the relevant expression is:
\begin{equation}
2\,g^4\,\cC_4^{\prime~\,b_1b_2a_1a_2}\,W_4(x,x;x_1,x_2)
\label{boxsub}
\end{equation}
where the color factor precisely corresponds to the trace combination:
\begin{equation}
\cC_4^{\prime~\,b_1b_2a_1a_2} = \trp \big(T^{b_1}T^{a_1}T^{b_2}T^{a_2}\big)~,
\end{equation}
that we encounter in the matrix model computations. We explicitly evaluate it again for the SCQCD theory:
\begin{equation}
C_4^{\,b_1b_2a_1a_2}\big|_A=-\frac{1}{2}\big(\delta^{b_1a_1}\,\delta^{b_2a_2}+\delta^{b_1 b_2}\,\delta^{a_1 a_2}
+\delta^{b_1a_2}\,\delta^{b_2a_1}\big)~,
\label{C4uncontracted}
\end{equation}
The superspace integral can be computed as a two-loops contribution of ladder diagrams to the four-point function in a $\phi^3$-theory, explicitly done in \cite{Usyukina:1993ch} and reproduced in Appendix B of \cite{Billo:2017glv}. It is always finite and leads to:
\begin{equation}
W_4(x,x;x_1,x_2)=\frac{6\, \zeta(3)}{(16 \pi^2)^2}\,\Big[
\frac{1}{4\pi^2(x-x_{1})^2}\,\frac{1}{4\pi^2(x-x_{2})^2}\Big]~.
\label{W4configspace}
\end{equation}
As is clear from the expression in square brackets, we still recover the same space dependence of two 
scalar propagators as in the tree-level computation, even if the color structure of the $C^\prime_4$ tensor is
different. Exploiting conformal invariance to set $x=0$ and recalling the parametrization (\ref{circle})
for points on a circle, the above square brackets simply becomes $1/(2\pi R)^4$;
thus the path-ordering and the integration over the Wilson loop become trivial to perform, just
as they were in the tree-level amplitude. Putting everything together and replacing any pair of external
scalar propagators with this effective two-loop vertex in all possible ways, we obtain
\begin{equation}
\begin{aligned}
J_{\vec{n}}&=g^4\,\frac{3\,\zeta(3)}{(8\pi^2)^2}\,
\Big[\frac{g^n}{N\,2^{\frac{n}{2}}}\,R_{\vec{n}}^{\,b_1\dots b_n}\,
\tr \big(T^{a_1}\dots T^{a_n}\big)\Big]
\\
&~~~\times 2\!\!\!
\sum_{p \in S_{n-1}} \!\!\!C_4^{\,b_1b_2 a_{p(1)} a_{p(2)}}
\, \delta^{b_3a_{p(3)}}
\dots\delta^{b_{n-1}a_{p(n-1)}}\,\delta^{b_n a_n}
\end{aligned}
\label{Jn}
\end{equation}
where $p\in S_{n-1}$ are the permutations of $(n-1)$ elements.
We observe that the $1/n!$ coming from the expansion of the Wilson loop operator at order $g^n$
is compensated by a factor of $n!$ that arises when we take into account the complete symmetry
of the tensor $R_{\vec{n}}$ and the cyclic symmetry of the trace factor in the square bracket. 
Furthermore the factor of 2 in the last line of (\ref{Jn}) is a combinatorial factor due to the 
multiplicity of the two-loop box diagram of Figure~\ref{fig:scalbox2l}.

Summing $I_{\vec{n}}$ and $J_{\vec{n}}$, we get
\begin{align}
\delta A_{\vec{n}}\Big|_{\mathrm{2-loop}}&= -g^4\,\frac{3\, \zeta(3)}{(8 \pi^2)^2}\,
\Big[\frac{g^n}{N\,2^{\frac{n}{2}}}\,R_{\vec{n}}^{\,b_1\dots b_n}\,
\tr \big(T^{a_1}\dots T^{a_n}\big)\Big]\phantom{\bigg|}\label{deltaAn2loopfin}\\
&\hspace{-0.75cm}
\times \Big[n\,(N^2+1)\,\delta^{b_1a_1}\dots\delta^{b_na_n}-2\!\!\!
\sum_{p \in S_{n-1}} \!\!\!C_4^{\,b_1b_2 a_{p(1)} a_{p(2)}}
\, \delta^{b_3a_{p(3)}}
\dots\delta^{b_{n-1}a_{p(n-1)}}\,\delta^{b_n a_n}\,\Big]~.\phantom{\bigg|}\notag
\end{align}
This is the final result of our diagrammatic computation of the two-loop correction to the
amplitude $A_{\vec{n}}$ in the $\cN=2$ superconformal theory.

As an example, we work out the explicit expression for the lowest dimensional operator $O_{(2)}$. In this
case, we simply have
\begin{equation}
R_{(2)}^{b_1b_2}=\tr\big(T^{b_1}T^{b_2}\big)=\frac{1}{2}\,\delta^{b_1b_2}~.
\end{equation}
Thus, the contribution from the diagram $\mathbf{(i)}$ is (see (\ref{In})):
\begin{equation}
I_{(2)}= -2\,g^4\,\frac{3\,\zeta(3)}{(8\pi^2)^2}\Big[\frac{g^2}{2\,N}
\frac{(N^2-1)}{4}\Big](N^2+1)~,
\label{I2a}
\end{equation}
while from the diagram $\mathbf{(j)}$ we get (see (\ref{Jn})):
\begin{equation}
J_{(2)}= -g^4\,\frac{3\,\zeta(3)}{(8\pi^2)^2}\Big[\frac{g^2}{2\,N}\frac{(N^2-1)}{4}\Big]
(N^2+1)~.\label{J2}
\end{equation}
Note that in this case both diagrams $\mathbf{(i)}$ and $\mathbf{(j)}$ provide color contributions with the same leading power of $N$. This is a specific property of this operator and it does not hold for higher dimensional operators unless they contain a factor of $\tr \phi^2$. We are going to give a further evidence in a moment. This fact will have important consequences 
for the planar limit as we will see in the following Subsection.
Summing (\ref{I2a}) and (\ref{J2}), we finally get
\begin{equation}
\delta A_{(2)}\Big|_{\mathrm{2-loop}}= -g^6\,\frac{\zeta(3)}{(8\pi^2)^2}\,
\frac{9(N^2-1)(N^2+1)}{8N}~,
\label{deltaA22loopft}
\end{equation}
in perfect agreement with the matrix model result (\ref{deltaA22loop}).

We have explicitly performed similar checks for many operators of higher dimension. We report the final results for the $(4)$ and $(2,2)$ cases. See Appendix B of \cite{Billo:2018oog} for the full computations. 
\begin{align}
I_{(4)}&= -g^8\,
\frac{\zeta(3)}{(8\pi^2)^2}\,\frac{(N^2-1)(N^2+1)(N^4-6N^2+18)}{32 N^3}~,\notag \\
J_{(4)}&= - g^8\,\frac{\zeta(3)}{(8 \pi^2)^2}\,\frac{(N^2-1)(7N^4-30N^2+63)}{32 N^3}~.
\label{IJ4}
\end{align} 
Notice that in the large-$N$ limit, $J_{(4)}$ is subleading with respect to $I_{(4)}$. 
Summing the two contributions, we find that 
the total amplitude $\delta A_{(4)}\big|_{\mathrm{2-loops}} $ is
\begin{equation}
\delta A_{(4)}\Big|_{\mathrm{2-loops}} = I_{(4)} + J_{(4)}
=-g^8\frac{\zeta(3)}{(8\pi^2)^2}
\frac{(N^2-1)(N^6+2N^4-18N^2+81)}{32N^3}
\end{equation}
which exactly matches the matrix model expression reported in the last-but-one row of Tab.~(\ref{tab2}).\\
We report the same results for the $(2,2)$ case:
\begin{align}\label{IJ22}
I_{(2,2)}&= -g^8\,
\frac{\zeta(3)}{(8\pi^2)^2}\,\frac{(N^2-1)(N^2+1)(2N^2-3)}{16 N^2}~,\notag \\
J_{(2,2)}&=-g^8\,
\frac{\zeta(3)}{(8\pi^2)^2}\,\frac{(N^2-1)(N^2+7)(2N^2-3)}{32 N^2}~.
\end{align}
We explicitly notice that in this case both $I_{(2,2)}$ and $J_{(2,2)}$ contribute to the leading order in the large-N limit. In total we get:
\begin{equation}
\delta A_{(2,2)}\Big|_{\mathrm{2-loops}} = I_{(2,2)} + J_{(2,2)}
=-g^8\frac{\zeta(3)}{(8\pi^2)^2}
\frac{3\,(N^2-1)(2N^2-3)(N^2+3)}{32N^2}
\end{equation}
which matches the matrix model expression reported in the last row of Tab.~(\ref{tab2}).\\
These checks has been done for many other operators, confirming the validity of (\ref{checkN2N4})
up to two loops.

\subsection{Planar limit}
All the above checks are easily extended in the planar limit by keeping the highest power of $N$ and performing the substitution $g^2N=\lambda$. In this limit the number of diagrams which contribute to the correlator is drastically reduced, and thus such checks can be pushed to higher orders in perturbation theory without much effort. Let us first review the well-known
$\cN=4$ case \cite{Semenoff:2001xp,Pestun:2002mr,Semenoff:2006am}.

\subsubsection*{The $\cN=4$ theory}
At leading order, using the tree-level result (\ref{Antree0}) that corresponds to the diagram
of Figure~\ref{fig:WOntree}, one easily finds
\begin{equation}
g^{n-2\ell}\,\widehat A_{\vec n}\Big|_{\mathrm{tree-level,planar}} 
=\lim_{N\to \infty}
\frac{g^{2n-2\ell}}{N\,2^{\frac{n}{2}}}\,R_{\vec{n}}^{\,b_1\dots b_n}\,\tr \big(T^{b_1}\dots T^{b_n}\big)
=c_{\vec{n},0}\,\lambda^{n-\ell}
\end{equation}
where $c_{\vec{n},0}$ are numerical coefficients which can be deduced from Tab.~\ref{tab1}. In particular
we have:
\begin{equation}
c_{(2),0}=\frac{1}{8}~,~~c_{(3),0}=\frac{1}{32\sqrt{2}}~,~~c_{(4),0}=\frac{1}{384}
~,~~c_{(2,2),0}=\frac{1}{96}~.
\end{equation}

\begin{figure}[ht]
	\begin{center}
		\includegraphics[scale=0.7]{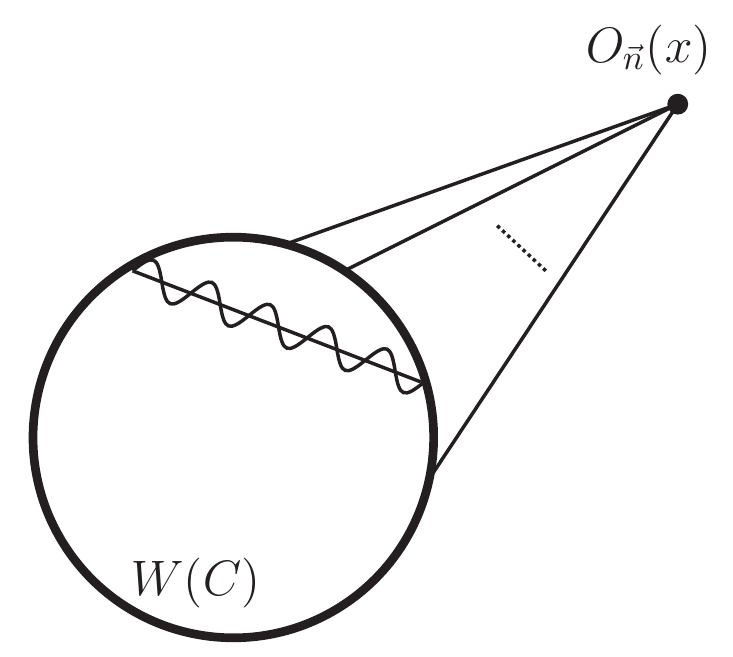} \hspace{1cm}\includegraphics[scale=0.7]{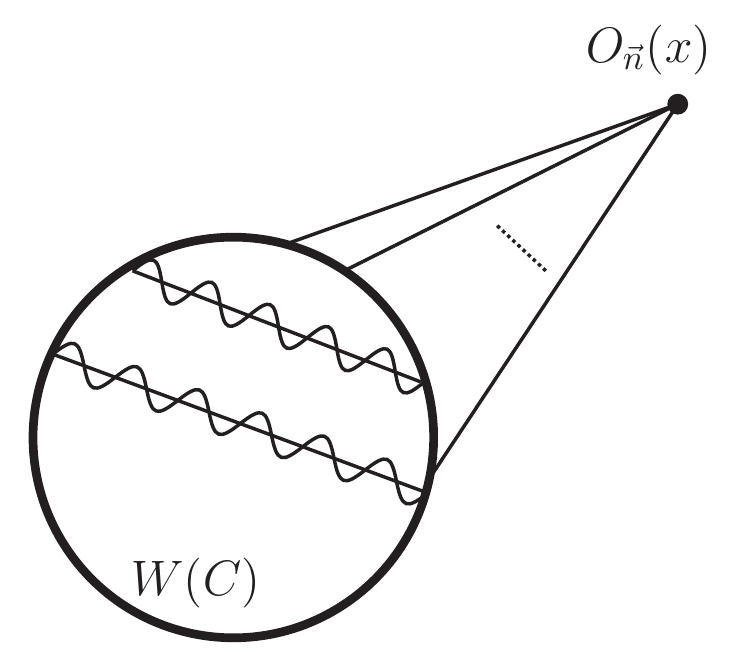}
	\end{center}
	\caption{In the planar limit of the $\cN=4$ theory, the tree-level expression encoded in 
Figure~\ref{fig:WOntree} gets corrected only by the so-called ``rainbow'' diagrams, the first two of which are
represented here. We have used the usual straight/wiggle line to denote the sum of the gluon and the scalar propagator, which always occur together when attached to the Wilson loop and yield the simple expression given in (\ref{ww}).}
\label{fig:WOnN4rain}
\end{figure}

In \cite{Semenoff:2001xp} it was argued that all diagrams with internal vertices cancel at the next order 
and it was conjectured that analogous cancellations should occur at all orders in perturbation theory. Thus,
only the ``rainbow'' diagrams of the type represented in Figure~\ref{fig:WOnN4rain} contribute to the
amplitude $\widehat{A}_{\vec{n}}$ in the planar limit.

The evaluation of these ``rainbow'' diagrams is particularly simple in the case of a circular Wilson loop, see the discussion in Section \ref{sec3:N4WL}. Indeed, 
if we denote by $w^a(x)$ the combination of gluons and scalars that appears inside the 
path-ordered exponential in (\ref{WLdefchapter5}), namely
\begin{equation}
w^a(x)=
\ii \,A^a_{\mu}(x)\,\dot{x}^{\mu}
+\frac{R}{\sqrt{2}}\Big(\varphi^a(x) + \bar\varphi^a(x)\Big)
\end{equation}
with $x$ being a point on the circle $C$, then we have
\begin{equation}
\big\langle
w^a(x_1)\,w^b(x_2)\big\rangle=\frac{\delta^{ab}}{4\pi^2}\,\frac{1-\dot{x_1}\cdot\dot{x_2}}{(x_1
-x_2)^2}=\frac{\delta^{ab}}{8\pi^2R^2}
\label{ww}
\end{equation}
where in the last step we have used the circular parameterization (\ref{circle}). Thus, the contribution of the
internal propagators, represented by straight/wiggle lines in Figure~\ref{fig:WOnN4rain}, is constant (see Section \ref{sec3:N4WL}) so that 
only combinatorial coefficients have to be computed. For example, the first diagram of Figure \ref{fig:WOnN4rain} yields a contribution of the form
\begin{equation}
c_{\vec{n},1}\,\lambda^{n-\ell+1}
\end{equation}
with
\begin{equation}
c_{(2),1}=\frac{1}{96}~,~~c_{(3),1}=\frac{1}{512\sqrt{2}}~,~~c_{(4),1}=\frac{1}{7680}
~,~~c_{(2,2),1}=\frac{1}{1536}~.
\end{equation}
Similarly, the second diagram of Figure \ref{fig:WOnN4rain} leads to
\begin{equation}
c_{\vec{n},2}\,\lambda^{n-\ell+2}
\end{equation}
with
\begin{equation}
c_{(2),2}=\frac{1}{3072}~,~~c_{(3),2}=\frac{1}{24480\sqrt{2}}~,~~c_{(4),2}=\frac{1}{368640}
~,~~c_{(2,2),2}=\frac{1}{61440}~.
\end{equation}
{From} these results it is possible to infer the following resummed expression
\begin{equation}
g^{n-2\ell}\,\widehat{A}_{\vec{n}}\Big|_{\mathrm{planar}} =
\sum_{j=0}^\infty c_{\vec{n},j}\,\lambda^{n-\ell+j}
= \frac{\big(\sqrt{\lambda}\big)^{n-\ell-1}}{2^{\frac{n}{2}+\ell-1}\phantom{\big|}}
\,I_{n-\ell+1}\big(\sqrt{\lambda}\big)\,\prod_{i=1}^\ell n_i~
\label{hatAnplanarQFT}
\end{equation}
which agrees with the matrix model result (\ref{hatAnplanar}).

\subsubsection*{The $\cN=2$ theory}
In this case we focus on the planar limit of the difference $\delta A_{\vec{n}}$ and in particular
on the terms proportional to $\zeta(3)$. To obtain the result at the lowest order, one simply has to take 
the two-loop result (\ref{deltaAn2loopfin}) and evaluate it in the large-$N$ limit. As we have seen
in the previous Subsection, there are two types of terms, corresponding to the diagrams 
$\mathbf{(i)}$ and $\mathbf{(j)}$ of Figure~\ref{fig:WOn2loop}. The correction to the scalar propagator
gives rise to a contribution that always survives in the planar limit; in fact in (\ref{scalprop2la}) 
it was proved to be proportional to $g^4(N^2+1)$, which in the planar limit reduces to $\lambda^2$.
On the other hand, the two-loop effective vertex does not always contribute in the planar limit, since
it is leading for $N\to \infty$ only when it is attached to $\tr \varphi^2$. This can be realized by noticing that in this case such a diagram, because of (\ref{C4uncontracted}), always produces the structure
\begin{equation}
\tr\big(T^{b_1}T^{b_2}\big)\delta^{b_1 b_2}\,\delta^{a_1 a_2}
=\frac{1}{2}(N^2-1)\delta^{a_1 a_2}~,
\end{equation}
with the $N^2$ factor making the contribution leading. Thus, the diagrams of type $\mathbf{(i)}$
always contribute in the planar limit, while the diagrams of type $\mathbf{(j)}$ are sub-leading unless
some of the components of the vector $\vec{n}$ are equal to 2.
This fact can be checked in the explicit computations for $O_{(2)}$ (see (\ref{I2}) and (\ref{J2})) 
and for $O_{(4)}$ and $O_{(2,2)}$ reported in \eqref{IJ4} and \eqref{IJ22}. These simple considerations 
give a nice field theory interpretation to some of the matrix model results presented in Section \ref{secn:largeN}.

Building on the idea that all diagrams with internal vertices cancel at all orders in perturbation theory, like in the $\cN=4$ model \cite{Semenoff:2001xp}, one can construct a class of $\zeta(3)$-proportional diagrams, starting from the $\cN=4$ ``rainbow'' diagrams and performing on them one of the aforementioned 
planar two-loop corrections. This can be done either by correcting one of the external scalar propagators, 
or by correcting one of the internal double-line propagators\,\footnote{Since these internal 
propagators
and the scalar propagators are proportional to each other (see (\ref{ww}) and (\ref{scalprop})), also their 
planar two-loop corrections are proportional.} or by including the two-loop effective vertex if $O_{\vec{n}}$ contains at least a factor $\tr \varphi^2$.
The result of performing any of these corrections is always equal to the original $\cN=4$ ``rainbow'' 
diagram multiplied by $-\frac{3\,\zeta(3)\,\lambda^2}{(8\pi^2)^2}$. This analysis tells us how to get the
$\cN=2$ correction proportional to $\zeta(3)$ in the planar limit starting from the
$\cN=4$ amplitude. In fact, expanding (\ref{hatAnplanarQFT}) for small $\lambda$, 
the term of order $k$ corresponds to a sum over ``rainbow'' diagrams with $(k-n+\ell)$ 
internal propagators and $n$ external ones. Using the method we just described, any such diagram can be corrected once for every internal propagator, once for every external propagator and once for 
every factor $\tr \varphi^2$ appearing in $O_{\vec{n}}$, giving a total of
\begin{equation}
(k-n+\ell)+n+\sum_{i=1}^\ell \delta_{n_i,2}=k+\ell+\sum_{i=1}^\ell \delta_{n_i,2}
\end{equation}
corrections proportional to $-\frac{3\,\zeta(3)\,\lambda^2}{(8\pi^2)^2}$.
This result precisely matches the matrix model expression (\ref{shiftn}) and suggests that this class of diagrams reconstructs the full $\zeta(3)$-term of the $\cN=2$ correlator at all orders in perturbation theory, just like the ``rainbow'' diagrams make up the full $\cN=4$ correlator.

\chapter{Emitted radiation in Superconformal field theories}\label{chap:6}
The present Chapter is a summary of the many results that have been recently conjectured and/or achieved in the context of the energy radiated by an accelerating quark in super-conformal theories in four dimensions with extended supersymmetry. The interest for this topic resides in the fact that it represents an old story for electrodynamics \cite{Dirac:1938nz}, with some particularly debated outcomes \cite{Fulton,Boulware:1979qj}, and  its generalizations have brought many new results and several connections with many areas of theoretical physics, from perturbative QCD to AdS/CFT duality. In context of the present thesis, the problem of the emitted radiation represents a perfect example of a physical observable that can be evaluated exactly using a consistent implementation of both conformal invariance and supersymmetry.  

The emitted radiation is proportional to the so called Bremsstrahlung function, which arises naturally in the study of vacuum expectation values of cusped Wilson loops, as a special limit of the cusp anomalous dimension. Therefore we devote Section \ref{sec6:Cusp} to an introduction to these physical quantities in quantum field theory, with some references to their relevance in other contexts. In Section \ref{sec:6N4Brem} we review the exact computation of the Bremsstrahlung function in $\cN=4$ SYM theory, performed by \cite{Correa:2012at} using a localization approach.

The radiated energy of an accelerated particle can be ``measured'' by computing a flux of the stress tensor of the theory. Therefore in a conformal theory the Bremsstrahlung function can be related to the one-point function of the stress energy tensor in presence of the probe particle represented by a Wilson loop. Section \ref{sec6:BhWCD} is devoted to justifying and elaborating this concept: we explain the problem through some examples given by free conformal theories (free Maxwell, free conformally coupled scalar, $\cN=2$ abelian gauge theory), then we discuss its generalization to non-abelian gauge theories, in presence of extended supersymmetry. It turns out that the crucial quantity for
computing the emitted radiation in superconformal theories is the stress tensor one-point function.

In Section \ref{sec:6.4} we provide a recipe for computing the one-point function of the stress tensor in presence of a line defect for any $\cN=2$ SCFT, in terms of a small variation of the background geometry represented by a four-dimensional ellipsoid. Finally in Section \ref{sec:6.5} we carry out a careful analysis of the perturbative structure of the result, using the supersymmetric
localization achievements in the presence of a Wilson loop.

\section{Cusp anomalous dimension and Bremsstrahlung function}\label{sec6:Cusp}
\begin{figure}[ht]
	\begin{center}
		\includegraphics[scale=0.7]{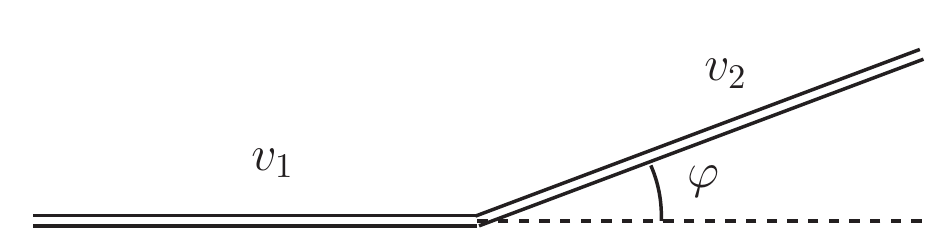}
	\end{center}
	\caption{Cusped Wilson loop with a physical angle $\varphi$. In the full Chapter the Wilson operator will be represented by a double line notation.}
\label{fig:Cuspdef}
\end{figure}

The cusp anomalous dimension represents a crucial quantity in theories of strong interactions. It has been introduced in the context of renormalization properties of Wilson loops \cite{Gervais:1979fv, Polyakov:1980ca, Dotsenko:1979wb, Brandt:1981kf} and especially in the scattering of a heavy quark off an external potential. The physical process is the following (see \cite{Grozin:2004yc} for a complete review): the heavy quark behaves as a classical charged particle, it moves with a velocity $v_1^\m$ that changes to $v_2^\m$ after the scattering with the external source. Due to instantaneous acceleration, the heavy quark starts emitting gluons, generating both infrared (IR) and ultraviolet (UV) divergences. The dipendence of the scattering apmplitude on both the IR and UV cut-offs is controlled by the Cusp Anomalous Dimension $\Gamma_{\mathrm{cusp}} (g, \varphi)$ \cite{Korchemsky:1985xj, Korchemsky:1987wg}, which explicitly depends on the angle $\varphi$ created by the change of velocities \footnote{$v_1\cdot v_2 = \cosh \varphi$ in a Minkowskian space, where $\varphi$ represents a change of rapidity.}
\begin{align}\label{physicalanglephi}
v_1\cdot v_2 = \cos \varphi~.
\end{align} 
The full physical process can be effectively described in terms of Wilson loops, in particular considering a Wilson loop in presence of a cusp, see Figure \ref{fig:Cuspdef}. 

After introducing a UV regulator $\Lambda_{\mathrm{UV}}$, shielding the tip of the contour and a IR regulator $\Lambda_{\mathrm{IR}}$ cutting the infinite length of the line, the cusped Wilson loop develops a logarithmic divergence of the form:
\begin{equation}\label{Wcusped1}
\vev{W[C_{\mathrm{cusp}}]} \propto e^{-\Gamma_{\mathrm{cusp}} \log \frac{\Lambda_{\mathrm{UV}}}{\Lambda_{\mathrm{IR}}}}~.
\end{equation}
In the next Section we will work out an explicit computation in a $\mathcal{N}=4$ theory, where the cusped Wilson loop behavior \eqref{Wcusped1} will clearly arise.\\
The observable $\Gamma_{\mathrm{cusp}}$ has several important properties. For a detailed discussion of QCD results see and references therein
\begin{itemize}
\item
At large $\varphi$ in Minkowski the cusp anomalous dimension is proportional to $\varphi$:
\begin{align}
\Gamma_{\mathrm{cusp}}(g,\ii \varphi) \xrightarrow{\varphi\to \infty}\varphi \gamma_{\mathrm{cusp}}~.
\end{align}
It was computed at weak coupling in \cite{Belitsky:2003ys}. It is also related to the anomalous dimensions of twist-two conformal operators with large spin \cite{Makeenko:1980bh, Korchemsky:1988si,Gubser:2002tv, Makeenko:2006ds}, computed also in supersymmetric theories using integrability techniques \cite{Beisert:2005tm, Beisert:2006ez, Drukker:2006xg, Hofman:2007xp}.
\item
Again for large values of the scattering angle $\varphi$, it parametrizes infrared divergences in gluon scattering amplitudes\cite{Anastasiou:2003kj,Bern:2005iz}, for which nice duality properties in the AdS/CFT context arise \cite{Alday:2007hr,Alday:2007he,Drummond:2007aua}.
\item
In conformal theories, $\Gamma_{\mathrm{cusp}}$ corresponds to the quark-antiquark potential \cite{Forini:2010ek,Drukker:2011za}, after performing a plane to cylinder map, see Figure \ref{fig:CuspCylinder}.
\begin{figure}[ht]
	\begin{center}
		\includegraphics[scale=0.5]{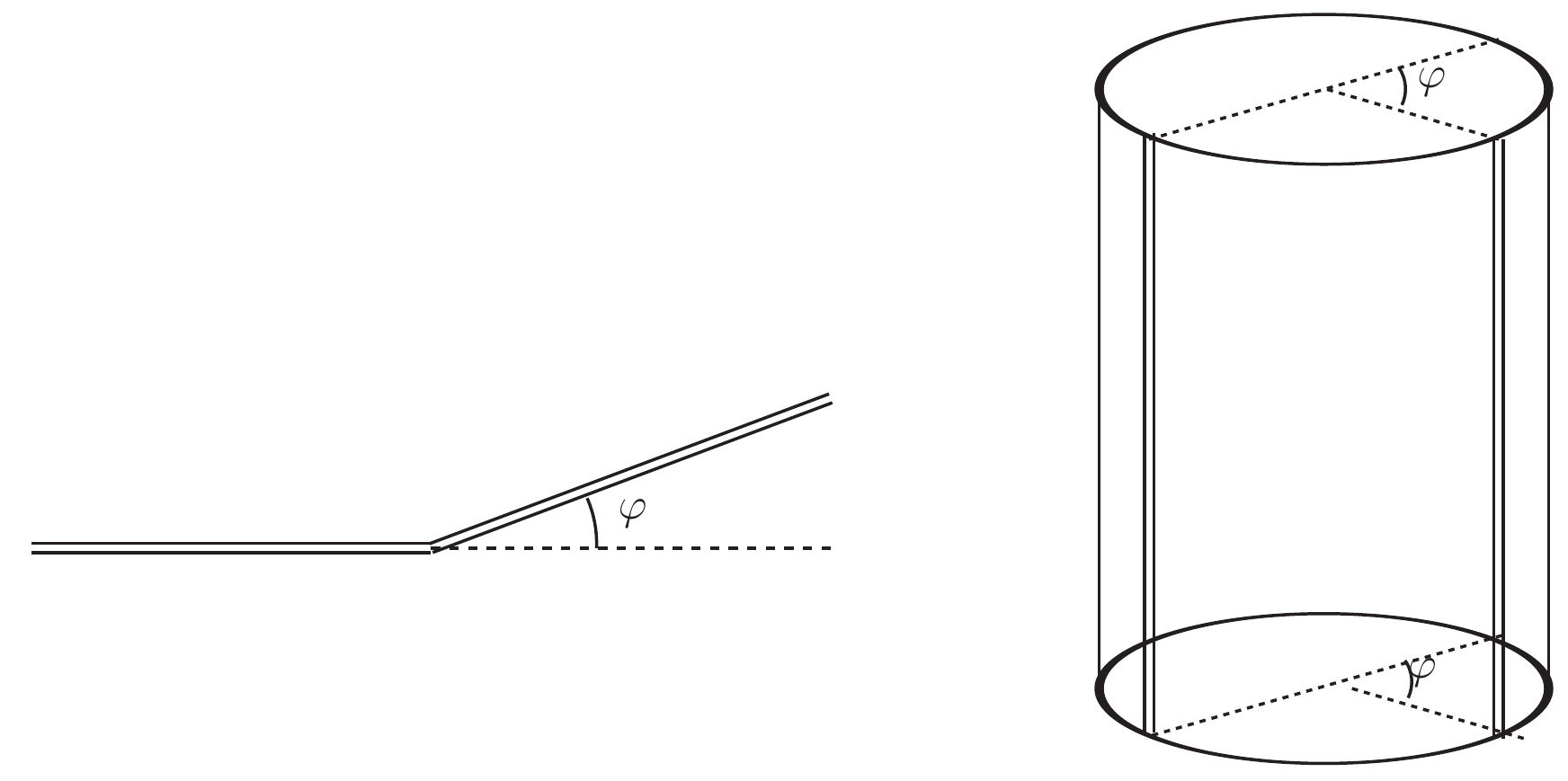}
	\end{center}
	\caption{Plane to cylinder map and correspondence with the $q\bar q$ potential}
\label{fig:CuspCylinder}
\end{figure}
 In this configuration, the $q\bar q$ pair sits on a pure spatial 3-sphere, at an angle $\pi-\varphi$. We have:
\begin{equation}\label{CuspV}
\Gamma_{\mathrm{cusp}} (g,\varphi) \xrightarrow{\varphi\to\pi} -\frac{V(g)}{\pi-\varphi}~,
\end{equation}
where $V(g)$ is the quark-antiquark potential for a $q\bar q$ pair, that was defined in Section \ref{sec1:WL}\footnote{In Section \ref{sec1:WL} we wrote the quark-antiquark potential for a $q\bar q$ pair at a distance $R$. In \eqref{CuspV} the scale distance $R$ is fixed to 1 for conformal symmetry.}.
\item
In the limit $\varphi\to 0$ the cusped Wilson line reduces to a straight line, the cusp divergence disappears and the cusp anomalous dimension vanishes as:
\begin{equation}\label{GammacuspBrem}
\Gamma_{\mathrm{cusp}} (g,\varphi) \xrightarrow{\varphi\to 0} -B(g)~\varphi^2~,
\end{equation}
where $B(g)$ is a positive definite function, dubbed Bremsstrahlung function. 
\end{itemize}
In the present Chapter we explicitly concentrate on this last case, the small angle limit \eqref{GammacuspBrem}. In particular the Bremsstrahlung function will be one of the crucial quantities for our computations.

\section{Exact Bremsstrahlung function in $\mathcal{N}=4$}\label{sec:6N4Brem}
We review the computation of the Bremsstrahlung function defined in \eqref{GammacuspBrem} in a $\cN=4$ SYM theory. After defining our set up and producing a simple 1-loop computation, we infer the exact result obtained in \cite{Correa:2012at} using a localization procedure.
\subsection{Perturbative computation at leading order}\label{subsec6:perturb}

We now compute the leading order coefficient of the small angle expansion of the cusp anomalous dimension (see equation (\ref{GammacuspBrem})). This quantity arises from the expectation value of a cusped Wilson line $W_\cusp$ which we take in the fundamental representation of SU($N$). 
Its contour is made of two semi-infinite rays parametrized as follows
\begin{equation}
\begin{aligned}
x^\mu&=v_1^\mu\,\t_1 \quad\mbox{for}\quad-\infty < \t_1 < 0 ~,\\
x^\mu&=v_2^\mu\,\t_2 \quad\mbox{for}\quad 0 < \t_2 < +\infty ~.
\end{aligned}
\label{contour}
\end{equation}
The velocity vectors $v_1^\mu$ and $v_2^\mu$ are such that $v_1\cdot v_1=v_2\cdot v_2=1$. They define the cusp angle $\varphi$ (see Figure~\ref{fig:Cuspdef}) by the relation $ v_1\cdot v_2 = \cos \varphi~.$

\noindent
The cusped Wilson line is explicitly defined by
\begin{equation}
W_\cusp=\frac{1}{N}\,\tr\mathcal{P} \exp\Big(g\!\int_{-\infty}^0\!\!d\t_1\,L_1(\t_1)+
g\!\int_0^{+\infty}\!\!d\t_2\,L_2(\t_2)\Big)~,
\label{Wloop}
\end{equation}
where we introduced the generalized connections
\begin{equation}
\label{defL1L2}
\begin{aligned}	
		L_1(\t_1)&= \ii \,v_1\cdot A(v_1\t_1)+\frac{1}{\sqrt{2}}\Big(\rme^{+\ii\,\vartheta/2}\,\phi(v_1\t_1)
		+\rme^{-\ii\,\vartheta/2}\,\bar\phi(v_1\t_1)\Big)~,\\
		L_2(\t_2)&= \ii \,v_2\cdot A(v_2\t_2)+\frac{1}{\sqrt{2}}\Big(\rme^{-\ii\,\vartheta/2}\,\phi(v_2\t_2)
		+\rme^{+\ii\,\vartheta/2}\,\bar\phi(v_2\t_2)\Big)~.
\end{aligned}
\end{equation}
Here $\vartheta$ is an ``internal'' angular parameter that can be defined at the cusp \cite{Drukker:1999zq,Correa:2012nk}; it represents an additional freedom due to the coupling with the scalars of the vector multiplet. It is defined as
\begin{equation}\label{defvartheta}
\cos\vartheta= \vec n\cdot \vec n~,
\end{equation}
where $n^u$ is the six-dimensional unit vector parametrizing the direction of the scalars inside the internal space $S^5$, see eq. \eqref{WLsusy}. \\
In general the Wilson loop defined by \eqref{Wloop} and \eqref{defL1L2} is not BPS, but there exists a nontrivial configuration where the loop becomes BPS as well, and it is when $\vartheta=\pm \varphi$. In this case we can extract the Bremsstrahlung function from an extended formula, starting from \eqref{GammacuspBrem}, which reads:
\begin{align}
	\label{GtoB}
		\Gamma_\cusp\simeq
		-\big(\varphi^2-\vartheta^2\big)\,B~.
\end{align}
This means that we can compute $B(g)$ either by taking $\varphi \ll 0$ with $\vartheta=0$, or $\vartheta \ll 0$ with $\varphi=0$. This property will be exploited in the following.

\begin{figure}[ht]
\vspace{0.3cm}
	\begin{center}
		\includegraphics[scale=0.7]{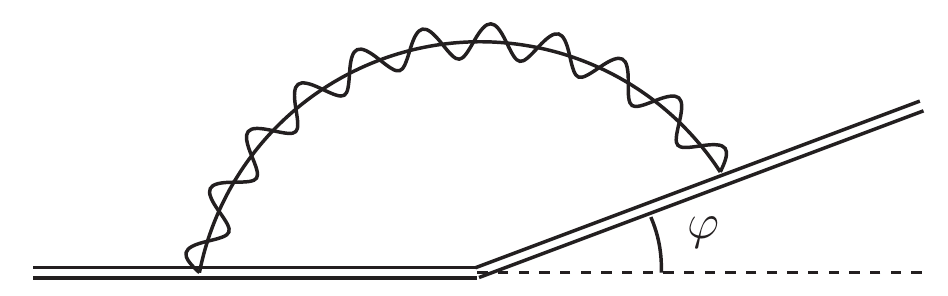}
	\end{center}
	\caption{The $g^2$-contribution to the vacuum expectation value of a cusped Wilson line.
		The double straight/wiggled line stands for the sum of the gluon and scalar propagators.}
	\label{fig:cusp1}
\end{figure}

Expanding $W_\cusp$ in the coupling constant $g$, we find that its vacuum expectation value at order 
$g^2$ is given by the diagram represented in Figure~\ref{fig:cusp1}. 

Using the explicit expression of the Wilson line and the propagators \eqref{propagators}, this 
leads to write
\begin{align}
	\label{cWL1}
		\big\langle W_\cusp \big\rangle =
		1 + \,g^2\,\frac{N^2-1}{2N}\,\big(\cos\varphi-\cos\vartheta\big)
		\,I(\varphi) + O(g^4)~,
\end{align}
where\,%
\footnote{Following \cite{Grozin:2015kna}, we regulate the IR divergence of the $\t_1$ and $\t_2$ 
	integrals by introducing a dumping factor $\rme^{-\ii\delta(\t_1-\t_2)}$ with $\mathrm{Im}\,\delta>0$
	which suppresses the contributions from the large $(-\t_1+\t_2)$ region
	and introduces the dependence on the IR cut-off $\delta$.} 
\begin{equation}
I(\varphi) = \int\!\frac{d^Dk}{(2\pi)^D}\,\frac{1}{k^2\,(k\cdot v_1-\delta)\,(k\cdot v_2-\delta)}~.
\label{Iis}
\end{equation}
This integral is evaluated in Appendix~\ref{app:ft}. Substituting the result
\begin{equation}
I(\varphi)=\frac{1}{\varepsilon}\left(-\frac{1}{8\pi^2}\,\frac{\varphi}{\sin\varphi}\right)+ O(\varepsilon^0)~.
\label{Int5}
\end{equation}
in (\ref{cWL1}), we get
\begin{equation}
\label{cWL2}
\big\langle W_\cusp \big\rangle = 1 - \frac{1}{\varepsilon}\,
\Big(\frac{g^2}{8\pi^2}\Big)\,\frac{N^2-1}{2N}\,
\frac{\varphi\,(\cos\varphi-\cos\vartheta)}{\sin\varphi}~.
+O(g^4)
\end{equation}
The cusp anomalous dimension $\Gamma_\cusp$ using dimensional regularization is defined by \cite{Polyakov:1980ca}\,% 
\footnote{Before we wrote the definition of $\Gamma_\cusp$ within a cut-off regularization scheme, in which case $1/(2\epsilon)$ gets replaced by $\log \left(\Lambda_\tmb{UV}/\Lambda_\tmb{IR}\right)$, see equation \eqref{Wcusped1}}
\begin{equation}
	\label{defGcusp}
\big\langle W_\cusp \big\rangle = \exp\Big(\!-\frac{1}{2\epsilon}\,\Gamma_\cusp\Big)~.
\end{equation}
Taking the logarithm of (\ref{cWL2}), expanding for small angles and comparing with \eqref{GtoB} we find:
\begin{equation}\label{Bfttree}
B=\Big(\frac{g^2}{8\pi^2}\Big)\,\frac{N^2-1}{2N}+O(g^4)~.
\end{equation}

This one-loop computation will be useful in the following, Let us notice that, in the same way as the Wilson loop vev, it contains the coupling and color dependence only. This observable is indeed protected and can be computed using a localization approach. This was done in a paper by Correa, Henn, Maldacena and Sever \cite{Correa:2012at}, which we are going to review.

\subsection{Bremsstrahlung function from localization}\label{subsec6:localiz}
The authors of \cite{Correa:2012at} computed the Bremsstrahlung function starting from the vacuum expectation value of a latitude Wilson loop \cite{Drukker:2006ga,Drukker:2006zk,Drukker:2007dw,Drukker:2007qr,Drukker:2007yx,Pestun:2009nn,Giombi:2009ms,Giombi:2009ds}, which we mentioned in Section \ref{sec1:WL}. This is a generalization of the circular loop, since it allows the contour to be on a non-maximal circle of a $S^2$ sphere embedded in $\mathbb{R}^4$. A latitude is parametrized by an angle $\theta_0$ inside the scalar profile:
\begin{equation}
n^u_{\theta_0}=(\sin\theta_0 \cos\t, \sin\theta_0\sin\t, \cos\theta_0,0,0,0)~,
\end{equation}
where we see that for $\theta_0=0$ the usual $n^u=\delta^{u3}$ typical of 1/2 BPS configurations (circular or straight) is restored.\\
Following the analysis described in Section \ref{sec1:WL}, we could see that the supersymmetry conditions for this contour impose two independent contraints on the SUSY parameter (differently from the single condition \eqref{susyconditionline} for the straight/circular case). So the latitude preserves 1/4 of the supercharges. The vacuum expectation value of the latitude Wilson loop has been computed by \cite{Drukker:2007dw}, and is given by the same expression as the circular one, up to a redefinition of the coupling constant in terms of $\theta_0$:
\begin{equation}
\vev{W_{\theta_0}(\lambda)}=\vev{W_{\mathrm{eq}}(\tilde\lambda)}~,~~~~~~~~\tilde{\lambda}=\lambda\,\sin^2\theta_0~.
\end{equation}
The result is given in terms of the 't Hooft coupling $\lambda=g^2N$.
Expanding the equation around $\theta_0=0$ up to $\theta_0^2$, $\tilde{\lambda}\sim \lambda(1-\theta_0^2)$ we have:
\begin{equation}\label{6.2a}
\frac{\vev{W_{\theta_0}}-\vev{W_{\mathrm{eq}}}}{\vev{W_{\mathrm{eq}}}} = -\theta_0^2 \, \lambda\partial_\lambda \log \vev{W_{\mathrm{eq}}(\lambda)}
\end{equation}
The left hand side of \eqref{6.2a} can be evaluated by expanding the functional integral in terms of the difference $n^u_{\theta_0}-n^u$. Expanding again around $\theta_0=0$, we obtain correlation functions of scalars along the line. Since conformal invariance is preserved along the line, we only have a non-vanishing two-point function, which can be written as:
\begin{equation}\label{f5}
\frac{\vev{W_{\theta_0}}-\vev{W_{\mathrm{eq}}}}{\vev{W_{\mathrm{eq}}}} = \frac{\theta_{0}^{2}}{2}\int_{0}^{2\pi} d\tau \int_{0}^{2\pi} d\tau'\: \hat{n}^{x}(\tau)\;\hat{n}^{y}(\tau')\;\left\langle \phi^{x}(\tau)\phi^{y}(\tau')\right\rangle_{W} +\cO(\theta_0^3)~,
\end{equation}
where $\hat{n}^{x}$ represents a 2-dimensional unit vector (here $x,y=1,2$) and the subscript $ W $ denotes by notation an expectation value evaluated along the Wilson loop, following the definition:
\begin{equation} \label{f10}
 \vev{\phi^{a}(\tau)\phi^{b}(\tau')}_{W} = \dfrac{\left\langle \Tr \bigg[ \mathcal{P}\, \phi^{a}(\tau)\, e^{\int_{\tau}^{\tau'} \left( \ii dx^{\mu} A_{\mu} + \norm{dx}  \phi^{u} n_{u}\right)}\phi^{b}(\tau') e^{\int_{\tau'}^{\tau} \left( \ii dx^{\mu} A_{\mu} + \norm{dx} \phi^{u} n_{u}\right)}\bigg] \right\rangle}{\big\langle W \big\rangle}, 
\end{equation}
which represents an explicit realization of the definition of defect correlation function (\ref{d3}). We evaluate this correlator without using the explicit form in terms of fundamental fields. We simply use the achievements of Chapter \ref{chap:4} to constrain the kinematics, in particular Equation \eqref{4.2.31}. Then we impose the circular parametrization of the loop and we find
\begin{equation}
\frac{\vev{W_{\theta_0}}-\vev{W_{\mathrm{eq}}}}{\vev{W_{\mathrm{eq}}}} = \theta_{0}^{2}\, a_\phi\,\frac{\pi}{2}\int_{\epsilon}^{2\pi-\epsilon} d\t \frac{\cos \t}{1-\cos \t}  = -\theta_{0}^{2}\,\pi^{2}\,a_\phi~,
\end{equation}
where $a_\phi$ is the 1-point coefficient. Now comparing this result with (\ref{6.2a}) we derive:
\begin{equation}\label{f13}
a_\phi = \frac{1}{\pi^{2}}\lambda\partial_{\lambda} \log \langle W_{C}(\lambda)\rangle,
\end{equation}
which is an explicit expression for the coefficient $ a_\phi $ in terms of the known expectation value of the 1/2 BPS circular Wilson loop.

Using a similar approach it is possible to relate $a_\phi$ with the second derivative of the cusp anomalous dimension, namely the Bremsstrahlung function. 
It is convenient to see the cusp configuration coming from the energy of two static quarks, sitting at opposite points on $S_3$, see Figure \ref{fig:CuspCylinder}. The cusp anomalous dimension arises from the vev of a cusped Wilson loop, but in this situation we switch off the physical angle $\varphi$, keeping the internal angle $\vartheta$ between the scalars, see equation \ref{defvartheta}. We proceed as before, varying with respect to the $\vartheta$ angle, and we obtain again a relation with the scalar two-point function:
\begin{equation}
\Gamma_{\textrm{cusp}}(\varphi=0,\theta, g) = -\theta^{2}\frac{1}{2} \int_{-\infty}^{\infty}d\tau \; \langle \phi(\tau)\bar\phi(0)\rangle_{W} + \mathcal{O}(\theta^{3})~.
\end{equation}
Exploiting again the DCFT achievements and solving the integral, we find the relation between $ \Gamma_{\textrm{cusp}} $ and the coefficient $ a_\phi $:
\begin{equation}\label{f11}
\Gamma_{\textrm{cusp}}(\theta,g) = \theta^{2}\frac{a_\phi}{2} + \mathcal{O}(\theta^{3})~.
\end{equation}
Now comparing (\ref{f11}) with (\ref{GtoB}), and inserting the value of $ a_\phi $ computed in (\ref{f13}) we obtain a formula for the Bremsstrahlung function in term of the vacuum expectation value of the circular Wilson loop, which we reviewed in Chapter \ref{chap:3}. We write this formula in terms of the Yang-Mills coupling $g =\sqrt{\lambda/N}$:
\begin{equation}\label{f17}
B(g,N) = \dfrac{1}{4\pi^{2}}g\,\frac{\partial}{\partial g}\log\vev{ W_{C}(g,N)},
\end{equation}
where $\vev{ W_{C}(g,N)}$ is given by \eqref{WN4exact}.
The result \eqref{f17} is remarkable, since it represents an exact formula, valid for any values of the coupling constant $g$. In particular it matches previous results at weak and strong coupling \cite{Drukker:2011za,Correa:2012hh,Correa:2012nk}.

\section{Bremsstrahlung, stress tensor and displacement operator}\label{sec6:BhWCD}
It is possible to infer a deeper meaning of the Bremsstrahlung function in terms of correlation function of important local operators. This is related to some of the basic questions in any field theory, namely the reaction to the presence of a source. We start by reviewing some simple theories, where the simplicity of the problem allows us to concentrate on the physical aspects.
\subsection{Radiation in free theories}\label{subsec6:freetheories}
We consider the problem of study the radiation induced by a probe coupled to a free field theory, in order to perform a purely classical analysis. The goal is to describe the Bremsstrahlung radiation, and how we can ``measure'' it using field theory instruments.

\subsubsection{Maxwell theory}
\noindent Given the Maxwell theory action
\begin{equation}
S_{\mathrm{Maxwell}} = \frac{1}{4} \int d^4x F_{\m\n}F^{\m\n}~,
\end{equation}
we consider a cusped Wilson line representing the world line of an external electron which emits energy along its trajectory, which in this case reads:
\begin{equation}\label{WLMaxwell}
\mathcal{W}=\exp\left[\ii\,e\!\int_{-\infty}^0\!\!dt\,\,v_1\cdot A(v_1t)+
\ii\,e\!\int_0^{+\infty}\!\!ds\,\,v_2\cdot A(v_2s)\right]~,
\end{equation}
where $e$ denotes the coupling constant ({\it{i.e.}} the electric charge).
Repeating the same computation as Subsection \ref{subsec6:perturb} (note that here the tree level contribution provides the full result) we find the Bremsstrahlung function for a charge $e$ in Maxwell theory \footnote{The subscript $v$ stands for vector, since here the vector field is the only contribution}
\begin{align}\label{BMaxwell}
B_{\mathrm{v}}=\frac{e^2}{12\pi^2}~.
\end{align}
We can relate this quantity to a correlation function of local operators in presence of the external probe.

Since this theory has no scales, a Wilson loop can be considered a conformal defect. We are particularly interested in the one-point function of the stress-energy tensor in presence of $W$ defined by $\vev{T_{\m\n}}_W$ and fixed in terms of a unique coefficient $A_T$ (see eq. \eqref{d7}) which in all this Chapter will be denoted $h_W$ and can be considered as the scaling weight of $W$ and in general depends on the couplings of the theory. Following \cite{Kapustin:2005py} we define $h_W$ as
\begin{equation}\label{definitionhW}
\vev{T_{00}}_W= \frac{h_W}{r^4}~,
\end{equation}
where $r$ represents the distance from $W$. The other components of $\vev{T_{\m\n}}_W$ are fixed analogously to \eqref{d7}.\\
The stress tensor for a Maxwell theory explicitly reads:
\begin{equation}\label{stressTmax}
T_{\m\n} = \frac{1}{4\pi} \left( F_{\m \l} F^\l_\n-\frac{1}{4} \delta_{\m\n} F_{\l\s}F^{\l\s}\right)~.
\end{equation}
It is traceless, even off-shell. It is possible to compute the explicit solution for the gauge potential $A_\m (x)$ in presence of the source, it corresponds to the Lienard-Wiechert retarded potential at a generic point $x^\m$, see \cite{Schild:1960,Teitelboim:1979px} and \cite{Rohrlich} for a complete analysis.
Evaluating \eqref{stressTmax} on the retarded solution \cite{Schild:1960} provides an explicit expression for $\vev{T_{\m\n}}_W$, from which it is possible to extract the value of the one-point coefficient:
\begin{equation}\label{hWmaxwell}
h_{W_{\mathrm{v}}}=\frac{e^2}{32\pi^2}~.
\end{equation}
To summarize, we showed that two separate computations determine both the Bremsstrahlung function $B$ and the stress tensor coefficient $h_W$ for a Maxwell theory in presence of an external electron. These quantities turn out to be functions of the coupling, \textit{i.e.} the electric charge $e$, and a simple relation between the two quantities arises:
\begin{equation}\label{BhWmaxwell}
B_{\mathrm{v}} = \frac{8}{3}h_{W_{\mathrm{v}}}~.
\end{equation}
At the end of this Subsection we will discuss the physical meaning of this relation. We now want to understand to what extent \eqref{BhWmaxwell} is general.

\subsubsection{Conformally coupled scalar}
A similar analysis can be performed for a pure conformally coupled scalar theory. The action and the line operator read:
\begin{align}
&S_{\mathrm{scalar}} = \frac{1}{2} \int d^4x \,\left(\partial_\m \phi \partial^\m \phi + \frac{R}{6}\phi^2  \right)~, \notag \\
\label{WLscalar}&\mathcal{W}=\exp\left[e\!\int_{-\infty}^0\!\!dt\,\,\phi(v_1t)+
e\!\int_0^{+\infty}\!\!ds\,\,\phi(v_2s)\right]
\end{align}
where $R$ is the Ricci scalar and the stress tensor is the following:
\begin{equation}\label{stressTscal}
T_{\m\n} = \frac{1}{4\pi} \left( \partial_\m \phi \partial_\n \phi+ \frac{1}{2}\delta_{\m\n} \partial_\l \phi \partial^\l \phi -\frac{1}{6} (\partial_\m\partial_\n +\delta_{\m\n} \Box)  \phi^2 \right)~.
\end{equation}
Performing the same computations as the Maxwell case we find:
\begin{equation}\label{BhWscalar}
B_{\mathrm{s}} = \frac{e^2}{24\pi^2}~, ~~~~~h_{W_{\mathrm{s}}}=\frac{e^2}{96\pi^2}~,~~~~~B_{\mathrm{s}} = 4\,h_{W_{\mathrm{s}}}~.
\end{equation}
We see that the coefficient of proportionality between $B$ and $h_W$ is different with respect to Maxwell theory.

\subsubsection{Abelian $\cN=2$ theory}
As a byproduct, we compute the same quantities in a $\cN=2$ theory with abelian gauge group U$(1)$. The Wilson loop in this case is simply the combination of \eqref{WLMaxwell} and \eqref{WLscalar}, so the result for the Bremsstrahlung function and the stress tensor coefficient are simply a sum of \eqref{BMaxwell}, \eqref{hWmaxwell} with \eqref{BhWscalar}:
\begin{equation}
B_{\mathrm{v}+\mathrm{s}} = \frac{e^2}{8\pi^2}~, ~~~~~h_{W~\mathrm{v}+\mathrm{s}}=\frac{e^2}{24\pi^2}~,~~~~~B_{\mathrm{v}+\mathrm{s}} = 3\,h_{W~\mathrm{v}+\mathrm{s}}~.
\end{equation}

We can state that the relation between $B$ and $h_W$ is not universal, but in general depends on the theory. However, the presence of a simple relation in terms of a numerical coefficient is a clear sign of a deeper physical meaning.

\subsubsection{Physical meaning}
The Bremsstrahlung function $B$ parametrizes the radiated energy of the probe particle in an accelerated motion. The formula reads:
\begin{align} 
  \D E_{\text{tot}}= 2\pi B \int \!d\tau \,a^2~,
 \end{align}
 where $a$ is the four-acceleration of the particle and $\t$ is the proper time, parametrizing its world-line.
 At a differential level, one finds the Larmor formula for a probe particle with momentum $p^\m$ and velocity $u^\m$:
 \begin{align}\label{Larmor}
 \cP= \frac{dp^0}{dt}= -2\pi B\, a^\m a_\m~,
 \end{align}
which is a well known result from classical electrodynamics \cite{Dirac:1938nz}. From its definition, this quantity is not Lorentz invariant, but it is valid under the assumption that 
the initial and final accelerations are equal (and in particular whenever they are equally vanishing, 
{\it{i.e.}} when the particle velocity is asymptotically constant). 
The subtleties related to this definition generated a strong debate in the past \cite{Fulton,Boulware:1979qj}.

One can define a different quantity, the invariant radiation rate (see Chapter 5 of \cite{Rohrlich}) as $\cR = u_\m \frac{dp^\m}{d\tau}$.\\
This power rate is not integrated along the world-line and is manifestly Lorentz invariant, so it constitutes the proper relativistic generalization of the Larmor formula. In \cite{Fiol:2019woe} it was found that $\cR$ is related to $h_W$ in a simple way:
\begin{equation}\label{Radiationrate}
\cR = -\frac{16\,\pi}{3} h_W a^\m a_\m~,
\end{equation}
which represents the proper Lorentz invariant quantity to measure the emitted radiation of the charged particle. We summarize all these results in Table \ref{tab:BhW}.

\begin{table}[ht]
	\begin{center}
		{\small
		\begin{tabular}{c|c|c|c|c}
			\hline
			\hline
			\,CFT \phantom{\bigg|}&  $B$ & $h_W$ &$B$ vs $h_W$
			&$\mathcal{R}$ \\
			\hline
			Maxwell$\phantom{\bigg|}$ & $\frac{e^2}{12\pi^2}$ & $\frac{e^2}{32\pi^2}$  & $B=\frac{8}{3}\,h_W$ & $-2\pi\Big(\frac{e^2}{12\pi^2}\Big)\,a^2$
			\\[7mm]
			Conformal scalar  
			& $\frac{e^2}{24\pi^2}$ & $\frac{e^2}{96\pi^2}$ & $B=4\,h_W$&  $-2\pi\Big(\frac{e^2}{36\pi^2}\Big)\,a^2$
			\\[7mm]
			$\mathcal{N}=2$ U(1) & 
			$\frac{e^2}{8\pi^2}$ & $\frac{e^2}{24\pi^2}$ & $B=3\,h_W$&  $-2\pi\Big(\frac{e^2}{9\pi^2}\Big)\,a^2$ \\[5mm]
			\hline
			\hline
		\end{tabular}
		}
	\end{center}
	\caption{The relevant quantities for three different free conformal theories.}
	\label{tab:BhW}
\end{table}

The physical meaning of this analysis is very clear. The relation between $B$ and $h_W$ derives from the fact that $B$ determines the emitted energy of the charged particle, which can be captured by the radiation rate $\cR$, proportional to the stress tensor coefficient $h_W$. We expect such a relation to exist for any conformal theory, but, as we see from Table \ref{tab:BhW}, there is no universal relation between $B$ and $h_W$.

We would like to generalize these concepts to non-abelian gauge theories.
Things get harder in this case, since two main problems arise:
\begin{itemize}
\item
It is difficult to find a relation between $B$ and $h_W$, since for non-abelian Yang-Mills theories conformal invariance is broken at the quantum level; this determines the problem of separating the radiation component from the self-energy part of the radiating particle.
\item
The explicit computation of $B$ and $h_W$ in terms of the couplings of the theory becomes very complicated in general.
\end{itemize}
We analyze these problems in theories that preserve conformal invariance, and are further constrained by extended supersymmetry.

\subsection{$\cN=4$ case}
For non abelian theories the relation between $B$ and $h_W$ is achieved with an intermediate step: following Section 4 of \cite{Correa:2012nk}, we derive the correspondence between the Bremsstrahlung function defined as the second derivative of the cusp and the two-point function of the displacement operator inserted along a Wilson loop, defined in Subsection \ref{subsec4:WLdispl}. This correspondence contributes to enrich the physical meaning discussed in the previous Subsection. Then, following the analysis of Section \ref{X14}, we will be able to discuss the relation with the stress tensor coefficient $h_W$.
\subsubsection*{Displacement two-point function}
From a conceptual point of view, the relation between the Bremsstrahlung function and the displacement operator comes from the idea that a small deformation of a defect can correspond to some operator insertions along its profile. It is possible to relate $B$ to the two-point coefficient $C_D$, defined through the two-point function:
\begin{equation} \label{displ2point}
\vev{\mathbb D_i(\t) \mathbb D_j (0)}_W = \frac{C_D\, \delta_{ij}}{\t^4}~,
\end{equation}
by following a reasoning which is very similar to what was done in Subsection \ref{subsec6:localiz}. Starting again from the cusped Wilson loop defined on the cylinder, see Figure \ref{fig:CuspCylinder} and the definitions \eqref{Wloop} and \eqref{defL1L2}, we vary $W_{\mathrm{cusp}}$ with respect to the physical angle $\varphi$. Differently from Section \ref{sec:6N4Brem} (when we performed the variation with respect to the angle $\vartheta$ defining the scalar coupling), $\varphi$ enters in both the arguments of the gauge connection and the scalars through the velocities, see eq. \eqref{defL1L2}. Thus from the variation we get precisely the displacement two-point function:
\begin{equation}
\Gamma_{\mathrm{cusp}} \xrightarrow{\varphi\to 0} -\frac{\varphi^2}{2} \int d\t\, \vev{\mathbb D_i(\t) \mathbb D_i (0)}_W = -\frac{\varphi^2}{2} \int d\t\, \frac{C_D}{4(\cosh\tau-1)^2} + O(\varphi^3)~.
\end{equation}
In the second step we used the Poincar\`e section specified by $(P^+,P^-,P^1,P^2,P^3,P^4)$ $= (e^\t, -e^\t,1,0,0,0)$.
Note that again $\cO(\varphi)$ terms are not present due to the vanishing of one-point functions along the line. Performing the integral and comparing with \eqref{GtoB} we get:
\begin{equation}\label{CDandB}
C_D = 12 B~.
\end{equation}
The same quantity $C_D$ also determines the total energy $\Delta E_{\mathrm{tot}}$ emitted by an accelerated particle, as discussed in the previous Subsection:
 \begin{align}
  \D E_{\text{tot}}=\frac{\pi}{6} \,C_D \int \!d\tau \,a^2~.
  \label{DEis}
 \end{align}
As we stressed before, this is true under the assumption that the initial and final acceleration are equal.
The proof of \eqref{DEis} can be found again in Section 4 of \cite{Correa:2012nk}. See also \cite{Mikhailov:2003er,Athanasiou:2010pv} for further considerations about properties of the radiation emitted by a moving quark in theories with a gravity dual.

\subsubsection*{Stress tensor one-point function}
In Section \ref{X14} it was pointed out that the relation between the bulk stress energy tensor and the displacement operator, which is evident from the conservation law \eqref{d8}, is not universal in a generic conformal field theory. However, the further constraints from supersymmetry determine a simple relation between the stress tensor one-point coefficient $h_W$ and the displacement two-point coefficient $C_D$.
The relation in $\cN=4$ arises from a simple observation \cite{Fiol:2012sg} : the exact formula for the Bremsstrahlung function in terms of the logarithmic derivative a circular Wilson loop \eqref{f17} corresponds (up to a numerical coefficient) to the one-point function of a chiral operator (see  eq. \eqref{A21}):
\begin{equation}\label{vevO2equaltoB}
\vev{O_2 (x)}_W= \frac{A_{2}(g,N)}{\big(2\pi\|x\|_C\big)^2\phantom{\Big|}}~, ~~~~~~ A_{2}(g,N) = \frac{1}{8\pi^2}g\partial_g \log \vev{W}~.
\end{equation}
See Section \ref{secn:mmcwl} for a complete derivation. In $\cN=4$ theory the $\Delta=2$ CPO belongs to the same supermultiplet as the stress tensor (see for example \cite{Dolan:2002zh}), so it is clear a direct correspondence between the Bremsstrahlung function $B$ and the stress tensor coefficient $h_W$. \\
The first attempt to fix the numerical coefficient between $B$ and $h_W$ has been done by Lewkowycz and Maldacena \cite{Lewkowycz:2013laa}, exploiting the connection with the energy radiated by an accelerated quark \eqref{DEis}. Indeed they were able to integrate the one-point function of the stress tensor over a hypersurface (in a similar way as \cite{Kapustin:2005py}), relating this result to the energy emitted by the quark. They found the result for a four-dimensional $\cN=4$ theory:
\begin{equation}\label{BandhW}
B = 3  \,h_W~.
\end{equation}
\subsubsection*{Resume}
We conclude the $\cN=4$ analysis with a recap. Also in non-abelian theories, but in presence of maximal supersymmetry, there exists an exact correspondence among the following physical observables:
\begin{itemize}
\item
the second derivative of the cusp anomalous dimension, also denoted as Bremsstrahlung function $B$;
\item
the two-point coefficient of the displacement operator $C_D$;
\item
the one-point coefficient of the stress-energy tensor $h_W$;
\item
the total energy emitted by an accelerated heavy particle $\D E_{\mathrm{tot}}$.
\end{itemize}
Such relations are displayed in eqs. \eqref{CDandB}, \eqref{DEis} and \eqref{BandhW}. \\
The second crucial point is that $\cN=4$ symmetry is powerful enough to provide an efficient way to compute all these quantities: exploiting the supersymmetric localization result for the 1/2 BPS Wilson loop, $B$, $C_D$ and $h_W$ are computed in terms of the logarithmic derivative of the Wilson loop vev, see \eqref{f17}. Since $\vev{W_C(g,N)}$ yields an exact result, in this $\cN=4$ case \eqref{f17} represents an exact formula for any values of the coupling $g$ and the rank $N$. This is similar to what happens in the free theories in Subsection \ref{subsec6:freetheories}, where $B$ and $h_W$ were simple functions of the coupling as well.

\subsection{$\cN=2$ case}
The natural question to ask is to what extent the previous relations hold when we decrease the degree of supersymmetry. We will see that the $\cN=2$ analysis enriches the physical understanding of the relations described above for the free theories and the $\cN=4$ case.

First of all, we can state that the relation between $B$ and $C_D$ is valid for any gauge theory: the argument of \cite{Correa:2012nk} lies in the fact that the introduction of an angle along the contour is equivalent to the slight modifications from the displacement operator insertions along the line. Furthermore, the relation between the displacement operator and the stress tensor has been proven in a $\cN=2$ theory in \cite{Bianchi:2018zpb}. Starting from the conformal analysis of Section \ref{X14}, the authors of \cite{Bianchi:2018zpb} built the stress tensor supermultiplet and found that the conservation law $\partial_{\mu} T^{\mu i} = - \delta_{W}(x) \mathbb D^{i}$  of the full supermultiplet gives rise to a corresponding displacement supermultiplet. Using the superconformal Ward identities they were able to prove that the series of equalities
\begin{equation}\label{allequalities}
C_D=12\,B=36\, h_W
\end{equation}
holds for any $\cN=2$ SCFT in presence of a line defect \footnote{A similar analysis has recently determined an analogous result $C_D = 48 h_W$ for a surface defect, see \cite{Bianchi:2019sxz}, proving the complete generality of this argument.}.\\

The crucial point then is to understand whether supersymmetric localization can still be the right tool to compute the observables $B,~C_D,~h_W$ in terms of the couplings of the theory. It turns out that the $\Delta=2$ CPO is still in the same multiplet as the Lagrangian density, so it has a well defined one-point function $\vev{O_2 (x)}_W$ in terms of the localized Wilson loop vev: eq \eqref{vevO2equaltoB} still holds in $\cN=2$, as we saw in Chapter \ref{chap:5}. However, this one-point function is now unrelated to the insertion of the stress tensor, which sits in a short supermultiplet of the $\cN=2$ superconformal group which does not contain any chiral primaries. The way to overcome this issue has been suggested by \cite{Fiol:2015spa}, where the authors conjectured a formula
\begin{align}
	\label{conj1}
	h_W=\frac{1}{12\pi^2}	\partial_b\ln \big\langle W_b\big\rangle\Big|_{b=1} ~.
\end{align}
Here $\big\langle W_b\big\rangle$ is the expectation value of the Wilson loop on the ellipsoid
with squashing parameter $b$ \cite{Hama:2012bg} and the value $b=1$ corresponds to the round sphere. The left hand side of this relation localizes and can be expressed in terms of a matrix model. The proposal \eqref{conj1} was motivated by the fact that a first order deformation of $S^4$ corresponds to the insertion of an integrated stress tensor supermultiplet. The proof of the formula represents the main result of \cite{Bianchi:2019dlw}.

That derivation only uses general properties of the geometric background and of defect CFTs, thus extending the relation \eqref{conj1} to any superconformal line defect. Furthermore, it provides a general recipe to extract exact results for the stress tensor one-point function by perturbing the background geometry. We stress that this is a peculiar feature of defect CFTs, where there is a non-vanishing one-point function and the first-order derivative gives a non-trivial result.

The relation (\ref{conj1}), together with the series of equalities \eqref{allequalities} discussed above, implies that all these apparently distinct observables are captured by the localization of a non-local operator on a deformed geometry. In particular, this provides a recipe to extract an exact prediction for a non-chiral scalar operator, such as the superprimary of the stress tensor multiplet. Indeed, after proving the relation \eqref{conj1}, we carry out a careful analysis of the perturbative structure of the result: we study the constraints imposed by the matrix model expansion on the structure of the diagrams. We find that a limited class of diagrams contribute to the final result and that the matrix model provides a precious organizing principle, grouping different diagrams according to their color structure in a clever way.

\section{Emitted radiation in $\cN=2$ SCFTs}\label{sec:6.4}
The proof of the formula \eqref{conj1} follows from the application of all the tools we introduced throughout the present thesis. Indeed, we consider a $\mathcal{N}=2$ SCFT on four-dimensional ellipsoids preserving rigid supersymmetry, as we reviewed in Section \ref{sec2:ellips}.

We recall that a four-dimensional ellipsoid can be described 
by the equation
\begin{equation}
	\label{defellipsoid2}
		\frac{x_1^2+x_2^2}{\ell^2}+\frac{x_3^2+x_4^2}{\widetilde{\ell}^{\,2}}+\frac{x_5^2}{r^2}=1~.
\end{equation}
and we introduce the squashing parameter $b^2=\ell/\widetilde{\ell}$ so that is it convenient to use the following parametrization 
\begin{equation}
\label{parametrization}
\ell=l(b)\,b~,\quad \widetilde{\ell}=\frac{l(b)}{b}~,\quad
r=r(b)~,
\end{equation}
where $l(b)$ and $r(b)$ are such that $l(1)=r(1)=\mathsf{r}$. In this way, 
the limit $b\to 1$ corresponds to the sphere limit.

We want to analyze how the vacuum expectation values of gauge invariant operators in the conformal $\cN=2$ SYM theory respond to a deformation of the ellipsoid geometry, and specifically how they depend on the squashing parameter $b$ in the
vicinity of the sphere limit. The goal is to find a direct relation between the quantity $h_W$ defined in the previous Section and the vacuum expectation value of half-BPS Wilson loops to prove the conjecture (\ref{conj1}).

Let us consider a gauge invariant operator $X_b$ which may depend on the ellipsoid squashing parameter. Its vacuum expectation
value is
\begin{equation}
\big\langle X_b\big\rangle =\frac{1}{Z_b}\,\int \!D A \,\, \rme^{-S_b}\, X_b~,
\label{vevO}
\end{equation}
where $A$ here denotes schematically all fields in the conformal $\cN=2$ SYM theory whose
action is $S_b$, and $Z_b$ is the partition function\footnote{Throughout this Section we will denote the ellipsoid metric as $G_{\m\n}$ to distinguish it from the gauge coupling $g$.}
\begin{equation}
Z_b=\int \!D A \,\, \rme^{-S_b}~, \hspace{1cm} S_b=\frac{1}{g^2} \int d^4\xi \sqrt{\det G}L~.
\label{Zb}
\end{equation}
{From} this definition it easily follows that
\begin{equation}
\begin{aligned}
\partial_b\,\ln \big\langle X_b\big\rangle\Big|_{b=1} &=
\frac{-\big\langle \partial_bS_b\,X_b\big\rangle
+\big\langle \partial_bS_b\big\rangle\,\big\langle X_b\big\rangle +\big\langle
\partial_b X_b\big\rangle}{\big\langle X_b\big\rangle}\Big|_{b=1}\\[2mm]
&=
-\frac{\big\langle \!:\!\partial_b S_b\!: X_b\big\rangle}{\big\langle X_b\big\rangle}\,
\Big|_{b=1}
+
\frac{\big\langle
\partial_b X_b\big\rangle}{\big\langle X_b\big\rangle}\,\Big|_{b=1}
\end{aligned}
\label{dbO}
\end{equation}
where the $:\,:$'s indicate the normal ordering, namely the subtraction of all possible
self-interactions. This expression should not depend on the parametrization (\ref{parametrization})
of the scales of the ellipsoid.

Since the action $S_b$ depends on $b$ only through the background supergravity fields, we have
\begin{equation}
\label{dbSbop}
\begin{aligned}
		\partial_b S_b
		& = \int \!d^4 \xi\,\sqrt{\det G}  \bigg[\frac{1}{\sqrt{\det G}}\frac{\pa (\sqrt{\det G}\,L) }{\pa G^{\mu\nu}}\, \pa_b G^{\mu\nu}+
		\frac{\pa L }{\pa (V^{\mu})^{\mathcal{J}}{}_{\mathcal{I}}}
		\,\pa_b (V^{\mu})^{\cJ}{}_{\cI}\\[2mm]
		&\qquad~~+ \frac{\pa L }{\pa \mathsf{T}^{\mu \nu}} \,\pa_b \mathsf{T}^{\mu \nu}
		+  \frac{\pa L }{\pa \bar{\mathsf{T}}^{\mu \nu}} \,
		\pa_b \bar{\mathsf{T}}^{\mu \nu}
		+  \frac{\pa L}{\pa \widetilde{M}} \,\pa_b 
		\widetilde{M}\bigg]~, 
\end{aligned}
\end{equation}
where the supergravity multiplet was defined in \eqref{offsm}.
We are interested in evaluating this expression at $b=1$.
By definition, the variation of the action  with respect to the metric at $b=1$
yields the stress-energy tensor $T_{\mu\nu}$ on the sphere. More precisely, we have:
\begin{align}
	\label{defT}
		\frac{\pa (\sqrt{\det G}\,L) }{\pa G^{\mu\nu}}\,\Big|_{b=1} 
		= - \frac 12 \sqrt{\det G^0}\,\,T_{\mu\nu}
\end{align}
where $G^0_{\mu\nu}$ is the metric on the round sphere $S^4$, namely
\begin{equation}
G^0_{\mu\nu}= \lim_{b\to1} G_{\mu\nu}~.
\label{G0}
\end{equation}
Similarly, the variations of the action with respect to the other background 
fields of the supergravity multiplet 
yield the other bosonic components of the stress-energy tensor supermultiplet, known also as the supercurrent multiplet. With the conventions given in Appendix~\ref{app:SUSYtransf}, we have
\begin{equation}
\label{defoperators}
\begin{aligned}
\frac{\pa L }{\pa (V^{\mu})^{\mathcal{J}}{}_{\mathcal{I}}}\,\Big|_{b=1} 
		&= -\frac{\ii}{2} (t_{\mu})_{\mathcal{J}}{}^{\mathcal{I}}~,~~~~~~~~ &
\frac{\pa L }{\pa \mathsf{T}^{\mu \nu}}\,\Big|_{b=1} 
		&= -16H_{\mu\nu}~,\\[2mm]
\frac{\pa L }{\pa \bar{\mathsf{T}}^{\mu \nu}}\,\Big|_{b=1} 
		&= -16\bar H_{\mu\nu}~, &
\frac{\pa L }{\pa \widetilde{M}}\,\Big|_{b=1} 
	     &= -O_2~.
\end{aligned}
\end{equation}
Using the Lagrangian $L = L_{\tmb{YM}} +  L_{\text{matter}}$ reviewed Subsection \ref{subsecn:susy}, we find 
\begin{equation}
\label{varscmultiplet}
\begin{aligned}
(t_{\mu})_{\mathcal{J}}{}^{\mathcal{I}}&= 
		 4 \ii \tr[\l^{\mathcal{I}} \s^{\mu} \bar \lambda_{\mathcal{J}}]-2\ii \tr[\l^{\mathcal{K}} \s^{\mu} \bar \lambda_{\mathcal{K}}]\,\delta_{\mathcal{J}}^{\mathcal{I}}
		+ q^{\mathcal{I}}\overset{\leftrightarrow}{D}_{\mu} q_{\mathcal{J}}+\frac{1}{2} q^{\mathcal{K}}\overset{\leftrightarrow}{D}_{\mu} q_{\mathcal{K}}\,\delta_{\mathcal{J}}^{\mathcal{I}}~,\\
		H_{\mu\nu}&= -\tr[F^+_{\mu\nu} \,\bar \phi]-\frac{\ii}{32} \psi \s_{\mu\nu} \psi~,\\
		{\bar H}_{\mu\nu}&=-\tr[F^-_{\mu\nu} \,\phi]+\frac{\ii}{32} \bar \psi \bar \s_{\mu\nu} \bar \psi ~,\\
		O_2&= -2 \tr[\bar \phi \phi] -\frac18 q^{\mathcal{I}} q_{\mathcal{I}}
\end{aligned}
\end{equation}
where $F_{\mu\nu}^+$ and $F_{\mu\nu}^-$ are the self-dual and anti self-dual parts of the
gauge field strength.
As a matter of fact, in the following we will not really need these explicit expressions, but we quoted them here to allow the check that the coefficients relating them to the variations of the Lagrangian as given in (\ref{defoperators}) are consistent with the supersymmetry transformations 
reported in Appendix~\ref{app:SUSYtransf} -- indeed, these coefficients will be important for our results. 

With these definitions, we can rewrite (\ref{dbSbop}) as
\begin{equation}
\begin{aligned}
		\partial_b S_b\,\big|_{b=1}&=- \int \!d^4 \xi \,\sqrt{\det G^0} \,
		\bigg[\frac{1}{2}\,T_{\mu\nu}\, \pa_b G^{\mu\nu}\big|_{b=1}+\frac{\ii}{2}\,          
		(t_{\mu})_{\mathcal{J}}{}^{\mathcal{I}}
		\,\pa_b (V^{\mu})^{\cJ}{}_{\cI}\big|_{b=1}
		\\[2mm]
		&\qquad~~+ 16 H_{\mu\nu}\,\pa_b \mathsf{T}^{\mu \nu}\big|_{b=1}
		+ 16 \bar{H}_{\mu\nu}\,
		\pa_b \bar{\mathsf{T}}^{\mu \nu}\big|_{b=1}
		+ O_2 \,\pa_b 
		\widetilde{M}\big|_{b=1}\bigg]~. 
\end{aligned}
\label{partialSb}
\end{equation}
In the following we will use this set-up to study how a half-BPS Wilson loop responds to
a deformation of the ellipsoid.

\subsection{Half-BPS Wilson loops}
\label{subsecn:WL}

On the ellipsoid there are two possible half-BPS Wilson loop defects. 
One wraps the circle of radius $\ell$ in the $x^1,x^2$ plane, the other wraps the circle of radius 
$\tilde\ell$ in the $x^3,x^4$ plane. The two configurations can be exchanged by sending
$b \leftrightarrow 1/b$. 
\begin{figure}[H]
	\begin{center}
		\includegraphics[scale=0.7]{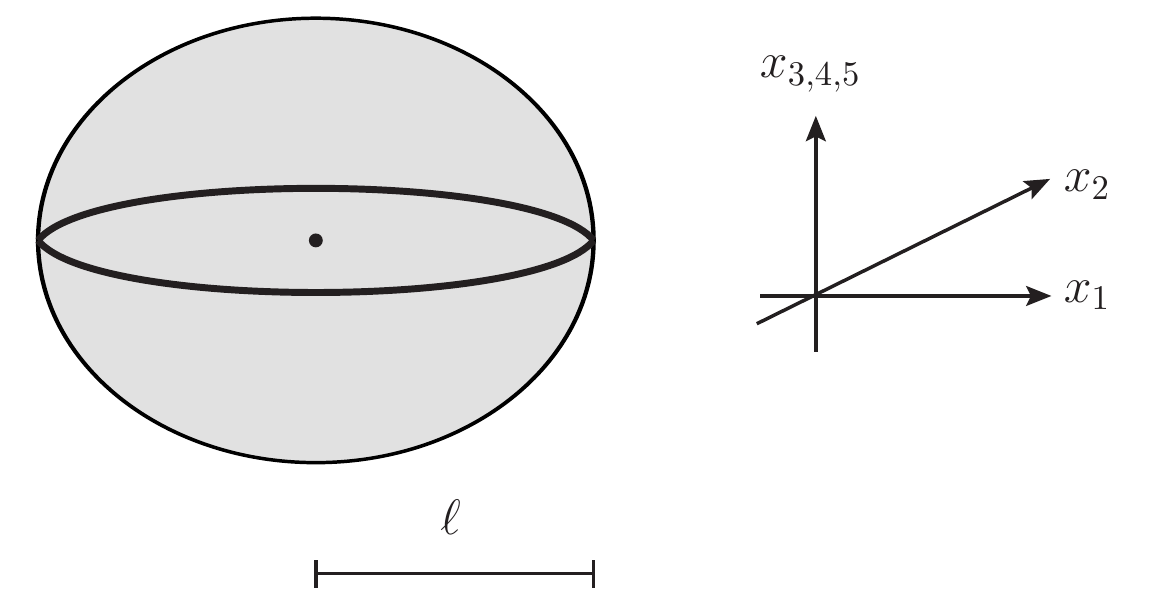}
	\end{center}
	\caption{Wilson loop wrapped around the circle of radius $\ell$ in the $x^1,x^2$ plane on the ellipsoid.}
	\label{fig:ellipsoid}
\end{figure}
Without loss of generality we can choose to wrap the circle of radius $\ell$, see 
Figure~\ref{fig:ellipsoid}. Hence, in the polar coordinates (\ref{coords}), the
Wilson loop locus $\mathcal{C}$ is defined by $\chi=\theta=0$, $\rho=\pi/2$. The explicit expression of
this Wilson loop is \cite{Hama:2012bg}
\begin{equation}
\label{WL}
W_b =\frac{1}{d_\mathcal{R}}\,\tr_{\mathcal{R}}\,\mathcal{P}
\exp\bigg[\ii\!\int_\mathcal{C}	\!d\varphi\,\Big(A_\varphi-\ell(\phi+\bar\phi)\Big)\bigg]
\end{equation}
where $d_\mathcal{R}$ is the dimension of the representation $\mathcal{R}$ in which the
Wilson loop transforms. Notice that this operator may explicitly depend on $b$ through the
coefficient $\ell$ of the scalar part, once the parametrization (\ref{parametrization}) is used.

{From} the formul\ae~(\ref{dbO}) and (\ref{partialSb}), we obtain
\begin{equation}
\begin{aligned}
\partial_b\ln \big\langle W_b\big\rangle\Big|_{b=1} &=
\int \!d^4 \xi \,\sqrt{\det G^0} \,
		\bigg[\frac{1}{2}\,\big\langle T_{\mu\nu}\big\rangle_W \, \pa_b G^{\mu\nu}\big|_{b=1}+\frac{\ii}{2}\, \big\langle
		(t_{\mu})_{\mathcal{J}}{}^{\mathcal{I}}\big\rangle_W
		\,\pa_b (V^{\mu})^{\cJ}{}_{\cI}\big|_{b=1}
		\\[2mm]
		&\qquad~~+ 16 \,\big\langle H_{\mu\nu}\big\rangle_W
		\,\pa_b \mathsf{T}^{\mu \nu}\big|_{b=1}
		+ 16\, \big\langle \bar{H}_{\mu\nu}\big\rangle_W \,
		\pa_b \bar{\mathsf{T}}^{\mu \nu}\big|_{b=1}\\[2mm]
		&\qquad~~+ \big\langle O_2\big\rangle_W \,\pa_b 
		\widetilde{M}\big|_{b=1}\bigg]
		+
\frac{\big\langle
\partial_b W_b\big\rangle}{\big\langle W_b\big\rangle}\,\Big|_{b=1}
\end{aligned}
\label{dlogW}
\end{equation}
where we have adopted the short-hand notation $\big\langle X \big\rangle_W$ to denote the normalized one-point function of $:\!X\!:$ in the presence of the Wilson loop on the sphere, namely
\begin{equation}
\big\langle X \big\rangle_W \equiv \frac{\big\langle \!:\!X\!: W_b \big\rangle}{\big\langle W_b \big\rangle}
\Big|_{b=1} =\frac{\big\langle X W \big\rangle}{\big\langle W \big\rangle}
-\big\langle X  \big\rangle 
\label{XW1}
\end{equation}
with $W$ denoting the Wilson loop on the sphere.
Our goal is to explicitly calculate the integrals in (\ref{dlogW}).

\subsection{Non-vanishing one-point functions}
\label{subsec:onepoint}

The half-BPS Wilson line in a $\mathcal{N}=2$ SCFT preserves an 
$\mathfrak{osp}(4^*|2)$ sub-algebra of the full $\mathfrak{su}(2,2|2)$ superconformal algebra and, in particular, it preserves the one-dimensional conformal group. 
We compute the one-point functions of \eqref{dlogW} using the machinery of Chapter \ref{chap:4}. In particular, we will use the embedding formalism (introduced in Section \ref{X4}), implementing a projection on the sphere, which is conformally equivalent to a plane. There is a very natural choice to make
for the light-cone section, namely
\begin{align}
\label{P}
P^\cM=\big(\mathsf{r},x_M\big|_{b=1}\big)
\end{align}
with $x_M\big|_{b=1},~ M=1,\dots,5$ are the coordinates given in (\ref{coords}) evaluated 
on the sphere of radius $\mathsf{r}$.
The coordinate $P^0$ is determined by the condition 
$P^\cM\eta_{\cM\cN}P^\cN=0$, while the two coordinates along which the defect stretches, 
{\it{i.e.}} $x_1$ and $x_2$, are the parallel coordinates in embedding space. 
To sum up, in our case $P^0=\mathsf{r}$ and $P^{1,2}=x_{1,2}\big|_{b=1}$ are the 
parallel coordinates, while $P^{3,4,5}=x_{3,4,5}\big|_{b=1}$ are the orthogonal ones. 

With this assignment, the extraction of the orthogonal and parallel scalar products is a trivial exercise:
\begin{equation}
\begin{aligned}
P \circ P &=\big( x_3^2+x_4^2+x_5^2\big)\big|_{b=1}=\mathsf{r}^2\big(\cos^2\rho+
\sin^2\theta\sin^2\rho\big)~,\\[1mm]
 P \bullet P &= -\mathsf{r}^2+\big(x_1^2+x_2^2\big)\big|_{b=1}=-\mathsf{r}^2\big(1-\cos^2\theta\sin^2\rho \big)=-P\circ P~.
\end{aligned}
\label{PP}
\end{equation}

A further ingredient that is needed to write the expression of the one-point functions
is the projection of indices using the auxiliary $z$-variables, see Section \ref{sec::4.2}.
For a symmetric traceless tensor, like the stress-energy tensor $T^{\mu\nu}$, one can contract 
all indices with a complex vector $z_\mu$, 
such that $z\cdot z\,\equiv \, z^\mu G^0_{\mu\nu} z^{\nu}=0$. 
Then the one-point function of this tensor in the presence of a defect is a polynomial in $z$. 
If one needs the one-point function with open indices, one can apply to this polynomial the Todorov operator \eqref{todorov}.
The strategy to extend this prescription to the light-cone is to introduce
a vector $Z$ in the embedding space given by
\begin{align}\label{Zlightcone}
Z^{\cM}=z^{\mu} \pa_\mu P^\cM~.
\end{align}
Using the relation
\begin{align}
\pa_\mu P^\cM\,\eta_{\cM\cN} \,\pa_\nu P^\cN=G^0_{\mu\nu}~,
\end{align}
which can be easily verified in our case, one can check that
\begin{equation}
P^\cM\eta_{\cM\cN} Z^\cN=Z^\cM\eta_{\cM\cN} Z^\cN=0
\end{equation}
if $z\cdot z=0$. 

In Section \ref{sec::4.2} we also included the case of tensors that are not symmetric or traceless, specifying the case of the anti-symmetric two-index tensors. In this case two different $z$-vectors,
$z^{(1)}$ and $z^{(2)}$, need to be introduced, see eqs. \eqref{4.2.18} and \eqref{4.2.19} and following.

\paragraph{The relevant one-point functions:}
In the presence of a conformal line defect, only operators with even spin can acquire an expectation value \cite{Billo:2016cpy} (the situation may be different for special cases where parity odd structures are available, but this is not the case for a line defect in four dimensions). Therefore, in
our case, the one-point function of $(t_{\mu})_{\mathcal{J}}{}^{\mathcal{I}}$ vanishes:
\begin{equation}
\big\langle (t_{\mu})_{\mathcal{J}}{}^{\mathcal{I}}\big\rangle_W=0~,
\label{vevt}
\end{equation}
and the only non-zero one-point functions are those of the stress-tensor $T_{\mu\nu}$, of the two anti-symmetric tensors $H_{\mu\nu}$ and $\bar{H}_{\mu\nu}$, and the scalar operator $O_2$.\\
The one-point function of the stress-energy tensor can be extracted from \eqref{4.2.33}
and reads
\begin{equation}
\label{zzT0}
z^{\mu} z^{\nu} \big\langle T_{\mu\nu}\big\rangle_W
= 4 h_W \frac{(P\circ Z)^2 - (Z\circ Z)\,(P\circ P)}{(P\circ P)^3}
\end{equation}
where $h_W$ is the one-point coefficient discussed before. Using the explicit expressions of $P$ and $Z$ given in (\ref{P}) and (\ref{Zlightcone}), we find
\begin{align}
z^{\mu} z^{\nu} \big\langle T_{\mu\nu}\big\rangle_W
\!=\!\frac{h_W}{\mathsf{r}^4 \big(\cos ^2\rho+\sin ^2\theta
		   \sin ^2\rho\big)^3}& \bigg(z_\chi^2\sin^2\theta \sin ^2\rho 
		 \big(\! \cos 2 \theta\!-\!2\cos ^2\theta \cos 2 \rho\! -\!3\big)\notag \\
		 &-4\big(z_\rho\,\sin\theta +z_\theta\,\cos \theta\sin \rho \cos \rho\big)^2\bigg)~.
		   \label{zzT}
\end{align}
Applying the Todorov operator we can open the indices and easily obtain the explicit
expression of $\big\langle T_{\mu\nu}\big\rangle_W$ in our coordinate system, namely
\begin{equation}
\big\langle T_{\mu\nu}\big\rangle_W=\cD_\mu \cD_\nu\Big(
z^{\lambda} z^{\kappa} \big\langle T_{\lambda\kappa}\big\rangle_W\Big)~.
\end{equation}
For the one-point function of $H_{\mu\nu}$ and $\bar{H}_{\mu\nu}$ we need \eqref{4.2.38}, which now becomes:
\begin{align}
 z_1^{\mu} z_2^{\nu} \big\langle H_{\mu\nu} +
 \bar{H}_{\mu\nu} \big\rangle_W
 = k_1\, \frac{\e_{IJK}P^I Z_1^J Z_2^K }{(P\circ P)^2}
 +k_2\,\frac{\e_{ABC} P^A Z_1^B Z_2^C}{(P\circ P)^2}~,
 \label{HH0}
\end{align}
where $I,J,K$ run over the orthogonal directions and $A,B,C$ run over the parallel directions.
To determine the constants $k_1$ and $k_2$, we use the supersymmetric Ward identities
that allow us to relate these coefficients to the prefactor $h_W$ appearing in the one-point function
of the stress-energy tensor.
This calculation 
is described in Appendix C of \cite{Bianchi:2019dlw} and the result is
\begin{equation}
k_1=0~,\qquad k_2=\frac{3h_W}{8}~.
\label{k1k2are}
\end{equation}
Inserting this in (\ref{HH0}), we then obtain
\begin{align}
 z_1^{\mu} z_2^{\nu} \big\langle H_{\mu\nu}\! +\!\bar{H}_{\mu\nu} \big\rangle_W
\! =\! \frac{3h_W}{8}\cos^2 \theta \sin^2 \rho\,\frac{(z_{1 \phi} z_{2 \theta}-z_{2 \phi} z_{1 \theta})\tan \theta+ (z_{1 \rho} z_{2 \phi}-z_{2 \rho} z_{1 \phi})\cot \rho}{\mathsf{r}^3
 \big(\cos ^2\rho+\sin ^2\theta \sin ^2\rho\big)^2}~. 
\end{align}
Opening the indices and projecting onto the self-dual and anti self-dual parts, we find
\begin{equation}
\begin{aligned}
\big\langle H_{\a}{}^{\b}\big\rangle_W \,&\equiv\,  \big\langle H_{\mu\nu}\big\rangle_W (\sigma^{\mu\nu})_{\a}{}^{\b} =\frac{3\ii h_W}{4}\,\frac{\cos\theta \cos\rho \,(\tau^1)_{\a}{}^{\b}
-\sin\theta \,(\tau^2)_{\a}{}^{\b}}{\mathsf{r}^3
 \big(\cos ^2\rho+\sin ^2\theta \sin ^2\rho\big)^2}~,\\[2mm]
\big\langle \bar{H}^{\dot \a}{}_{\dot\b}\big\rangle_W \,&\equiv\,  \big\langle \bar{H}_{\mu\nu}\big\rangle_W (\bar{\sigma}^{\mu\nu})^{\dot \a}{}_{\dot\b}
=-\frac{3\ii h_W}{4}\,\frac{\cos\theta \cos\rho \,(\tau^1)^{\dot\a}{}_{\dot\b}
+\sin\theta \,(\tau^2)^{\dot\a}{}_{\dot\b}}{\mathsf{r}^3
 \big(\cos ^2\rho+\sin ^2\theta \sin ^2\rho\big)^2}~,
\end{aligned}
\label{HHbar}
\end{equation}
where $\tau^i$ are the usual Pauli matrices.\\
The last one-point function, that of the scalar superprimary operator
$O_2$, is the easiest one. Its functional form can be extracted from \eqref{4.2.32} and, in our
coordinate system, reads
\begin{align}
	\label{vevO2}
		\braket{O_2}_W=\frac{3h_W}{8}\,\frac{1}{P\circ P}=\frac{3h_W}{8}\,
		\frac{1}{\mathsf{r}^2 \big(\cos^2 \rho+ \sin^2 \theta \sin^2 \rho\big)}~.
\end{align}
The coefficient $3h_W /8$ has been fixed from the superconformal Ward identities (see also
\cite{Fiol:2015spa}).

\paragraph{Absence of anomalies:} The functional form of the one-point functions
\eqref{zzT}, \eqref{HHbar} and \eqref{vevO2} on $S^4$ has been obtained from that of the
corresponding one-point functions on $\mathbb{R}^4$ by performing a conformal transformation.
However, this transformation is affected by a Weyl anomaly and thus we have to make sure that
this anomaly will not plague our results. To show this, we can use a simple argument inspired
by \cite{Bianchi:2016xvf}.

Let us recall that the one-point function of the stress-energy tensor on $S^4$ is not vanishing, even in the absence of a defect, and that it contains a contribution proportional to the anomaly coefficient $a$ \cite{Brown:1977sj}\,%
\footnote{For conformally flat manifolds there is no contribution from the B-type anomalies.}. 
For a supersymmetric field theory in the presence of additional background fields, like the $\cN=2$ SYM theory we are considering, the conformal anomaly is constructed out of the full 
Weyl supergravity multiplet and not just out of the background metric \cite{Kuzenko:2013gva,Gomis:2015yaa}. As a consequence, we expect non-vanishing one-point functions for the various components of the stress tensor multiplet. These would all be proportional to the anomaly coefficient $a$. This anomalous contribution is a local feature of the stress tensor multiplet, which is not affected by the presence or absence of a defect. This is very natural since one never expects that bulk CFT data, like the anomaly coefficients, are modified by a defect. 
Therefore, under a Weyl transformation $\widehat{G}_{\mu\nu}\to G_{\mu\nu}=\e^{2\s} \widehat{G}_{\mu\nu}$ of a flat metric $\widehat{G}_{\mu\nu}$, 
the stress tensor one-point function in the presence of a Wilson line $W$ changes as follows
\begin{align}
\frac{\big\langle \widehat{T}_{\mu\nu} W \big\rangle}{\big\langle W \big\rangle}~~\to~~
\frac{\big\langle {T}_{\mu\nu} W \big\rangle}{\big\langle W \big\rangle}
=\rme^{-2\s}\,\frac{\big\langle \widehat{T}_{\mu\nu} W \big\rangle}{\big\langle W \big\rangle}+
\big\langle {T}_{\mu\nu} \big\rangle
\end{align}
where $\widehat{T}_{\mu\nu}$ is the stress tensor in flat space. The
last term in the right hand side is the anomalous contribution, while the
term proportional to $\rme^{-2\s}$ is the result of the conformal transformation
applied to the one-point function in the flat space. In the case where the conformal transformation
maps $\mathbb{R}^4$ to $S^4$, this term is just what we have denoted by 
$\big\langle {T}_{\mu\nu} \big\rangle_W$ in the previous Subsection. 
Indeed, from (\ref{XW1}) we have
\begin{align}
\big\langle {T}_{\mu\nu} \big\rangle_W= 
\frac{\big\langle {T}_{\mu\nu} W \big\rangle}{\big\langle W \big\rangle}-
\big\langle {T}_{\mu\nu} \big\rangle=
\rme^{-2\s}\,\frac{\big\langle \widehat{T}_{\mu\nu} W \big\rangle}{\big\langle W \big\rangle}~.
\end{align}

This argument, which applies of course to all other components of the stress tensor multiplet,
shows that the sphere one-point functions that appear in (\ref{dlogW}) are precisely those that are
obtained by performing the conformal transformation on those in flat space, as we have done
to write \eqref{zzT}, \eqref{HHbar} and \eqref{vevO2}. Thus, our result is not affected
by the anomaly. 
Actually, this argument is rather general and holds for an arbitrary line defect in any $\mathcal{N}=2$ SCFT. For the specific case we consider in this paper though, {\it{i.e.}} 
$\mathcal{N}=2$ SYM theory, we know that the anomaly coefficient $a$ does not depend on the coupling and the absence of anomalous contributions can also be ascertained from a simple 
free theory computation.

\subsection{Explicit integration}
Using the one-point functions of Subsection \ref{subsec:onepoint}, together with the explicit results of the background values of the bosonic fields of the supergravity multiplet reported in Subsection \ref{subsecn:background}, we have all the ingredients that are necessary to perform the integrations in (\ref{dlogW}). 
Let us begin by considering the integral involving the one-point function of the stress-energy tensor.
This has to be regularized by introducing a cutoff $\e$ to keep the integration away from the location of the defect; the result is
\begin{align}
	\label{res1}
		\int \!d^4 \xi \,\sqrt{\det G^0} \,
		\Big[\frac{1}{2}\,\big\langle T_{\mu\nu}\big\rangle_W \, \pa_b G^{\mu\nu}\big|_{b=1}
		\Big]= \Big(\frac{3l^{\prime}-3r^{\prime}
		-3}{\e^3}-\frac{l^{\prime}-r^{\prime}-5}{\e}\Big)2\pi h_W
		+\mathcal{O}(\e)
\end{align}
where
\begin{equation}
l^{\prime}=\partial_b l(b)\big|_{b=1}~,~~~r^{\prime}=\partial_b r(b)\big|_{b=1}
\end{equation}
with $l(b)$ and $r(b)$ being the functions used in (\ref{parametrization}) to parametrize the scales
of the ellipsoid. The expression \eqref{res1} is purely divergent and does not contain any finite contribution. The divergent part is clearly a feature of the regularization procedure since there is no universal logarithmic term. In particular, if we computed the integral \eqref{res1} in dimensional regularization we would simply find zero. For this reason the contribution \eqref{res1} can be discarded.

The other terms in (\ref{dlogW}), instead, yield finite contributions. 
In fact, we find
\begin{subequations}
\begin{align}
\int \!d^4 \xi \,\sqrt{\det G^0} \,
		\Big[16 \,\big\langle H_{\mu\nu}\big\rangle_W
		\,\pa_b \mathsf{T}^{\mu \nu}\big|_{b=1} \Big] &=
		\int \!d^4 x \,\sqrt{\det G} \,
		\Big[\!-2\ii \,\big\langle H_{\a}{}^{\b} \big\rangle_W
		\,\pa_b \mathsf{T}_{\b}{}^{\a}\big|_{b=1}\Big]
		\notag\\
		&=\big(14+4l^{\prime}
		-4r^{\prime}\big)\pi^2 h_W-\frac{3}{2}\pi^4h_W~,\label{resH}\\[2mm]
\int \!d^4 \xi \,\sqrt{\det G^0} \,
		\Big[16 \,\big\langle \bar{H}_{\mu\nu}\big\rangle_W
		\,\pa_b \bar{\mathsf{T}}^{\mu \nu}\big|_{b=1} \Big] &=
		\int \!d^4 \xi \,\sqrt{\det G^0} \,
		\Big[\!-2\ii \,\big\langle \bar{H}^{\dot\a}{}_{\dot\b} \big\rangle_W
		\,\pa_b \bar{\mathsf{T}}^{\dot\b}{}_{\dot\a}\big|_{b=1}\Big]
		\notag\\
		&=\big(14+4l^{\prime}
		-4r^{\prime}\big)\pi^2 h_W-\frac{3}{2}\pi^4h_W~,\label{resHbar}\\[2mm]
\int \!d^4 \xi \,\sqrt{\det G^0} \,
		\Big[\big\langle O_2\big\rangle_W \,\pa_b 
		\widetilde{M}\big|_{b=1}\Big]&=-\big(16+8l^{\prime}
		-8r^{\prime}\big)\pi^2 h_W+3\pi^4h_W~.\label{resO2}
\end{align}
\end{subequations}
It is interesting to observe that, while the individual integrals depend on the constants
$l^{\prime}$ and $r^{\prime}$ that are related to the chosen parametrization of the
ellipsoid scales, remarkably their sum is independent of such a choice. 
Indeed, all terms involving
$l^{\prime}$ and $r^{\prime}$ exactly cancel when we add (\ref{resH}), (\ref{resHbar}) and (\ref{resO2}). Notice that also the terms proportional to $\pi^4$
cancel in the sum. Therefore, discarding the unphysical divergent terms (\ref{res1}) for the aforementioned reasons and 
collecting all the finite contributions, we can rewrite (\ref{dlogW}) as follows
\begin{equation}
\partial_b\ln \big\langle W_b\big\rangle\Big|_{b=1} =12\pi^2 h_W
		+
\frac{\big\langle
\partial_b W_b\big\rangle}{\big\langle W_b\big\rangle}\,\Big|_{b=1}~.
\label{dlogW1}
\end{equation}
The quantity in the left hand side is independent of the parametrization of the ellipsoid, and so also the last term the right hand side must be independent of this parametrization. We can then evaluate it choosing $l(b)=\mathsf{r}/b$, which according to (\ref{parametrization})
implies that $\ell=\mathsf{r}$. In this case the Wilson loop (\ref{WL}) does not explicitly depend on
$b$ and thus $\big\langle \partial_b W_b\big\rangle=0$. 
On the other hand, if we choose a different parametrization for the ellipsoid scales,
we still get this same result. Indeed, as one can see from (\ref{WL}) 
the Wilson loop may explicitly depend on $b$ only through the coefficient $\ell$ in front of the scalar term in the exponent, and the derivative
 $\big\langle \partial_b W_b\big\rangle\big|_{b=1}$ would lead to the integral of a defect one-point function, which clearly vanishes if the defect preserves conformal invariance along its profile. This fact can also be easily checked perturbatively at leading order, as shown in Appendix~D of \cite{Bianchi:2019dlw}.

In conclusion the result of our calculation is
\begin{equation}
\partial_b\ln \big\langle W_b\big\rangle\Big|_{b=1} =12\pi^2 h_W~,
\label{dlogW2}
\end{equation}
which proves the conjecture of \cite{Fiol:2015spa}.

\paragraph{Independence on $c_1$, $c_2$ and $c_3$:}
The supergravity background of the ellipsoid given in (\ref{HHsol}) depends on three arbitrary functions $c_1$, $c_2$ and $c_3$ that parametrize the ambiguity in the solution of the Killing spinor equations. These arbitrary functions appear in the $\Delta$-terms given in 
(\ref{DeltaM}) and (\ref{Deltas}). However, our result (\ref{dlogW2}) is robust and 
does not depend on these arbitrary functions. Here we would like to explain why this happens.

The $\Delta$-terms in the supergravity background give rise to the following contribution
\begin{equation}
\begin{aligned}
\int \!\!d^4 \xi \sqrt{\det G^0} \,
		\bigg[\big\langle O_2\big\rangle_W \,\pa_b 
		\Delta\widetilde{M}\big|_{b=1}\!\!-\!2\ii \,\big\langle H_{\a}{}^{\b} \big\rangle_W
		\,\pa_b \Delta\mathsf{T}_{\b}{}^{\a}\big|_{b=1}
		\!\!-2\ii \,\big\langle \bar{H}^{\dot\a}{}_{\dot\b} \big\rangle_W
		\,\pa_b \Delta\bar{\mathsf{T}}^{\dot\b}{}_{\dot\a}\big|_{b=1}\!
		\bigg]
\end{aligned}
\label{DeltalogW}
\end{equation}
Let us first observe that the terms proportional to $c_i^2$ in $\Delta\widetilde M$ do not contribute
since their $b$-derivative at $b=1$ vanishes because of (\ref{ci}). Similarly, the dependence 
on $c_3$ disappears because in $\pa_b \D  \mathsf{T}_{\a}{}^{\b}$ and 
$\pa_b \D  \bar{\mathsf{T}}^{\dot\a}{}_{\dot\b}$ it multiplies the diagonal matrix $\t^3$, 
while, as one can see from (\ref{HHbar}), the one-point functions 
$\big\langle H_{\a}{}^{\b} \big\rangle_W$ and $\big\langle \bar{H}^{\dot\a}{}_{\dot\b} \big\rangle_W$ are proportional to $\t^1$ and $\t^2$ and hence are anti-diagonal. 

We then remain with the terms proportional to $c_1$ and $c_2$. Evaluating them, we find that
they vanish because they can be recast as total derivatives. Indeed, (\ref{DeltalogW}) becomes
\begin{equation}
\begin{aligned}
3h_W\int \!d^4 \xi \,
		\bigg[\partial_\rho\Big(\frac{\sin\theta \cos\theta\sin^3\rho}{\cos ^2\rho+\sin ^2\theta \sin ^2\rho}\,c_1^\prime\Big)
		+\partial_\theta\Big(\frac{\sin\theta \cos\theta\sin^2\rho}{\cos ^2\rho+\sin ^2\theta \sin ^2\rho}\,c_2^\prime\Big)
		\bigg]=0~.
\end{aligned}
\label{DeltalogW1}
\end{equation}
This proves that the ambiguity in the background solutions does not affect our result (\ref{dlogW2}).

Using (\ref{allequalities}) we can find analogous expressions
in any $\cN=2$ conformal SYM theory for the coefficient $C_D$ of the two-point function of the
displacement operator:
\begin{equation}
C_D=\frac{3}{\pi^2}\,\partial_b\ln \big\langle W_b\big\rangle\Big|_{b=1}
\label{CDW}
\end{equation}
and equivalently for the Bremsstrahlung function $B$:
\begin{equation}
B=\frac{1}{4\pi^2}\,\partial_b\ln \big\langle W_b\big\rangle\Big|_{b=1}
\label{BW}
\end{equation}
as conjectured in \cite{Fiol:2015spa}.

\section{Perturbative structure of the result}\label{sec:6.5}
The relation (\ref{dlogW2}) between the coefficient $h_W$ in the stress tensor one-point function and the $b$-derivative of the ellipsoid Wilson loop, which also implies the relations (\ref{CDW}) and (\ref{BW}) for $C_D$ and $B$, relies on the superconformal symmetry of the gauge theory on the ellipsoid constructed in \cite{Hama:2012bg}. In that same reference, supersymmetric localization was applied to this theory to express its partition function and the expectation value of circular Wilson loops in terms of a matrix model, as we reviewed in Subsection \ref{subsec2:MMellips}. This makes it possible to explicitly evaluate $h_W$ using matrix model techniques.

\subsection{Stress tensor coefficient $h_W$ in the localized matrix model}
\noindent The Wilson loop (\ref{WL}) on the saddle point locus reads:
\begin{align}
	\label{wlspm}
		\cW_b = \frac{1}{N}\tr \exp \Big(\frac{b\,g}{\sqrt{2}}\, a\Big)~,
\end{align}
and its expectation value is (see eq. \eqref{partb})
\begin{align}
	\label{derWbmm}
		\big\langle \cW_b\big\rangle = \frac{1}{\cZ_b} \int da~ 
		\cW_b\, \rme^{-\tr a^2} \,\big|\cZ^\text{1-loop}_b\big|^2
		\, \big|\cZ^\text{inst}_b\big|^2~. 
\end{align}
Using the special property $\partial_b \cZ_b\big|_{b=1}=0$ of the ellipsoid partition function, in computing $\partial_b \big\langle \cW_b\big\rangle$ at $b=1$ we get a contribution only when the derivative is applied to the operator $\cW_b$ itself. Thus, we obtain
\begin{align}
	\label{derbB}
		\partial_b \ln \big\langle\, \cW_b\,\big\rangle \Big|_{b=1}
		= \frac{\big\langle\partial_b \cW_b\big|_{b=1}\big\rangle\phantom{\Big|}}{\big\langle \cW\big\rangle}\,
		\equiv \,\frac{\big\langle\cW^{\,\prime}\big\rangle}{\big\langle \cW\big\rangle}~.   
\end{align}
Here $\cW$ stands for $\cW_{b=1}$, that is
\begin{equation}
	\label{Wdef}
		\cW = \frac{1}{N}\tr \exp \Big(\frac{g}{\sqrt{2}}\, a\Big) = 
		1 + \frac{g^2}{4N}  \tr a^2 + O(g^3) \ldots~,
\end{equation} 
while 
\begin{align}
	\label{derbW}
		\cW^{\,\prime} =\partial_b \cW_b\big|_{b=1}=  \frac{g}{\sqrt{2}} \frac 1N 
		\tr \Big(a\,\exp\Big(\frac{g a}{\sqrt{2}}\Big)\Big)
		= \frac{g^2}{2N} \tr a^2 + O(g^3)~.
\end{align}
Note that we have the identity
\begin{equation}
\cW^{\,\prime} = g\,\frac{\partial \cW}{\partial g}~.
\label{Wprimeg}
\end{equation}
In (\ref{derbB}), both expectation values in the right hand side
are given by expressions analogous to (\ref{derWbmm})
but at $b=1$, {\it{i.e.}} they are expectation values in the matrix model on the round sphere. 

Inserting (\ref{derbB}) into (\ref{dlogW2}) expresses $h_W$ in terms of expectation values of operators in the sphere matrix model:
\begin{align}
	\label{fs4mm}
		h_W = \frac{1}{12\pi^2} \frac{\big\langle\cW^{\,\prime}\big\rangle}{\big\langle \cW\big\rangle}~.
\end{align}
Let us observe that in the matrix model it is convenient to choose a strategy, implemented through the rescaling (\ref{defa}), such that the $b$-derivative acts on the operator only. This is the opposite of what happened in the field theory proof of Section \ref{sec:6.4}, where the $b$-dependence occurred only through the action.

\paragraph{The $\cN=4$ case:} In the $\cN=4$ SYM theory,  
the matrix model is purely gaussian as both the one-loop determinant and the instanton factor reduce to $1$. Then, after using (\ref{Wprimeg}) in (\ref{fs4mm}), the $g$-derivative commutes with the expectation value and thus, as already derived in \cite{Correa:2012nk}, one has
\begin{equation}
	\label{N4B}
		h_W\Big|_{\cN=4}= \frac{1}{12\pi^2}\, g \frac{\partial \ln \big\langle\cW\big\rangle }{\partial g}~.
\end{equation}
This big simplification no longer occurs in the $\cN=2$ case, due to the non-trivial 1-loop determinant and instanton factors. Nevertheless the quantity in (\ref{derbW}), and then through eq. (\ref{dlogW2}) the value of $h_W$ and $B$, can be computed in a standard fashion in the interacting $\cN=2$ matrix model on $S^4$. In particular,  we will employ the techniques of \cite{Billo:2019fbi} to describe its perturbative expansion in $g$. 

\subsubsection{Perturbative expansion}

We now want to explicitly evaluate $h_W$ in a $\cN=2$ superconformal gauge theory using (\ref{fs4mm}).
We consider the perturbative limit in which the coupling $g$ is small and the instanton contributions
become trivial, namely we set $\cZ_\tmb{inst}=1$. 
The one-loop determinant can instead be expanded
as follows:
\begin{align}
	\label{interactiveaction}
		|\cZ_{\mathrm{1-loop}}|^2 =~\rme^{-\cS_\text{int}}~,~~~~~	
		\cS_{\mathrm{int}}= \sum_{n=2}(-1)^{n}\left(\frac{g^2}{8\pi^2}\right)^{n}\,\frac{\zeta(2n-1)}{n}\trp a^{2n}~.
\end{align}
See Subsection \ref{subsec2:MMsphere} for a complete description of all the properties of this matrix model.
%Notice that the absence of the $g^2$ term in this expansion is due to the fact that we are considering a conformal theory for which the $\beta$-function vanishes. In the right hand side of (\ref{intact2}) we used the trace combination introduced in \eqref{deftrp} where again $\cR$ is the representation in which the matter hypermultiplets transform. In the $\cN=4$ SYM theory, where $\cR$ is the adjoint, we easily see that $\cS_\tmb{int}=0$.  For $\cN=2$ models, instead, this combination accounts for the matter content of the ``difference  theory'', see Section \ref{subsec::difference}.

The vacuum expectation value of any observable $f$ in the interacting matrix model 
can be expressed in terms of vacuum expectation values computed in the Gaussian matrix model, which we distinguish by a subscript $0$. In particular, we can rewrite (\ref{fs4mm}) as
\begin{align}
	\label{hvev0}
		h_W=\frac{1}{12\pi^2}
	 \frac{\big\langle\cW^{\,\prime}\,\rme^{-\cS_\text{int}}\big\rangle_0}{
	 \big\langle \cW\,\rme^{-\cS_\text{int}}\big\rangle_0}~.
\end{align}
Expanding $\cW$ and $\cW^{\,\prime}$, as well as $\cS_\text{int}$, in series of $g$ we obtain the perturbative expansion of $h_W$ in terms of expectation values of multi-traces of powers of the matrix $a$ in the Gaussian model. Such quantities can be easily computed in a recursive way, see Section \ref{sec:mmvev}.

\paragraph{Transcendentality driven expansion:}
It is interesting to organize the computation in terms of the Riemann zeta-values appearing in (\ref{interactiveaction}). Expanding (\ref{hvev0}) in powers of $g$, we get an expression of the form
\begin{equation}
\label{hexp}
\begin{aligned}
		h_W & = g^2 \,x_1 \big(1 + O(g^2)\big) + g^6 \,\zeta(3)\, x_3  \big(1 + O(g^2)\big)
		+ g^8\, \zeta(5)\, x_5 \big(1 + O(g^2)\big)\\[1mm]
		& ~~~+ g^{10} \Big[\zeta(7)\, x_7 \big(1 + O(g^2)\big)+ \zeta(3)^2\, x_{3,3}
		\big(1 + O(g^2)\big)\Big] + \ldots
\end{aligned}
\end{equation}
where the coefficients $x_{n_1,n_2,\ldots}$ can be explicitly computed in terms of the color dependence of the theory.

Let us then introduce the quantity $\tilde h_W$ obtained by keeping, for each Riemann zeta-value, only the lowest term in $g$, namely
\begin{align}
	\label{htexp}
		\tilde h_W & = g^2\, x_1  + g^6\, \zeta(3)\, x_3 
		+ g^8\, \zeta(5)\, x_5
		+ g^{10}\, \big[\zeta(7)\, x_7  + \zeta(3)^2\, x_{3,3}\big]
	    + \ldots~.
\end{align}
This quantity is interesting for the comparison with explicit field-theoretic perturbative computations that we will carry out in the next Section.

Considering the expression of $h_W$ given in (\ref{hvev0}), we see that it reduces to $\tilde h_W$
if we keep only the lowest term in the perturbative expansions of both $\cW$ and $\cW^{\,\prime}$ given in (\ref{Wdef}) and (\ref{derbW}). Thus we can formally resum (\ref{htexp}) and write  
\begin{align}
	\label{htis}
		\tilde h_W = \frac{1}{12\pi^2}\frac{g^2}{2N} 
		\frac{\big\langle\tr a^2\,\rme^{-\cS_\text{int}}\big\rangle_0}{
	 \big\langle \rme^{-\cS_\text{int}}\big\rangle_0}= 
	 \frac{1}{12\pi^2}\frac{g^2}{2N} \,\big\langle\tr a^2\big\rangle
\end{align}
to express $\tilde h_W$ in terms of the propagator of the interacting 
matrix model. This corresponds to what has been found in Section \ref{sec:propWilson}, see in particular \eqref{corrbcres1} . Using this in (\ref{htis}), we find that $\tilde h_W$ is given by
\begin{align}
	\label{htisPi}
		\tilde h_W = \frac{1}{12\pi^2}\,\frac{g^2(N^2-1)}{4N} \,\big(1 + \Pi\big)~.
\end{align}
The perturbative corrections $\Pi$ were computed in Section \ref{subsec:propagator} with the result \eqref{Piis} which we display here for convenience:
\begin{align}
	\label{Pisecondcall}
		\Pi &= \zeta(3) \left(\frac{g^2}{8\pi^2}\right)^2 \cC^{\prime}_{4}
		- \zeta(5) \left(\frac{g^2}{8\pi^2}\right)^3 \cC^{\prime}_{6} + \cO(g^8)~,
\end{align}
where $\cC_{2n}^\prime$ is the fully symmetric contraction of the tensor $\cC^\prime_{c_1\ldots c_{2n}} = \trp T_{c_1}\ldots T_{c_{2n}}$, already encountered in \eqref{defC} (see Table \ref{tab:C4C6conf} for all the results in the various conformal theories). From the explicit expressions of the tensors $\cC_{2n}^\prime$ we can read the  coefficients $x_{n_1,n_2,\ldots}$ introduced in \eqref{hexp}.

Exploiting these methods and using the relations (\ref{CDW}) and (\ref{BW}), one can derive the
perturbative expansion of the coefficient $C_D$ in two-point function of the displacement operator and the Bremsstrahlung function $B$, at any desired order.

\subsection{Field theory interpretation}
\label{sec:ft}
We now compare the results of the previous Sections 
to the computation of the Bremsstrahlung function $B$, of the normalization 
$C_D$ in two-point function of the displacement operator, and of the normalization $h_W$ of the stress-energy one-point function using ordinary perturbative field theory in flat space. 
This comparison is not meant as a check of the relation (\ref{conj1}) of these quantities to the Wilson loop on the ellipsoid, since this is no longer conjectured but proven. Rather, it is meant to illustrate 
how the matrix model results based on this relation suggest how to organize the diagrammatic computations. These suggestions might be useful in the future for studying related quantities and/or different theories. 

We will focus on the lowest order contributions in $g$ for each given structure of 
Riemann zeta values. In the matrix model we introduced the notation $\tilde h_W$ for the sum 
of all such contributions to $h_W$ given in (\ref{htexp}) and (\ref{htis}); 
analogously we will use the notations $\tilde B$ and $\tilde C_D$.
As shown in (\ref{htisPi}), in the matrix model $\tilde h_W$ is proportional to the 
propagator. This fact suggests that also on the field-theory side the diagrams contributing to 
$\tilde h_W$, $\tilde B$ and $\tilde C_D$ are given by propagator corrections. 
We will see that for the Bremsstrahlung and for the displacement two-point function this is indeed natural. It is instead much less obvious for the one-point function of operators in the stress-energy multiplet.

\paragraph{Notations and conventions:}
In order to rely on standard literature, we perform a change of conventions with respect to Section \ref{sec2:ellips}. We redefine the adjoint scalar fields of the vector multiplet
by
\begin{align}
	\label{redphi}
		\phi \to \frac{\ii \,g}{\sqrt{2}} \,\phi~,
		~~~
	\bar\phi \to \frac{\ii\, g}{\sqrt{2}} \,\bar\phi~,
\end{align}
while all other components of the gauge multiplet are rescaled by $g$, namely 
$A_\mu\to g A_\mu$, etc. In this way, the sum of the YM and matter Lagrangians given in (\ref{LYM}) and (\ref{Lhm}), in flat space and with all supergravity background fields set to zero, reduces to the Lagrangian described in Section \ref{sec:FTactions}. This Lagrangian 
yields canonical (super) propagators.
In particular, at tree level we have
\begin{equation}
\label{propagators}
		\big\langle A_\mu^c(x)\,A_\nu^d(y)\big\rangle_0 
		= \delta^{cd}\,\delta_{\mu\nu}\, \Delta(x-y)~,\phantom{\Big|}~~~~~
		\big\langle \phi^c(x)\,\bar\phi^{\,d}(y)\big\rangle_0 
		= \delta^{cd}\,\Delta(x-y)~,
\end{equation}
where 
\begin{equation}
	\label{Delta}
		\Delta(x) = \int\!\frac{d^Dk}{(2\pi)^D}\,\frac{\rme^{\ii\,k\cdot x}}{k^2}~.
\end{equation}
with $D=4-2\epsilon$.\\
In the $\cN=4$ SYM theory, the tree-level propagators (\ref{propagators}) receive no corrections. 
In the $\cN=2$ case, instead, they are corrected in perturbation theory, and take the form already encountered in \eqref{propsAphicorr}:
\begin{equation}
\label{propagatorscorr}
		\big\langle A_\mu^c(x)\,A_\nu^d(y)\big\rangle = (1 + \Pi)\, \delta^{cd}\,\delta_{\mu\nu}\, \Delta(x-y)~,~~~
		\big\langle \phi^c(x)\,\bar\phi^{\,d}(y)\big\rangle = (1 + \Pi)\,\delta^{cd}\,\Delta(x-y)~.
\end{equation}
In Section \ref{secn:fieldtheory} it has been shown explicitly up to three loops that the correction factor $\Pi$ introduced above coincides with the factor $\Pi$ appearing in the matrix model given in (\ref{Pisecondcall}). 

\subsubsection{Bremsstrahlung function}
\begin{figure}[ht]
	\begin{center}
		\includegraphics[scale=0.83]{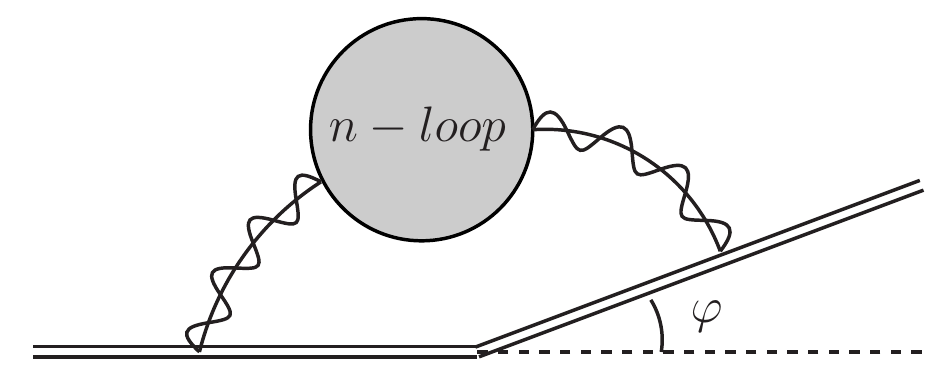}
	\end{center}
	\caption{The contribution to the vacuum expectation value of the cusped Wilson line arising
		from the a single, loop corrected, propagator - of the gluon or of the scalar.}
	\label{fig:cuspn}
\end{figure}

We already computed the leading order coefficient of the small angle expansion of the cusp anomalous dimension (see equation (\ref{Bfttree})). The pure $\cN=2$ corrections follow from that result: the form of (\ref{htisPi}) indicates that the sum of all perturbative corrections contributing to the lowest order for each transcendentality weight, which we denoted by $\tilde B$, 
can be obtained by replacing in the derivation of Section \ref{sec:6N4Brem} the tree level propagators 
(\ref{propagators}) with their loop-corrected counterparts (\ref{propagatorscorr}). In other words, at $n$ loops, we just have to consider the diagram represented in Figure~\ref{fig:cuspn}. 

Indeed, it is not difficult to realize that considering diagrams with more propagators attached to the Wilson line increases the order in $g$ without giving rise to higher transcendentality. 
The only difference in the explicit expression of the diagrams in Figure \ref{fig:cuspn} with respect to the tree-level case of Figure \ref{fig:cusp1}, is an overall factor of $(1+\Pi)$. In this way we get
\begin{equation}
	\label{Bftcomplete}
		\tilde B=\Big(\frac{g^2}{8\pi^2}\Big)\,\frac{N^2-1}{2N}\,(1+\Pi)~,
\end{equation}
in perfect agreement with (\ref{htisPi}), since $\tilde B = 3 \tilde h_W$.

\subsubsection{The displacement two-point function}
\label{subsec:disp}
We now consider the field-theory computation of the coefficient $C_D$ of the 
displacement two-point function, introduced in (\ref{Displ2points}).
From \eqref{CDandB} this quantity was shown to be related to the Bremsstrahlung function by $C_D=12 B$ in the $\cN=4$ SYM case. This relation holds as well in any $\cN=2$ superconformal theory, and it is understandable at the diagrammatic level in a simple way. 
 
We take a circular Wilson loop\,%
\footnote{We could have chosen as well a straight Wilson line.} in the fundamental representation
given by
\begin{align}
\label{Wline}
W &= \frac{1}{N} \tr \cP \exp \Big(\ii\,g \!\int_0^{2\pi}\! \!d\tau L(\tau)
\Big)~,
\end{align}
where
\begin{equation}
L(\tau)=A_\mu \dot{x}^\mu -\ii\, \frac{|\dot{x}|}{\sqrt{2}}(\phi +\bar{\phi})
\end{equation}
with the circular contour being parametrized as $x^\mu(\tau) = (R\cos \tau, R\sin\tau,0,0)$ 
for $\tau\in [0,2\pi]$.
Rather than the displacement operator $\mathbb{D}^i$, in this case 
it is easier to consider its scalar superpartner $\mathbb{O}$. While $\mathbb{D}^i$ arises from the breaking of the conservation of the stress-energy tensor by the Wilson loop defect, the scalar operator $\mathbb{O}$ arises from the breaking of the conservation law for the SO$(1,1)_R$ R-symmetry current. {From} this fact, 
following the prescription in \cite{Bianchi:2018zpb}, one can determine its explicit expression 
finding
\begin{align}
\label{Displscalar}
\mathbb{O}(\tau) = \frac{\ii\, g\,R}{\sqrt{2}} \big(\phi(\tau) -\bar{\phi}(\tau)\big)
\end{align}
where $\phi(\tau) \equiv \phi(x(\tau))$ and similarly for $\bar{\phi}$.

The functional form of the defect two-point function of this operator is fixed by the residual conformal symmetry, see eq. \eqref{4.2.31}, and its coefficient is related to the one of the displacement two-point function by supersymmetric Ward identities. For the circular Wilson loop we are considering, 
this amounts to 
\begin{align}
	\label{Displ2points}
		\big\langle \mathbb{O}(\tau_1)\,\mathbb{O}(\tau_2)\big\rangle_W 
		= \frac{C_D}{12}\,\frac{1}{(1-\cos\tau_{12})^2}
\end{align}
where $\tau_{12}=\tau_1-\tau_2$.
Using \eqref{Wline} and \eqref{Displscalar}, at the lowest order in $g$, we find
\begin{equation}
\label{Displpert}
\begin{aligned}
		\big\langle \mathbb{O}(\tau_1)\,\mathbb{O}(\tau_2)\big\rangle_W &= 
		\frac{1}{N} \tr \cP \,\Big\langle
		\rme^{\ii \,g \int_{0}^{\tau_1} \!d\tau L(\tau)}~
		\mathbb{O}(\tau_1) ~\rme^{\ii\, g \int_{\tau_1}^{\tau_2} \!d\tau L(\tau)}
		~\mathbb{O}(\tau_2) ~\rme^{\ii\, g \int_{\tau_2}^{2\pi}\! d\tau L(\tau)}
		\Big\rangle \\
		&= -\frac{g^2 R^2}{4N} \Big\langle
		\big(\phi^c(\tau_1) -\bar{\phi}^c(\tau_1)\big)\,\big(\phi^c(\tau_2) -\bar{\phi}^c(\tau_2)
		\big)	\Big\rangle + O(g^4)~.
\end{aligned}
\end{equation}
Using the tree-level scalar propagator (\ref{propagators}) and the explicit parametrization 
$x(\tau)$, we find
\begin{align}
	\label{Displpert2}
		\big\langle \mathbb{O}(\tau_1)\,\mathbb{O}(\tau_2)\big\rangle_W &
		=\frac{g^2(N^2-1)}{16\pi^2N} \frac{1}{(1-\cos\tau_{12})^2}+ O(g^4)~.
\end{align}
Thus, comparing with (\ref{Displ2points}), we obtain
\begin{align}\label{Cdtree}
C_D=12\, \Big(\frac{g^2}{8\pi^2}\Big)\,\frac{N^2-1}{2N}+ O(g^4)~,
\end{align}
which agrees with (\ref{Bfttree}) since $C_D = 12 B$. 
This tree-level computation of $C_D$ is based on the insertion of a scalar propagator 
attached to the defect, and is strictly analogous to what we have done in the previous Subsection
for the calculation of $B$; the only difference is that in that case both the scalar and the gluon propagator contribute.

\begin{figure}[ht]
	\begin{center}
		\includegraphics[scale=0.83]{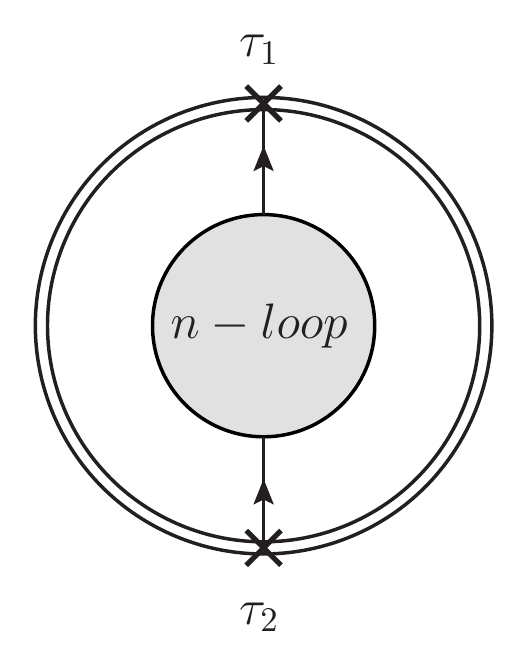}
	\end{center}
	\caption{The contribution to the two-point function of the scalar partner of the displacement operator arising from the $n$-loop correction of the scalar propagators.}
	\label{fig:displacement}
\end{figure}

The matrix model result (\ref{htisPi}) tells us that the contributions at the lowest order for each transcendentality are simply obtained by replacing the tree-level scalar propagator with the full propagator (\ref{propagatorscorr}), as represented in Figure~\ref{fig:displacement}.

By summing all these contributions, we produce an extra factor of $(1+\Pi)$ so that
\begin{align}\label{Cdcorr}
\tilde C_D=12\, \Big(\frac{g^2}{8\pi^2}\Big)\,\frac{N^2-1}{2N} \,(1+\Pi) = 12 \tilde B~.
\end{align}

\subsubsection{The stress tensor one-point function}
\label{subsec:sto}
We finally consider the direct diagrammatic computation of the $h_W$ appearing in the defect
one-point functions of the operators of the stress-energy tensor multiplet on the sphere. To do so
we consider the scalar component of this multiplet, namely the operator $O_2$ defined in the
last line of (\ref{varscmultiplet}), which in terms of the rescaled adjoint scalar fields becomes\,%
\footnote{Notice that here we do not include the factor of $g$ in the rescaling of $\phi$ and
$\bar{\phi}$, to avoid introducing in the operator an explicit dependence on the coupling constant.}
\begin{equation}\label{O2ft}
O_2 (x)=\tr[\bar \phi \phi] (x)-\frac18 q^{\mathcal{I}} q_{\mathcal{I}}(x)~. 
\end{equation}
As before, we take the defect to be the circular Wilson loop (\ref{Wline}).

The one-point function of $O_2$ in the presence of $W$ is fixed by the conformal symmetry and
depends on the orthogonal scalar product $P\circ P$, as shown in \eqref{vevO2}. 
While in Section~\ref{sec:6.4} we used the sphere projection, here we project on $\mathbb{R}^4$. Then, we exploit the residual conformal symmetry to place $O_2$ in the origin, where 
$P\circ P = R^2/4$. In this way we have
\begin{align}
\label{vevO2ft}
\big\langle \,O_2\,\big\rangle_W = \frac{3 h_W}{2 R^2}~.
\end{align}
Using (\ref{Wline}), at the lowest order we find
\begin{align}
\label{O2ft1}
\big\langle \,O_2\,\big\rangle_W&\!=\! \frac{g^2}{2N} \frac{R^2}{2} \!
\oint d\tau_1 d\tau_2 \, \Big\langle \tr[\bar \phi(0) \phi(0)] \,
\tr \!\big[ (\phi +\bar{\phi})(x(\tau_1))\, (\phi +\bar{\phi})(x(\tau_2)) \big]\Big\rangle \!
+\! O(g^4)~.
\end{align}
Inserting the tree-level scalar propagator (\ref{propagators}) and taking into account that 
$x(\tau_i)^2 =R^2$, we get
\begin{align}
\label{O2ft2}
\big\langle \,O_2\,\big\rangle_W&= \frac{g^2(N^2-1)}{8N}\frac{1}{4\pi^2 R^2} + O(g^4)~, 
\end{align}
from which it follows that 
\begin{equation}
\label{aOtree}
h_W=\frac{1}{3}\, \Big(\frac{g^2}{8\pi^2}\Big)\,\frac{N^2-1}{2N}  + O(g^4)~, 
\end{equation}
in agreement with the lowest order term in the matrix model result (\ref{htisPi}), and the relations
$C_D=12B=36h_W$.

\begin{figure}[ht]
	\begin{center}
		\includegraphics[scale=0.83]{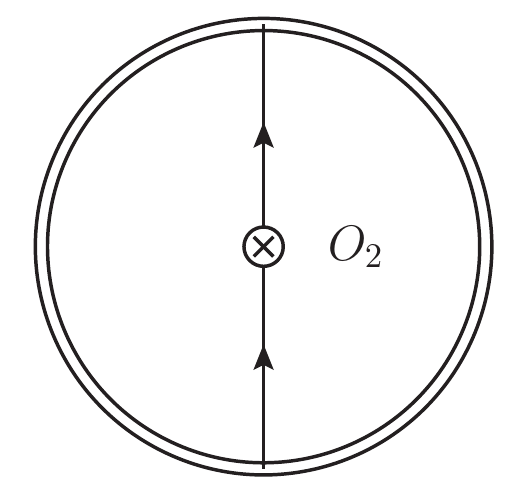}
	\end{center}
	\caption{Tree level contribution to the one-point function of $O_2$}
	\label{fig:O2}
\end{figure}

We note, however, that already at tree level the diagrammatic expansion of this observable 
differs significantly from that of the Bremsstrahlung function $B$ and of the normalization constant $C_D$ in displacement two-point function, because it involves two propagators, and not just one,
as is clear from Figure~\ref{fig:O2}. 

Despite this fact, the matrix model result (\ref{htisPi}) for $\tilde h_W$ suggests that the loop diagrams that correct the result at leading order in each transcendentality should organize themselves in terms of loop corrections to a single scalar propagator. This is far from obvious from the point of view of the Feynman diagrams, which are not so easy to compute beyond one loop. 
Indeed, $O_2$ does not belong to the class of chiral operators which enjoy nice cancellation properties due to superconformal symmetry (see for example \cite{Billo:2017glv,Billo:2018oog,Gerchkovitz:2016gxx,Baggio:2014ioa}). In this case, the matrix model could therefore provide non-trivial suggestions on how one should organize the higher loop diagrams 
contributing to the correlators of non-chiral operators.

\chapter*{Conclusions}
\addcontentsline{toc}{chapter}{Conclusion}
\chaptermark{Conclusion}
In the present thesis we have presented a series of relevant results in $\cN=2$ Lagrangian superconformal field theories, in presence of a Wilson loop operator as a conformal defect. The general philosophy resides in the interplay between defect conformal field theory, which allows to constrain the kinematic factor of many observables of the theory, and supersymmetric localization, which enables the computation of such observables in their coupling dependence, and represents a guideline for the perturbative computations which have been performed using a modern approach to the $\cN=1$ superspace formalism. We now make some comments about the original results mainly presented in Chapters \ref{chap:3}, \ref{chap:5} and \ref{chap:6} and discuss possible further developments.

In Chapter \ref{chap:3} we have considered the perturbative part of the matrix model, derived from localization, written for a generic conformal $\cN=2$ theory. Using this general approach we have described the color structure of the matrix model interactions in terms of the difference between the $\cN=2$ and the $\cN=4$ theories. This organization has allowed us to determine the four-loop terms of order $g^8$. This is in itself a significant progress with respect to the checks previously available, but the relevance of this computation stays mainly in the fact that we have shown how the perturbative computations are made more efficient and tractable by organizing them in the way suggested by the matrix model, namely by focusing on the color factors corresponding to traces of adjoint generators inserted on a loop of hypermultiplets. We think that such an organization is potentially useful also for different theories, for example non conformal ones or, maybe, even theories with less supersymmetry for which localization techniques are not presently available.\\ Furthermore we singled out some theories for which the Wilson loop vev approaches the $\mathcal{N}=4$ result in the planar limit. Beside the circular Wilson loop, it would be interesting to study other observables in the various families of $\cN=2$ superconformal theories described in this paper and analyze their behavior in the large-$N$ limit to gain some insight on their holographic dual counterparts. This direction is along the line of finding new holographic properties of $\cN=2$ theories \cite{Gaiotto:2009gz}.

In Chapter \ref{chap:5} we have verified up to two loops in $\cN=2$ superconformal theories
that the one-point amplitude $\cA_{\vec{n}}$ of a chiral operator in presence of a circular Wilson loop computed using the matrix model exactly matches
the amplitude $A_{\vec{n}}$ computed using standard field theory methods with (super) Feynman diagrams. This is a further proof of the power of supersymmetric localization. \\
Several extensions and generalizations are possible. One could the two-point functions in presence of a Wilson loop (as in \cite{Buchbinder:2012vr}) and see what kind of information could be extracted from the matrix model in this case. These correlation functions are no longer fully constrained by residual conformal symmetry, therefore additional techniques should enter in the game, such as the conformal bootstrap \cite{Liendo:2016ymz,Liendo:2018ukf,Gimenez-Grau:2019hez}.\\
Another direction, unexplored at the moment, would be investigating if Pestun's matrix model still contains information on flat space correlation functions in presence of a Wilson loop also in a \emph{non-conformal case} \cite{Billo:2019job}, up to a proper renormalization group analysis from the perturbative side. It would allow to shed some light on conformal anomaly which arises moving from $\mathbb{R}^4$ to $S^4$, as well as to work out a direct analysis reducing the amount of extra symmetries.

In Chapter \ref{chap:6} we proved that the insertion of a stress tensor operator in presence of a Wilson loop is directly related to the variation of the geometry of the spacetime. Starting from a consistent $\mathcal{N}=2$ SCFT on a four dimensional ellipsoid, the squashing acts on the background profiles of the $\mathcal{N}=2$ off-shell supergravity multiplet, while Defect CFT fixes the one-point functions of the full stress tensor supermultiplet. This one-point function captures the energy emitted by an accelerated heavy particle represented by the Wilson loop, and it represents one of the most important observables of the theory.\\
This result only uses general properties of the geometric background and the residual conformal symmetry, thus our derivation provides a general recipe to extract exact results for the stress tensor one-point function in presence of a superconformal defect by perturbing the background geometry. Another reason of interest lies in the fact that the ellipsoid matrix model formula shows an interesting way to approach correlation functions of non-chiral operators, whose computation is no longer protected by supersymmetry.

Finally it is necessary to point out many other interesting results that have been recently achieved, following the interpretation of the Wilson operator as a conformal defect in supersymmetric theories \cite{Beccaria:2017rbe,Giombi:2017cqn,Giombi:2018qox,Giombi:2018hsx,Beccaria:2019dws}. An intriguing connection with integrability (also in the $\cN=2$ case \cite{Pomoni:2013poa,Mitev:2014yba, Mitev:2015oty, Pittelli:2019ceq, Pomoni:2019oib}) has been suggested, and a special version of holographic correspondence (AdS$_2$/CFT$_1$) has been realized. The study of correlation functions in presence of a Wilson loop has recently contributed also to the exploration of new sectors of gauge theories, such as the \emph{large-charge} sector of $\cN=2$ SCFTs \cite{Hellerman:2017sur,Hellerman:2018xpi,Bourget:2018obm,Beccaria:2018xxl,Beccaria:2018owt,Grassi:2019txd,Beccaria:2020azj}. It is clear that the approach pursued in the present thesis shows a remarkable capacity of standing at the crossroad of many powerful techniques to explore non-perturbative physics.

\begin{appendices}
\part{Appendices} 

\chapter{Conventions on spinors, Grassmann variables and SUSY transformations}\label{app:Notations}

\iffalse
\section{Notations for indices}\label{app:Notations}
\begin{itemize}
\item
$d=4$ vector indices (ellipsoid): $\mu,\nu,\dots = 1,\dots,4$;

\item
$d=4$ vector indices (flat space): $m,n,\dots = 1,\dots,4$;

\item
spatial flat space indices: $i,j = 1,2,3$;

\item
$d=5$ embedding space indices (flat space): $M,N\dots = 1,\dots,5$;

\item
$d=6$ light-cone embedding coordinates: $\mathcal{M},\mathcal{N}= 0,\dots,5$;

\item
$d=6$ ``parallel'' indices: $A,B,C = 0,1,2$, and ``orthogonal'' indices:  $I,J,K = 3,4,5$;

\item
$d=4$ chiral and anti-chiral spinor indices: $\alpha,\beta$ and $\dot{\alpha},\dot{\beta}$;

\item
SU$(2)_R$ symmetry indices: $\mathcal{I},\mathcal{J},\dots = 1,2$;

\item
Sp$(r)$ indices: $\mathcal{A},\mathcal{B}= 1,\dots,2r$;

\item
SU$(N)$ adjoint indices: $c,d,\dots = 1,\dots,N^2-1$.
\end{itemize}
\fi
\section{Spinor notations}
We denote by $\psi$ a chiral spinor of components $\psi_\alpha$ with $\alpha=1,2$, 
and by $\bar \psi$ an anti-chiral one of 
components $\bar{\psi}^{\dot{\alpha}}$, with $\dot\alpha=1,2$.
The spinor indices are raised and lowered with the following rules:
\begin{align}
	\label{raislow}
		\psi^\alpha = \epsilon^{\alpha\beta}\,\psi_\beta~,~~~
		\psi_\alpha = \epsilon_{\alpha\beta}\,\psi^\beta~,~~~
		\bar\psi^{\dot{\alpha}} =\epsilon^{\dot{\alpha}\dot{\beta}}\,\bar\psi_{\dot{\beta}}~,~~~
		\bar\psi_{\dot{\alpha}} =\epsilon_{\dot{\alpha}\dot{\beta}}\,\bar\psi^{\dot{\beta}}~,
\end{align}
where
\begin{equation}
	\label{epsilons}
		\epsilon^{12} = \epsilon^{\dot{1}\dot{2}} =\epsilon_{21} = \epsilon_{\dot{2}\dot{1}} = 1~.
\end{equation}
We contract indices according to
\begin{align}
	\label{contractions}
		(\psi\chi) &\equiv \psi^\alpha\,\chi_\alpha =
		\epsilon^{\alpha\beta}\,\psi_\beta\,\chi_\alpha 
		= \psi^\alpha\,\chi^\beta\,\epsilon_{\alpha\beta}~,\\[1mm]
		(\bar\psi\bar\chi) &\equiv \bar\psi_{\dot{\alpha}}\,\bar\chi^{\dot{\alpha}} 
		=\epsilon_{\dot{\alpha}\dot{\beta}}\,\bar\psi^{\dot{\beta}}\,\bar\chi^{\dot{\alpha}}
		=\bar\psi_{\dot{\alpha}}\,\bar\chi_{\dot{\beta}}\,\epsilon^{\dot{\alpha}\dot{\beta}}~.
\end{align}
For the ``square'' of spinors, we use the notation
\begin{align}
	\label{t2tb2}
		\psi^2 \equiv (\psi\psi)~,~~~
		\bar{\psi}^2 \equiv (\bar{\psi}\bar{\psi})~.
\end{align}
{From} the previous relations, it is straightforward to obtain the Fierz identities
\begin{align}
	\label{2tist2}
		\psi^\alpha\psi^\beta = - \frac 12\, \epsilon^{\alpha\beta}\,\psi^2~,~~~
		\bar{\psi}^{\dot\alpha} \bar{\psi}^{\dot\beta} 
		= +\frac 12 \,\epsilon^{\dot\alpha\dot\beta}\,\bar{\psi}^{\,2}~.
\end{align}		
	
\section{Clifford algebra}	
We realize the Euclidean Clifford algebra
\begin{equation}
	\label{cliff4}
		\sigma_\mu\bar\sigma_\nu + \sigma_\nu\bar\sigma_\mu =
		-2\,\delta_{\mu\nu}\,\mathbf{1}
\end{equation}
by means of the matrices $(\sigma^\mu)_{\alpha\dot\beta}$ and
$(\bar\sigma^{\mu})^{\dot\alpha\beta}$ that can be taken to be
\begin{equation}
	\label{sigmas}
		\sigma^\mu =
		(\vec\tau,-\ii\mathbf{1})~,\qquad
		\bar\sigma^\mu =
		-\sigma_\mu^\dagger = (-\vec\tau,-\ii\mathbf{1})~,
\end{equation}
where $\vec\tau$ are the ordinary Pauli matrices. They are such that
\begin{equation}
	\label{traspsigma}
		(\bar\sigma^{\mu})^{\dot\alpha\alpha}=\epsilon^{\alpha\beta}\,\epsilon^{\dot{\alpha}\dot{\beta}}(\sigma^\mu)_{\beta\dot\beta}~.
\end{equation}
With these matrices we can write the 4-vectors as bispinors:
\begin{equation}
	\label{bispinork}
		k_{\alpha\dot{\beta}} = k_\mu \, (\sigma^\mu)_{\alpha\dot{\beta}}~,~~~
		\bar k^{\alpha\dot\beta} = k^\mu \, (\bar\sigma_\mu)^{\dot\alpha\beta}~.		
\end{equation}
We will often use the notations $k$ and $\bar{k}$ to indicate the matrices $k_{\alpha\dot{\beta}}$ and  $\bar k^{\alpha\dot\beta}$
and form spinor bilinears of the type
\begin{equation}
	\label{tptb}
	\theta\, k\, \bar{\theta}= \theta^\alpha\, k_{\alpha\dot{\beta}} \,\bar{\theta}^{\dot{\beta}}~.
\end{equation}
The Clifford algebra, together with the property (\ref{traspsigma}),  allows to evaluate traces of $\sigma$ and $\bar\sigma$ matrices, which we can also write in terms of traces of matrices of the type (\ref{bispinork}). In our computations we will need the following traces:
\begin{align}
	\label{sigmatraces}
		\tr \big(k_1 \bar k_2\big)  = & - 2\, k_1\!\cdot\! k_2 ~,\nonumber\\[1mm]
		\tr \big(k_1 \bar k_2 k_3 \bar k_4\big) = &  
        +2 \,\Big[(k_1\!\cdot\!  k_2)\, (k_3\!\cdot\!  k_4) - (k_1\!\cdot\!  k_3)\,(k_2\!\cdot\!  k_4)
		+ (k_1\!\cdot\!  k_4)\, (k_2\!\cdot\!  k_3)\Big] + \ldots~,\nonumber\\[1mm]
		\tr \big(k_1 \bar k_2 k_3 \bar k_4 k_5 \bar k_6\big) = &  
		- 2\,k_1\!\cdot\!  k_2\, \Big[(k_3\!\cdot\!  k_4)\, (k_5\!\cdot\!  k_6) - (k_3\!\cdot\!  k_5)\, (k_4\!\cdot\!  k_6) + (k_3\!\cdot\!  k_6) \,(k_4\!\cdot\!  k_5)\Big]	\nonumber\\
		& +  2\, k_1\!\cdot\! k_3\, \Big[(k_2\!\cdot\!  k_4) \,(k_5\!\cdot\!  k_6) - (k_2\!\cdot\!  k_5)\, (k_4\!\cdot\!  k_6) + (k_2\!\cdot\!  k_6)\, (k_4\!\cdot\! k_5)\Big] \nonumber\\
		& -  2 \,k_1\!\cdot\! k_4\,\Big[(k_2\!\cdot\!  k_3) \,(k_5\!\cdot\!  k_6) - (k_3\!\cdot\! k_5)\, (k_3\!\cdot\!  k_6) + (k_2\!\cdot\!  k_6)\, (k_3\!\cdot\! k_5)\Big] \nonumber\\
		& +  2 \,k_1\!\cdot\! k_5\, \Big[(k_2\!\cdot\!  k_3)\, (k_4\!\cdot\!  k_6) - (k_3\!\cdot\!  k_4)\, (k_3\!\cdot\!  k_6) + (k_2\!\cdot\! k_6) \,(k_3\!\cdot\! k_4)\Big] \nonumber\\
		& -  2\, k_1\!\cdot\! k_6\,\Big[(k_2\!\cdot\! k_3)\, (k_4\!\cdot\!  k_5) - (k_3\!\cdot\! k_4)\, (k_3\!\cdot\!  k_5) + (k_2\!\cdot\! k_5) \,(k_3\!\cdot\!  k_4)\Big] \nonumber\\
		& + \ldots~,
\end{align}
where the ellipses in the second and last line stand for parity-odd terms containing contractions with a space-time $\varepsilon$-tensor that do not enter in our computations.

\section{Grassmann integration formul\ae}
The basic integration formul\ae~ for Grassmann variables are
\begin{align}
	\label{intt2}
		\int d^2\theta\, \theta^2 = 1~,~~~
		\int d^2\bar{\theta}\, \bar{\theta}^2 = 1~.
\end{align} 
These imply that the $\theta^2$ and $\bar{\theta}^2$ act as fermionic $\delta$-functions; 
more in general, writing $\theta_{ij} = \theta_i - \theta_j$, we have
\begin{align}
	\label{deltat}
		\theta_{ij}^2 = \delta^2(\theta_{ij})~,~~~
		\bar\theta_{ij}^{\,2} = \delta^2(\bar\theta_{ij})~;
\end{align} 
we also use the notation
\begin{align}
	\label{delta4t}
		\theta_{ij}^2\, \bar\theta_{ij}^{\,2} = \delta^4(\theta_{ij})~.
\end{align}

\subsection*{Spinor derivatives} 
Writing $\partial_\alpha\equiv 
{\displaystyle{\frac{\partial}{\partial\theta^\alpha}}}$ and $\bar\partial_{\dot{\alpha}}\equiv{\displaystyle{\frac{\partial}{\partial\bar\theta^{\dot{\alpha}}}}}$, we have
\begin{equation}
\begin{aligned}
	\label{spdert2}
		&\partial_\alpha\,\theta^2 = 2\,\theta_\alpha~,~~~
		\partial\partial\,\theta^2 = - 4\\[1mm]
		&\bar\partial_{\dot{\alpha}}\,\bar\theta^{\,2} = -2\,\bar\theta_{\dot{\alpha}}~,~~~
		\bar\partial\bar\partial\,\bar\theta^{\,2}= - 4~.
\end{aligned}
\end{equation}
The covariant spinor derivatives are defined as 
\begin{align}
	\label{covspinder}
		D_\alpha=\partial_\alpha+\ii\,(\sigma^\mu)_{\alpha\,\dot{\alpha}}\,
		\bar\theta^{\dot{\alpha}}\,\partial_\mu
		\quad\mbox{and}\quad
		\bar D_{\dot{\alpha}}=-\bar\partial_{\dot{\alpha}}-
		\ii\,\theta^\alpha\,(\sigma^\mu)_{\alpha\,\dot{\alpha}}\,\partial_\mu~.
\end{align}
In momentum space, they become
\begin{align}
	\label{covspink}
		D_\alpha=\partial_\alpha-(k\,\bar\theta)_{\alpha}
		\quad\mbox{and}\quad
		\bar D_{\dot{\alpha}}=-\bar\partial_{\dot{\alpha}}+
		(\theta\,k)_{\dot{\alpha}}~,
\end{align}
where $k$ is the momentum flowing outward from the space-time point $x$, {\it{i.e.}} 
the Fourier transform is taken with the phase $\exp(+\ii \,k\!\cdot\! x)$.

\section{SUSY transformations}
\label{app:SUSYtransf}
Let us start by listing the on-shell SUSY transformations of the fields in the vector multiplet.
We follow \cite{Hama:2012bg}, but consider the SUSY parameters $\xi$ as Grassmann odd.
\begin{equation}
\label{susygauge}
\begin{aligned}	
		\d A_{\mu}&=\ii\xi^{\mathcal{I}} \s_{\mu} \bar \l_{\mathcal{I}}-\ii\bar \xi^{\mathcal{I}} \bar \s_{\mu} \l_{\mathcal{I}}~,\\
		\d \phi &= - \ii \xi^{\mathcal{I}} \l_{\mathcal{I}}~,\\
		\d \bar \phi &= + \ii \bar \xi^{\mathcal{I}} \bar \l_{\mathcal{I}}~,\\
		\d \l_{\mathcal{I}}&=\frac12 \s^{\mu\nu}\xi_{\mathcal{I}} (F_{\mu\nu}+8\bar \phi T_{\mu\nu}) + 2\s^{\mu} \bar \xi_{\mathcal{I}} D_{\mu} \phi+\s^{\mu} D_{\mu} \bar \xi_{\mathcal{I}} \phi+ 2\ii \xi_{\mathcal{I}} [\phi,\bar \phi]~,\\
		\d \bar \l_{\mathcal{I}}&=\frac12 \bar \s^{\mu\nu}\bar \xi_{\mathcal{I}} (F_{\mu\nu}+8 \phi \bar T_{\mu\nu}) + 2\bar \s^{\mu} \bar \xi_{\mathcal{I}} D_{\mu} \bar \phi+\bar \s^{\mu} D_{\mu}  \xi_{\mathcal{I}} \bar \phi- 2\ii \bar \xi_{\mathcal{I}} [\phi,\bar \phi]~.
\end{aligned}
\end{equation}
This algebra closes on the following field equations
\begin{align}
	\label{eomgauge}
		\bar \s^{\mu}D_{\mu} \l_{\mathcal{I}}=2\ii[\phi,\bar \l_{\mathcal{I}}]~,~~~  
		\s^{\mu}D_{\mu} \bar \l_{\mathcal{I}}=2\ii[\bar\phi, \l_{\mathcal{I}}]~.
\end{align}
For the hypermultiplet the on-shell SUSY transformations are 
\begin{align}
	\label{susyhyper}
		\d q_{\mathcal{I}}&=-\ii\xi_{\mathcal{I}} \psi + \ii \bar \xi_{\mathcal{I}} \bar \psi~,\notag\\
		\d \psi &=2 \s^{\mu} \bar \xi_{\mathcal{I}} D_{\mu} q^{\mathcal{I}}+ \s^{\mu} D_{\mu} \bar \xi_{\mathcal{I}} q^{\mathcal{I}}-4\ii \xi_{\mathcal{I}} \bar \phi q^{\mathcal{I}}~,\notag\\
		\d \bar \psi &=2 \bar \s^{\mu} \xi_{\mathcal{I}} D_{\mu} q^{\mathcal{I}}+ \bar \s^{\mu} D_{\mu}  \xi_{\mathcal{I}} q^{\mathcal{I}}-4\ii \bar\xi_{\mathcal{I}}  \phi q^{\mathcal{I}}~.
\end{align}

Now we consider the stress tensor multiplet. In flat space, the on-shell SUSY transformations are
\begingroup
\allowdisplaybreaks
\begin{align}
	\label{susystm}
		\d O_2&=\ii\bar \chi_{\dot \a \mathcal{I}} \bar \xi^{\dot \a \mathcal{I}} +\ii \xi_{\mathcal{I}}^{\a} \chi_{\a}^{\mathcal{I}}~,\notag\\
		\d\chi_{\a}^{\mathcal{I}}&=H_{\a}{}^{\b} \xi^{\mathcal{I}}_\b+\frac12 j_{\a \dot \a} \bar \xi^{\dot \a \mathcal{I}}+\frac12 t_{\a \dot \a \mathcal{J}}{}^{\mathcal{I}} \bar \xi^{\dot \a \mathcal{J}}+\pa_{\a \dot \a} O_2 \bar \xi^{\dot \a \mathcal{I}}~,\notag\\
		\d\bar \chi_{\dot \a \mathcal{I}}&=-\bar H^{\dot \b}{}_{\dot \a} \bar \xi_{\dot \b \mathcal{I}}+\frac12 j_{\a \dot \a} \xi_{\mathcal{I}}^{\a}+\frac12 t_{\a \dot \a \mathcal{I}}{}^{\mathcal{J}}  \xi_\mathcal{J}^{\a}-\pa_{\a \dot \a} O_2 \xi_{\mathcal{I}}^{\a}~,\notag\\
		\d H_{\a}{}^{\b}&=\frac{\ii}{2} J_{\a \dot \a}{}^{\b}{}_{\mathcal{I}} \bar \xi^{\dot \a \mathcal{I}}+\frac{2 \ii}{3} \big(\pa_{\a \dot \a} \chi^\b_{\mathcal{I}} +\pa^{\b}{}_{\dot \a} \chi_{\a \mathcal{I}}\big)\bar \xi^{\dot \a \mathcal{I}}~,\notag\\
		\d \bar H^{\dot \b}{}_{\dot \a}&=-\frac{\ii}{2} \bar J_{\a \dot \a}{}^{\dot \b \mathcal{I}} \xi^{\a}_{\mathcal{I}}-\frac{2 \ii}{3} \big(\pa_{\a \dot \a} \bar \chi^{\dot \b \mathcal{I}} +\pa_{\a}{}^{\dot \b} \bar \chi_{\dot \a}^{\mathcal{I}}\big) \,\xi_{\mathcal{I}}^{\a}~,\notag\\
		\d j_{\a \dot \a}&=- \frac{\ii}{2} J_{\a \dot \a \b}{}^{\mathcal{I}} \xi_{\mathcal{I}}^{\b} -\frac{\ii}{2} \bar J_{\a \dot \a \dot \b \mathcal{I}} \bar \xi^{\dot \b \mathcal{I}}+\frac{4\ii}{3} \xi_{\mathcal{I}}^\b \big(2 \pa_{\b \dot \a} \chi_{\a}^{\mathcal{I}} -\pa_{\a \dot \a} \chi_{\b}^{\mathcal{I}}\big)+\frac{4\ii}{3} \bar\xi^{\dot \b \mathcal{I}} \big(2 \pa_{\a \dot \b} \bar\chi_{\dot \a \mathcal{I}} -\pa_{\a \dot \a} \bar \chi_{\dot \b \mathcal{I}}\big)~,\notag\\
		\d t_{\a \dot \a \mathcal{I}}{}^{\mathcal{J}}&=\ii J_{\a \dot \a \b}{}^{\mathcal{J}} \xi_{\mathcal{I}}^\b+\ii \bar J_{\a \dot \a \b \mathcal{I}} \bar \xi^{\dot \b \mathcal{J}}+\frac{4\ii}{3} \xi_{\mathcal{I}}^\b \big(2 \pa_{\b \dot \a} \chi_{\a}^{\mathcal{J}} -\pa_{\a \dot \a} \chi_{\b}^{\mathcal{J}}\big)+\frac{4\ii}{3} \bar\xi^{\dot \b \mathcal{J}} \big(2 \pa_{\a \dot \b} \bar\chi_{\dot \a \mathcal{I}} -\pa_{\a \dot \a} \bar \chi_{\dot \b \mathcal{I}}\big)\notag\\
		-\frac12&\d^{\mathcal{J}}_{\mathcal{I}} \Big[\ii J_{\a \dot \a \b}{}^{\mathcal{K}} \xi_{\mathcal{K}}^\b+\ii \bar J_{\a \dot \a \b {\mathcal{K}}} \bar \xi^{\dot \b {\mathcal{K}}}+\frac{4\ii}{3} \xi_{\mathcal{K}}^\b \big(2 \pa_{\b \dot \a} \chi_{\a}^{\mathcal{K}} -\pa_{\a \dot \a} \chi_{\b}^{\mathcal{K}}\big)+\frac{4\ii}{3} \bar\xi^{\dot \b {\mathcal{K}}} \big(2 \pa_{\a \dot \b} \bar\chi_{\dot \a {\mathcal{K}}} -\pa_{\a \dot \a} \bar \chi_{\dot \b {\mathcal{K}}}\big)\Big]~,\notag\\
		\d J_{\a \dot \a \b}{}^{\mathcal{I}}&=2 T_{\a \dot \a \b \dot \b} \bar \xi^{\dot \b \mathcal{I}}  +\frac23 \big(\pa_{\a \dot \a} H_{\b}{}^{\g}+\pa_{\b \dot \a} H_{\a}{}^{\g}\big)\xi^{\mathcal{I}}_\g-2 \pa_{\g \dot \a} H_{\b}{}^{\g} \xi_{\a}^{\mathcal{I}}-2 \pa_{\g \dot \a} H_{\a}{}^{\g} \xi_{\b}^{\mathcal{I}}\notag\\
		&-\bar \xi^{\dot \b \mathcal{I}} \Big(\frac23 \pa_{\a \dot \a} j_{\beta \dot \beta}- \frac13 \pa_{\b \dot \a} j_{\a \dot \b}- \pa_{\a \dot \b} j_{\b \dot \a}\Big) +2\bar \xi^{\dot \b \mathcal{J}} \Big(\frac23 \pa_{\a \dot \a} t_{\beta \dot \beta \mathcal{J}}{}^{\mathcal{I}}- \frac13 \pa_{\b \dot \a} t_{\a \dot \b \mathcal{J}}{}^{\mathcal{I}}- \pa_{\a \dot \b} t_{\b \dot \a \mathcal{J}}{}^{\mathcal{I}}\Big)~,\notag\\
		\d \bar J_{\a \dot \a \dot \b \mathcal{I}}&=-2 T_{\a \dot \a \b \dot \b} \xi_{\mathcal{I}}^{ \b}  -\frac23 \big(\pa_{\a \dot \a}\bar  H^{\dot \g}{}_{\dot \b}+\pa_{\a \dot \b} \bar H^{\dot \g}{}_{\dot\a}\big)\bar \xi_{\dot \g \mathcal{I}}+2 \pa_{\a \dot \g} \bar H_{\dot \b}{}^{\dot \g} \bar \xi_{\dot \a \mathcal{I}}+2 \pa_{\a \dot \g} \bar H_{\dot \a}{}^{\dot \g} \bar \xi_{\dot \b \mathcal{I}}\notag\\
		&- \xi_{\mathcal{I}}^{\b} \Big(\frac23 \pa_{\a \dot \a} j_{\beta \dot \beta}- \frac13 \pa_{\a \dot \b} j_{\b \dot \a}- \pa_{\b \dot \a} j_{\a \dot \b}\Big) +2 \xi_\mathcal{J}^{\b} \Big(\frac23 \pa_{\a \dot \a} t_{\beta \dot \beta \mathcal{I}}{}^{\mathcal{J}}- \frac13 \pa_{\a \dot \b} t_{\b \dot \a \mathcal{I}}{}^{\mathcal{J}}- \pa_{\b \dot \a} t_{\a \dot \b \mathcal{I}}{}^{\mathcal{J}}\Big)~,\notag\\
		\d T_{\a \dot \a \b \dot \b}&=\frac{\ii}4 \xi_{\mathcal{I}}^\g \big(2\pa_{\g \dot \a} J_{\b \dot \b \a}{}^{\mathcal{I}}-\pa_{\a \dot \a} J_{\b \dot \b \g}{}^{\mathcal{I}}\big)-\frac{\ii}4 \bar \xi^{\dot \g \mathcal{I}} \big(2\pa_{\a \dot \g} \bar J_{\b \dot \b \dot \a \mathcal{I}}-\pa_{\a \dot \a} \bar J_{\b \dot \b \dot \g \mathcal{I}}\big) + \big( \{\a,\dot \a\} \leftrightarrow \{\b,\dot \b\} \big)~.
\end{align}
\endgroup
These transformations obey the commutation relations
\begin{align}
	\label{susyalg}
		\Big[\big[\d_{\xi_1}, \d_{\xi_2}\big] , \bullet\Big]
		=-2\ii (\xi_{1 \mathcal{K}}^\a \bar \xi_2^{\dot \a \mathcal{K}}-\xi_{2 \mathcal{K}}^\a \bar \xi_1^{\dot \a \mathcal{K}}) \pa_{\a \dot \a} \bullet~.
\end{align}
It is possible to verify that the normalization factors of the operators listed in (\ref{varscmultiplet})
are consistent with these SUSY transformations.

\chapter{Group theory conventions}
\label{app:group}
\section{Useful group theory formul\ae~for SU$(N)$}
We denote by $T_a$, with $a=1,\ldots, N^2-1$, a set of Hermitean generators satisfying 
the $\mathfrak{su}(N)$ Lie algebra
\begin{align}
	\label{sunalgebrar}
		\big[\,T_a\,,\,T_b\,\big] = \ii f_{abc}\, T_c~.
\end{align} 
We indicate by $t_a$ the representative of $T_a$ in the fundamental representation; they are Hermitean, traceless $N\times N$ matrices that we normalize by setting 
\begin{equation}
	\label{norm}
		\tr\,t_a t_b = \frac{1}{2}\,\delta_{ab}~.
\end{equation}
In the conjugate fundamental representation the generators are
\begin{align}
	\label{genconj}
		\bar t_a = - t_a^{\,T}~.
\end{align}
The generators $t^a$ are such that the following fusion/fission identities hold
\begin{align}
	\label{fussion}
		\tr\left(t_a M_1 t_a M_2\right) & = \frac{1}{2}\,\tr\, M_1\, \tr\,M_2
		-\frac{1}{2N}\,\tr\left(M_1 M_2\right)~,\\
		\tr\left(t_a M_1\right) ~\tr\left(t_a M_2\right) & = \frac{1}{2}\,\tr\left(M_1 M_2\right) -\frac{1}{2N}\,\tr\,M_1 \,\tr\,M_2~,
\end{align}
for arbitrary $(N\times N)$ matrices $M_1$ and $M_2$.

In the enveloping matrix algebra, we have
\begin{align}
	\label{tatb}
		t_a\,t_b =\frac{1}{2}\,\left[\frac{1}{N}\,\delta_{ab}\,\mathbf{1}
		+\left(d_{abc}+\ii\,f_{abc}\right)\,t^c\right]~,
\end{align}
where $d^{abc}$ is the totally symmetric $d$-symbol of su$(N)$.
Using (\ref{norm}) and (\ref{tatb}), we obtain
\begin{align}
	\label{key}
		\tr\left(\big\{\,t_a\,,\,t_b\big\}\,t_c\right) = \frac{1}{2}\,d_{abc}~,\quad
		\tr\left(\big[\,t_a\,,\,t_b\big]\,t_c\right)= \frac{\ii}{2}\,f_{abc}~,
\end{align}
from which it follows that $d_{aac}=0$. 
We can write the $d$- and $f$-symbols as $(N^2-1)\times (N^2-1)$ matrices
\begin{equation}
\ii f^{abc} = (F^a)^{bc} , \hspace{0.5cm} d^{abc} = (D^a)^{bc}
\end{equation}
and derive the following useful identities:
\begin{equation}\label{adjtraces}
\begin{split}
&\Tr F^a=\Tr D^a=\Tr F^a D^b=0~,\\
&\Tr F^a F^b=N \delta^{ab}~,
\hspace{0.6cm}\Tr D^a D^b=\frac{N^2-4}{N} \delta^{ab}~,\\
&\Tr F^a F^b F^c=\frac{\ii N}{2} f^{abc}~,
\hspace{0.6cm}\Tr D^a F^b F^c=\frac{N}{2} d^{abc}~,\\
&\Tr F^a F^b F^c D^d=\frac{\ii N}{4}(d^{ade}f^{bce}-f^{ade}d^{bce})
\end{split}
\end{equation}
where $\Tr$ denotes the trace in the adjoint representation.

\section{Traces of generators}
In any representation $\cR$ we have
\begin{equation}
	\label{indexR}
		\Tr_\cR T_a T_b = i_\cR\, \delta_{ab}~,
\end{equation}
where $i_\cR$ is the index of $\cR$, and is fixed once the generators have been normalized in
the fundamental representation (see (\ref{norm})). 
The quadratic Casimir operator in the representation $\cR$ is defined by 
\begin{equation}
	\label{casimirR}
		T_a \,T_a = c_\cR\, \mathbf{1}~.
\end{equation}  
By tracing this equation and comparing to (\ref{indexR}), we have
\begin{equation}
	\label{cRtoiR}
		c_\cR = \frac{N^2-1}{d_\cR} \, i_\cR~,
\end{equation}   
with $d_\cR$ being the dimension of the representation $\cR$. 

The traces of products of generators  define a set of cyclic tensors 
\begin{align}
	\label{defCapp}
		C_{a_1\ldots a_n} = \Tr_{\cR} T_{a_1}\ldots T_{a_n}
\end{align}
whose contractions are higher order invariants characterizing the representation $\cR$. 
Let us note that we can switch the order of any two consecutive indices using the Lie algebra relation (\ref{sunalgebrar}); indeed:
\begin{align}
	\label{switchC}
		C_{\ldots a b \ldots} = C_{\ldots b a \ldots} + \ii\, f_{abc}\, C_{\ldots c \ldots}~.
\end{align}

In our computations we encounter the particular combination of traces introduced in 
(\ref{defC}), namely
\begin{align}
	\label{defCagain}
		C^\prime_{a_1\ldots a_n} \,=\, \trp T_{a_1}\ldots T_{a_n} \,=\, 
		\Tr_{\cR} T_{a_1}\ldots T_{a_n}- 
		\Tr_{\mathrm{adj}} \,T_{a_1}\ldots T_{a_n}~.
\end{align}
These are of course also cyclic, and the relation (\ref{switchC}) applies to them as well.

If $\cR$ is the representation in which the matter hypermultiplets of a superconformal theory transform, one can prove that
\begin{equation}
C^\prime_{ab}= 0~,
\end{equation}
since $C^\prime_{ab}$ is proportional to the one-loop $\beta$-function coefficient. Therefore,
using this property and the relation (\ref{switchC}) one can easily show that for conformal theories
\begin{align}
\label{C3confsym}
C^\prime_{abc} = C^\prime_{acb} + \ii f_{abe} C^\prime_{ec} = C^\prime_{acb}
\end{align}
which, together with cyclicity, implies that the tensor $C^\prime_{abc}$ is totally symmetric. Thus, 
it is proportional to $d_{abc}$. Finally, one can prove that
\begin{equation}
C^\prime_{abcc} = C^\prime_{(abcc)}~.
\end{equation}
Indeed, if we exchange the two free indices we have
\begin{align}
	\label{C4sw1}
		C^\prime_{abcc} = C^\prime_{bacc} + \ii f_{abe} C^\prime_{ecc} = C^\prime_{bacc}~,
\end{align}
where the last step follows from the fact that $C^\prime_{ecc}=0$ since $d_{ecc} =0$. 
If instead we switch the position of a free and a contracted index, we have
\begin{align}
	\label{C4sw12}
		C^\prime_{abcc} = C^\prime_{acbc} + \ii f_{bce} C^\prime_{aec} = C^\prime_{acbc}~,
\end{align}
where have used the fact 
that $C^\prime_{aec}$, being symmetric, vanishes when contracted with $f_{bce}$. 

\section{Some particular representations}
The generators in the direct product representation $\cR =  \Yfund \otimes \Yfund$ are given by
\begin{equation}
	\label{Tafunfun}
		T_a = t_a \otimes \mathbf{1} \oplus \mathbf{1} \otimes t_a~.
\end{equation} 
This representation is reducible into its symmetric and anti-symmetric parts:
\begin{equation}
	\label{funfundec}
		\Yfund \otimes \Yfund = \Ysymm \oplus \Yasymm~.	
\end{equation}
In the symmetric representation one has 
\begin{equation}
	\label{trsymm}
		\Tr_{\Ysymm}\big( X\otimes Y \big)=
		\frac{1}{2}\Big(\tr\,X\,\,\tr\,Y 
		+\tr\left(X\, Y\right)\Big)~,
\end{equation}
while in the anti-symmetric representation one has
\begin{equation}
	\label{trasymm}
		\Tr_{\Yasymm} \big(X\otimes Y\big) =
		\frac{1}{2}\Big(\tr\,X\,\,\tr\,Y 
		-\Tr\left(X\, Y\right)\Big)~.
\end{equation}

The adjoint representation is contained in the direct product of a fundamental and an anti-fundamental:
\begin{equation}
	\label{funbfun}
		\Yfund \otimes \overline{\Yfund} = \mathrm{singlet} \oplus \adj~.
\end{equation}
The generators in the adjoint can thus represented simply\,%
\footnote{They should be thought of as acting on the $N^2-1$-dimensional subspace orthogonal to the invariant vector $\sum_i e_i\otimes \bar e_i$, where $e_i$ and $\bar e_i$, for $i=1,\ldots N$, are basis vectors in the carrier spaces of the fundamental and anti-fundamental representations. This however makes no difference for the computation of the traces we are interested in.} by 
\begin{equation}
	\label{genadj}
		T_a = t_a \otimes \mathbf{1} + \mathbf{1} \otimes \overline{t}_a~.
\end{equation}
Using these relations it is easy to obtain the well-known results collected in table~\ref{tab:saad}. 
\begin{table}[ht]
	\begin{center}
		{\small
			\begin{tabular}{c|c|c}
				\hline
				\hline
				$\cR$ \phantom{\bigg|}& $d_\cR$ & $i_\cR$ \\
				\hline
				$\phantom{\bigg|}\Yfund\phantom{\bigg|}$ & $~~N~~$ & $~~\frac{1}{2}~~$\\
				$\phantom{\bigg|}\Ysymm\phantom{\bigg|}$ & $~~\frac{N(N+1)}{2}~~$ & $~~
				\frac{N+2}{2}~~$\\
				$\phantom{\bigg|}\Yasymm\phantom{\bigg|}$ & $~~\frac{N(N-1)}{2}~~$ & $~~
				\frac{N-2}{2}~~$\\
				$\phantom{\bigg|}\adj\phantom{\bigg|}$ & $~~N^2-1~~$ & $~~N~~$ \\
				\hline
				\hline
			\end{tabular}
		}
	\end{center}
	\caption{Dimensions and indices of the fundamental, symmetric, anti-symmetric and adjoint representations of SU($N$).}
	\label{tab:saad}
\end{table}

If we consider a representation $\cR$ made of $N_F$ fundamental, $N_S$ symmetric and $N_A$ anti-symmetric representations, namely
\begin{equation}
\cR = N_F\, \Yfund \oplus N_S\, \Ysymm \oplus N_A\, \Yasymm
\end{equation}
as in (\ref{RNFNASNA}), we immediately see that 
\begin{align}
	\label{RFSAb0}
		\trp T^a T^b =\big(N_F + N_S(N+2) + N_A(N-2) - 2N\big)\,\tr t^a t^b 
		= -{\beta_0}\,\tr t^a t^b 
	\end{align}
where 
$\beta_0$ is the one-loop $\beta$-function coefficient of the $\mathcal{N}=2$ SYM theory
(see (\ref{b0is})).

With a bit more work, but in a straightforward manner, one can compute traces of more 
generators. In particular, one can evaluate
\begin{align}
	\label{colfact1}
		\trp a^n & = 
		N_F\, \tr a^n 
		+ N_S\, \Tr_{\Ysymm}\big(a\otimes \bone + \bone \otimes a\big)^n
		+ N_A\, \Tr_{\Yasymm}\big(a\otimes \bone + \bone \otimes a\big)^n\nonumber\\[1mm]
		& ~~~- \Tr_\adj \big(a\otimes \bone + \bone \otimes (-a^{\,T})\big)^n~,
\end{align}
with the result
\begin{align}
	\label{colfact2}
		\trp a^n & = \Big[(N_F + 2^{n-1} \big(N_S - N_A\big) 
		+ N \big(N_S + N_A -(1 + (-1)^n)\big) \Big]\, \tr a^n
	\nonumber\\
	& ~~~+ \sum_{p=1}^{n-1} \binom{n}{p} \left(\frac{N_S + N_A}{2} - (-1)^{n-p}\right)\tr a^p\, \tr a^{n-p}~.		
\end{align}
In particular, when $n=2k$, this expression can be rewritten as in (\ref{S2n}) of the main text.
 
\subsection{Traces in a generic representation} 
\label{secn:frob}
A representation $\cR$ is associated to a Young diagram $Y_R$; let $r$ be the number of boxes
in the tableau. Traces in the representation $\cR$ can be evaluated in terms of traces in the fundamental representation using the Frobenius theorem. 
For any group element $U$ in SU$(N)$, this theorem theorem states that 
\begin{equation}
	\label{frob}
		\Tr_{\cR} U = \sum_{M} \frac{1}{|M|}\, \chi^R(M)\,(\tr U)^{m_1}\, (\tr U^2)^{m_2}\,
		\ldots (\tr U^r)^{m_r}~.
\end{equation}  
We denote by $M$ a conjugacy class\,%
\footnote{$M$ is associated to a Young diagram with $r$ boxes, containing $m_j$ columns of length $j$.} of $S_r$ containing permutations made of $m_j$ cycles of length $j$, with $j=1,\ldots r$; the number of elements in the class is $r!/|M|$, with 
\begin{equation}
	\label{orderclassM}
		|M| = \prod_{_j=1}^{r} m_j!\, j^{m_j}~. 
\end{equation}  
With $\chi^R(M)$ we denote the character of the conjugacy class $M$ in the representation $R$ of the group $S_r$ associated to the tableau $Y_R$.
If we write $U = \rme^a$, with $a\in \mathfrak{su}(N)$, equation (\ref{frob}) reads
\begin{equation}
	\label{frob2}
		\Tr_{\cR} \rme^a = \sum_{M} \frac{1}{|M|} \,\chi^R(M)\,(\tr \rme^a)^{m_1}\, (\tr \rme^{2a})^{m_2}\,
		\ldots (\tr \rme^{ra})^{m_r}~,
\end{equation}  
and expanding it in powers of $a$, one can obtain the expression of all traces of the form
$\Tr_{R} a^k$ in terms of products of traces of powers of $a$ in the fundamental representation,
generalizing what we have seen before for the symmetric, anti-symmetric and adjoint representations.

\chapter{Field theory computations}

\section{Grassmann integration in superdiagrams}
\label{app:grass-super}
We discuss a method to carry out the Grassmann integrations appearing in $\cN=1$ superdiagrams 
involving chiral/anti-chiral multiplet and vector multiplet lines.

\subsection*{Diagrams with only chiral/anti-chiral multiplet lines}
As we can see from the Feynman rules in Figure~\ref{fig:Feynmatter}, the three-point 
vertex with incoming chiral lines carries a factor of $\theta^2$ and thus in the
integration over the fermionic variables associated to the vertex, one remains with
only an integral over $\bar{\theta}$. For the three-point vertex with outgoing anti-chiral 
lines, we remain instead with an integration over $\theta$ only.

\begin{figure}[ht]
	\begin{center}
	\includegraphics[width=0.92\textwidth]{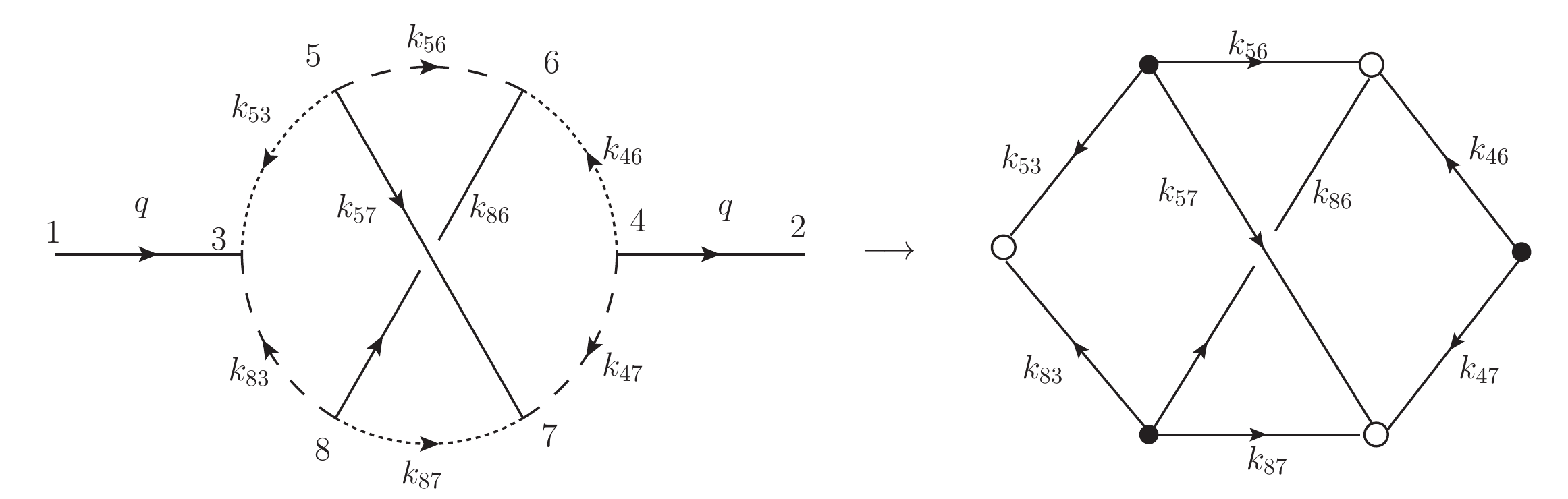}
	\end{center}
	\caption{On the left, a Feynman super-diagram involving only chiral/anti-chiral lines. 
	On the right, the corresponding $\theta$-graph encoding the Grassmann integrals. 
	The two ``external'' propagators with momentum $q$
	do not play a r\^ole in the bipartite graph because 
	the external states are the lowest components of the chiral and 
	anti-chiral superfields, and so the corresponding Grassmann variables are set to $0$.}
	\label{fig:1}
\end{figure}

We will use a graphical notation in which a black dot represents a $\theta$ variable and 
a white circle represents a $\bar{\theta}$ variable. {From} the point of view of the 
Grassmann integrations, superdiagrams with only hypermultiplet lines reduce to bipartite 
graphs, which we call ``$\theta$-graphs''. 
In these graphs a solid line connecting the $i$-th dot to the $j$-th circle corresponds to 
the factor 
\begin{equation}
	\label{exp2}
		\exp\Big(2\, \theta_i \,k_{ij} \,\bar{\theta}_j\Big)
		= 1 + 2\,\theta_i\, k_{ij}\, \bar{\theta}_j + \frac 12 \,\Big(2\,\theta_i\, k_{ij}\, 
		\bar{\theta}_j
		\Big)^2
\end{equation} 
coming from the chiral superfield propagator connecting two vertices at points $i$ and $j$ 
in a Feynman superdiagram. An example of a $\theta$-graph associated to a 
superdiagram is illustrated in Figure~\ref{fig:1}, where the momenta respect momentum 
conservation at each node.

\begin{figure}[ht]
	\begin{center}
		\includegraphics[width=0.85\textwidth]{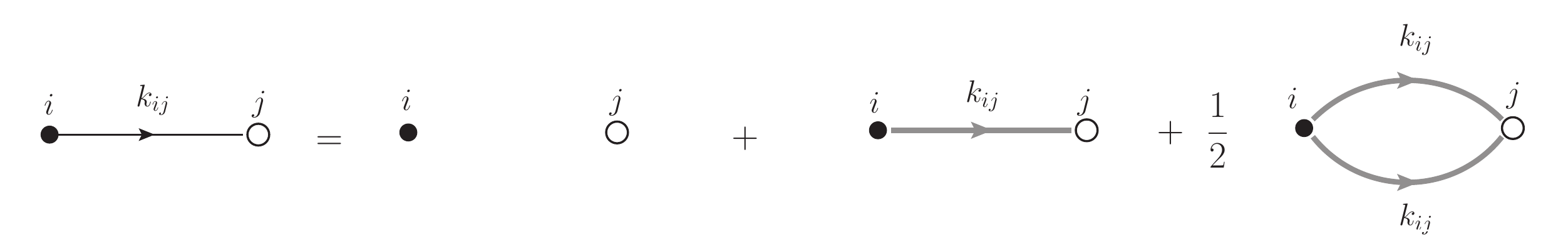}
	\end{center}
	\caption{Expansion of the exponential factor corresponding to a black line in 
	the $\theta$-graph. In the right hand side, each grey line corresponds to a 
	$\theta_i\, k\, \bar{\theta}_j$ term.}
	\label{fig:2}
\end{figure}

To compute the diagram we have to integrate over all $\theta_i$ and $\bar{\theta}_j$ 
variables. To do so, we expand the exponential factor corresponding to each line as in 
(\ref{exp2}); we graphically represent this expansion in Figure~\ref{fig:2}.

\begin{figure}[ht]
	\begin{center}
		\includegraphics[width=\textwidth]{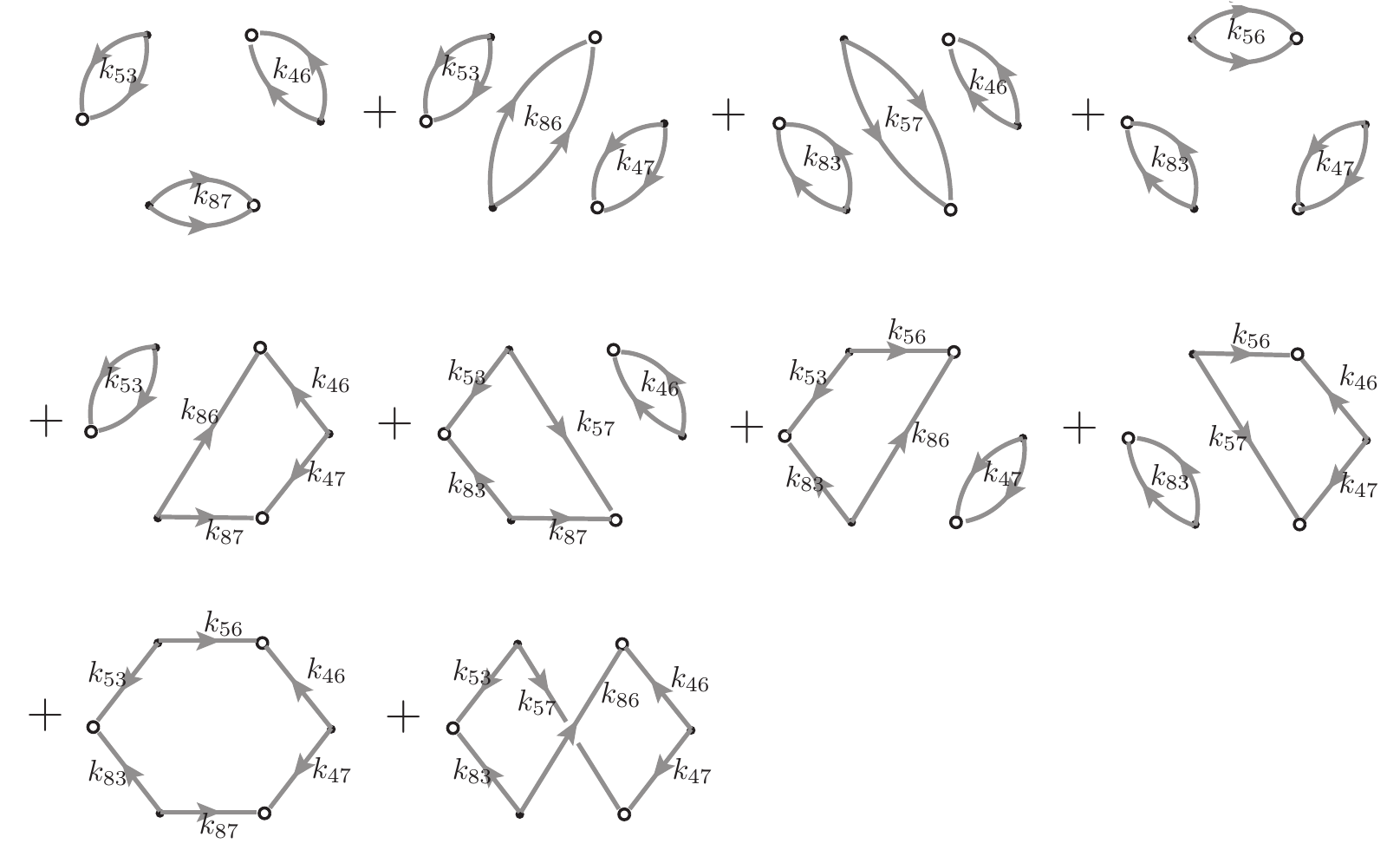}
	\end{center}
	\caption{The paths corresponding to non-vanishing contributions to the integral 
	encoded in the 
	diagram of Figure \ref{fig:1}. Note that all cycles of length two are actually accompanied 
	by a factor of $1/2$ which, however, we did not write in the figure to avoid clutter.}
	\label{fig:3}
\end{figure}

Once this is done, it is easy to realize that one gets a non-zero contribution from the 
Grassmann integration if and only if in each black (or white) node one selects exactly two 
incoming (or outgoing) lines. As a consequence, one gets a contribution for each possible 
non-self-intersecting path passing through all the nodes that uses the edges present in the 
diagram. Such paths are collections of closed cycles. In the example of Figure~\ref{fig:1} there are 
ten such paths, which are drawn in Figure~\ref{fig:3}.

We can now integrate over all Grassmann variables belonging to a cycle. By using the Fierz 
identities (\ref{2tist2}) and the integration rules (\ref{intt2}), it is possible to show the following relation:
\begin{align}
\label{loopnruleg}
\parbox[c]{.3\textwidth}{\includegraphics[width = .3\textwidth]{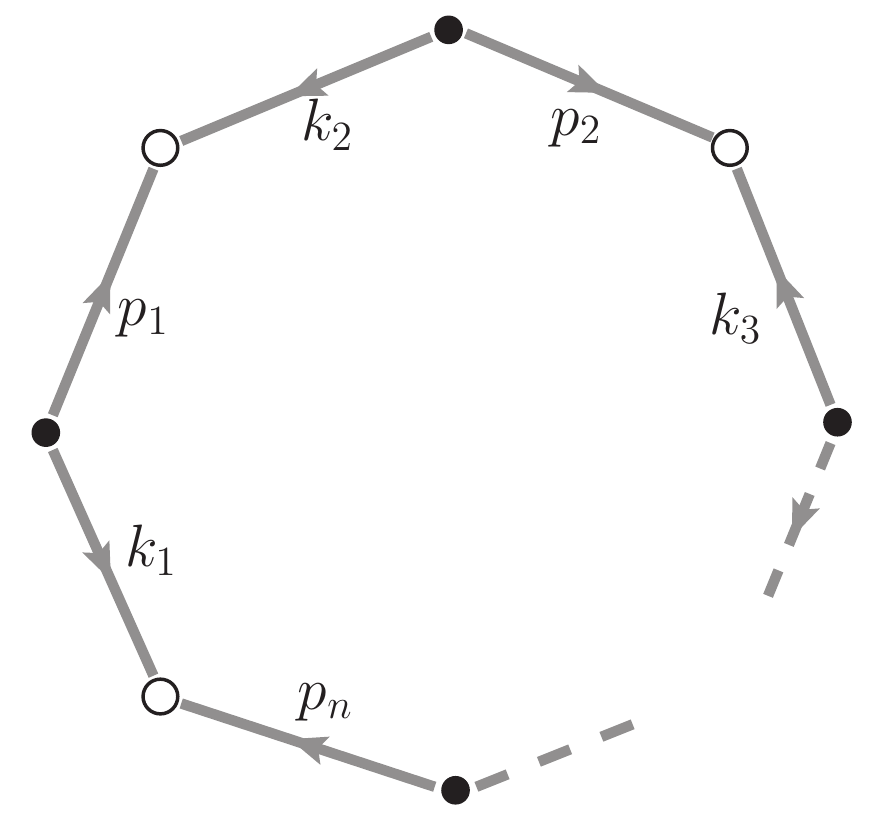}}
& = \int d^2\theta_1\, d^2\bar{\theta}_1 \ldots d^2\theta_n\,d^2\bar{\theta}_n\,
\big(2\,\theta_1\,k_1\,\bar{\theta}_1)\,\big(2\,\theta_1\,p_1\,\bar{\theta}_1\big)
\ldots
\nonumber\\[-15mm]
& 
= (-1)^{n+1} \, \tr\big(k_1\,\bar{p}_1\,k_2\,\bar{p}_2\,\ldots k_n\,\bar{p}_n\big)
\\
\nonumber
\end{align}
where the traces can be computed using (\ref{sigmatraces}) - or analogous formul\ae
~for $n>3$. This is the key Grassmann integration formula for the calculation of 
Feynman superdiagrams.

Applying this procedure to the $\theta$-graph of Figure~\ref{fig:1}, we obtain
\begin{align}
\label{Fdefinition}
\parbox[c]{.35\textwidth}{\includegraphics[width = .35\textwidth]{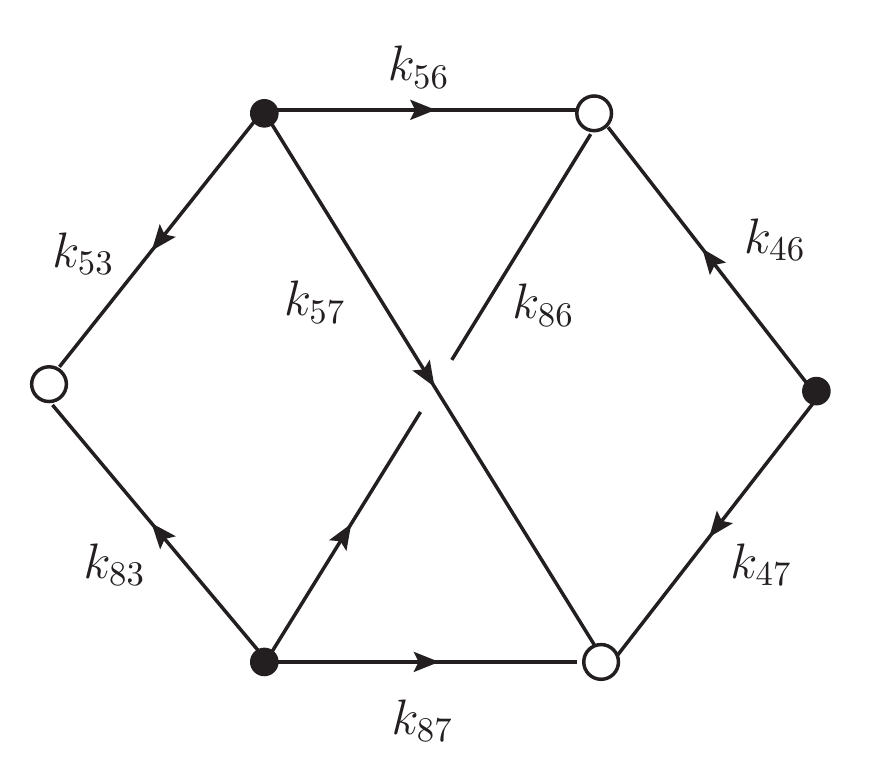}}
= F(k_{83},k_{87},k_{86},k_{53},k_{57},k_{56},k_{47},k_{46})~,
\end{align}
where we have introduced the function $F$ defined by
\begin{align}
F(p_1,p_2,p_3,p_4,p_5,p_6,p_7,p_8)&=
 -\,p_1^2 \,p_6^2 \,p_7^2 -\, p_2^2\,p_8^2\, p_4^2
 -\,p_3^2\, p_4^2\,p_7^2 - \,p_1^2 \,p_5^2\,p_8^2\nonumber\\[1mm]
& ~~~\,+ p_4^2 \,\tr\big(p_8\,\bar p_7\, p_2\, \bar p_3\big)
+ p_8^2\, \tr\big(p_2\,\bar p_1\, p_4\, \bar p_5\big) \nonumber\\[1mm]
&~~~\,+ p_7^2 \,\tr\big(p_1\,\bar p_4\, p_6\, \bar p_3\big) 
+ p_1^2 \,\tr\big(p_6\,\bar p_8\, p_7\, \bar p_5\big)\nonumber\\[1mm]
&~~~\,+ \tr\big(p_6\, \bar p_8\,p_7\, \bar p_2\, p_1\, \bar p_4\big)
+ \tr\big(p_8\, \bar p_7\, p_5 \,\bar p_4\, p_1 \bar p_3\big)~.
\label{Fisp}
\end{align}
With the momentum assignments as in (\ref{Fdefinition}), the ten terms in the right hand side
of (\ref{Fisp}) precisely reproduce the ten terms represented in Figure~\ref{fig:3}. 
Computing the traces with the help of (\ref{sigmatraces}), one obtains in the end a polynomial 
of order six in the momenta entirely made of scalar products. 

We have explicitly worked out this example because this $\theta$-graph 
actually describes the prototypical example for the Grassmann factor associated to 
many of the Feynman superdiagrams that we will consider in detail in 
Appendix~\ref{app:diagrams}, the only difference being in the different assignments 
of the momenta to the various lines.

\subsection*{Vector multiplet lines}

For Feynman superdiagrams containing vector multiplet lines, the most convenient strategy
to handle the Grassmann integration is first to eliminate the vector lines, so that one 
remains with graphs containing hypermultiplet lines only, which can then be computed as 
we have previously described. 

\begin{figure}[ht]
	\begin{center}
		\includegraphics[width=0.8\textwidth]{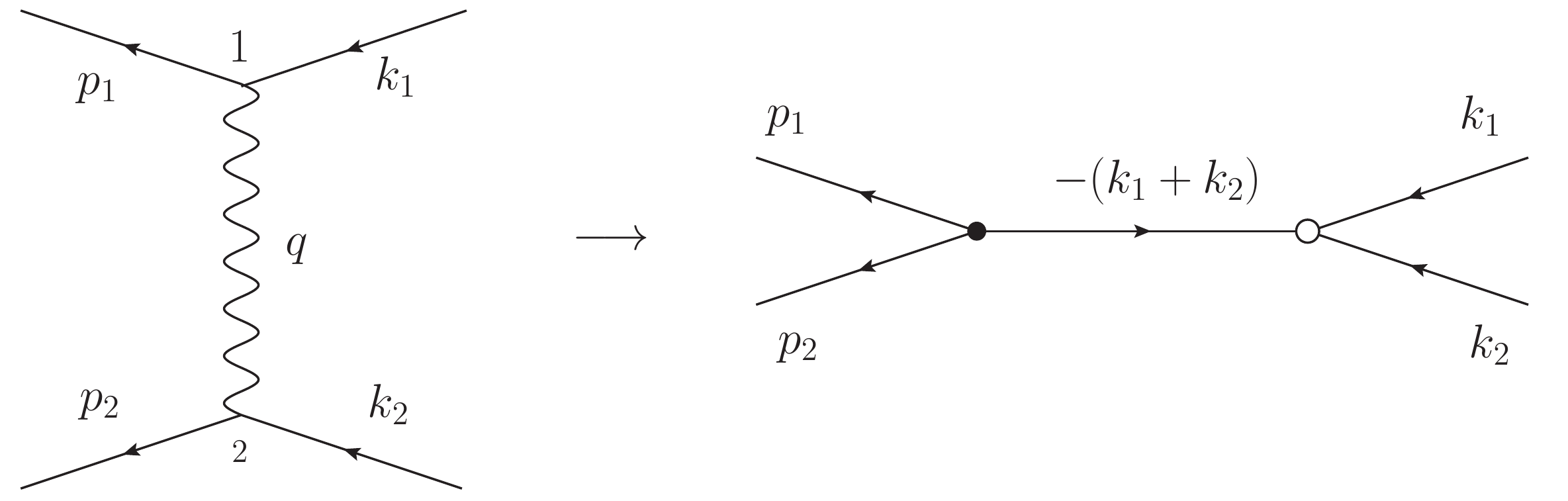}
	\end{center}
	\caption{How to associate a $\theta$-graph to a diagram with a vector line attached to matter current.}
	\label{fig:5}
\end{figure}

Let us first consider the graphs in which all vector lines are attached at both ends to a 
hypermultiplet line. In this case, for every vector line we have a sub-graph of the form 
described on the left of Figure \ref{fig:5}, where the solid oriented lines indicate a generic 
chiral/anti-chiral multiplet propagator. 

As one can see from the Feynman rules listed in Section~\ref{sec:FTactions}, at each cubic vertex,
labeled by 1 and 2, both $\theta_1$ and $\bar{\theta}_1$, and $\theta_2$
and $\bar{\theta}_2$ are present and have to be integrated.
However, the vector propagator contains a factor of $\theta_{12}^2\, \bar{\theta}_{12}^2$ 
which acts as a $\delta$-function identifying $\theta_2$ and $\bar{\theta}_2$ with 
$\theta_1$ and $\bar{\theta}_1$, respectively. Therefore, we remain with two 
Grassmann variables, say $\theta_1$ and $\bar{\theta}_1$, to be integrated. 
The hypermultiplet lines attached to these variables provide the factor
\begin{equation}
\label{expt1t1b}
\exp\Big[ -\theta_1 \big(k_1+p_1+k_2+p_2\big)\,\bar{\theta}_1\Big]
= \exp\Big[ - 2 \,\theta_1 \big(k_1 + k_2\big)\, \bar{\theta}_1\Big]
\end{equation}
where in the second step we have used momentum conservation. 
This is exactly the same type of exponential factor
that in a $\theta$-graph we associate to a solid line from the black dot 
representing $\theta_1$ to the white dot representing $\bar{\theta}_1$ (see (\ref{exp2})).  
Thus, we deduce the rule of Figure \ref{fig:5} which allows us to write the portion of a 
$\theta$-graph corresponding to a vector line attached to matter lines.

\begin{figure}[ht]
	\begin{center}
	\vspace{0.3cm}
		\includegraphics[width=0.8\textwidth]{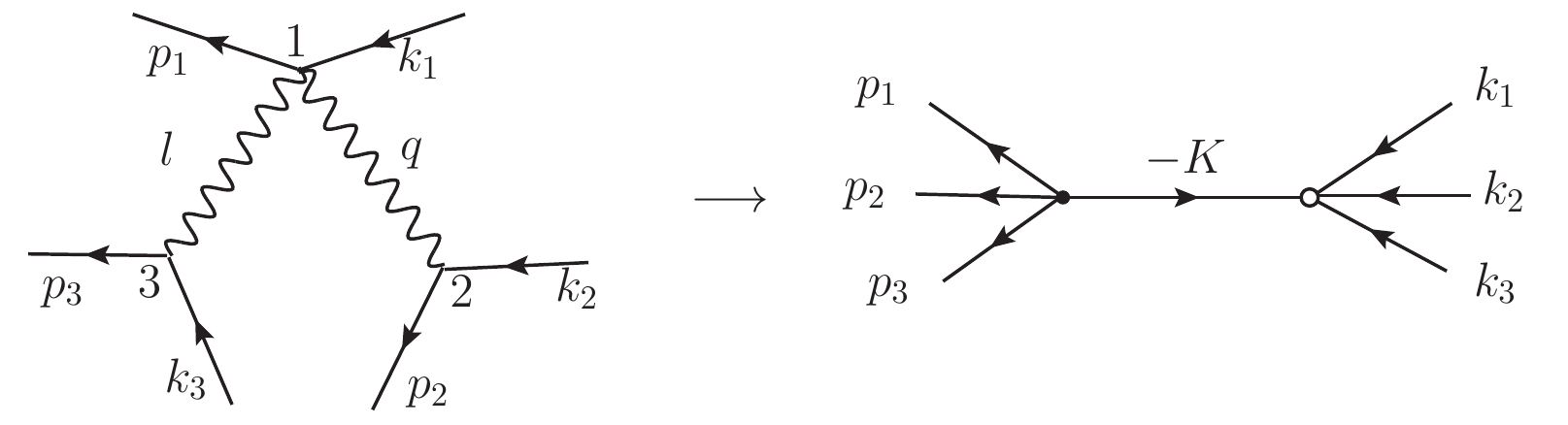}
	\end{center}
	\caption{The rule to replace a quartic vertex with two vector lines with the corresponding 
	$\theta$-graph. Here $K= k_1+k_2+k_3$.}
	\label{fig:6}
\end{figure}

Analogous rules can be worked out when there are vertices with the simultaneous emission of two vector lines from a scalar current line. The simplest case is the one represented in Figure \ref{fig:6}.

Things proceed in a perfectly analogous way if there are more quartic vertices. In the end, the subdiagram gives rise to a $\theta$-subgraph with the same ``external'' lines. However now the outgoing lines are all attached to a single black dot - corresponding to an integration variable 
$\theta$ - and the incoming lines are all attached to a single white circle - corresponding to a variable $\bar{\theta}$. The dot and the circle are connected by a line, associated with 
the exponential factor $\exp\big(-2\,\theta\, K\,\bar{\theta}\,\big)$, where $K$ is the sum of the incoming momenta. 

When the diagram contains interaction vertices with three or more vectors, things are slightly 
more involved because of the presence of covariant spinor derivatives in such vertices.
We will not describe the procedure in general, because only one diagram with a three-vector vertex
is needed in our computations. Indeed, we find more convenient to deal directly with this case, 
in which it is again possible to rewrite the Grassmann integrals in terms of a $\theta$-graph of the type introduced above.

\section{Evaluation of the relevant superdiagrams}
\label{app:diagrams}
We report the computation of the Feynman superdiagrams that yield a contribution 
proportional to $\zeta(5)$ in the three-loop corrections to the propagator 
of the scalar field in the $\cN=2$ vector multiplet. 

Any diagram of this kind, with external adjoint indices $b$ and $c$, external 
momentum $q$ and $s$ internal lines, is written as
\begin{align}
	\label{gen-diag}
		\cW_{bc}(q) = \cN\times \cT_{bc}\times \!\int \!\prod_s \frac{d^d k_s}{(2\pi)^d}
		~\delta^{(d)}(\mathrm{cons}) ~\frac{\cZ(k)}{\prod_s k_s^2}~.
\end{align}
Here $\cN$ is the product of the symmetry factor of the diagram and all the factors (like 
the powers of the coupling constant $g$) appearing in the vertices - except for the color 
factors which give rise to the tensor $\cT_{bc}$. We have then the scalar integral over the 
internal momenta $k_s$ which we perform using dimensional regularization setting 
$d=4-2\varepsilon$. The momenta are subject to the appropriate momentum 
conservation relations enforced by the $\delta$-functions $\delta^{(d)}(\mathrm{cons})$. 
Beside the denominator coming from the massless propagators, the integrand 
contains also a numerator $\cZ(k)$ which is the result of the integration over all the 
Grassmann variables of the $\theta$-dependent expressions present in the superdiagram.

The massless scalar integrals at three loops with cubic or quartic vertices can be evaluated 
by various means; in particular, we use the FORM version of the program Mincer discussed 
in \cite{Larin:1991fz}, which classifies them according to different ''topologies'' 
described by diagrams in which a solid line indicates a massless scalar propagator, and 
momentum conservation is enforced at each vertex. 

\subsection{Diagrams with six insertions on the hypermultiplet loop}
We start by considering the diagrams with six insertions of an adjoint generator on the hypermultiplet loop. The color factor of these diagrams is proportional to a doubly contracted $\cC^\prime$ tensor with six indices defined in (\ref{defC}).

The first diagram we consider is the following
\begin{align}
	\label{LA}
	\cW_{bc}^{(1)}(q) =
	\parbox[c]{.5\textwidth}{\includegraphics[width = .5\textwidth]{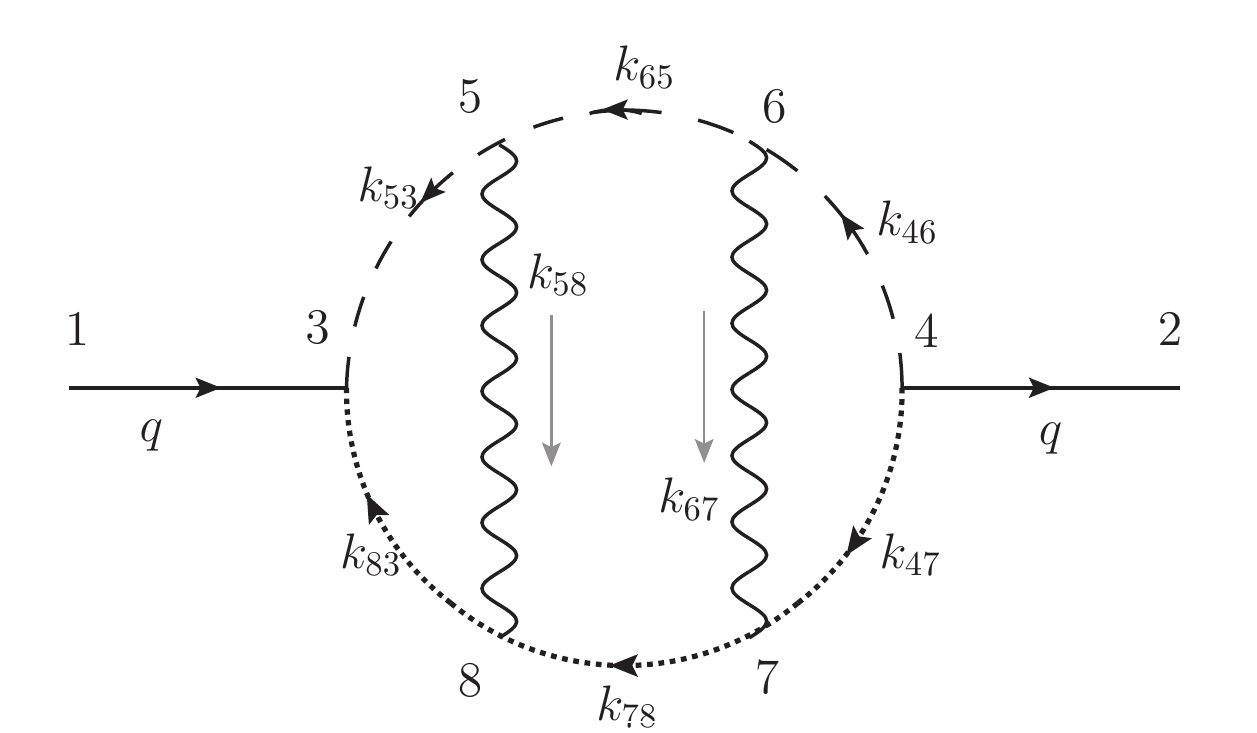}}
\end{align}
In this first diagram we set up the notation that we will use also in all subsequent ones. 
The external momentum is always denoted as $q$. Regarding the labeling of internal momenta,
we label the internal vertices (from $3$ to $8$ in this case) and we denote as $k_{ij}$ the 
momentum flowing in a propagator from the vertex $i$ to the vertex $j$, which is also 
the same convention introduced in (\ref{exp2}). Assuming it, from now on we will display in the 
figures only the labels of the vertices and not of the internal momenta.
The Feynman rules for propagators and vertices are given in Section~\ref{sec:FTactions}. 
Using them, we get
\begin{align}
	\label{LAbis}
		\cW_{bc}^{(1)}(q)
		= 8 g^6 \times C^\prime_{bdeced} \times 
		\parbox[c]{.35\textwidth}{\includegraphics[width = .35\textwidth]{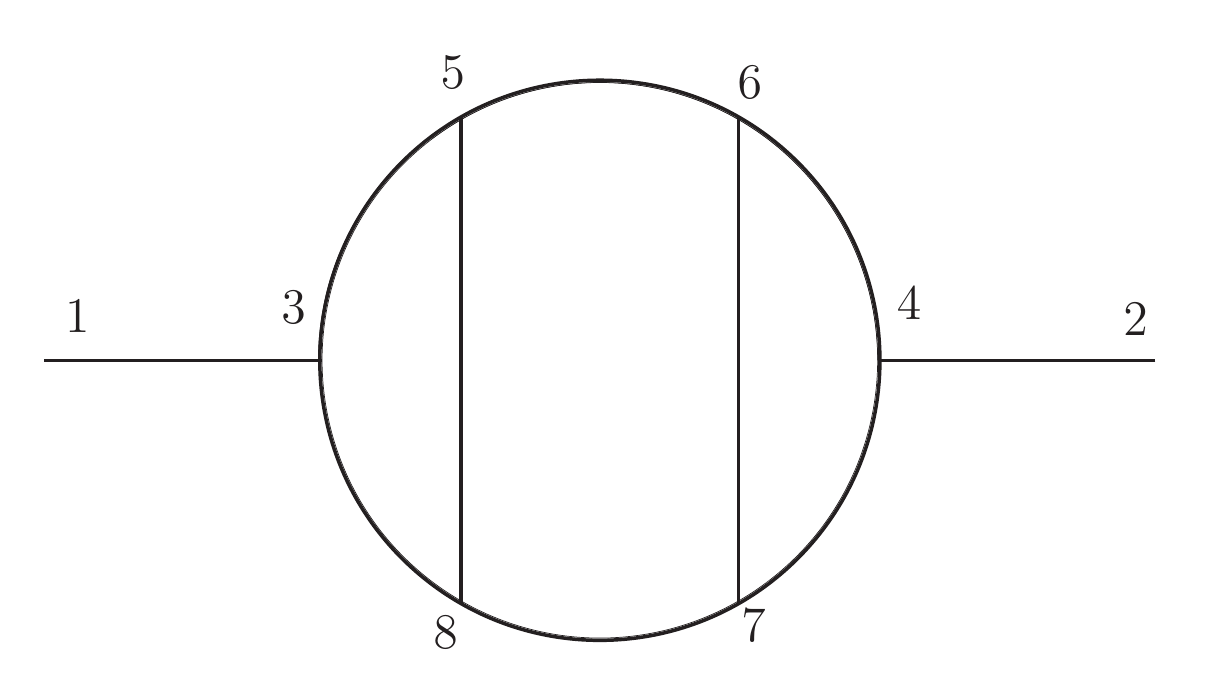}}
		\cZ^{(1)}(k).
\end{align}
The scalar diagram has the ladder topology denoted as LA in \cite{Larin:1991fz}.
The Grassmann factor $\cZ^{(1)}(k)$ is obtained integrating over $d^4\theta_i$ for $i=3,\ldots,8$ and is easily determined using the rule described in Figure \ref{fig:5}. It is given by the following
$\theta$-diagram
\begin{align}
	\label{Z1is}
		\cZ^{(1)}(k) & = 
		\parbox[c]{.5\textwidth}{\includegraphics[width = .5\textwidth]{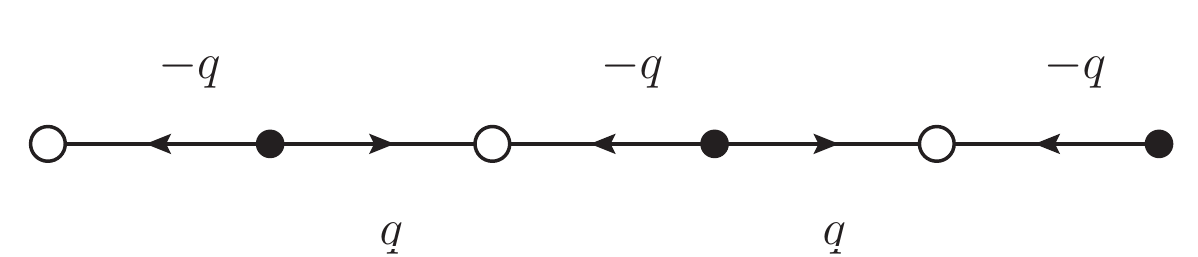}}
		= - q^6~.
\end{align}
The evaluation of this $\theta$-diagram by means of its cycle expansion, as explained after (\ref{exp2}) and illustrated in Figure \ref{fig:1}, is immediate using (\ref{loopnruleg}).  
A factor of $q^4$ removes the two external propagators in the scalar diagram, so that it reduces to
\begin{align}
	\label{scal1}
		- q^2\,
		\parbox[c]{.12\textwidth}{\includegraphics[width = .12\textwidth]{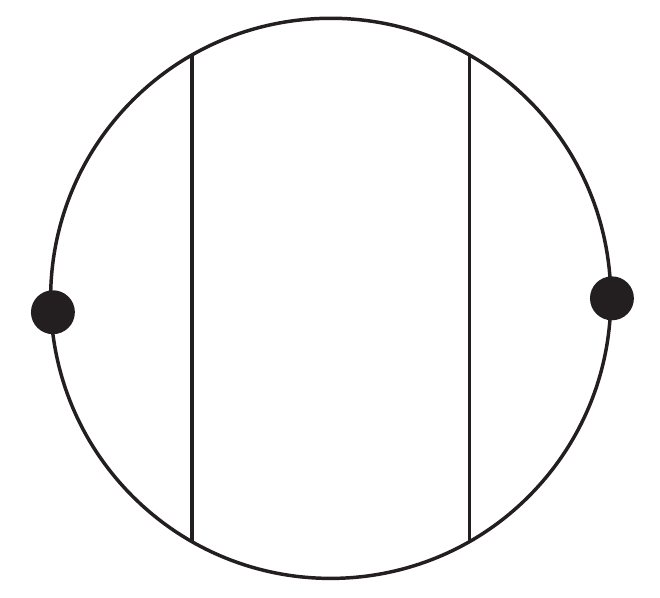}}
		= - \frac{20 \zeta(5)}{(4\pi)^6} \frac{1}{q^2}~.
\end{align}
Here we have employed the standard graphical notation for diagrams with canceled external propagators and we have given the value of this scalar integral, which is finite, directly in $d=4$. Altogether we get thus
\begin{align}
	\label{W1res}
		\cW_{bc}^{(1)}(q) = -\frac{1}{q^2}\left(\frac{g^2}{8\pi^2}\right)^3 \zeta(5) \times \left(20
		\,C^\prime_{bdeced}\right)~. 
\end{align}

The next diagram is 
\begin{align}
	\label{W2}
		\cW_{bc}^{(2)}(q) &=
		\parbox[c]{.40\textwidth}{\includegraphics[width = .40\textwidth]{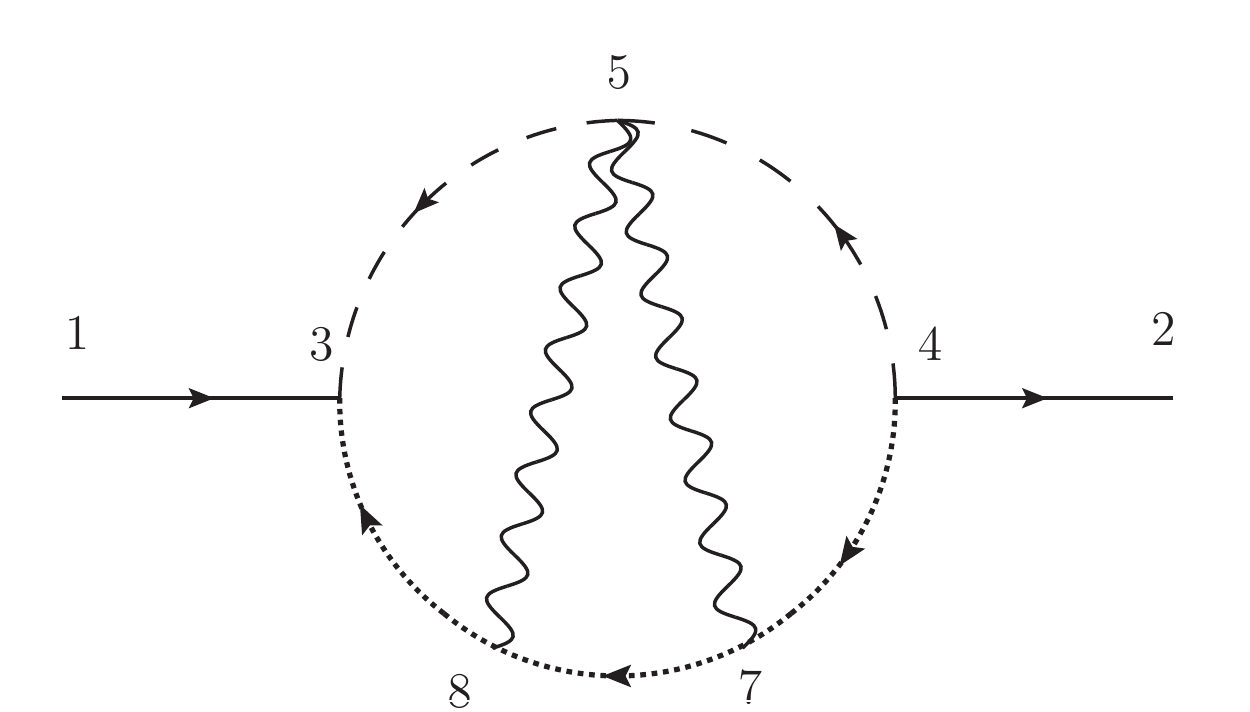}}
		\nonumber\\
		&=   4 g^6 \times 2 \,\cT^ {(2)}_{bc} \times 
		\parbox[c]{.33\textwidth}{\includegraphics[width = .33\textwidth]{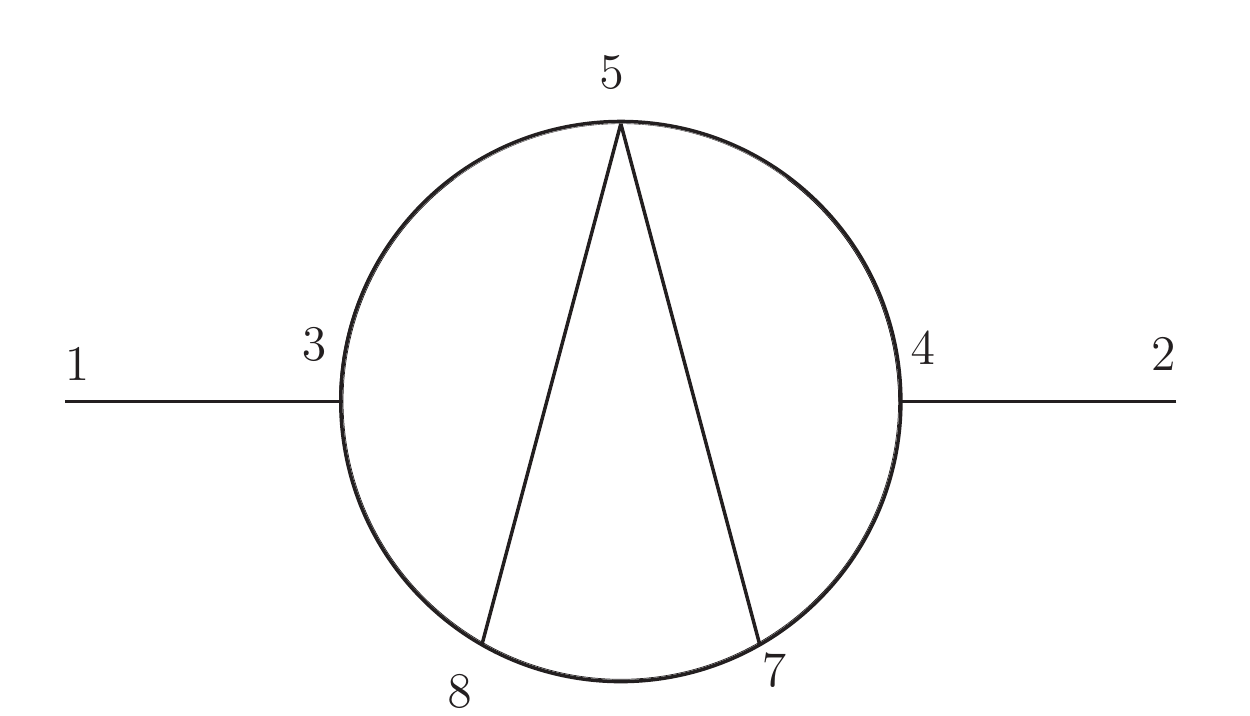}}
		\cZ^{(2)}(k)~.
\end{align}
Here the color tensor reads
\begin{equation}
	\label{T2bc}
		\cT_{bc}^{(2)} = C^\prime_{bdecde} + C^\prime_{bdeced}~, 
\end{equation}
the two terms stemming from the two ways to attach the gluon lines to the quartic vertex. This expression comes with a factor of $2$ in (\ref{W2}) to account for 
the diagram in which the dashed and dotted parts of the hypermultiplet loop are switched.    
The scalar diagram has the fan topology denoted as FA in \cite{Larin:1991fz}. 
The Grassmann factor can be determined using the rule described in Figure \ref{fig:6} and 
it is given by
\begin{align}
	\label{Z2is}
		\cZ^{(2)}(k) =  
		\parbox[c]{.33\textwidth}{\includegraphics[width = .33\textwidth]{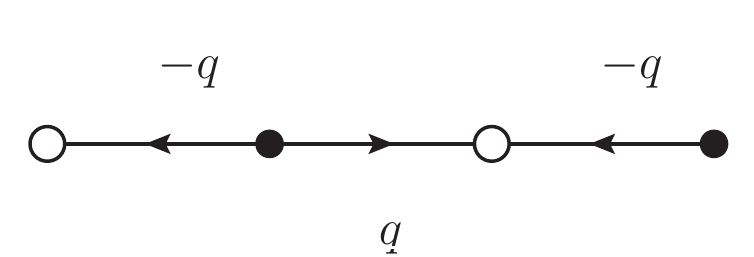}}
		= q^4~.
\end{align}
This factor removes the two external propagators in the scalar diagram, so that it reduces to
\begin{align}
	\label{scal2}
		\parbox[c]{.12\textwidth}{\includegraphics[width = .12\textwidth]{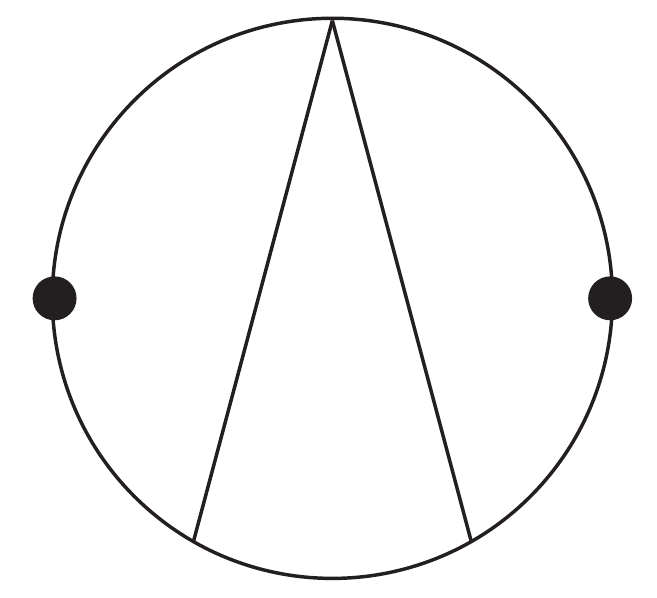}}
		=  \frac{20 \zeta(5)}{(4\pi)^6} \frac{1}{q^2}~.  
\end{align}
Altogether we find thus
\begin{align}
	\label{W2res}
		\cW_{bc}^{(2)}(q) = -\frac{1}{q^2}\left(\frac{g^2}{8\pi^2}\right)^3 \zeta(5) \times 
		\left(-20\,C^\prime_{bdecde} -20\,C^\prime_{bdeced}\right)~. 
\end{align}

The third diagram that contributes is 
\begin{align}
	\label{W3}
		\cW_{bc}^{(3)}(q) &=
		\parbox[c]{.4\textwidth}{\includegraphics[width = .4\textwidth]{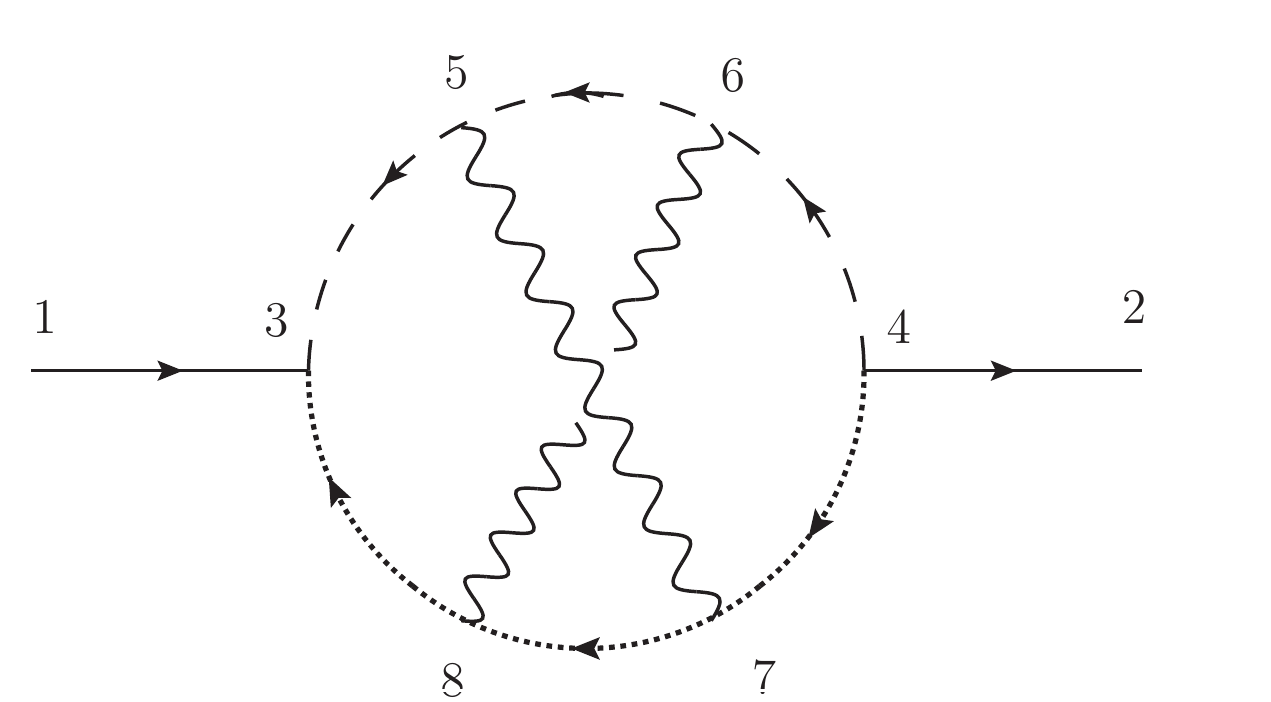}}   
		\nonumber\\
		&= 8 g^6 \times C^\prime_{bdecde} \times 
		\parbox[c]{.33\textwidth}{\includegraphics[width = .33\textwidth]{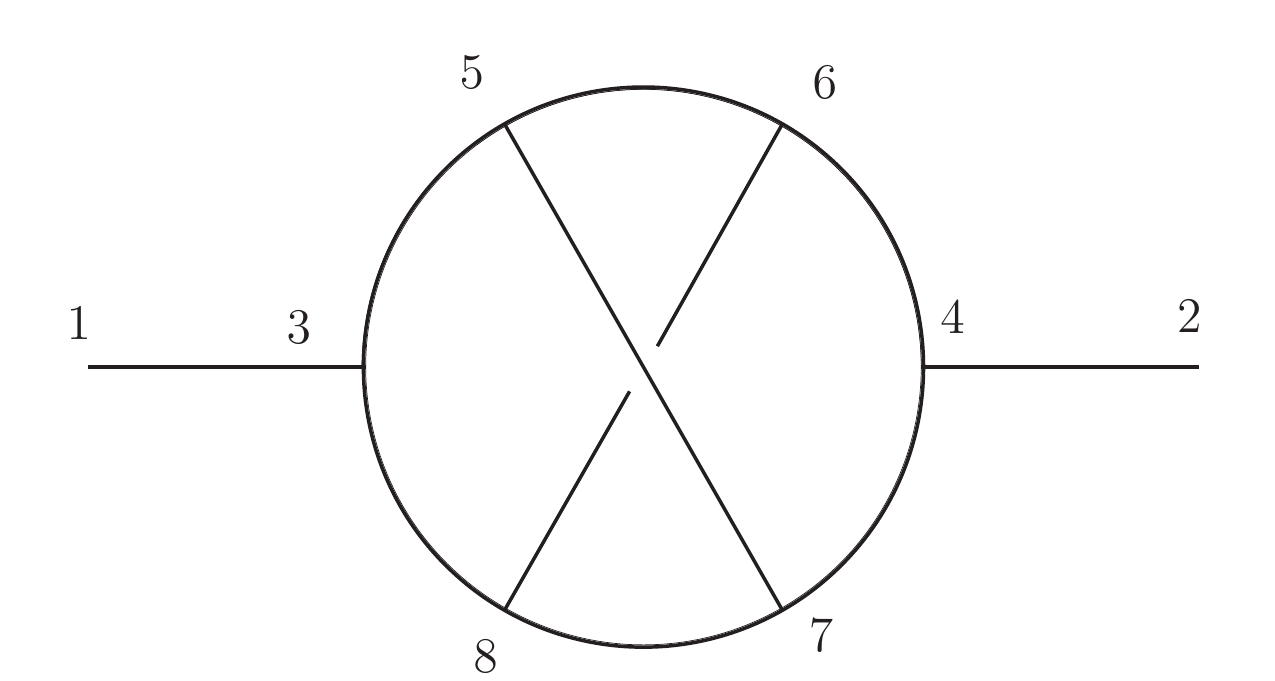}}
		\cZ^{(3)}(k)~.
\end{align}
The scalar diagram has the non-oriented topology denoted as NO in \cite{Larin:1991fz}. 
The Grassmann factor is found applying the rule of Figure \ref{fig:5} and it is given by
a $\theta$-diagram of the type depicted in \ref{Fdefinition}, but with a different assignment of momenta. In particular, one has 
\begin{align}
	\label{Z3is}
		\cZ^{(3)}(k) = F\big(k_{83},-(k_{46} + k_{78}),k_{65},k_{53},k_{78},-(k_{47}+ k_{65}),k_{46},k_{47}\big)~.
\end{align}
Evaluating this and inserting it in the scalar momentum integral, we find that the results 
contains a $\zeta(5)$-contribution. Indeed we have
\begin{align}
\label{scal3}
		\parbox[c]{.33\textwidth}{\includegraphics[width = .33\textwidth]{NO.pdf}}
		\cZ^{(3)}(k)
		=  - \frac{10 \zeta(5)}{(4\pi)^6} \frac{1}{q^2} + \ldots
\end{align}
where the ellipses stand for terms that do not contain $\zeta(5)$.
Putting together the various factors, we find
\begin{align}
	\label{W3res}
		\cW_{bc}^{(3)}(q) = -\frac{1}{q^2}\left(\frac{g^2}{8\pi^2}\right)^3 \zeta(5) \times 
		\left(10 \,C^\prime_{bdecde}\right) + \ldots~. 
\end{align}

Next we consider
\begin{align}
	\label{W4}
		\cW_{bc}^{(4)}(q) &=
		\parbox[c]{.4\textwidth}{\includegraphics[width = .4\textwidth]{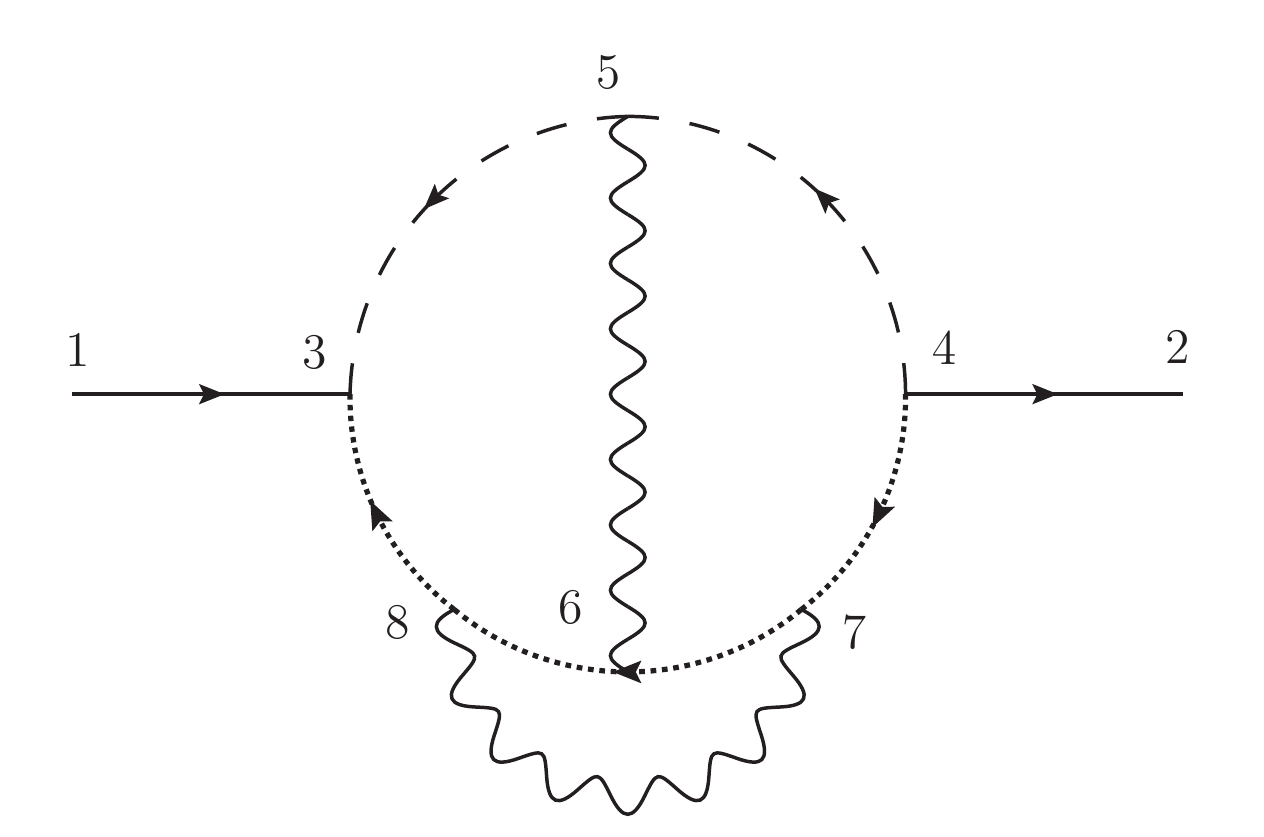}}
		\nonumber\\
		&= - 8 g^6 \times \cT_{bc}^{(4)}\times 
		\parbox[c]{.33\textwidth}{\includegraphics[width = .33\textwidth]{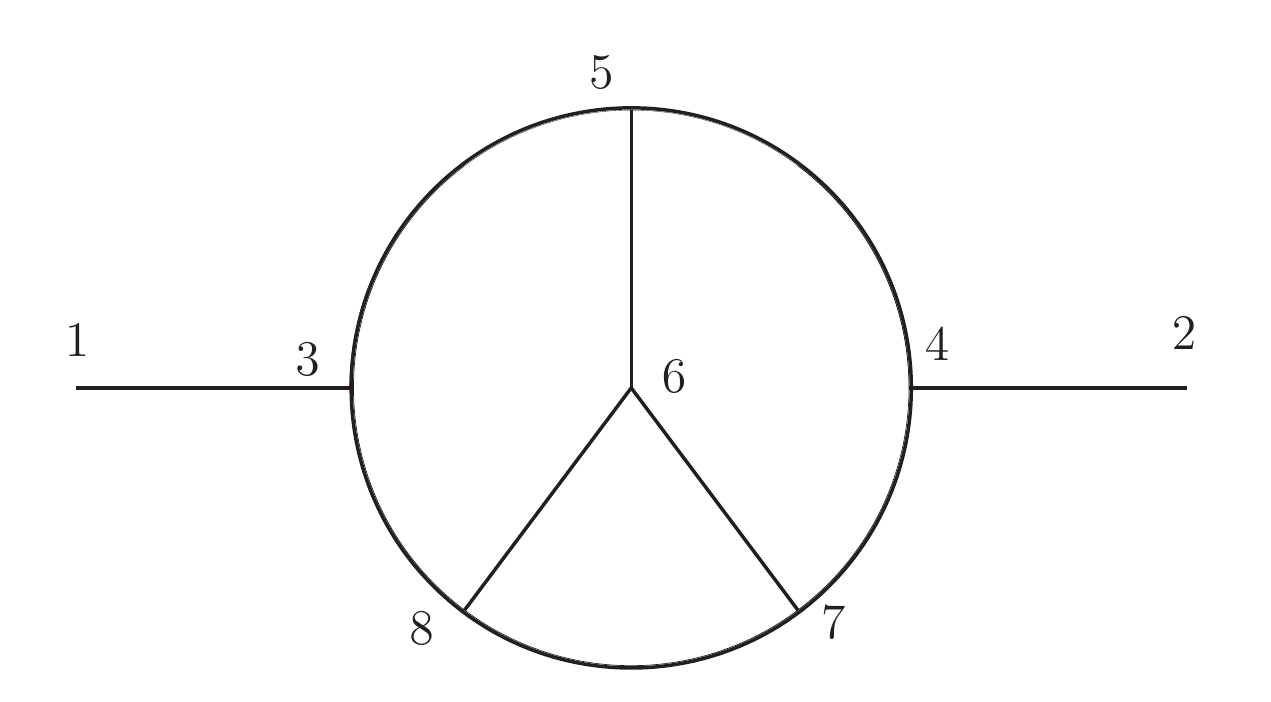}}
		\cZ^{(4)}(k)~.
\end{align}
where the color tensor reads
\begin{equation}
	\label{T4bc}
		\cT_{bc}^{(4)} = C^\prime_{bdedce} + C^\prime_{bdcede}~.
\end{equation}
Here the second term comes from the diagram where the dashed and dotted parts of the 
hypermultiplet loop are exchanged. The scalar diagram has the ``Benz'' topology denoted as BE 
in \cite{Larin:1991fz}. The Grassmann factor is found using the rule of Figure \ref{fig:5} and it is given by
\begin{align}
	\label{Z4is}
		\cZ^{(4)}(k) = F\big(k_{83},-(k_{46} + k_{68}),k_{67},k_{53},k_{68},-(k_{45}+ k_{67}),
		k_{47},k_{45}\big)~.
\end{align}
The corresponding scalar momentum integration contains a $\zeta(5)$ contribution; indeed
\begin{align}
	\label{scal4}
		\parbox[c]{.33\textwidth}{\includegraphics[width = .33\textwidth]{BE.pdf}}
		\cZ^{(4)}(k)
		=   \frac{20 \zeta(5)}{(4\pi)^6} \frac{1}{q^2} + \ldots~.  
\end{align}
Altogether we have thus
\begin{align}
	\label{W4res}
		\cW_{bc}^{(4)}(q) = -\frac{1}{q^2}\left(\frac{g^2}{8\pi^2}\right)^3 \zeta(5) \times 
		 \left( 20\,C^\prime_{bdedce} + 20\,C^\prime_{bdcede}\right) + \ldots~. 
\end{align}

\subsection{Diagrams with five insertions on the hypermultiplet loop}
We now consider the diagrams with five insertions of an adjoint generator on the hypermultiplet loop. The first diagram of this kind we consider is
\begin{align}
	\label{W5}
		\cW_{bc}^{(5)}(q) &=
		\parbox[c]{.4\textwidth}{\includegraphics[width = .4\textwidth]{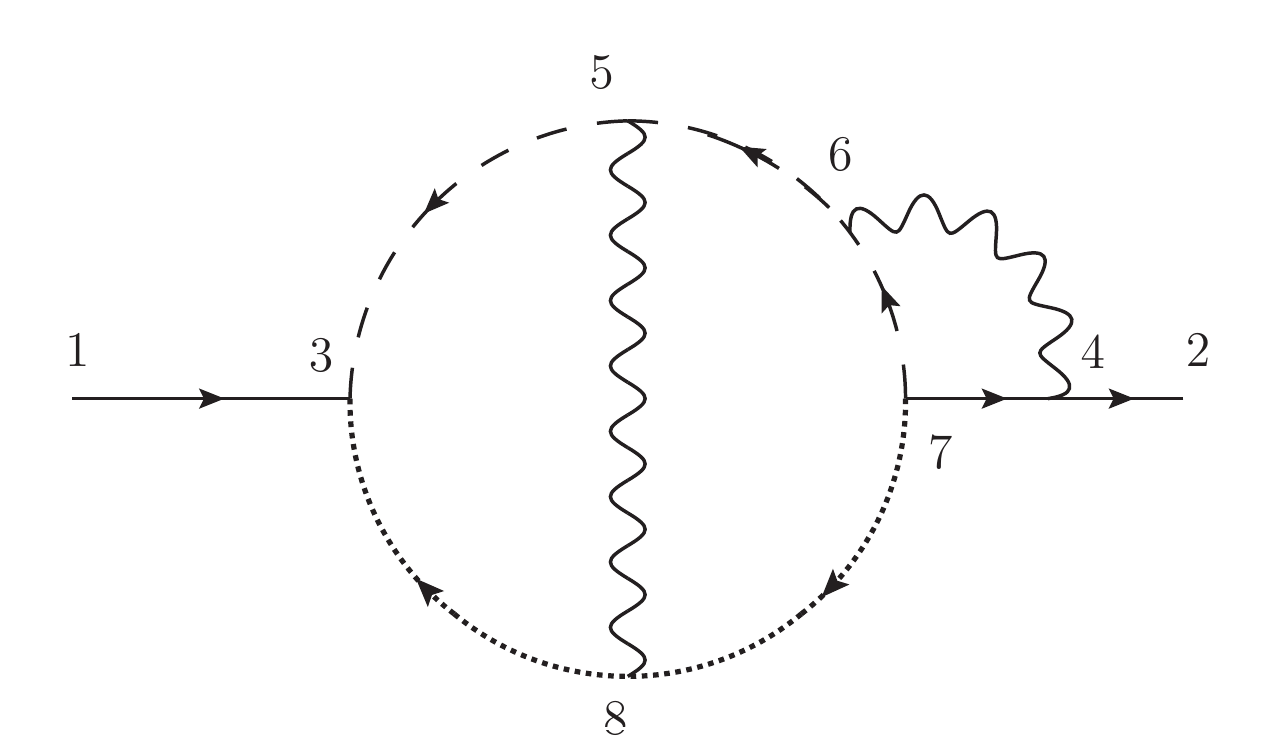}}
		\nonumber\\
		&= -8 g^6 \times \cT_{bc}^{(5)} \times 
		\parbox[c]{.33\textwidth}{\includegraphics[width = .33\textwidth]{LA.pdf}}
		\cZ^{(5)}(k)~.
\end{align}
The color factor is given by 
\begin{align}
	\label{T5bc}
		\cT_{bc}^{(5)} & = 
		\ii f_{cef} C^\prime_{bdefd} - \ii f_{cef} C^\prime_{bdfed}
		+ \ii f_{bef} C^\prime_{cdefd} - \ii f_{bef} C^\prime_{cdfed}
		\nonumber\\[1mm]
		& = 2 \,\ii f_{cef} C^\prime_{bdefd} + 2 \,\ii f_{bef} C^\prime_{cdefd}~,
\end{align}
where the four terms that appear in the first line correspond to the four possible ways to 
attach the ``external'' vector multiplet line. The Grassmann factor is again found using 
the rule of Figure \ref{fig:5} and it is given by
\begin{align}
	\label{Z5is}
		\cZ^{(5)}(k) = F\big(0,-k_{78},k_{78},-q,0,q,k_{78},-(k_{78}+ q)\big)~.
\end{align}
Using this result inside the scalar momentum integral, which has the LA topology, one finds
\begin{align}
	\label{scal5}
		\parbox[c]{.33\textwidth}{\includegraphics[width = .33\textwidth]{LA.pdf}}
		\cZ^{(5)}(k)
		=   - \frac{20 \zeta(5)}{(4\pi)^6} \frac{1}{q^2} + \ldots~.  
\end{align}
The final result for this diagram is then
\begin{align}
	\label{W5res}
		\cW_{bc}^{(5)}(q) = -\frac{1}{q^2}\left(\frac{g^2}{8\pi^2}\right)^3 \zeta(5) \times 
		\left(-40\,\ii f_{cef} C^\prime_{bdefd} - 40\,\ii f_{bef} C^\prime_{cdefd} \right) + \ldots~. 
\end{align}

Another diagram in this class is
\begin{align}
\label{W6}
	\cW_{bc}^{(6)}(q) &=
		\parbox[c]{.4\textwidth}{\includegraphics[width = .4\textwidth]{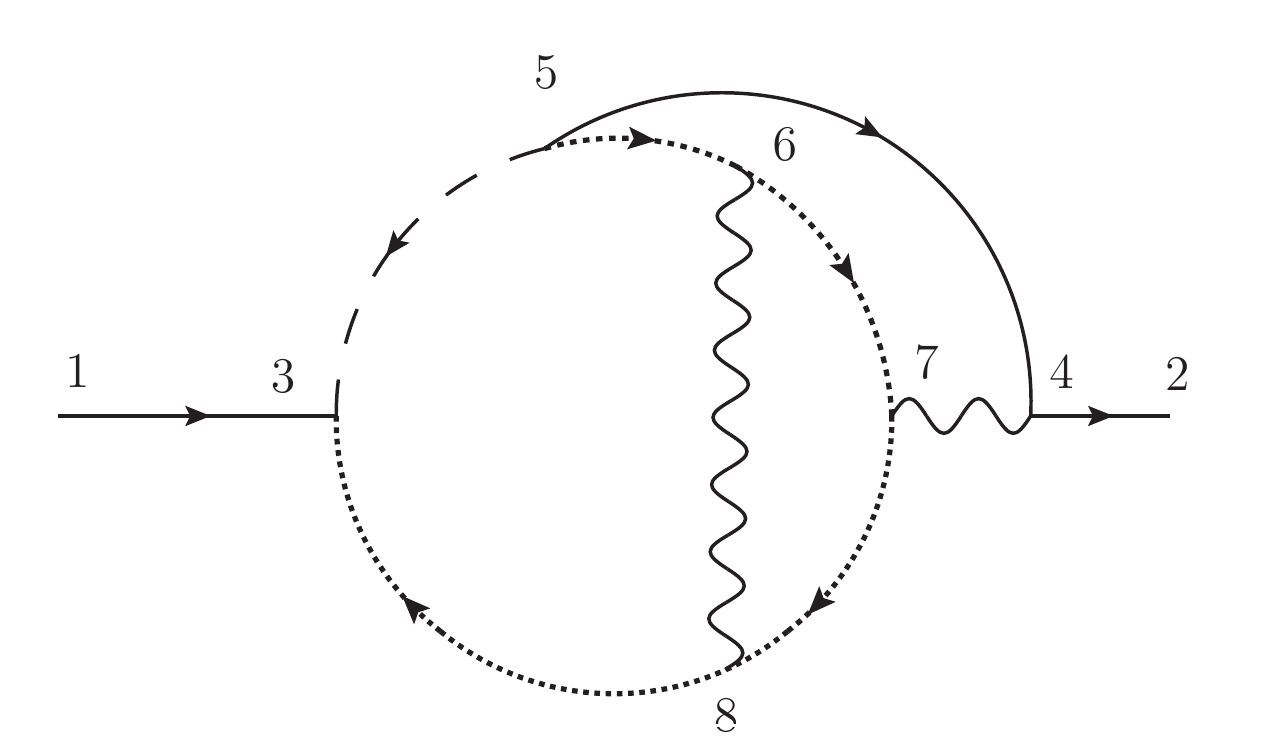}}
		\nonumber\\		
		&= -8 g^6 \times \cT_{bc}^{(6)} \times 
		\parbox[c]{.33\textwidth}{\includegraphics[width = .33\textwidth]{BE.pdf}}
		\cZ^{(6)}(k)~,
\end{align}
where the color factor is
\begin{align}
	\label{T6bc}
		\cT_{bc}^{(6)} & = 
		\ii f_{ced} C^\prime_{bfdfe} - \ii f_{ced} C^\prime_{befdf}
		+ \ii f_{bed} C^\prime_{cfdfe} - \ii f_{bed} C^\prime_{cefdf}~.
\end{align}
Here the four terms correspond to the four possible ways to attach the ``external'' adjoint 
chiral multiplet line. Using the by-now familiar procedure, the Grassmann factor is found to be
\begin{align}
	\label{Z6is}
		\cZ^{(6)}(k) = F\big
		(k_{73},-(k_{56} + k_{87}),k_{68},k_{53},k_{56},k_{54},k_{87},-(k_{87}+ q)\big)~.
\end{align}
The scalar integral, which has the BE topology, yields the result
\begin{align}
	\label{scal6}
		\parbox[c]{.33\textwidth}{\includegraphics[width = .33\textwidth]{BE.pdf}}
		\cZ^{(6)}(k)
		=   - \frac{20 \zeta(5)}{(4\pi)^6} \frac{1}{q^2} + \ldots~.  
\end{align}
The total result is thus
\begin{align}
	\label{W6res}
		\cW_{bc}^{(6)}(q) & = -\frac{1}{q^2}\left(\frac{g^2}{8\pi^2}\right)^3 \zeta(5)\nonumber\\
		& ~~~\times 
		\left(	-20\,\ii f_{ced} C^\prime_{bfdfe} + 20\,\ii f_{ced} C^\prime_{befdf}
		-20\, \ii f_{bed} C^\prime_{cfdfe} + 20\,\ii f_{bed} C^\prime_{cefdf}\right) + \ldots~. 
\end{align}

Among the diagrams with five insertions that give a $\zeta(5)$ contribution, there is one
whose Grassmann factor cannot be computed simply by using the rules illustrated in 
Appendix~\ref{app:grass-super}. It is the following:
\begin{align}
	\label{W7}
		\cW_{bc}^{(7)}(q) &=
		\parbox[c]{.4\textwidth}{\includegraphics[width = .4\textwidth]{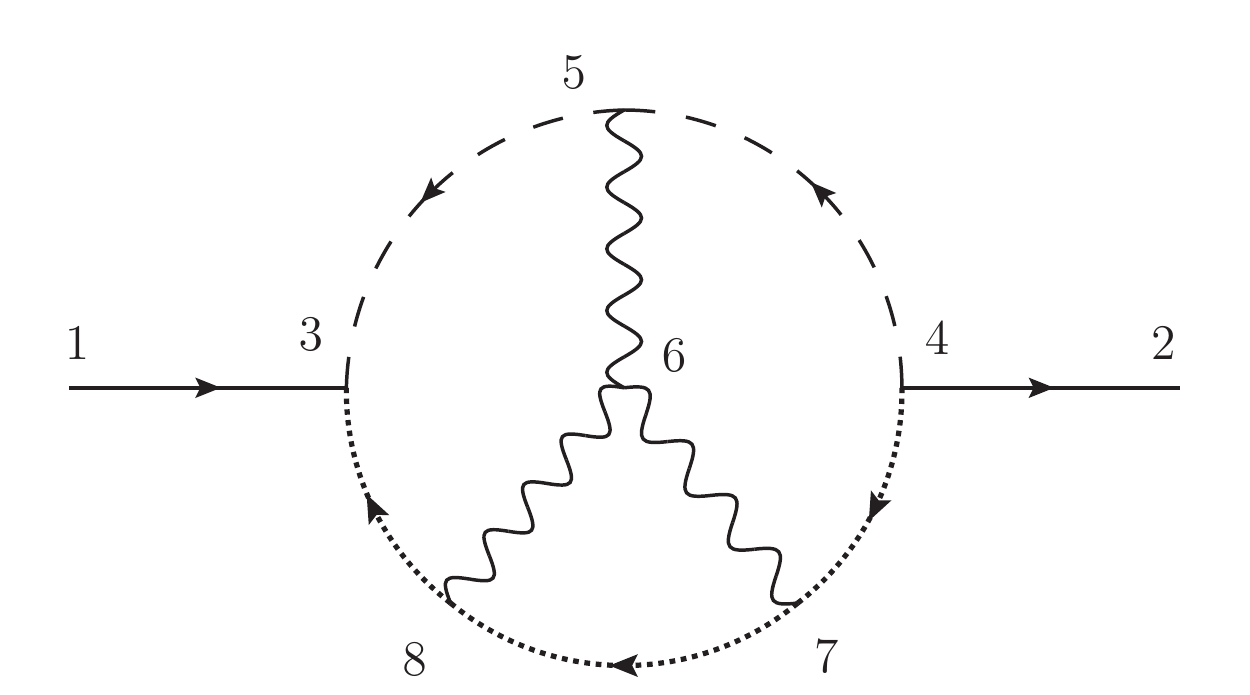}}
		\nonumber\\
		&= -\frac{1}{16}\, (8 g^6) \times \cT_{bc}^{(7)} \times 
		\parbox[c]{.33\textwidth}{\includegraphics[width = .33\textwidth]{BE.pdf}}
		\cZ^{(7)}(k)~.
\end{align}
The color factor reads
\begin{align}
	\label{T7bc}
		\cT_{bc}^{(7)} & =
		\ii f_{def} C^\prime_{bfecd} + \ii f_{def} C^\prime_{cfebd}~,
\end{align}
with the two terms corresponding to the fact that in the hypermultiplet loop the dashed or dotted parts can be exchanged. Since the cubic vector vertex contains covariant spinor derivatives and is not symmetric in the three vector lines that it contains, the diagram gets six distinct contributions 
arising from the six different ways it is contracted with the other vertices of the diagram. We write
these six terms as follows
\begin{align}
	\label{Z7def}
		\cZ^{(7)} = \cZ^{(7)}_{578} + \cZ^{(7)}_{758} + \cZ^{(7)}_{785} + \cZ^{(7)}_{875}
		+ \cZ^{(7)}_{857} + \cZ^{(7)}_{587}~.
\end{align}
The first term above is
\begin{align}
	\label{Z578}
		\cZ^{(7)}_{578}(k) & = 
		\Big[\big(\overline{D}_6\big)^2 D_6^\alpha\, \delta^4(\theta_{65}) \Big]
		\, \delta^4(\theta_{67})\, \Big[ D_{6,\alpha} 
		\, \delta^4(\theta_{68})\Big]\,\exp\big[\cA(k)\big]~.
\end{align}
Here we have denoted by $D_{6,\alpha}$ and $\overline{D}_{6,\dot{\alpha}}$ 
the covariant spinor derivatives defined in (\ref{covspink}) with respect to $\theta_6$ and
$\bar\theta_6$.  The last exponential factor $\exp\big[\cA(q,k)\big]$ 
contains all other contributions which amount to
\begin{align}
	\label{CA3g}
		\cA(k) & 
		= 2 \,\theta_4 \,k_{45}\, \bar{\theta}_5 + 2 \,\theta_5 \,k_{53} \,\bar{\theta}_3
		- \theta_5\,\big(k_{45} + k_{53}\big)\, \bar{\theta}_5
		+ 2 \,\theta_4\, k_{47}\, \bar{\theta}_7
		+ 2 \,\theta_7\, k_{78}\, \bar{\theta}_8 \nonumber\\
		&~~~ - \theta_7 \,\big(k_{47} + k_{78}\big)\, \bar{\theta}_7
		+ 2 \,\theta_8\, k_{83} \,\bar{\theta}_3
		- \theta_8 \,\big(k_{78} + k_{83}\big)\, \bar{\theta}_8~. 		 
\end{align}
Using the identity
\begin{align}
	\label{D6toD8}
		D_{6,\alpha} \,\delta^4(\theta_{68}) = 
		\big(\partial_{6,\alpha} - k_{68} \bar{\theta}_6\big)\, \delta^4(\theta_{68}) 
		= - \big(\partial_{8,\alpha} + k_{68} \bar{\theta}_6\big)\, 
		\delta^4(\theta_{68})
\end{align}
and then integrating by parts with respect to $\theta_8$, we can rewrite (\ref{Z578})
as follows 
\begin{align}
	\label{Z578bis}
		\cZ^{(7)}_{578}(k) & = 
		\delta^4(\theta_{67})\, \delta^4(\theta_{68})
		\Big[\big(\overline{D}_6\big)^2 D_6^\alpha\, \delta^4(\theta_{65}) \Big]
		\big(\partial_{8,\alpha} - (k_{68}\,\bar{\theta}_8)_\alpha\big)
		\exp\big[\cA(k)\big]~.
\end{align} 
By direct evaluation one can show that
\begin{align}
	\label{D3id}
		\big(\overline{D}_6\big)^2 D_6^\alpha\, \delta^4(\theta_{65}) = -4\,
		\rme^{-\theta_6\, k_{65}\, \bar{\theta}_{65}} \,
		\left[2 \,\theta_{65}^\alpha + (k_{65}\,\bar{\theta}_5)^\alpha 
		 \left(\theta_{65}\right)^2\right]~,
\end{align}
and
\begin{align}
	\label{DA}
			\big(\partial_{8,\alpha} - (k_{68}\,\bar{\theta}_8)_\alpha\big)\, 
		\exp\big[\cA(k)\big]= 2 (k_{83} \,\bar{\theta}_{38})_\alpha\,
			\exp\big[\cA(k)\big]
\end{align}
where in the last step we used momentum conservation.
Substituting (\ref{D3id}) and (\ref{DA}) into (\ref{Z578bis}), after a Fierz rearrangement we 
arrive at 
\begin{align}
	\label{Z578tris}
		\cZ^{(7)}_{578}(k) & = 
		-16\, \delta^4(\theta_{67})\, \delta^4(\theta_{68}) \,
		\left(\theta_{65}\,k_{83}\,\bar{\theta}_{38}\right)
		\left(1 + \theta_{65}\, k_{65}\, \bar{\theta}_5\right)
		\exp\big[\cA(k) - \theta_6\, k_{65}\,\bar{\theta}_{65}\big]
		\nonumber\\[1mm]
		& = -16 \,\delta^4(\theta_{67})\, \delta^4(\theta_{68}) \,
		\left(\theta_{65}\,k_{83}\,\bar{\theta}_{38}\right)
		\exp\big[\cA(k) - \theta_6\, k_{65}\,\bar{\theta}_{65}
		+ \theta_{65}\, k_{65}\, \bar{\theta}_5\big]~,
\end{align}	
where in the second step we could replace the factor $\big(1 + \theta_{65}\, k_{65} \,
\bar{\theta}_5\big)$ with $\exp\big[\theta_{65}\, k_{65} \,\bar{\theta}_5\big]$ because 
it is multiplied by $\theta_{65}$. 

We now perform the $\theta$-integrations using the $\delta$-functions present in 
(\ref{Z578tris}) and keep as remaining independent variables $\theta_4$, 
$\bar{\theta}_{63}$, $\theta_{65}$, $\theta_6$ and $\bar{\theta}_6$; with straightforward manipulations, involving also the use of momentum conservation, we rewrite
$\big[\cA(k) - \theta_6\, k_{65}\,\bar{\theta}_{65}
		+ \theta_{65}\, k_{65}\, \bar{\theta}_5\big]$ as
\begin{align}
	\label{exprepl}
-2\,\theta_4 \,q \,\bar{\theta_6} - 2\, \theta_4\, k_{45}\,\bar{\theta}_{65}
+2\,\theta_5 \,q \,\bar{\theta}_{63}+ 2 \,\theta_{65}\, k_{53}\,\bar{\theta}_{63}
+2\,\theta_6\, k_{45}\, \bar{\theta}_{65}
-2\,\theta_{65}\, k_{53}\, \bar{\theta}_{65}~.
\end{align}	
We also have
\begin{align}
	\label{floorrepl}
		2\,\theta_{65}\,k_{83}\,\bar{\theta}_{38}
		= -2 \,\theta_{65}\,k_{83}\,\bar{\theta}_{63}\, 
		\equiv\, \exp\big[-2 \lambda \,\theta_{65}\,k_{83}\,\bar{\theta}_{63}\big]
		\Big|_{\lambda}~
\end{align}		
where the notation $X\big|_\lambda$ means the term of $X$ that is linear in $\lambda$.
Altogether we have managed to express $\cZ^{(7)}_{578}(k)$ as an exponential:
\begin{align}
	\label{Z578quater}
		\cZ^{(7)}_{578}(k)  = 
		& -8\, \exp\big[\!-2 \,\theta_4 \,q\, \bar{\theta}_6
		- 2 \,\theta_4\, k_{45}\,\bar{\theta}_{65} 
		+ 2 \,\theta_5\, q \,\bar{\theta}_{63} \nonumber\\
		&\qquad + 2\, \theta_{65}\, (k_{53}-\lambda k_{83})\, \bar{\theta}_{63}
		+ 2 \,\theta_6\, k_{45}\, \bar{\theta}_{65} 
		- 2 \,\theta_{65}\, k_{53}\, \bar{\theta}_{65}\big]
		\Big|_{\lambda}~.
\end{align}
This exponential can be interpreted as a $\theta$-graph\,%
\footnote{Since we use as Grassmann variables the differences $\bar{\theta}_{63}$
and $\theta_{65}$ of original variables, in the resulting $\theta$-graph momentum conservation is not realized at each node. However, this is does not cause any problem.}:
\begin{align}
	\label{Z578fifth}
		\cZ^{(7)}_{578}(k)  =-8
		\parbox[c]{.32\textwidth}{\includegraphics[width = .32\textwidth]{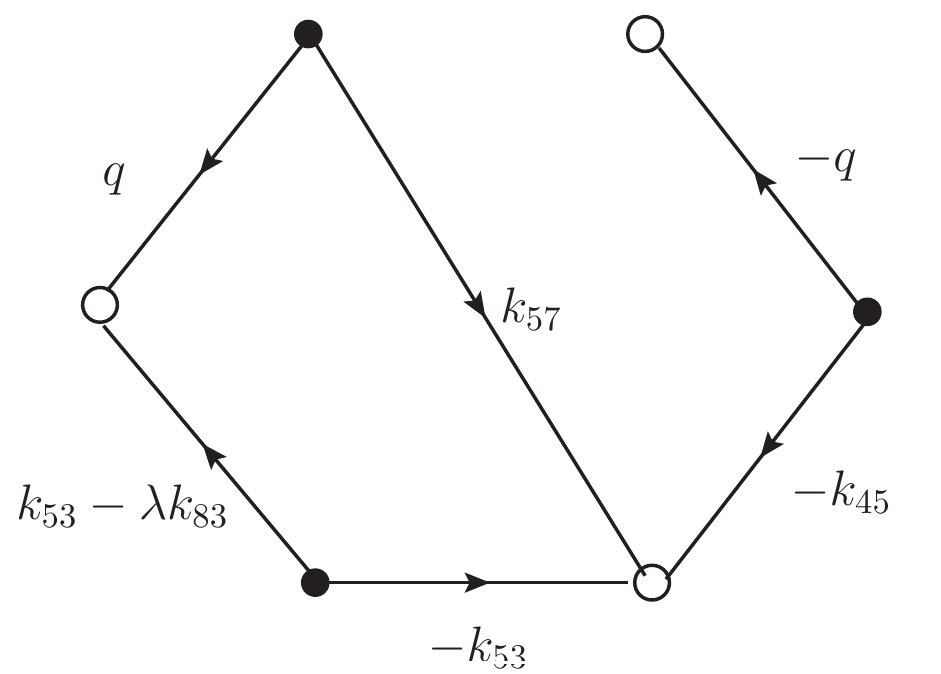}}\Bigg|_{\lambda}
		=  - 8 \,F\big(k_{53}-\lambda k_{83},-k_{53},0,q,k_{45},0,-k_{45},-q\big)\Big|_\lambda~.
\end{align}		 

We can apply this same procedure to evaluate the other five terms in (\ref{Z7def}) and obtain
\begin{align}
	\label{Z758}
		\cZ^{(7)}_{758}(k) & = 
		\delta^4(\theta_{65})\,
		\Big[\big(\overline{D}_6\big)^2 D_6^\alpha\, \delta^4(\theta_{67}) \Big]\,  
		\Big[ D_{6,\alpha}\, \delta^4(\theta_{68})\Big]\,
		\exp\big[\cA(k)\big]\nonumber\\
		&= - 8 \, F\big(-\lambda k_{83},-k_{78},0,q,k_{47},0,-k_{47},-q\big)\Big|_\lambda~,
		\\[1mm]
		\cZ^{(7)}_{785}(k) & = 
		\Big[\big(\overline{D}_6\big)^2 D_6^\alpha \,\delta^4(\theta_{67}) \big]\, 
		\delta^4(\theta_{68})\,		
		\big[ D_{6,\alpha} \,\delta^4(\theta_{65})\big]\,
		\exp\big[\cA(k)\big]\nonumber\\
		& = - 8 \, F\big(-\lambda k_{53},-k_{78},0,q,k_{47},0,-k_{47},-q\big)\Big|_\lambda~,
		\label{Z785}\\[1mm]
		\cZ^{(7)}_{875}(k) & = 
		\delta^4(\theta_{67})\,		
		\Big[\big(\overline{D}_6\big)^2 D_6^\alpha\, \delta^4(\theta_{68}) \big]\, 
		\Big[ D_{6,\alpha} \,\delta^4(\theta_{65})\Big]\,
		\exp\big[\cA(k)\big]\nonumber\\
		& = - 8 \, F\big(k_{83}-\lambda k_{53},-k_{83},0,q,0,0,0,-q\big)\Big|_\lambda = 0~,
		\label{Z875}\\[1mm]
		\cZ^{(7)}_{857}(k) & = 
		\delta^4(\theta_{65})\,		
		\Big[\big(\overline{D}_6\big)^2 D_6^\alpha\, \delta^4(\theta_{68}) \Big]\, 
		\Big[ D_{6,\alpha} \,\delta^4(\theta_{67})\Big]\,
		\exp\big[\cA(k)\big]\nonumber\\
		& = - 8 \, F\big(-k_{83},-k_{83}-\lambda k_{78},k_{83},-q,0,q,0,-q\big)\Big|_\lambda~,
		\label{Z857}\\[1mm]
		\cZ^{(7)}_{587}(k) & = 
		\Big[\big(\overline{D}_6\big)^2 D_6^\alpha\, \delta^4(\theta_{65}) \Big]\, 
		\Big[ D_{6,\alpha} \,\delta^4(\theta_{67})\Big]\,
		\delta^4(\theta_{68})\,		
		\exp\big[\cA(k)\big]\nonumber\\
		& = 0~.	\label{Z587}
\end{align}
The vanishing of the last contribution is due to the fact that in the step analogous to the one in
(\ref{DA}) we compute
\begin{align}
	\label{DAzero}
		\Big(\partial_{7,\alpha} - (k_{67}\,\bar{\theta}_7)_\alpha\Big) \exp\big[\cA(k)\big] 
		= 2 \big(k_{78}\, \bar{\theta}_{87}\big)_\alpha\,		\exp\big[\cA(k)\big] = 0~;
\end{align}
indeed in presence of $\delta^4(\theta_{68})\, \delta^4(\theta_{67})$, the difference 
$\bar{\theta}_{87}$ is null. The vanishing of this factor makes zero the entire expression. 

Now that we have computed all six terms of (\ref{Z7def}), we can insert the resulting expression 
for $\cZ^{(7)}(k)$ in the momentum integration, which has the BE topology, obtaining 
\begin{align}
	\label{scal7}
		\parbox[c]{.33\textwidth}{\includegraphics[width = .33\textwidth]{BE.pdf}}
		\cZ^{(7)}(k)
		=    \frac{160 \zeta(5)}{(4\pi)^6} \frac{1}{q^2} + \ldots~.  
\end{align}
Putting everything together, we finally get
\begin{align}
	\label{W7res}
		\cW_{bc}^{(7)}(q) & = -\frac{1}{q^2}\left(\frac{g^2}{8\pi^2}\right)^3 \zeta(5)
		\times 
		\left( 10\,\ii\, f_{def} C^\prime_{bfecd} + 10\,\ii\, f_{def}
		C^\prime_{cfebd} \right) + \ldots~. 
\end{align}

We have made a thorough analysis of all diagrams that can contribute to the propagator at order
$g^8$ and the ones we have listed above are the only ones that yield a term proportional to 
$\zeta(5)$ in the difference theory for a generic superconformal matter content. 
Other diagrams, indeed, either vanish due their color structure or give contributions that do not
contain $\zeta(5)$.

\section{Cusp integral}
\label{app:ft}
\noindent
In the following we will make use of the following integrals:
\begin{itemize}
\item Feynman parametrizations:
\begin{subequations}
\begin{align}
\frac{1}{A^\alpha\,B^\beta}&=\frac{\Gamma(\alpha+\beta)}{\Gamma(\alpha)\,\Gamma(\beta)}\int_0^1 \!dx\,\frac{x^{\alpha-1}(1-x)^{\beta-1}}{\big(x A+(1-x)B\big)^{\alpha+\beta}}
\label{Fx}\\[2mm]
\frac{1}{A^\alpha\,B^\beta}&=
\frac{\Gamma(\alpha+\beta)}{\Gamma(\alpha)\,\Gamma(\beta)}
\int_0^\infty \!dy \,\frac{y^{\beta-1}}{\big(A+y B\big)^{\alpha+\beta}}
\label{Fy}
\end{align}
\end{subequations}
\item The one-loop momentum integral (with Euclidean signature):
\begin{equation}
\int\!\frac{d^Dq}{(2\pi)^D}\,\frac{1}{\big(q^2+M^2\big)^n}=
\frac{\Gamma\big(n-\frac{D}{2}\big)}
{(4\pi)^\frac{D}{2}\,\Gamma(n)}\,\big(M^2\big)^{\frac{D}{2}-n}
\label{intq}
\end{equation}
\item The integral:
\begin{equation}
\int_0^\infty\!dy\,y^\alpha(A y+ B)^\beta =
\frac{\Gamma(-\alpha-\beta-1)\Gamma(\alpha+1)}{\Gamma(-\beta)}\, \frac{B^{\alpha+\beta+1}}{A^{\alpha+1}}~.
\label{Inty}
\end{equation}
\end{itemize}
With these ingredients, we can now perform the calculation of the following integral
\begin{equation}
I(\varphi) = \int\!\frac{d^Dk}{(2\pi)^D}\,\frac{1}{k^2\,(k\cdot v_1-\delta)\,(k\cdot v_2-\delta)}
\label{Iis1}
\end{equation}
where $D=4-2\varepsilon$, and $v_1$ and $v_2$ are two 4-vectors such that
\begin{equation}
v_1\cdot v_1=v_2\cdot v_2=1\quad\mbox{and}\quad v_1\cdot v_2 = \cos \varphi~.
\label{v1v2}
\end{equation}
We follow essentially the procedure outlined in \cite{Grozin:1992yq} (correcting a few typos).

We first use the Feynman parametrization (\ref{Fx})
to combine the two factors that are linear in $k$, obtaining
\begin{align}
I(\varphi)
=\int_0^1\!dx\int\!\frac{d^Dk}{(2\pi)^D}\,\frac{1}{k^2\,\big[\big(x v_1+(1-x) v_2\big)\cdot k-\delta\big]^2}~.
\end{align}
Then, we use the alternative Feynman parametrization (\ref{Fy}) and get
\begin{align}
I(\varphi)&=\int_0^1\!dx\int_0^\infty\!dy
\int\!\frac{d^Dk}{(2\pi)^D}\,\frac{2y}{\big[k^2+ y\big(x v_1+(1-x) v_2\big)\cdot k-y\delta\big]^3}
\label{I1}
\end{align}
Evaluating the integral over $k$, we obtain
\begin{align}
I(\varphi)&=2\int_0^1\!dx\int_0^\infty\!dy\,y
\int\!\frac{d^Dq}{(2\pi)^D}\,\frac{1}{(q^2+M^2)^3}
\label{I2}
\end{align}
with
\begin{align}
M^2
=-y\Big[\frac{y}{4}\big(x^2+(1-x)^2+2x(1-x)\cos\varphi\big)+\delta\Big]~.
\end{align}
Now we can use (\ref{intq}) and get
\begin{align}
I(\varphi)
=-(-1)^{-\varepsilon}\,\frac{\Gamma(1+\varepsilon)}{(4\pi)^{2-2\varepsilon}}
\int_0^1\!dx\int_0^\infty\!dy\,y^{-\varepsilon}\Big[\frac{y}{4}\big(x^2+(1-x)^2+2x(1-x)\cos\varphi\big)+\delta\Big]^{-1-\varepsilon}~.
\end{align}
The integral over $y$ can be computed using (\ref{Inty}), and the result is
\begin{align}
I(\varphi)
=-(-1)^{-\varepsilon}\,\frac{\Gamma(2\varepsilon)\,\Gamma(1-\varepsilon)\,\delta^{-2\varepsilon}}{(2\pi)^{2-2\varepsilon}}
\int_0^1\!dx\,
\frac{1}{\big(x^2+(1-x)^2+2x(1-x)\cos\varphi\big)
^{1-\varepsilon}}~.
\label{Int3}
\end{align}
{From} this expression we explicitly see the UV divergence signaled by the pole for 
$\varepsilon\to 0$.
Since we are ultimately interested in the coefficient of this divergence, we have
\begin{align}
I(\varphi)&=\frac{1}{\varepsilon}\left[-\frac{1}{8\pi^2}
\int_0^1\!dx\,
\frac{1}{\big(x^2+(1-x)^2+2x(1-x)\cos\varphi\big)}\right]+ O(\varepsilon^0)~.
\label{Int4}
\end{align}
The integral over $x$ can be evaluated by setting
\begin{equation}
x=\frac{1}{2}\Big(1+\cot\frac{\varphi}{2}\,z\Big)~.
\end{equation}
In this way we find
\begin{equation}
I(\varphi)=\frac{1}{\varepsilon}\left(-\frac{1}{8\pi^2}\,\frac{\varphi}{\sin\varphi}\right)+ O(\varepsilon^0)~.
\label{Int6}
\end{equation}
\end{appendices}

\providecommand{\href}[2]{#2}\begingroup\raggedright\endgroup

\end{document}